\documentclass[11pt,twoside]{report}
\usepackage[nosort]{cite}
\usepackage{anysize,feynmp,epsfig,graphics}
\marginsize{2.5cm}{2.5cm}{2.5cm}{2.5cm}
\makeatletter
  \newcommand{\FMslash}[2][0mu]{%
    \mathchoice
      {\fsl@sh\displaystyle{#1}{#2}}%
      {\fsl@sh\textstyle{#1}{#2}}%
      {\fsl@sh\scriptstyle{#1}{#2}}%
      {\fsl@sh\scriptscriptstyle{#1}{#2}}}
  \newcommand{\fsl@sh}[3]{%
    \m@th\ooalign{$\hfil#1\mkern#2/\hfil$\crcr$#1#3$}}
\makeatother

\def\tr{{\rm tr} \,}

\def\pslash{\FMslash p}
\def\barpslash{\FMslash{\bar{p}}}
\def\lslash{\FMslash l}
\def\m0{m^{\!\!\!\!^o}}
\def\qslash{\FMslash q}
\def\barqslash{\FMslash{\bar{q}}}

\def\wslash{\FMslash w}
\def\barwslash{\FMslash{\widetilde{w}}}
\def\kslash{\FMslash k}
\def\barkslash{\FMslash{\bar{k}}}

\def\partialslash{\FMslash \partial}

\begin{document}


\centerline{\bf \Large The relativistic nuclear dynamics }
\vskip1.3cm
\centerline{\bf \Large for the SU(3) Goldstone bosons of chiral QCD}

\vskip2.5cm

\centerline{Schrift zur Habilitation}

\vskip1.6cm
\centerline{dem}
\vskip0.6cm
\centerline{Fachbereich Physik der}
\vskip0.6cm
\centerline{Technischen Universit\"at Darmstadt}

\vskip2.5cm

\centerline{vorgelegt von}
\vskip0.4cm
\centerline{Matthias Friedrich Michael Lutz}
\vskip0.4cm
\centerline{aus Br\"uhl bei K\"oln}

\pagenumbering{roman}

\cleardoublepage

\setcounter{page}{1}
\pagestyle{headings}
\tableofcontents

\cleardoublepage

\pagenumbering{arabic}
\pagestyle{myheadings}

\chapter{Introduction}
\label{k1}
\markboth{\small CHAPTER \ref{k1}.~~~Introduction}{}

The question whether and to what extent hadrons change their properties
in a dense nuclear environment is an exciting subfield of nuclear and particle physics
which attracts considerable attention in the community \cite{Brown:rep,Migdal:1990vm,Oset:rep,Weise,Rapp:Wambach,Chanfray,Mosel}. Several
complementary approaches exist to study this question empirically. There is a long history for the study of
pion- and kaon-nucleus scattering. Pion and kaon beams have been available at many facilities and a
wealth of data has been collected. Also there exist data on level shifts of pionic and kaonic atoms.
Phenomenological optical potentials were constructed that successfully describe the data
set \cite{Hufner:1975ys, Dover, Gal:report}.
The optical potential reflects the in-medium properties of the mesons in cold nuclear
matter for nuclear densities smaller or equal the nuclear saturation density. Furthermore, photon induced meson
production off the nucleus offer a promising probe of in-medium effects \cite{Metag1,Metag2}.

A further important tool for the study of hadron properties in hot and dense matter is offered by the heavy
ion facilities at GSI, CERN and BNL. A heavy ion reaction can be visualized
as the creation of an initial hot and dense fireball, which after some evolution is assumed to freeze out to release
the observed particles \cite{PBM:1,PBM:2,Redlich:1,Redlich:2}. In this picture the measured
hadron multiplicities are characterized by a freeze-out surface,  temperature and baryon chemical
potential. An alternative picture, avoiding any
equilibrium assumption, is offered by transport model calculations which provide the
bridge to a more microscopic understanding of heavy ion reactions
\cite{Fuchs, Gaitanos:1999bu, Bratkovskaya, Bass:1998ca, Ko, Aichelin, Cassing:review,
Effenberger:1999uv, Leupold, Knoll, Cassing, Schaffner}.
The properties of the fireball may change dramatically if the energy of the heavy ion beam is varied.
In the SPS (CERN) experiments, with typical energies per nucleon of 40 GeV  $< E/N <$ 200 GeV, and more recently
in the RHIC (BNL) experiments, with even larger energies, one attempts to create a
fireball which is in the quark gluon plasma state \cite{QGP:theory}. There are indications that the
temperatures reached are indeed sufficiently high, exceeding 200 MeV, as to melt the initial heavy ions into
their quark and gluon constituents \cite{CERN}.

A complementary heavy ion program is pursued at GSI with
typical energies of up to 2 GeV per nucleon. Here the initial fireball reaches nuclear densities
of up to three times the density of nuclear saturation. The associated temperatures are typically
smaller than 100 MeV. A spectacular finding of the KaoS collaboration at GSI was the enhanced subthreshold
production of negatively charged kaons as compared to positively charged kaons in heavy ion reactions \cite{Senger}.
This may hint at a substantially
decreased antikaon mass in a dense nuclear environment. Consequently there has been much theoretical
effort to access the properties of kaons in nuclear matter
\cite{cpth, njl-lutz, Koch, Kolomeitsev, Pethick, Min, Brown, Waas1, Waas2, ml-sp, ramossp,
Cieply, Tolos, Florkowski}.
A possible exciting consequence of a significantly reduced effective $K^-$ mass could be that kaons
condense in the interior of neutron stars \cite{K:condensation:1,K:condensation:2}.

Quantitative empirical constraints from heavy ion reactions, on the in-medium antikaon spectral
functions require rather involved transport model calculations which are performed
extensively by various groups \cite{Fuchs,Gaitanos:1999bu,Bratkovskaya,Bass:1998ca,Ko,Aichelin,Schaffner}. The next generation of
transport codes which are able to incorporate particles with finite width more consistently
are being developed \cite{Leupold,Knoll,Cassing,Schaffner}. This is of considerable
importance when dealing with antikaons which are poorly described by a quasi-particle
ansatz \cite{ml-sp,ramossp}. It is an open challenge to find a microscopic understanding of the
particle production data in heavy ion reactions. The challenges are manyfold. In particular one
needs to derive improved microscopic input like in-medium reaction rates and spectral functions
for hadrons to be used in transport model calculations.

In this work we will focus on the properties of antikaons in cold nuclear matter.
We do not address the question how hadronic modes behave in a quark-gluon plasma environment.
From an experimental point of view the pions, kaons and etas offer a particularly unique tool
to study the nuclear many-body system since their in-medium properties are accessible in
different types of experiments. Besides measuring pion, kaon and eta multiplicities in
heavy-ion reactions one may scatter pions or kaons off nuclei or study the systems of
negatively charged pions or antikaons bound at a nucleus via $\gamma$-spectroscopy. There is
even the possibility that a eta meson may be bound at a nucleus forming an eta-atom \cite{eta-atom, Sokol}.

\vskip1.5cm \section{Fundamental aspects}

The foundation of all theoretical attempts to describe and possibly predict hadron properties
in a dense nuclear environment should be Quantum Chromo Dynamics (QCD), the fundamental theory of
strong interactions. For an exposition of the basic concepts of QCD we refer to standard
textbooks like \cite{QCD:textbook}. The QCD Lagrangian predicts all hadron interactions in terms
of quark and gluon degrees of freedom. At large momentum transfers QCD is extremely successful and predictive,
because observable quantities can be evaluated perturbatively in a straightforward manner.
At small momentum transfers, in contrast, QCD is less predictive so far due to its non-perturbative
character. It is still an outstanding problem to systematically unravel the properties of QCD in its low-energy phase
where the effective degrees of freedom are hadrons rather than quarks and gluons. That is a challenge since
many of the most exciting phenomena of QCD, like the zoo of meson and baryon resonances, manifest themselves
at low-energies. Also the fundamental question addressed in this work 'how do hadrons change their properties
in dense nuclear matter' requires a solid understanding of QCD at small momentum transfer.

There are two promising paths along which one achieved and expects further significant progress. A direct evaluation
of many observable quantities is possible by large scale numerical simulations where QCD is put on a finite
lattice \cite{lattice:1}. Though many static properties, like hadron ground-state properties, have been successfully
reproduced by lattice calculations, the description of the wealth of hadronic scattering data is still outside the
scope of that approach \cite{lattice:1}. Also the determination of hadron properties at finite nuclear densities is still
not feasible on the lattice \cite{lattice:2,lattice:3}. Here a complementary approach, effective field theory, is more
promising at present. Rather than solving low-energy QCD  in terms of quark and gluon degrees of freedom,
inefficient at low energies, one aims at constructing an effective quantum field theory in terms of hadrons
directly. The idea is to constrain that theory by as many properties of QCD as possible. This leads to a significant
parameter reduction and a predictive power of the effective field theory approach. In this spirit many effective field
theory models, not all of which fully systematic, have been constructed and applied to the data set.
Given the many empirical data points to be described, it is not always necessary to build in all
constraints of QCD. Part of QCD's properties may enter the model indirectly once it successfully
describes the data and provided that the model is not in conflict with the most basic properties
of quantum field theory. The most difficult part in constructing a systematic effective field theory
is the identification of its applicability domain and its accuracy level.

An effective field theory, which meets the above criteria, is the so called Chiral Perturbation Theory ($\chi$PT)
applicable in the flavor $SU(2)$ sector of low-energy QCD. This effective field theory is based on a simple
observation, namely that QCD is chirally symmetric in the limiting case where the up and down current quark masses
vanish $m_{u,d} =0$. This implies in particular that the handedness of quarks is a conserved property in
that limit. There is mounting empirical evidence that the QCD ground state breaks the chiral $SU(2)$ symmetry
spontaneously in the limit $m_{u,d} =0$. For instance the observation that hadrons
do not occur in parity doublet states directly reflects that phenomenon. Also the smallness of the pion masses
with $m_\pi \simeq $ 140 MeV much smaller than the nucleon mass $m_N \simeq $ 940 MeV, naturally fits
into this picture once the pions are identified to be the Goldstone bosons of that spontaneously broken
chiral symmetry. The merit of standard $\chi$PT is that first it is based on an effective Lagrangian density
constructed in accordance with all chiral constraints of QCD and second that it permits a systematic
evaluation applying formal power counting rules \cite{book:Weinberg}. In $\chi$PT the finite but small values of
the up and down quark masses $m_{u,d} \simeq $ 10 MeV are taken into account as a small perturbation defining the
finite masses of the Goldstone bosons. The smallness of the current quark masses on the typical chiral scale
of $1$ GeV explains the success of standard $\chi$PT. For applications of $\chi$PT to pion-nucleon scattering see
\cite{Gasser, Bernard, Meissner, pin-q4, pin-em}.

It is of course tempting to generalize the chiral $SU(2)$ scheme to the $SU(3)$ flavor group which
includes the strangeness sector. To construct the appropriate chiral $SU(3)$ Lagrangian is mathematically straightforward
and has been done long ago (see e.g. \cite{Krause}). The mass $m_s \simeq 10\,m_{u,d}$ of the strange quark, though
much larger than the up and down quark masses, is still small on the typical chiral scale of
1 GeV \cite{GL85}. The required approximate Goldstone boson octet is readily found with the pions, kaons and the eta-meson.
Nevertheless, important predictions of standard $\chi $PT as applied to the $SU(3)$ flavor group are in stunning
conflict with empirical observations. Most spectacular is the failure of the Weinberg-Tomozawa theorem \cite{WT:1,WT:2}
which predicts an attractive $K^-$-proton scattering length, rather than the observed large and repulsive
value. Progress can be made upon accepting a crucial observation that the power counting rules
must not be applied to a certain subset of Feynman diagrams. Whereas for irreducible diagrams
the chiral power counting rules are well justified, this is no longer necessarily the case for the irreducible diagrams
\cite{Weinberg}. The latter diagrams are enhanced as compared to irreducible diagrams and therefore
may require a systematic resummation scheme in particular in the strangeness sectors. First intriguing works taking
up this idea in the chiral context are given in  \cite{Kaiser,Ramos,Hirschegg,Oller-Meissner}.

It is amusing to observe that this development may be viewed as a justification of the many $K$-matrix analyses
in the 60's and the 70's which in spirit were already rather close to modern effective field theories only that
chiral constraints were not considered. A severe
drawback of those early attempts, where the inverse $K$-matrix is systematically expanded in a Taylor series, is
that the number of free parameters increases excessively along with the number of terms kept in that Taylor series.
For instance, the $K$-matrix analyses for the antikaon-nucleon scattering process, quite systematic close to threshold,
involve up to 44 parameters which all need to be adjusted to the data set \cite{A.D.Martin, martsakit, kim, sakit, gopal}.
Here modern effective field theories constitute an important advance once they successfully incorporate chiral
constraints of QCD  and therewith reduce the number of unknown parameters significantly.

\vskip1.5cm \section{Goldstone bosons in nuclear matter}

We return to the main theme of this work to study the properties of Goldstone bosons in nuclear matter.
Not only from an experimental  but also from a theoretical point of view the pions and kaons, the
lightest excitation of the QCD vacuum with masses of $140$ MeV and $495$ MeV respectively, are outstanding probes
for exciting many-body dynamics.
The Goldstone bosons are of particular interest since their in-medium properties reflect the structure of the
nuclear many-body ground state. For example at high baryon densities one expects the chiral symmetry to be
restored. One therefore anticipates that the Goldstone bosons change their properties substantially as
one compresses nuclear matter.

Even though in the $SU(3)$ limit of QCD with degenerate current quark
masses $m_u=m_d=m_s$ the pions and kaons have identical properties with respect to the strong
interactions, they provide very different means to explore the nuclear many-body system.
This is because the chiral $SU(3)$ symmetry is explicitly broken  by a nuclear matter
state with strangeness density zero, a typical property of matter produced in the
laboratory. A pion, if inserted into isospin degenerate nuclear matter, probes rather
directly the spontaneously broken or possibly restored chiral $SU(2)$ symmetry.
A kaon, propagating in strangeness free nuclear matter, looses its chiral $SU(3)$ symmetry since
the matter by itself explicitly breaks the $SU(3)$ symmetry. It is subject to three
different phenomena: the spontaneously broken chiral $SU(3)$ symmetry, the explicit symmetry
breaking of the small current quark masses and the explicit symmetry breaking of the
nuclear matter bulk. The various effects are illustrated by recalling the effective pion
and kaon masses in a dilute isospin symmetric nuclear matter gas. The low-density
theorem \cite{dover,njl-lutz} predicts mass changes $\Delta m_\Phi^2$ for any meson $\Phi $ in
terms of its isospin averaged s-wave meson-nucleon scattering length $a_{\Phi N}$
\begin{eqnarray}
\Delta m_\Phi^2 =-4\,\pi \left(1+\frac{m_\Phi}{m_N}\right) a_{\Phi N}\,\rho
+ {\mathcal O} \left( \rho^{4/3} \right)
\label{LDT}
\end{eqnarray}
where $\rho $ denotes the nuclear density. According to the above arguments one expects that the
pion-nucleon scattering length $a_{\pi N} \sim m_\pi^2$ must vanish in the chiral $SU(2)$ limit since
isospin symmetric nuclear matter conserves the Goldstone boson character of the pions at least at
small densities. On the other hand, kaons loose their Goldstone boson properties and therefore one expects
$a_{K N} \sim m_K$ and in particular $a_{K^- N} \neq a_{K^+ N}$. This is demonstrated by the
Weinberg-Tomozawa theorem which predicts the s-wave scattering length in terms of the chiral order
parameter $f_\pi \simeq 93$ MeV :
\begin{eqnarray}
a_{\pi N} = 0 + {\mathcal O} \left( m_\pi^2\right) \,, \qquad
a_{K^\pm N} = \mp \frac{m_K}{4\pi \,f_\pi^2}+ {\mathcal O} \left( m_K^2\right) \,.
\label{WT-theorem}
\end{eqnarray}
In the pion sector the Weinberg-Tomozawa theorem (\ref{WT-theorem}) is beautifully confirmed by the
smallness of the empirical isospin averaged pion-nucleon scattering length $a_{\pi N} \simeq -0.01$ fm.
In the kaon sector the Weinberg-Tomozawa theorem misses the empirical scattering $K^+$ nucleon scattering length
$a_{K^+ N} \simeq - 0.3 $ fm by about a factor of three. Even more spectacular is the disagreement
of Weinberg-Tomozawa term in the $K^-$ case where (\ref{WT-theorem}) predicts $a_{K^- N} \simeq + 0.9 $ fm
while the empirical $K^-$ nucleon scattering length is $a_{K^- N} \simeq (- 0.6 +i\,1.1 ) $ fm.

Whereas  in conjunction with the low-density theorem the Weinberg-Tomozawa theorem predicts a decreased
effective $K^-$ mass, the empirical scattering length unambiguously states that there must be
repulsion in the $K^-$ channel at least at very small nuclear densities. The antikaon-nucleon scattering
process is complicated due to the open inelastic $\pi \Sigma$ and $\pi \Lambda $ channels. This is reflected
in the large imaginary part of the empirical $K^-$ nucleon scattering length. A quite rich variety of
phenomena arises from the presence of both the s-wave $\Lambda(1405)$ and p-wave $\Sigma(1385)$ resonances
just below, and the  d-wave $\Lambda(1520)$ resonance not too far above the antikaon-nucleon threshold.
In nuclear matter there exist multiple modes with quantum numbers of the $K^-$ resulting from the
coupling of the various hyperon states to a nucleon-hole state. As a consequence the $K^-$ spectral
function shows a rather complex structure as a function of baryon density, kaon energy and momentum.
This is illustrated by recalling the low-density theorem as applied for the energy dependence of the kaon
self energy $\Pi_{\bar K}(\omega , \rho)$.  The self energy is determined in terms of the
s-wave kaon-nucleon scattering amplitude $f^{\rm s-wave}_{\bar K N}(\sqrt{s})$ is
\begin{eqnarray}
\Pi_{\bar K}(\omega , \rho) =
- 4\,\pi \left( 1+ \frac{\omega }{m_N}\right) \,f^{\rm s-wave}_{\bar K N}(m_N + \omega )\,\rho
+{\mathcal O} \left(  \rho^{4/3}\right) \,,
\label{LDT-energy}
\end{eqnarray}
where we assume zero kaon momentum. A pole contribution to $\Pi_{\bar K}(\omega , \rho)$
from a hyperon state with mass $m_H$, if sufficiently strong, may lead to a $K^-$ like state of approximate energy
$ m_H-m_N$. Most important are the $\Lambda (1405)$ s-wave resonance and the $\Sigma(1385)$ p-wave
resonance. An attractive modification of the antikaon spectral function was already anticipated in the 70's by the many
K-matrix analyses of antikaon-nucleon scattering (see e.g. \cite{A.D.Martin}) which predicted
considerable attraction in the subthreshold s-wave $K^-$ nucleon scattering amplitudes. In
conjunction with the low-density theorem (\ref{LDT-energy}) this leads to an attractive antikaon spectral function
in nuclear matter. As was pointed out in \cite{ml-sp} the realistic evaluation of the antikaon self energy
in nuclear matter requires a self consistent scheme. In particular the feedback effect of an attractive
antikaon spectral function on the antikaon-nucleon scattering process was found to be important for
the $\Lambda(1405)$ resonance structure in nuclear matter.

The quantitative evaluation of the antikaon spectral function in nuclear matter is still an outstanding
problem. Further progress
requires an improved understanding of the antikaon-nucleon scattering process in free space. In particular reliable
subthreshold antikaon-nucleon scattering amplitudes are needed. The present data set for antikaon-nucleon scattering
leaves much room for different theoretical extrapolations to subthreshold energies
\cite{A.D.Martin,martsakit,kim,sakit,gopal,oades,Juelich:2,dalitz,Kaiser,Ramos,keil}.
As a consequence the subthreshold $\bar K N$ scattering amplitudes of different analyses may differ by as much as
a factor of two \cite{Kaiser,Ramos} in the region of the $\Lambda (1405)$ resonance. Also  the
antikaon spectral function should be evaluated in an improved many-body treatment including the effects of
p-wave interactions systematically.

\vskip1.5cm \section{Meson-baryon scattering}

The meson-baryon scattering processes are an important test for effective field theories
which aim at reproducing QCD at small energies, where the effective degrees of freedom are
hadrons rather than quarks and gluons. In this work we focus on the strangeness sector, because here
the acceptable effective field theories are much less developed and also the empirical data set still
leaves much room for different theoretical interpretations. In the near future the
new DA$\Phi$NE facility at Frascati could deliver new data on kaon-nucleon scattering \cite{DAPHNE} and therewith
help to establish a more profound understanding of the role played by the $SU(3)$ flavor symmetry in hadron
interactions. At present the low-energy elastic $K^+$-proton scattering data set leads to a rather well
established $K^+p$ scattering length with $a_{K^+p}\simeq 0.28-0.34$ fm \cite{Dover}. Uncertainties exist,
however, for the $K^+$-neutron cross sections which are typically extracted
 from the scattering data of the $K^+d\to K^+ p\,n , K^0 p \,p$ reactions. Since data are
available only for $p_{\rm lab} > 350$ MeV the model dependence of the deuteron wave function, the
final state interactions and the necessary extrapolations down to threshold lead to conflicting values
for the $K^+$-neutron scattering length \cite{Dumbrajs}. A recent analysis \cite{Barnes} favors
a repulsive and small value $a_{K^+n}\simeq 0.1$ fm. Since low-energy polarization
data are not available for $K^+$-nucleon scattering at present, the separate strength of the various
p-wave channels can only be inferred from theory. This leads to large uncertainties
in the p-wave scattering volumes \cite{Dumbrajs}.

The $K^-$-proton scattering length was only recently determined convincingly by a
kaonic-hydrogen atom measurement \cite{Iwasaki}. In contrast to that the $K^-$-neutron scattering
length remains model dependent \cite{A.D.Martin,martsakit}. This reflects the fact
that at low energies there are no $K^-$ deuteron scattering data available
except for some $K^-d$ branching ratios \cite{Veirs}, which however are commonly not included in theoretical
models of kaon-nucleon scattering. The rather complex multi-channel dynamics of the
strangeness $-1$ channel is constrained by not too good quality low-energy elastic and
inelastic $K^-p$ scattering data \cite{Landoldt} but rather precise $K^-p$
threshold branching ratios \cite{branch-rat}. Therefore the isospin one scattering amplitude
is  only indirectly constrained for instance by the $\Lambda \pi^0$ production data
\cite{mast-pio}. All this leaves much room for different theoretical extrapolations
\cite{A.D.Martin,martsakit,kim,sakit,gopal,oades,Juelich:2,dalitz,keil}.
As a consequence the subthreshold $\bar K N$ scattering amplitudes, which determine
the $\bar K$-spectral function in nuclear matter at leading order in the density expansion,
are poorly controlled.

Therefore it is desirable to make use of the chiral symmetry constraints of
QCD \cite{Kaiser,Ramos,Hirschegg,Oller-Meissner}. The reliability of the extrapolated subthreshold
scattering amplitudes can substantially be improved as compared to previous works \cite{Kaiser,Ramos,Oller-Meissner},
if s- {\it and} p-waves are considered in the analysis. This is important because the available data, in particular
in the strangeness $-1$ channel, are much more precise for $p_{\rm lab}>200$ MeV than for $p_{\rm lab}<200$ MeV where
one expects s-wave dominance.

\vskip1.5cm \section{Perspectives}

In this work  the relativistic chiral $SU(3)$ Lagrangian including
an explicit baryon decuplet resonance field with $J^P\!= \!\frac{3}{2}^+$ is applied.
The baryon decuplet field is an important ingredient, because it is part of the baryon
ground state multiplet which arises in the large-$N_c$ limit of QCD \cite{Hooft,Witten}.
Also the effects of a phenomenological baryon nonet d-wave resonance field with
$J^P\!=\!\frac{3}{2}^-$ is considered. This is necessary because some of the p-wave
strengths in the $\bar K N$-system can be extracted from the present day data set reliably
only via their interference effects with the d-wave resonance $\Lambda (1520)$. This is the
first application of the chiral $SU(3)$ Lagrangian density to the kaon-nucleon and
antikaon-nucleon system which systematically considers constraints from the pion-nucleon sector.
A convenient minimal chiral subtraction scheme for relativistic Feynman diagrams is
proposed which complies manifestly with the standard chiral counting rule
\cite{Becher,nn-lutz,Gegelia}.
Furthermore it is argued that due to the rather large kaon mass the kaon-nucleon
dynamics is non-perturbative in contrast to to the pion-nucleon system where standard chiral perturbation
theory ($\chi$PT) can be applied successfully \cite{Gasser, Bernard, Meissner, pin-q4, pin-em}. In the strangeness
sectors a partial resummation scheme is required \cite{Siegel,Kaiser,Ramos}.
The Bethe-Salpeter equation for the scattering amplitude is solved
with an interaction kernel truncated at chiral order $Q^3$. Only
those terms are included in the interaction kernel which are leading in the
large-$N_c$ limit of QCD \cite{Hooft,Witten,DJM,Carone,Luty}.
An important aspect to be addressed concerns the independence of the
on-shell scattering amplitude on the choice of chiral coordinates or the choice of
interpolating fields. If one solves the Bethe-Salpeter equation with an interaction
kernel evaluated in perturbation theory the resulting on-shell scattering amplitudes
depend on the choice of interpolating fields even though physical quantities should
be independent on that choice \cite{Lahiff}. It will be demonstrated that this problem can
be avoided by introducing an appropriate on-shell reduction scheme for the
Bethe-Salpeter equation.

The s-, p- and d-wave contributions with $J=\frac{1}{2},\frac{3}{2}$ in the
scattering amplitude
are considered. Detailed predictions for the poorly known meson-baryon
coupling constants of the baryon octet and decuplet states will be made. It is
found that the scattering amplitudes are quite sensitive to the
precise values of the meson-baryon coupling constants permitting
the determination of SU(3) symmetry breaking effects in the
3-point vertices. That observation justifies to include effects of
chiral order $Q^3$ in the interaction kernel. At chiral order
$Q^2$ one encounters 7 unknown symmetry conserving and 3 reasonably
well known symmetry breaking parameters. At chiral order $Q^3$ the relevance of 4
unknown parameters is derived refining the quasi-local two-body vertices. Additional
10 parameters are required to describe SU(3) symmetry breaking effects in the meson-baryon
coupling constants as well as those in the axial vector coupling constants of the
baryon octet states.
As a novel technical ingredient a covariant projector formalism is constructed.
It is supplemented by a subtraction scheme rather than a cutoff scheme as employed previously
\cite{Kaiser,Ramos}. The renormalization scheme is an essential input for the chiral
$SU(3)$ dynamics, because it leads to consistency with chiral counting rules and an approximate
crossing symmetry of the subthreshold kaon-nucleon scattering amplitudes. In particular, the
scheme avoids breaking the $SU(3)$ symmetry by channel dependent cutoff parameters as
suggested in \cite{Kaiser} and also an implicit sensitivity
of the $\Lambda(1405)$ resonance structure on the cutoff parameter \cite{Ramos}.

The set of parameters is adjusted successfully to describe the existing low-energy cross section
data on kaon-nucleon and antikaon-nucleon scattering including angular distributions to good
accuracy. At the same time a good description of the low-energy s- and p-wave
pion-nucleon phase shifts as well as the empirical axial-vector coupling constants of the
baryon octet states is achieved. Detailed predictions are given for the poorly known meson-baryon
coupling constants of the baryon octet and decuplet states. Furthermore the many $SU(3)$ reaction
amplitudes and cross sections like $\pi \Lambda \to \pi \Lambda, \pi \Sigma , \bar K \,N $
relevant for transport model simulations of heavy-ion reactions are presented. As a result
of the analysis, particularly important for any microscopic description of antikaon propagation
in dense nuclear matter, sizable contributions from the p-waves parts in the
subthreshold $K^-$-nucleon forward scattering amplitude are predicted. Besides deriving
a realistic antikaon spectral function first quantitative results on the
in-medium structure of the p-wave $\Sigma (1385)$ and the d-wave $\Lambda(1520)$ resonances
are obtained.

In chapter 2  a critical review of previous non-perturbative
applications of the chiral SU(3) Lagrangian in the meson-baryon sector is given. It follows
a discussion on the generic nature of baryon resonances in view of their role played
in chiral coupled channel theories. The chapter is closed by a series of more formal
sections in which the parts of the relativistic chiral Lagrangian relevant for this work
are constructed. All interaction terms are systematically analyzed in the $1/N_c$ expansion
of QCD. In chapter 3 a formalism is developed how to properly treat
the Bethe-Salpeter equation in a chiral coupled channel approach.
Details on the renormalization scheme will be given.
In chapter 4 the coupled channel effective interaction kernel in accordance with
the scheme presented in chapter 3 is derived.
A scheme-independent and covariant framework, in which
self consistency for the in-medium meson spectral functions is implemented in terms of the
vacuum meson-nucleon scattering amplitudes is developed in chapter 5.  For the first time the
scheme considers the in-medium mixing of s-, p- and d-waves in a self consistent manner. Any
reader not primarily interested in the numerous technical developments may directly jump to
the result chapter 6 which can be read independently.

The work presented here is based in large parts on collaborations with E.E. Kolomeitsev and
C.L. Korpa \cite{Lutz:Kolomeitsev,Lutz:Korpa}.

\cleardoublepage

\chapter{Chiral SU(3) symmetry and meson-baryon interactions}
\label{k2}
\markboth{\small CHAPTER \ref{k2}.~~~Chiral SU(3) symmetry and meson-baryon interactions}{}

In this chapter a detailed derivation of the constraints
imposed on the meson-baryon scattering processes by the chiral
SU(3) symmetry of QCD will be given. The discussion of the existing literature
will emphasize the strangeness sectors since there exist various
review articles on the pion-nucleon scattering process (see e.g. \cite{Bernard:Kaiser:Meissner}).
The latter process was studied in much more detail than the kaon and
anti-kaon nucleon scattering processes, the focus of this work.
After reviewing in some detail the more recent developments in the
strangeness sectors the scope of this work will be defined. For
earlier descriptions of the meson-baryon scattering data based on
meson-exchange phenomenology we refer to standard review articles (see e.g. \cite{EW}).
Before entering the many technical details on how to implement the
chiral constraints into the coupled channel dynamics an exposition about
the generic structure of baryon
resonances will be offered in a separate section. This is important since we will
encounter various baryon resonances in this work. As a result of
the discussion a conjecture will be formulated as to what the
nature of baryon resonances may be.

The first work by Kaiser, Siegel and Weise \cite{Kaiser}, who
applied the chiral SU(3) Lagrangian to the low-energy
antikaon-nucleon scattering data, described the elastic and
inelastic scattering process in terms of a coupled channel
Lippman-Schwinger equation. The scattering kernel was identified
with the leading and subleading quasi-local interaction vertices
of the chiral Lagrangian in a phenomenological manner. The
power-divergencies, which arise when solving the Lippman-Schwinger
equation, were tamed by phenomenological form factors. All
together the scheme has 5 free parameters characterizing the
strength of the leading and subleading quasi-local interaction
vertices. Three additional parameters were required to model the form
factors. The low-energy scattering data for laboratory momenta
$p_{\rm lab} < 200$ MeV were reproduced accurately in terms of
s-wave dynamics only. In particular the s-wave $\Lambda(1405)$
resonance was generated dynamically by coupled channels effects. A
criticism of this work was raised by Oset and Ramos \cite{Ramos}
who pointed out that in \cite{Kaiser} the $\eta \,\Sigma$, $\eta \,
\Lambda$ and $K\, \Xi $ channels, indispensable ingredients of a chiral
SU(3) theory, were not considered in the coupled channel
scattering equations. Superficially these additional channels may
appear unimportant since they are closed at small energies.
However, the effect of those channels may turn important since
they give rise to a particular renormalization of the leading open
$\bar K \,N , \pi \,\Sigma$ and $\pi \,\Lambda $ channels. Oset and
Ramos demonstrated that a reasonable fit of the low-energy
scattering data was possible in terms of the leading
Weinberg-Tomozawa interaction term and a suitable constructed
universal form factor which was modelled by one parameter
only. They argued that the form factor represents higher order
effects of the chiral Lagrangian in some way. From a fundamental
point of view not much was gained as compared to previous K-matrix
or potential model analyses of the scattering data. The interplay
of form factor effects and the scattering kernel, constrained by
the chiral SU(3) symmetry in a phenomenological manner, remained
puzzling in those early works. In particular, a fit to the
scattering data required a fine-tuning of the form factors
\cite{Ramos} which explicitly break the SU(3) symmetry
\cite{Kaiser}. Moreover, the leading Weinberg-Tomozawa interaction
was long known to be equivalent at low energies to an interaction
kernel approximated by a simple t-channel vector meson exchange
diagram if the vector-meson coupling constants were chosen in an
universal manner \cite{Siegel}. A successful description of the
low-energy scattering data was obtained previously by Siegel and
Weise \cite{Siegel} in terms of the t-channel vector meson
exchange diagram as the driving term in the coupled channel scattering equation.
Such models, which reproduce the threshold strength as predicted
by the leading chiral Weinberg-Tomozawa interaction, were originally proposed and studied
by Wyld  \cite{Wyld} and also by Dalitz, Wong and Rajasekaran \cite{dalitz-1}.
In those early computations \cite{Wyld,dalitz-1} the t-channel vector-meson
exchange was treated in the static approximation but already all
inelastic channels required by the flavor SU(3) symmetry were considered.
The resulting coupled channel Schr\"odinger equation does not require
any form factors to tame ultraviolet divergencies since the Yukawa type potential, which
arises in the static approximation of the t-channel vector meson exchange diagram, is
well behaved at short distances.

In order to make further progress and unravel the chiral SU(3)
structure of the meson-baryon scattering data a more systematic
treatment of the chiral Lagrangian was asked for. In particular a
more transparent scheme in which the interaction kernel and the
scattering equation are considered on the same footing was
desirable. A first important step in this direction was achieved
in \cite{Hirschegg} where it was suggested to use a subtraction scheme
rather than a phenomenological form factor. It was argued that the
subtraction point is strongly constrained by the requirement that
the associated relativistic loop integral should obey the chiral
counting rule. Once a subtraction scheme is applied there is no
need anymore for a fine-tuning of the parameters. Using a
universal subtraction point for all partial waves and channels it
was argued that the subtraction point should be identified with
the hyperon mass as to protect the s-channel hyperon exchange
contribution in the p-wave scattering amplitude. Then using the
physical strength of the Weinberg-Tomozawa interaction as
parameterized in terms of the pion decay constant $f_\pi \simeq
93$ MeV one obtained a parameter-free leading order description of
the antikaon-nucleon s-wave dynamics. An additional strong argument as
to why the subtraction point must be universal and close to the hyperon
mass follows from dimensional regularization and the requirement that the theory should lead to
crossing symmetric amplitudes, at least approximatively. This will be elaborated
on in this work in great detail. In the more recent work
Oller and Meissner \cite{Oller-Meissner} suggested a similar scheme where the value of
the pion-decay constant and the subtraction point were varied as to obtain a
quantitative description of
the low-energy s-wave antikaon-nucleon scattering data. The resulting
effective pion-decay constant was found to be significantly larger
than the value required by the pion-nucleon scattering data
\footnote{Though the general strategy advocated by Oller and Meissner \cite{Oller-Meissner} is
similar to that one applied previously in \cite{QM-lutz,Hirschegg} and used in this work, there
are important technical differences. At leading  order and in particular considering only s-wave
dynamics the differences are of minor importance. However, when
correction terms and higher partial waves, not studied in \cite{Oller-Meissner},
are considered, this is not anymore the case and important differences arise.
In the scheme of Oller and Meissner the crucial problem of
how strength coming from the subtraction coefficients
is constrained by the chiral SU(3) symmetry remains obscure. This will be
discussed in chapter 3.4 when presenting the technical details of our scheme.}.

The challenge, to obtain a simultaneous description of the
low-energy pion-, kaon- and antikaon-nucleon scattering data
remained. If different values of the chiral parameters are used in
the various sectors the chiral SU(3) structure remains unclear.
Obviously, one should consider chiral correction terms in a
systematic way and treat the pion-, kaon-, and antikaon-nucleon
scattering processes on equal footing. Also higher partial wave
components, so far not studied in the strangeness sectors, should
be evaluated. This defines the scope of this work. The analysis
becomes necessarily much more involved but at the same time more
rewarding since one may learn more about the chiral SU(3)
symmetry. Of particular interest are the explicit SU(3) symmetry
breaking effects which are so far only poorly understood.

It is emphasized that here the goal is not to describe pion-induced
strangeness production off the nucleon. A solid
understanding of such reactions $\pi N \to K \Lambda,K \Sigma,
...$ requires the incorporation of the correlated two-pion
production process which constitutes a major source of inelastic
strength at these energies. That lies clearly outside the scope of
this work which aims at understanding the scattering processes at
low energies only. In this respect the antikaon-nucleon
scattering process is better suited to unravel the chiral SU(3)
symmetry of QCD quantitatively. For instance, the inelastic
reactions $\bar K N \to \pi \Lambda, \pi \Sigma $ occur at
energies where the phase space for the complicated three-body
hadronic final states are still closed.

The incorporation of the correlated two-pion production into a chiral coupled
channel theory is a quite ambitious enterprize, which we believe, however,
to be feasible if the large-$N_c$ constraints of QCD are used.
The inclusion of vector mesons into the chiral Lagrangian as explicit degrees
of freedom is subtle as to the fact that rigorous power counting rules
do not exist. The effect of the implicit t-channel vector meson exchange
contribution was suggested in \cite{N-D-Oller-Oset,N-D-Meissner-Oller} to be
incorporated by the N/D method of Chew and Mandelstam \cite{Chew-Mandelstam}.
However, the question how to incorporate for instance the inelastic $\rho_\mu N$ or
$\omega_\mu N$ channels into a chiral description of pion-nucleon scattering remains an open
problem. We conjecture that this should be possible once the chiral expansion is
combined with the $1/N_c$ expansion. The $1/N_c$ expansion provides an intuitive and possibly
systematic guide as to why higher order loop diagrams involving multiple powers of
the vector meson propagators are suppressed as the number of independent
loop integrals in the diagram increases. Here we exploit the obvious observation that
hadronic loop functions if visualized in terms of quark-gluon diagrams are suppressed in the
$1/N_c$ expansion.

A phenomenological attempt by Caro Ramon et al. \cite{new-muenchen} to
describe pion induced strangeness production within the chiral
SU(3) Lagrangian but without the inclusion of the correlated
two-pion production is obviously plagued by severe short comings.
In particular, their description of differential cross sections,
which includes s- and p-wave terms, leaves much room for
improvements. It is unclear to what degree the SU(3) symmetry is
obscured by form factor effects or the fact that chiral parameters
were adjusted in the strangeness zero sector only.
The more recent work \cite{Nieves:Arriola},
which also describes pion induced $\eta N $ and $K \Lambda $ production without
the inclusion of the correlated two-pion production, is subject to similar objections.
In particular the strong SU(3) symmetry breaking allowed for in the subtraction constants is
difficult to understand. At this stage it may be
advantageous to postpone the fundamental description of such reactions in terms
of the chiral Lagrangian and rather improve the description of these reactions in terms of
phenomenological approaches which do however incorporate the
correlated two-pion production processes modelled in terms of
effective $\rho_\mu N $ and $\pi \Delta_\mu $ channels
\cite{Manley,Sauerman,Feuster:Mosel:1,Feuster:Mosel:2,Vrana:Dytman:Lee,QM-lutz,LWF}.
The solid understanding of these reactions requires necessarily a profound
theory for the structure of baryon resonances like for instance the s-wave N(1535)
and N(1650) states. Since we will have to make assumptions on
the nature of baryon resonances throughout this work we discuss
this aspect in the following section in some detail.

\vskip1.5cm \section{Baryon resonances and coupled channel
dynamics}


There have been many discussions on how to incorporate the
$\Delta(1232)$ isobar resonance into the chiral Lagrangian \cite{Hemmert:Holstein:Kambor,Fettes:Meissner}.
One legitimate point of view would be to insist that the two relevant
small scales $m_\pi$ and $m_\Delta -m_N$ are not correlated and
that one should insist on $m_\pi \ll m_\Delta -m_N$, at least in
the chiral limit where $m_\pi =0$ but $m_\Delta \neq m_N$. In more
physical terms the isobar degree of freedom would be integrated
out and therefore would not appear as a fundamental field in the
chiral Lagrangian. Obviously this point of view  necessarily restricts the
applicability of the chiral Lagrangian  to threshold
physics. For instance the well known rapid energy variation of the
$P_{33}$ pion-nucleon phase shift close to $\sqrt{s} \simeq 1232$
MeV can only be described in such a scheme if an infinite number
of interaction terms in the chiral Lagrangian are considered. A
more economical scheme arises if one accepts the physical values
of $m_\pi$ and $m_\Delta -m_N$ which suggest to set up an
effective chiral field theory where one introduced so-called power
counting rules with $m_\pi \sim m_\Delta -m_N \sim Q $
parametrically. With $Q$ we denote any typical small scale of the
system. A brief discussion of power counting rules will be given
in section 2.2. Such a scheme is well justified if one
incorporates an additional property of QCD into the effective
field theory. If the number of colors, $N_c$, in OCD is considered
as a free parameter a systematic expansion of hadron properties
in powers of $1/N_c$ is feasible. At a given order, the $1/N_c$
expansion selects a well defined infinite set of quark-gluon
diagrams which may be considered at the physical value $N_c=3$.
Even though at present it appears impossible to evaluate the sum
of these diagrams, at any given order in that expansion it is
possible to work out sum rules obeyed by hadronic observables. For
instance at leading order in the $1/N_c$ expansion the nucleon and
isobar masses are degenerate. This observation may serve as a
justification for the counting rule $m_\pi \sim m_\Delta -m_N \sim
Q $ in a combined chiral and $1/N_c$ expansion. The generalization
to the chiral SU(3) Lagrangian is straightforward. It necessarily
involves baryon octet and decuplet fields since their mass
parameters would be degenerate in the large-$N_c$ limit of QCD up
to small corrections from explicit SU(3) symmetry breaking
effects. The importance of the decuplet degrees of freedom for the
antikaon-nucleon scattering process was pointed out in
\cite{Hirschegg}.

The picture becomes more complicated when baryons of the second
generation are involved. Let us focus for a while on the s-wave
baryon resonance $\Lambda$(1405) which is known to play a dominant
role in the antikaon-nucleon scattering process. Since there
appears no fundamental reason at hand to insist on degenerate mass
parameters of the $\Lambda(1115)$ ground state and the resonance
state $\Lambda(1405)$ in the large-$N_c$ or chiral limit of QCD,
the incorporation of an explicit $\Lambda(1405)$ resonance field
into the chiral Lagrangian would lead to the breakdown of
power counting rules. As a consequence it would be unclear what
diagrams to evaluate at given order implying an uncontrolled
breaking of the chiral Ward identities. This is a disaster from a
fundamental point of view. Thus, if the $\Lambda (1405)$ is within
reach of the chiral SU(3) Lagrangian it preferably be generated
dynamically by coupled channel effects. Indeed, as is known from
the many K-matrix analyses of the 70's the $\Lambda (1405)$ can be
generated dynamically \cite{Dalitz:Tuan,Wyld,dalitz-1,martsakit,kim,sakit}.
This mechanism is confirmed beautifully by the recent applications of the chiral
SU(3) Lagrangian to the antikaon-nucleon scattering process
\cite{Kaiser,Ramos,Hirschegg,Oller-Meissner}.

An immediate question arises. To what extent is the dynamic
generation of a baryon resonance consistent with the quite
successful quark model picture of such states suggested by Isgur
and Karl \cite{Isgur-Karl,Capstick-Roberts,Riska-Glozman} where three
constituent quarks are bound together by a confining potential?
The answer of this question is nontrivial and we will not be able
to offer a final answer. What we can do here is, however, to
collect various arguments which at least illustrate the
complexity of this problem. We do not claim originality for all of
these arguments since they are in some way implicit in the literature
even though it is difficult to associate them to a given person. As a
result of this discussion we will formulate a conjecture as to what
the nature of the baryon resonances may be. Before entering the discussion we
emphasize that here we are concerned exclusively about baryon
resonances build up from the light quarks of QCD, the up, down and
strange quarks. For heavy quark systems there are convincing
derivations of a non-relativistic confining potential which
suffices to describe the interactions of heavy quarks \cite{confinement}. The
dynamics of the light quarks of QCD, on the other hand, is highly
non-trivial and subject to controversial opinions. The problem is
so complicated because for the light quarks a non-relativistic
treatment of the bound state problem cannot be justified.

To some degree the problem may be simply a question of economy.
What is the most natural description of the scattering data?
Suppose some quark model predicts a bare resonance state of given
mass. Then the coupling to the meson-baryon states gives rise to
a finite hadronic decay width but also to a mass shift of the resonance.
If the mass shift induced by the coupled channel
dynamics is sufficiently large it may be more economical or at
least legitimate to consider the resonance to be generated
dynamically. It is difficult to imagine that the bare mass parameter
is an observable quantity in this case. This suggests for instance
that a bare-pole term in the scattering kernel may be Taylor expanded
around the physical resonance mass provided that the bare pole
position is outside the applicability domain of the effective
field theory. The crucial question is therefore how important is
the coupling of the bare resonance states to the final hadronic
states \cite{qqqMB}. The fact that baryon resonances exhibit
large hadronic decay widths, not all of them well understood within the
quark model approach so far, demonstrates that this is a non-trivial question.
In this context it is important to realize, that channels which
open only at energies larger than a given resonance mass may still
give rise to a sizeable mass shift for that resonance.

It is instructive to return to the $1/N_c$ expansion strategy
of QCD already touched upon above and discuss its implications for the
baryon states. We would like to emphasize that the impressive
success of the quark model in describing the properties of the
baryon octet and decuplet ground states, like for instance their
magnetic moments, can equivalently be obtained applying the
$1/N_c$ expansion \cite{DJM}. Only at the level of the second
generation baryon resonances there appear important differences
between the quark model approach and the $1/N_c$ expansion method
\cite{Carone,Luty}. For the large-$N_c$ baryon ground state
properties the so called contracted SU(6) spin-flavor symmetry,
inherent in the quark model and also predicted by the $1/N_c$
expansion, is responsible for this remarkable success. A stringent
test of the quark model dynamics is provided by the second and
third generation of baryon resonances. It remains to be seen
whether the predictions of present day quark model calculations
\cite{Capstick-Roberts}, a large number of so far unobserved
baryon resonances, comes true. The challenge is to reproduce the
empirical baryon resonance spectrum and at the same time obtain a detailed
description of the resonance decay channels. The question to be
answered is whether it is efficient to classify baryon resonances
not part of the large-$N_c$ ground states as higher dimensional
representations of a O(3)$\otimes $SU(6) group or
whether it is more physical to consider those as a result of
coupled channel dynamics. For instance the baryon octet and decuplet states
form together the ground state ${\bf 56}$-plet. For the excited states
this classification is indicative but hampered by the fact that it requires
additional dynamical assumption on how to model the orbital excitation of a quark
inside a resonance state consistently with the contracted SU(6) spin-flavor symmetry
of QCD. For the first excited resonance ${\bf 56}$-plet with negative parity
there are hyperon resonance states which are so far not observed \cite{Carone-1,Carone-2}.
Similarly the first ${\bf 70}$-plet of positive parity is incomplete containing a number of
missing hyperon resonance states \cite{Schat-Goity-Scoccola}. The latter multiplet assumes
one quark to carry angular momentum $L=1$. Of course one may say that this is a beautiful
prediction anticipating that the missing states should eventually
be observed in experiment. However, one also may be worried by the fact that the
study of the excited resonance states reveals that operators
that are subsubleading in the $1/N_c$ analysis are in fact describing the main
splitting patterns observed in the multiplets, rather than the
subleading operators \cite{Carone-1,Carone-2,Schat-Goity-Scoccola}.
Thus, it remains unclear whether the $1/N_c$ expansion converges for the excited baryon
resonances. Moreover it was argued by Witten that the large-$N_c$ picture of excited
resonances is applicable for states with small total spin only \cite{Witten,Manohar,Georgi}.
In the Skyrme model for instance, a large-$N_c$ realization of QCD, only states
with $J \ll N_c$ are believed to be reliably described \cite{Witten}.

In the framework of the $1/N_c$ expansion the conjecture that resonances may be
generated dynamically is quite attractive since it amounts to a significant parameter
reduction. The interaction of the large-$N_c$ meson and baryon ground states may be
systematically constrained by the $1/N_c$ analysis and then would have to predict
all excited resonance states. On the other hand a classification
of the excited baryon resonance states as higher dimensional
representations of the contracted O(3)$\otimes $SU(6) group required
for each multiplet a new tower of operators with unknown
coefficients \cite{Carone-1,Carone-2,Schat-Goity-Scoccola}. Of
course any given quark model would predict the latter
coefficients. However, one should not forget that there is no
unique quark model so that strictly speaking the particular ansatz
used for the inter-quark potential represents already a particular
choice for an infinite set of parameters.

One may argue that the hadronic final state effect on a bare
resonance state is suppressed by $1/N_c$. This follows at a formal
level since any hadronic loop if visualized in terms of quark and
gluon lines gives rise to that suppression factor $1/N_c$. Though
this argument is certainly applicable in a world in which the
parameter $N_c\gg 1$ is really large, it is less immediate for the
physical case with $N_c=3$. If for some reason a particular subset
of hadronic loop diagrams is systematically enhanced by a factor
larger than the suppression factor $1/N_c$ for $N_c=3$ the above
argument breaks down and one must sum that infinite set of
hadronic diagrams. In this work we will argue that precisely the
set of diagrams representing the hadronic final state effect on a
bare resonance state are enhanced coherently by a typical factor
$2\pi$. Such diagrams are commonly called reducible in the sense
that they give rise to unitarity cuts as compared to irreducible
diagrams which build up the interaction kernel of the relativistic
Bethe-Salpeter two-body scattering equation. Therefore we would
expect that indeed the effect of the hadronic final state on a
bare resonance state is non-perturbative and potentially large.
That explains naturally why the chiral Lagrangian is able to
generate for instance the $\Lambda(1405)$ resonance in terms of
coupled channel dynamics. For a recent discussion of the competing
picture in which the $\Lambda (1405)$ resonance is considered as a
quark-model state we refer to \cite{Kimura}. The fact that the
existence of the $\Lambda (1405)$ resonance is predicted by the
chiral SU(3) Lagrangian basically parameter free \cite{Hirschegg}
we take as a strong indication that it is legitimate to consider
at least part of the baryon resonances of the second and third
generation as dynamically generated.

It is important to explore the consequence of this point of
view. Since in the quark model and also in the large-$N_c$
approach of QCD the excited resonances come in multiplets
classified by the contracted spin-flavor SU(6) group, the
statement that one member of a given multiplet should lie outside
the scope of the quark-model affects of course at the time the
remaining members of that multiplet \footnote{Of course one may
argue that the $\Lambda(1405)$ is not part of the ${\bf 70}$-plet. However,
we find that point of view not very convincing since that would
further increase the number of the so-called 'missing' resonance
states.}. For instance within the
quark-model the s-wave $\Lambda (1405)$ and d-wave $\Lambda(1520)$
resonance states belong to the same ${\bf 70}$-plet
\cite{Schat-Goity-Scoccola}. Incidentally, the quark model has a
notoriously difficult time to reproduce the empirical mass
splitting of those states \cite{Kimura,Schat-Goity-Scoccola}.
However, if the $\Lambda (1405)$ resonance is a consequence of
coupled channel dynamics than we would except the same to hold for
the $\Lambda (1520)$. Indeed there is an old speculation that the
latter resonance may be formed as consequence of a strong $\bar
K_\mu N $ channel \cite{Ball:Frazer,Aaron:1,Aaron:2}. Analogously, for the
SU(3) partner N(1520) of the $\Lambda(1520)$ the $\rho_\mu N,
\omega_\mu N$ would be the driving channels. For another member of
the same ${\bf 70}$-plet, the s-wave N(1535) resonance, there are
hints from the t-channel vector meson exchange model of Wyld \cite{Wyld} and
chiral coupled channel analyses \cite{Kaiser-Siegel-Weise,Inoue} that it may also be a
state that is generated dynamically.

It is then natural to expect
that the s-wave N(1650) resonance, another member of the same
${\bf 70}$-plet, is of dynamic origin also. This conjecture was tested in a
phenomenological coupled channel analysis of pion- and
photon-nucleon scattering data which considered in particular the
$\omega_\mu N$ and $\rho_\mu N$ channels \cite{QM-lutz,LWF}. The s-wave
N(1535) and N(1650) as well as the d-wave N(1520) resonance were
generated successfully by coupled channel effects. A similar chain
of arguments may be applied for the remaining multiplets. For
instance, the p-wave N(1440) resonance studied in great detail by
the J\"ulich group \cite{Roper-Juelich}, was claimed to be of
dynamic origin too. If that were true we would expect that the
remaining 55 states of that ${\bf 56}$-plet
\cite{Carone-1,Carone-2}, more precisely the subset of states
which exists, would also be of dynamic origin. Based on these
observations it is tempting to conjecture that in fact all excited
baryon resonance states, i.e. baryon states which do not belong to
the baryon octet and decuplet states, are a consequence of coupled
channel dynamics\footnote{This conjecture would imply that a similar
mechanism should explain excited meson resonances. These states would then
be generated dynamically in terms of the large-$N_c$
ground states, the pseudoscalar, the vector
and axial-vector mesons, $\pi, \rho_\mu, \omega_\mu, a_\mu$ etc.}.

There are two further interesting aspects we wish to discuss
in connection with the conjecture formulated above. First we would
like to comment on the possible parity doubling of resonance states
with large mass and
second on the Regge trajectory phenomena commonly used as a strong
argument in favor of the quark-model \cite{Regge-quark:1,Regge-quark:2}.
As was pointed out by Cohen and Glozman in recent papers
\cite{Cohen-Glozman-1,Cohen-Glozman-2}, the QCD operator product
expansion \cite{Wilson} suggests that correlation functions of
opposite parity should behave identically at high energies.
Asymptotically the non-perturbative structure is caused by gluon
dynamics which does not break the chiral symmetry. Therefore, if
the operator product expansion is valid at energies where there
are still resonance states visible, at least the high-energy tail
thereof, one should conclude that high-mass baryon resonances must
occur in parity doublets. Cohen and Glozman argue that indeed the
known high mass baryon resonance states confirm this behavior.
Though it is certainly premature to draw definite conclusion, if
this is true it would pose a severe challenge to the quark-model
picture \cite{Glozman-1}. Though by fine-tuning the quark model
parameters one may obtain the approximate degeneracy of some
states of opposite parity, it appears impossible to achieve this
property for the infinite tower of states typical predicted by any
quark model \cite{Glozman-1}. On the other hand in the coupled
channel picture of resonance states this property can be
implemented very easily by an appropriate constraint on the
scattering kernel. It is evident that if the scattering kernel in
channels of opposite parity is degenerate at high energies the
same would hold for resonances generated in those channels. In
fact that constraint would naturally lead to a significant
parameter reduction, a welcome effect. We turn to the
discussion of the baryon Regge trajectories. A quark model with a
linearly rising inter-quark potential can easily reproduce the
empirical linear Regge trajectory behavior of the baryon resonance
states \cite{Collins}. How can this be achieved in the competing
picture of resonance states? This is a serious
challenge for the coupled channel mechanism which may ultimately
favor the quark-model picture for highly excited resonances.
Common wisdom is that a simple t-channel exchange driving the scattering kernel of
a single channel problem can not produce a linear Regge trajectory
\cite{Collins}. However, we would conjecture that this may change
once more channels are considered. In this respect it is
encouraging that Diakonov and Petrov found a linear Regge trajectory for
baryon resonances in the Skyrme model \cite{Diakonov}. Even though
we do not want to advocate this model to be realistic for baryon
resonances it demonstrates that if sufficiently many inelastic
channels, in case of the Skyrme model multiple pion emissions of a rotating chiral
soliton, are considered one may have a chance to reproduce an approximatively
linear Regge trajectory. It would be important to investigate this
issue in more detail.

It is clear that in any case the rigorous study of baryon
resonances requires extensive coupled channel calculations.
On the one hand it is desirable to couple the bare three-quark states
of the quark-model to the meson-baryon final states
\cite{qqqMB} in order to arrive at a more realistic description of the baryon resonances.
Such calculation are similar in spirit to coupled channel analyses within generalized
isobar models \cite{Manley,Sauerman,Feuster:Mosel:1,Feuster:Mosel:2,Vrana:Dytman:Lee}
where preformed baryon resonance states are part of the meson-baryon scattering kernel.
On the other hand it is also useful to explore how far the competing picture where baryon
resonances are generated dynamically can be carried \cite{QM-lutz,LWF}. In this context it is
illuminating to review a particular aspect of coupled channel analyses. Most generalized
isobar model calculations so far employ the K-matrix approximation or extensions
thereof~\cite{Feuster:Mosel:1,Feuster:Mosel:2,Manley}. There is a conceptual
drawback of such computations. It is not guaranteed that the resulting scattering amplitudes
satisfy dispersion-integral relations predicted by the micro causality property
of quantum field theory. Analyticity is a fundamental property of the S-matrix and
the analytic structure of the scattering amplitude plays an important role e.g. close
to thresholds (see e.g.~\cite{Kondratyuk:Scholten}). It is to be emphasized that this
argument is
not necessarily a valid objection against such calculations. By fitting the data set a particular
model may satisfy the analyticity constraint approximatively. However, it would be useful
to study this problem in detail and demonstrate to what extent the resulting amplitudes are consistent
with their expected analyticity structure. A further property of the K-matrix approximation is
the fact that it suppresses potentially large resonance mass shifts \cite{Roper-Juelich,LWF}.
Therefore the study of the question whether a given resonance is primarily a three-quark bound
state or mainly a result of coupled channel dynamics should preferably be done in a framework
that does not inherently suppress the formation of resonances. Such schemes have been developed
\cite{N-D-Oller-Oset,N-D-Meissner-Oller,Vrana:Dytman:Lee,QM-lutz,Hirschegg,LWF}.
The question as to what is the nature of baryon resonances may be answered by a
detailed coupled channel study where the bare resonance mass parameters and bare resonance
coupling constants are considered as free parameters adjusted to the data set
\cite{Roper-Juelich,N-D-Meissner-Oller}. Of course the model should be general
enough as to avoid any bias of the analysis. If a given resonance is of
dynamic origin one would expect that the result of such a fit would give either an
extremely large bare resonance mass parameter or almost zero bare resonance coupling
constants. It may well be that such an analysis will reveal some type of mixed scenario.
Even though it would be preferable to perform such analyses for all resonances separately,
a tremendous challenge, it may be advantageous to allow for some minimal bias from large-$N_c$
QCD. As discussed in some detail above, if for a given member of a large-$N_c$ multiplet
there are convincing analyses demonstrating that this resonance is of dynamic origin
it is quite natural to 'give up' that multiplet and attempt a dynamic coupled
channel interpretation also of the remaining so far observed states.

In this work the resonance physics is not the main theme. Rather the goal is to
improve and better understand the chiral coupled channel approach elevating it onto
more rigorous grounds and at the same time extending its
applicability domain. Once the parameters of the chiral coupled
channel approach are fixed by a fit to the low-energy data we will, however,
investigate whether the scheme shows precursor effects of some
baryon resonances at the edge of its applicability domain. If that is the
case we take this as a hint or confirmation that those resonances may be of dynamic origin.
The above discussion on the nature of baryon resonances was necessary since
along the developments we will encounter various baryon resonances.
Therefore it is quite useful to have a reasonable working
hypothesis available that will guide us through the dark and dense
forest of technical and physical difficulties towards the sunny
glade of a profound understanding of hadronic interactions.

\vskip1.5cm \section{Relativistic chiral SU(3) interaction terms in large-$N_c$ QCD}

In the following sections we present all chiral interaction terms to be used in the remaining
chapters to describe the low-energy meson-baryon scattering data set.
Particular emphasis is put on the constraints implied by the large-$N_c$
analysis of QCD. The reader should not be discouraged by the many fundamental parameters
introduced in this chapter. The empirical data set includes many hundreds of
data points and will be reproduced rather accurately. Our scheme makes many predictions
for poorly known or not known observable quantities like for example the p-wave scattering volumes
of the kaon-nucleon scattering processes or the $SU(3)$ reactions like
$\pi \Lambda \to \pi \Sigma $. In a more conventional
meson-exchange approach, which lacks a systematic approximation scheme, many parameters
are implicit in the so-called form factors. In a certain sense the parameters used in the
form factors reflect the more systematically constructed and controlled quasi-local
counter terms of the chiral Lagrangian.

We recall the interaction terms of the relativistic chiral
$SU(3)$ Lagrangian density relevant for the meson-baryon scattering process.
For details on the systematic construction principle see for example \cite{Krause}. The basic
building blocks of the chiral Lagrangian are
\begin{eqnarray}
&& U_\mu = \frac{1}{2}\,e^{-i\,\frac{\Phi}{2\,f}} \left(
\partial_\mu \,e^{i\,\frac{\Phi}{f}}
+ i\,\Big[A_\mu , e^{i\,\frac{\Phi}{f}} \Big]_+
\right) e^{-i\,\frac{\Phi}{2\,f}}  \;,\qquad \!\!
B \;, \qquad \! \!\Delta_\mu \,, \qquad \! \! B^{*}_\mu \;,
\label{def-fields}
\end{eqnarray}
where we include the pseudo-scalar meson octet field
$\Phi(J^P\!\!=\!0^-)$, the baryon octet field $B(J^P\!\!=\!{\textstyle{1\over2}}^+)$, the baryon
decuplet field $\Delta_\mu(J^P\!\!=\!{\textstyle{3\over2}}^+)$ and the baryon nonet resonance
field $B^*_\mu(J^P\!\!=\!{\textstyle{3\over2}}^-)$ (see \cite{Tripp:1,Tripp:2,Plane}).

In (\ref{def-fields}) we consider
an external axial-vector source function $A_\mu $ which is required for the systematic
evaluation of matrix elements of the axial-vector current. A corresponding term
for the vector current is not shown in (\ref{def-fields}) because it will
not be needed in this work. The axial-vector source function
$A^\mu =\sum A_a^\mu\,\lambda^{(a)} $,
the meson octet field $\Phi=\sum \Phi_a\,\lambda^{(a)}$ and the baryon octet fields
$B= \sum B_a\,\lambda^{(a)}/\sqrt{2}$,
$B_\mu^*= B_{\mu,0}^*/\sqrt{3}+\sum B_{\mu,a}^*\,\lambda^{(a)}/\sqrt{2}$ are decomposed using the
Gell-Mann matrices $\lambda_a$ normalized by $\tr \lambda_a \,\lambda_b =
2\,\delta_{ab}$. The baryon decuplet field $\Delta^{abc} $ is completely symmetric
and related to the physical states by
\begin{eqnarray}
\begin{array}{llll}
\Delta^{111} = \Delta^{++}\,, & \Delta^{113} =\Sigma^{+}/\sqrt{3}\,, &
\Delta^{133}=\Xi^0/\sqrt{3}\,,  &\Delta^{333}= \Omega^-\,, \\
\Delta^{112} =\Delta^{+}/\sqrt{3}\,, & \Delta^{123} =\Sigma^{0}/\sqrt{6}\,, &
\Delta^{233}=\Xi^-/\sqrt{3}\,, & \\
\Delta^{122} =\Delta^{0}/\sqrt{3}\,, & \Delta^{223} =\Sigma^{-}/\sqrt{3}\,, &
& \\
\Delta^{222} =\Delta^{-}\,. & & &
\end{array}
\label{dec-field}
\end{eqnarray}
The parameter $f $ in (\ref{def-fields}) is determined by the weak decay
widths of the charged pions and kaons properly corrected for chiral $SU(3)$ effects. Taking
the average of the empirical decay parameters $f_\pi = 92.42 \pm 0.33 $
MeV  and $f_K \simeq 113.0 \pm 1.3$ MeV \cite{fpi:exp} one obtains the naive estimate
$f \simeq  104$ MeV. This value is still within reach of the more detailed analysis
\cite{GL85} which lead to $f_\pi/f = 1.07 \pm 0.12$. As was emphasized in \cite{MO01},
the precise value of $f$ is subject to large uncertainties.

Explicit chiral symmetry-breaking effects are included in terms
of scalar and pseudo-scalar source fields $\chi_\pm $ proportional to the quark-mass
matrix of QCD,
\begin{eqnarray}
\chi_\pm = \frac{1}{2} \left(
e^{+i\,\frac{\Phi}{2\,f}} \,\chi_0 \,e^{+i\,\frac{\Phi}{2\,f}}
\pm e^{-i\,\frac{\Phi}{2\,f}} \,\chi_0 \,e^{-i\,\frac{\Phi}{2\,f}}
\right) \,,
\label{def-chi}
\end{eqnarray}
where $\chi_0 \sim {\rm diag} (m_u,m_d,m_s)$.
All fields in (\ref{def-fields}) and (\ref{def-chi}) have identical properties under
chiral $SU(3)$ transformations. The chiral Lagrangian consists of all possible interaction
terms, formed with the fields $U_\mu, B, \Delta_\mu, B^*_\mu$ and $\chi_\pm$ and their
respective covariant derivatives. Derivatives of the fields must be included in compliance
with the chiral $SU(3)$ symmetry. This leads to the notion of a covariant derivative
${\mathcal D}_\mu$ which is identical for all fields in (\ref{def-fields}) and (\ref{def-chi}). For example,
it acts on the baryon octet fields as
\begin{eqnarray}
\Big[{\mathcal D}_\mu , B\Big]_- &=& \partial_\mu \,B +
\frac{1}{2}\,\Big[ e^{-i\,\frac{\Phi}{2\,f}} \left(
\partial_\mu \,e^{+i\,\frac{\Phi}{2\,f}}\right)
+e^{+i\,\frac{\Phi}{2\,f}} \left(
\partial_\mu \,e^{-i\,\frac{\Phi}{2\,f}}\right), B\Big]_-
\nonumber\\
&+& \frac{i}{2}\,\Big[ e^{-i\,\frac{\Phi}{2\,f}} \,
A_\mu \,e^{+i\,\frac{\Phi}{2\,f}}- e^{+i\,\frac{\Phi}{2\,f}} \, A_\mu
\,e^{-i\,\frac{\Phi}{2\,f}}, B\Big]_-  \,.
\label{}
\end{eqnarray}

The chiral Lagrangian is a powerful tool once it is combined with appropriate
power counting rules leading to a systematic approximation strategy.
One aims at describing hadronic interactions at low energy by constructing an expansion
in small momenta and the small pseudo-scalar meson masses. The infinite set
of Feynman diagrams are sorted according to their chiral powers. The minimal chiral
power $Q^{\nu }$ of a given relativistic Feynman diagram,
\begin{eqnarray}
\nu = 2-{\textstyle {1\over2}}\, E_B + 2\, L
+\sum_i V_i \left( d_i +{\textstyle {1\over2}}\, n_i-2 \right) \;,
\label{q-rule}
\end{eqnarray}
is given in terms of the number of loops, $L$, the number, $V_i$, of vertices of type $i$
with $d_i$ 'small' derivatives and $n_i$ baryon fields involved, and
the number of external baryon lines $E_B$ \cite{Weinberg}. Here one calls a derivative small
if it acts on the pseudo-scalar meson field or if it probes the virtuality of a baryon field.
Explicit chiral symmetry-breaking effects are perturbative and included in the counting scheme
with $\chi_0 \sim Q^2$.
For a discussion of the relativistic chiral Lagrangian and its required systematic regrouping
of interaction terms we refer to \cite{nn-lutz}. We will encounter explicit
examples of this regrouping subsequently. The relativistic chiral Lagrangian requires
a non-standard renormalization scheme. The $MS$ or $\overline{MS}$ minimal subtraction schemes
of dimensional regularization do not comply with the chiral counting rule \cite{Gasser}.
However, an appropriately modified subtraction scheme for relativistic Feynman diagrams leads
to manifest chiral counting rules \cite{Becher,nn-lutz,Gegelia}.
Alternatively one may work with the chiral
Lagrangian in its heavy-fermion representation \cite{J&M} where an appropriate frame-dependent
redefinition of the baryon fields leads to a more immediate  manifestation of the chiral
power counting rule (\ref{q-rule}). We will return to this issue in chapter 3.1 where we
propose a simple modification of the $\overline {MS}$-scheme which leads to
consistency with (\ref{q-rule}). Further subtleties of the chiral power counting rule (\ref{q-rule})
caused by the inclusion of an explicit baryon resonance field $B^*_\mu$ are addressed in chapter 4.1
when discussing the u-channel resonance exchange contributions.

In the $\pi N$ sector the $SU(2)$ chiral Lagrangian was successfully
applied \cite{Gasser,Bernard} demonstrating good convergence properties of the perturbative
chiral expansion. In the $SU(3)$ sector the situation is more involved due in part
to the rather large kaon mass $m_K \simeq m_N/2$. The perturbative
evaluation of the chiral Lagrangian cannot be justified and one must change
the expansion strategy. Rather than expanding directly the scattering amplitude one may
expand the interaction kernel according to chiral power counting rules \cite{Weinberg,LePage}.
The scattering amplitude then follows from the solution of a scattering equation like the
Lipmann-Schwinger or the Bethe-Salpeter equation. This is analogous to the treatment of the
$e^+\,e^-$ bound-state problem of QED where a perturbative evaluation of the interaction kernel
can be justified. The rational behind this change of scheme lies in the observation that
reducible diagrams are typically enhanced close to their unitarity threshold.
The enhancement factor $(2\pi)^n$, measured relative to a reducible diagram with the
same number of independent loop integrations, is given by
the number, $n$, of reducible meson-baryon pairs in the diagram, i.e. the number of unitary
iterations implicit in the diagram. In the $\pi N$ sector this enhancement factor does not
prohibit a perturbative treatment, because the typical expansion parameter of
\begin{eqnarray}
\frac{m^2_\pi}{8 \pi \,f^2} \simeq 0.1 \;,
\label{}
\end{eqnarray}
remains sufficiently small. In the $\bar K N$ sector,
on the other hand, the factor $(2\pi)^n$ invalidates a perturbative
treatment, because the typical expansion parameter would be
\begin{eqnarray}
\frac{m^2_K}{8 \pi\,f^2} \simeq 1 \;.
\label{}
\end{eqnarray}
This is in contrast to irreducible diagrams. They yield the typical expansion parameters
\begin{eqnarray}
\frac{m_\pi}{4 \pi \,f } \;,\qquad
\frac{m_K}{4\pi \,f} \;,
\label{}
\end{eqnarray}
which justifies the perturbative evaluation of the scattering kernels. This issue will
be taken up again later
when exemplifying the enhancement factor $(2\pi)$ in terms of the Weinberg-Tomozawa
interaction term.

In the next chapter a formalism will be developed which defines how to construct the
interaction kernel to be used in an 'on-shell reduced' Bethe-Salpeter
equation. In the remainder of this chapter all interaction terms are collected
that will be required for the construction of the leading orders
interaction kernel. All terms of chiral order $Q^2$ but
only the subset of chiral $Q^3$-terms, those which are leading in the large-$N_c$ limit,
will be considered. Loop corrections to the Bethe-Salpeter kernel are neglected, because
they carry minimal chiral order $Q^3$ and are $1/N_c$ suppressed. The chiral Lagrangian
\begin{eqnarray}
{\mathcal L} = \sum_n \,{\mathcal L}^{(n)}+\sum_n\,{\mathcal L}^{(n)}_\chi
\label{}
\end{eqnarray}
can be decomposed into terms of different classes ${\mathcal L}^{(n)}$ and ${\mathcal L}^{(n)}_\chi$.
With an upper index $n$ in ${\mathcal L}^{(n)}$ we indicate the number of fields
in the interaction vertex. The lower index $\chi $ signals terms with explicit chiral
symmetry breaking. Charge conjugation symmetry and parity invariance is assumed here.
At leading chiral order the following interaction terms are required:
\begin{eqnarray}
{\mathcal L}^{(2)} &=&
\tr \bar B \left(i\,\partialslash-\m0_{[8]}\right) \, B
+\frac{1}{4}\,\tr (\partial^\mu \,\Phi )\,
(\partial_\mu \,\Phi )
\nonumber\\
&+&\tr \bar \Delta_\mu \cdot \Big(
\left( i\,\partialslash -\m0_{[10]} \right)g^{\mu \nu }
-i\,\left( \gamma^\mu \partial^\nu + \gamma^\nu \partial^\mu\right)
+i\,\gamma^\mu\,\partialslash\,\gamma^\nu
+\,\m0_{[10]} \,\gamma^\mu\,\gamma^\nu
\Big) \, \Delta_\nu
\nonumber\\
&+&\tr \bar B^*_\mu \cdot \Big(
\left( i\,\partialslash -\m0_{[9]} \right)g^{\mu \nu }
-i\,\left( \gamma^\mu \partial^\nu + \gamma^\nu \partial^\mu\right)
+i\,\gamma^\mu\,\partialslash\,\gamma^\nu
+\,\m0_{[9]} \,\gamma^\mu\,\gamma^\nu
\Big) \, B^*_\nu
\nonumber\\
{\mathcal L}^{(3)} &=&\frac{F_{[8]}}{2\,f} \,\tr  \bar B
\,\gamma_5\,\gamma^\mu \,\Big[\left(\partial_\mu\,\Phi\right),B\Big]_-
+\frac{D_{[8]}}{2\,f} \,\tr  \bar B
\,\gamma_5\,\gamma^\mu \,\Big[\left(\partial_\mu\,\Phi\right),B\Big]_+
\nonumber\\
&-&\frac{C_{[10]}}{2\,f}\,
\tr \left\{
\Big( \bar \Delta_\mu \cdot
(\partial_\nu \,\Phi ) \Big)
\Big( g^{\mu \nu}-{\textstyle{1\over 2}}\,Z_{[10]}\, \gamma^\mu\,\gamma^\nu \Big) \,
B +\mathrm{h.c.}
\right\}
\nonumber\\
&+&\frac{D_{[9]}}{2\,f}\,
\tr \left\{  \bar B^*_\mu \cdot \Big[
(\partial_\nu \,\Phi ) \,
\Big( g^{\mu \nu}-{\textstyle{1\over 2}}\,Z_{[9]}\, \gamma^\mu\,\gamma^\nu \Big)\,\gamma_5 ,
B \Big]_++\mathrm{h.c.} \right\}
\nonumber\\
&+&\frac{F_{[9]}}{2\,f}\,
\tr \left\{
 \bar B^*_\mu \cdot \Big[
(\partial_\nu \,\Phi ) \,
\Big( g^{\mu \nu}-{\textstyle{1\over 2}}\,Z_{[9]}\, \gamma^\mu\,\gamma^\nu \Big)\,\gamma_5 ,
B \Big]_-+\mathrm{h.c.}
\right\}
\nonumber\\
&+&\frac{C_{[9]}}{8\,f}\,
\tr  \left\{ \bar B^*_\mu \,\tr   \Big[
(\partial_\nu \,\Phi ) \,
\Big( g^{\mu \nu}-{\textstyle{1\over 2}}\,Z_{[9]}\, \gamma^\mu\,\gamma^\nu \Big)\,\gamma_5 ,
B \Big]_++\mathrm{h.c.}
\right\} \;,
\nonumber\\
{\mathcal L}^{(4)}&=& \frac{i}{8\,f^2}\,\tr\bar B\,\gamma^\mu \Big[\Big[ \Phi ,
(\partial_\mu \,\Phi) \Big]_-,B \Big]_- \,,
\label{lag-Q}
\end{eqnarray}
where we use the notations $[A,B]_\pm = A\,B\pm B\,A$ for
$SU(3)$ matrices $A$ and  $B$. Note that the complete chiral interaction terms which
lead to the terms in (\ref{lag-Q}) are easily recovered by replacing
$i\,\partial_\mu \,\Phi /f \to U_\mu $. A derivative acting on a baryon field in (\ref{lag-Q})
must be understood as a covariant derivative with
$\partial_\mu \,B \to [{\mathcal D}_\mu ,B ]_- $  and
$\partial_\mu \,\Delta_\nu \to [{\mathcal D}_\mu ,\Delta_\nu ]_- $ .
The $SU(3)$ meson and baryon fields are
written in terms of their isospin symmetric components
\begin{eqnarray}
\Phi &=& \tau \cdot  \pi
+ \alpha^\dagger \!\cdot \! K +  K^\dagger \cdot \alpha  +
\eta \,\lambda_8 \;,
\nonumber\\
\sqrt{2}\,B &=&  \alpha^\dagger \!\cdot \! N(939)+\lambda_8 \,\Lambda(1115)+ \tau \cdot \Sigma(1195)
 +\Xi^t(1315)\,i\,\sigma_2 \!\cdot \!\alpha   \, ,
\nonumber\\ \nonumber\\
\alpha^\dagger &=&
 {\textstyle{1\over\sqrt{2}}}\left( \lambda_4+i\,\lambda_5 ,
\lambda_6+i\,\lambda_7 \right)
\;,\;\;\;\tau = (\lambda_1,\lambda_2,\lambda_3)\;,
\label{field-decomp}
\end{eqnarray}
with the isospin doublet fields $K =(K^+,K^0)^t $, $N=(p,n)^t$ and
$\Xi = (\Xi^0,\Xi^-)^t$. The isospin Pauli matrix $\sigma_2$ acts exclusively in the space of
isospin doublet fields $(K,N,\Xi)$ and the matrix valued isospin doublet $\alpha$
(see \cite{Lutz:Kolomeitsev}).
For this work we chose the isospin basis, because isospin breaking effects are important only
in the $\bar K N$ channel. Note that in (\ref{field-decomp})  the numbers in the parentheses
indicate the approximate mass of the baryon octet fields and $(...)^t$ means matrix transposition.
Analogously we write the baryon resonance field $B_\mu^*$ as
\begin{eqnarray}
\sqrt{2}\,B^*_\mu &=&
\Big( \sqrt{{\textstyle{2\over 3}}}\,\cos \vartheta- \lambda_8\,\sin \vartheta\Big) \,\Lambda_\mu(1520)
+\alpha^\dagger \!\cdot\!  N_\mu(1520)
\nonumber\\
&+& \Big( \sqrt{{\textstyle{2\over 3}}}\,\sin \vartheta+ \lambda_8\,\cos \vartheta\Big) \,\Lambda_\mu(1690)
+\tau \!\cdot \!\Sigma_\mu(1670)  +\Xi_\mu^t(1820)\,i\,\sigma_2 \!\cdot\! \alpha \, ,
\label{def-b-stern}
\end{eqnarray}
where we allow for singlet-octet mixing by means of the mixing angle $\vartheta $ (see \cite{Plane}).
The parameters $\m0_{[8]}$, $\m0_{[9]}$ and $\m0_{[10]}$ in (\ref{lag-Q}) denote the baryon masses
in the chiral $SU(3)$ limit. Furthermore the products of an anti-decuplet field $\bar
\Delta$ with a decuplet field $\Delta$ and an octet field $\Phi$ transform as  $SU(3)$ octets
\begin{eqnarray}
\Big(\bar \Delta \cdot \Delta \Big)^a_b &=&
\bar \Delta_{bcd}\,\Delta^{acd}\,, \qquad
\Big( \bar \Delta \cdot \Phi \Big)^a_b
=\epsilon^{kla}\,\bar \Delta_{knb}\,\Phi_l^n \,,
\nonumber\\
\Big( \Phi \cdot \Delta \Big)^a_b
&=&\epsilon_{klb}\,\Phi^l_n\,\Delta^{kna} \; ,
\label{dec-prod}
\end{eqnarray}
where $\epsilon_{abc}$ is the completely anti-symmetric pseudo-tensor. For the isospin
decomposition of $\bar \Delta \cdot \Delta$, $\bar \Delta \cdot \Phi$ and
$\Phi \cdot \Delta$ we refer to \cite{Lutz:Kolomeitsev}.

The parameters  $F_{[8]}\simeq 0.45$
and $D_{[8]}\simeq 0.80$ are constrained by the weak decay widths of the baryon octet states
\cite{Okun} (see also Tab. \ref{weak-decay:tab}) and $C_{[10]}\simeq 1.6$ can be
estimated from the hadronic decay width of the baryon decuplet states.
The parameter $Z_{[10]}$ in (\ref{lag-Q}) may be best determined in an $SU(3)$ analysis of
meson-baryon scattering. While in the pion-nucleon sector it can be absorbed into the
quasi-local 4-point interaction terms to chiral accuracy $Q^2$ \cite{Tang} (see also \cite{Lutz:Kolomeitsev})
this is no longer possible if the scheme is extended to $SU(3)$. Our detailed analysis reveals that
the parameter $Z_{[10]}$ is relevant already to order $Q^2$ if a simultaneous chiral analysis of
the pion-nucleon and kaon-nucleon scattering processes is performed. The resonance parameters may
be estimated by an update of the analysis \cite{Plane}. That leads to the values
\begin{eqnarray}
F_{[9]} \simeq 1.8\;,\qquad  D_{[9]} \simeq 0.84 \;,\qquad
C_{[9]} \simeq 2.5 \;.
\label{}
\end{eqnarray}
The singlet-octet mixing angle $\vartheta \simeq $ 28$^\circ$
confirms the finding of \cite{Tripp:2} that the $\Lambda(1520)$ resonance is predominantly a flavor singlet
state. The  value for the background parameter $Z_{[9]}$ of the $J^P\!\!=\!{\textstyle{3\over2}}^-$ resonance
is expected to be rather model dependent, because it is unclear so far how to incorporate the
$J^P\!\!=\!{\textstyle{3\over2}}^-$ resonance in a controlled approximation scheme.
As will be explained in detail $Z_{[9]}$ will drop out completely in our scheme (see chapters 4.1-4.2).

The empirical values of $F_{[8]},D_{[8]}$ and $C_{[10]}$ are consistent with the expectation from
large-$N_c$ counting rules which predict for example $C_{[10]}/D_{[8]} =2+{\mathcal O}(1/N_c )$
\cite{Jenkins}. The large-$N_c$ scaling of  a chiral interaction
term is easily worked out applying the operator analysis proposed in \cite{Dashen}.
Interaction terms involving baryon fields are represented by matrix
elements of many-body operators in the large-$N_c$ ground-state baryon
multiplet $| {\mathcal B}  \rangle $. A n-body operator is the product of
n factors formed exclusively in terms of the bilinear quark-field operators
$J_i, G_i^{(a)}$ and $T^{(a)}$. These operators are characterized fully by their commutation
relations,
\begin{eqnarray}
&&[ G^{(a)}_i\,,G^{(b)}_j]_- ={\textstyle{1\over 4}}\,i\,\delta_{ij}\,f^{ab}_{\;\;\;\,c}\,T^{(c)}
+{\textstyle{1\over 2}}\,i\,\epsilon_{ij}^{\;\;\;k}
\left({\textstyle{1\over 3}}\,\delta^{ab}\,J_{k}+d^{ab}_{\;\;\;\,c}\,G^{(c)}_k\right), \;\;
\nonumber\\
&&[ J_i\,,J_j]_- =i\,\epsilon_{ij}^{\;\;\;k}\,J_{k}\, , \quad
[ T^{(a)}\,,T^{(b)}]_- =i\,f^{ab}_{\;\;\;\,c}\,T^{(c)}\,,\quad
\nonumber\\
&& [ T^{(a)}\,,G^{(b)}_i]_- =i\,f^{ab}_{\;\;\;c}\,G^{(c)}_i \;,\quad \!
[ J_i\,,G^{(a)}_j]_- =i\,\epsilon_{ij}^{\;\;\;k}\,G^{(a)}_k \;,  \quad \!
 [ J_i\,,T^{(a)}]_- = 0\;.
\label{comm}
\end{eqnarray}
The algebra (\ref{comm}), which reflects the so-called contracted spin-flavor symmetry
of QCD, leads to a transparent derivation of the many sum rules implied by the
various infinite subclasses of QCD quark-gluon diagrams as collected to a given order in the
$1/N_c$ expansion. A convenient realization of the algebra (\ref{comm}) is obtained in
terms of non-relativistic, flavor-triplet and color $N_c$-multiplet field operators
$q$ and $q^\dagger$ as
\begin{eqnarray}
&& J_i = q^\dagger \Bigg( \,\frac{\sigma_i^{(q)}}{2} \otimes 1\Bigg) \,q \,, \qquad
T^{(a)} = q^\dagger \Bigg( 1 \otimes \frac{\lambda^{(a)}}{2} \Bigg) \,q \,, \;\,
\nonumber\\
&& G^{(a)}_i = q^\dagger \Bigg(\, \frac{\sigma_i^{(q)}}{2} \otimes \frac{\lambda^{(a)}}{2} \Bigg) \,q \,.
\label{}
\end{eqnarray}
If the fermionic field operators $q$ and $q^\dagger $ are assigned
anti-commutation rules the algebra (\ref{comm}) follows. The Pauli spin matrices
$\sigma^{(q)}_i$ act on the two-component spinors of the fermion fields $q, q^\dagger $
and the Gell-Mann matrices $\lambda_a$ on their flavor components. Here one needs to emphasize
that the non-relativistic quark-field operators $q$ and $q^\dagger $ should not be identified
with the quark-field operators of the QCD Lagrangian \cite{DJM,Carone,Luty}. Rather, they
constitute an effective tool to represent the operator algebra (\ref{comm}) which allows for an
efficient derivation of the large-$N_c$ sum rules of QCD. A systematic truncation scheme
results in the operator analysis, because a $n$-body operator is assigned the suppression factor
$N_c^{1-n}$ \cite{Witten}. The analysis is complicated by the fact that  matrix elements of
$ T^{(a)}$ and $G_i^{(a)}$ may be of order $N_c$ in the baryon
ground state $|{\mathcal B} \rangle$. That implies for instance that matrix elements of
the (2$n$+1)-body operator $(T_a\,T^{(a)})^n\,T^{(c)}$ are not suppressed relative to the matrix
elements of the one-body operator $T^{(c)}$. The systematic large-$N_c$ operator analysis
relies on the observation that matrix elements of the spin operator $J_i$, on the other hand,
are always of order $N_c^0$. Then a set of identities shows how to systematically represent the
infinite set of many-body operators, which one may write down to a given order in the $1/N_c$
expansion, in terms of a finite number of operators. This leads to
a convenient scheme with only a finite number of operators at given order \cite{Dashen}.
Typical examples of the required operator identities read
\begin{eqnarray}
&&[T_a,\, T^{(a)}]_+ -  [J_i,\,J^{(i)}]_+  ={\textstyle {1\over 6}}\,N_c\,(N_c+6)\;,
\quad [ T_{a}\,,G^{(a)}_i]_+ ={\textstyle {2\over 3}}\,(3+N_c)\,J_i \;,
\nonumber\\
&& 27\,[T_a,\, T^{(a)}]_+-12\,[G_i^{(a)},\, G_a^{(i)}]_+
= 32\,[J_i,\,J^{(i)}]_+ \;,
\nonumber\\
&& d_{abc}\,[T^{(a)},\,T^{(b)}]_+  -2\,[J_{i},\,G_c^{(i)}]_+
= -{\textstyle {1\over 3}}\,(N_c+3)\,T_c \;,
\nonumber\\
&& d^a_{\;\;bc}\,[G_a^{(i)},\,G_i^{(b)}]_+
+{\textstyle{9\over 4}}\, d_{abc}\,[T^{(a)},\,T^{(b)}]_+
= {\textstyle {10\over 3}}\,[J_{i},\,G_c^{(i)}]_+ \;,
\nonumber\\
&& d_{ab}^{\;\;\; c}\,[T^{(a)},\,G^{(b)}_i]_+ =  {\textstyle{1\over 3}}\,[J_{i},\,T^{(c)}]_+
-{\textstyle{1\over 3}}\,\epsilon_{ijk}\,f_{ab}^{\;\;\;c}\,[G^{(j)}_a\,,G^{(k)}_b]_+ \;.
\label{operator-ex}
\end{eqnarray}
For instance the first identity in (\ref{operator-ex}) shows how to avoid that
infinite tower $(T_a\,T^{(a)})^n\,T^{(c)}$ discussed above. Note that the 'parameter'
$N_c$ enters in (\ref{operator-ex}) as a mean to classify the possible realizations of
the algebra (\ref{comm}).

As a first and simple example recall the large-$N_c$ structure of the 3-point vertices.
One readily establishes two operators with parameters $g$ and $h$ at leading order
in the $1/N_c$ expansion \cite{Dashen}:
\begin{eqnarray}
\langle {\mathcal B}' |\, {\mathcal L}^{(3)}\,| {\mathcal B}  \rangle  =\frac{1}{f}\,
\langle {\mathcal B}' |\, g\,G_i^{(c)}+h\,J_i\,T^{(c)}| {\mathcal B}  \rangle \,
\tr \,\lambda_c\,\nabla^{(i)}\,\Phi + {\mathcal O}\left( \frac{1}{N_c}\right) \;.
\label{3-point-vertex}
\end{eqnarray}
Further possible terms in (\ref{3-point-vertex}) are either redundant or suppressed
in the $1/N_c$ expansion. For example, the two-body operator
$ i\,f_{abc}\,G_i^{(a)} \,T^{(b)}  \sim N_c^0$ is reduced by applying the relation
\begin{eqnarray}
&& i\,f_{ab}^{\;\;\;c}\,\Big[G_i^{(a)} \,,T^{(b)} \Big]_- = i\,f_{ab}^{\;\;\;c}\,i\,f^{ab}_{\;\;\;\,d}\,G_i^{(d)} =
-3\,G_i^{(c)} \;. \nonumber
\label{}
\end{eqnarray}
In order to make use of the large-$N_c$ result it is necessary to evaluate the matrix elements
in (\ref{3-point-vertex}) at
$N_c=3$ where one has a $\bf 56$-plet with
$| {\mathcal B}  \rangle= |B(a) ,\Delta(ijk) \rangle $. Most
economically this is achieved with the completeness identity
$1=|B\rangle \langle B|+ |\Delta\rangle \langle \Delta | $ in conjunction with
\begin{eqnarray}
&&T_c\,| B_a(\chi)\rangle = i\,f_{abc}\,| B^{(b)}(\chi)\rangle \;,\qquad
J^{(i)} \,| B_a(\chi )\rangle
= \frac{1}{2}\,\sigma^{(i)}_{\chi' \chi}| B_a(\chi')\rangle \;,
\nonumber\\
&& G^{(i)}_c\,| B_a(\chi)\rangle =
\left(\frac{1}{2}\,d_{abc}
+ \frac{1}{3}\,i\,f_{abc}\right)\,\sigma^{(i)}_{\chi'\chi}\,| B^{(b)}(\chi')\rangle
\nonumber\\
&& \qquad \qquad \quad \;
+ \frac{1}{\sqrt{2}\,2}\,\Big(\epsilon_{l}^{\;jk}\,\lambda^{(c)}_{mj}
\,\lambda_{nk}^{(a)}\Big)\,S^{(i)}_{\chi' \chi}
| \Delta^{\!(lmn)}(\chi') \rangle \;,
\label{matrix-el}
\end{eqnarray}
where $S_i\,S^\dagger_j=\delta_{ij}-\sigma_i\,\sigma_j/3$ and
$\lambda_a\,\lambda_b = {\textstyle{2 \over 3}}\,\delta_{ab}
+(i\,f_{abc}+d_{abc})\,\lambda^{(c)}$. In (\ref{matrix-el}) the baryon octet states
$| B_b(\chi)\rangle $ are labelled according to their $SU(3)$ octet index $a=1,...,8$ with
the two spin states represented by $\chi=1,2$. Similarly the decuplet states
$| \Delta_{lmn}(\chi') \rangle$ are listed with $l,m,n=1,2,3$ as defined in (\ref{dec-field}).
Note that the expressions (\ref{matrix-el}) may be
verified using the quark-model wave functions for the baryon octet and decuplet states. The result (\ref{matrix-el}) is
however much more general than the quark-model, because it follows from the structure of the ground-state baryons
in the large-$N_c$ limit of QCD only. Matching the interaction vertices of the relativistic chiral Lagrangian
onto the static matrix elements arising in the large-$N_c$ operator analysis requires a
non-relativistic reduction. It is standard to decompose the
4-component Dirac fields $B$ and $\Delta_\mu $ into baryon octet and
decuplet spinor fields $B(\chi)$ and $\Delta (\chi)$:
\begin{eqnarray}
\Big(B, \Delta_\mu \Big) \to  \left(
\begin{array}{c}
 \left(\frac{1}{2}+\frac{1}{2}\,\sqrt{1+\frac{\nabla^2}{M^2}}
\right)^{\frac{1}{2}} \Big(B(\chi ),S_\mu\,\Delta (\chi )\Big) \\
\frac{(\sigma \cdot \nabla )}{\sqrt{2}\,M}
\left(1+\sqrt{1+\frac{\nabla^2}{M^2} }\,\right)^{-\frac{1}{2}}
\Big(B(\chi ),S_\mu\,\Delta (\chi )\Big)
\end{array}
\right) \,,
\label{}
\end{eqnarray}
where $M$ denotes the baryon octet and decuplet mass in the large-$N_c$ limit. At
leading order one finds $S_\mu =(0, S_i) $ with the transition matrices $S_i$ introduced in
(\ref{matrix-el}). It is then straightforward to expand in powers of $\nabla/M$ and
achieve the desired matching. This leads for example to the identification $D_{[8]}=g $,
$F_{[8]}= 2\,g/3+h$ and $C_{[10]}=2\,g$. Operators at subleading order in
(\ref{3-point-vertex}) then parameterize the deviation from $C_{[10]}\simeq 2 \,D_{[8]}$.

\vskip1.5cm \section{Quasi-local interaction terms}

Consider the two-body interaction terms to chiral order $Q^2$.
From phase space consideration it is evident that at this order
there are only terms which contribute to the meson-baryon s-wave scattering lengths, the
s-wave effective range parameters and the p-wave scattering volumes. Higher partial
waves are not affected at this order. The various contributions are grouped according
to their scalar, vector or tensor nature as
\begin{eqnarray}
{\mathcal L}^{(4)}_2= {\mathcal L}^{(S)}+{\mathcal L}^{(V)}+{\mathcal L}^{(T)}
\,,
\label{l42}
\end{eqnarray}
where the lower index k in ${\mathcal L}^{(n)}_k$ denotes the minimal chiral order
of the interaction vertex. In the relativistic framework one observes mixing of the
partial waves in the sense that for instance ${\mathcal L}^{(S)}, {\mathcal L}^{(V)}$
contribute to the s-wave channels and ${\mathcal L}^{(S)}, {\mathcal L}^{(T)}$ to the
p-wave channels. We write
\begin{eqnarray}
{\mathcal L}^{(S)}&=&\frac{g^{(S)}_0}{8\!\,f^2}\,\tr\bar B\,B
\,\tr (\partial_\mu\Phi) \, (\partial^\mu\Phi)
+\frac{g^{(S)}_1}{8\!\,f^2}\,\tr \bar B \,(\partial_\mu\Phi)
\,\tr(\partial^\mu\Phi) \, B
\nonumber\\
&+&\frac{g^{(S)}_F}{16\!\,f^2} \,\tr\bar B \Big[
\Big[(\partial_\mu\Phi),(\partial^\mu\Phi)
\Big]_+ ,B\Big]_-
\!+\frac{g^{(S)}_D}{16\!\,f^2}\,\tr \bar B \Big[
\Big[(\partial_\mu\Phi),(\partial^\mu\Phi)
\Big]_+, B \Big]_+ \;,
\nonumber\\
{\mathcal L}^{(V)}&=&
\frac{g^{(V)}_0}{16\!\, f^2}\,
\Big(\tr \bar B \,i\,\gamma^\mu\,( \partial^\nu B) \,
\tr(\partial_\nu\Phi) \, ( \partial_\mu\Phi)
+\mathrm{h.c.}\Big)
\nonumber\\
&+&\frac{g^{(V)}_1}{32\!\,f^2}\,
\tr \bar B \,i\,\gamma^\mu\,\Big( ( \partial_\mu\Phi)
\,\tr(\partial_\nu\Phi) \, ( \partial^\nu B)
+ ( \partial_\nu\Phi) \,
\tr ( \partial_\mu \Phi) \, ( \partial^\nu B)
+\mathrm{h.c.}\Big)
\nonumber\\
&+&\frac{g_F^{(V)}}{32\!\,f^2}\,\Big(
\tr   \bar B \,i\,\gamma^\mu\,\Big[
\Big[(\partial_\mu \Phi) , (\partial_\nu\Phi)\Big]_+,
( \partial^\nu B) \Big]_-
+\mathrm{h.c.} \Big)
\nonumber\\
&+& \frac{g^{(V)}_D}{32\!\,f^2}\,\Big(\tr
\bar B \,i\,\gamma^\mu\,\Big[
\Big[( \partial_\mu\Phi) , (\partial_\nu\Phi)\Big]_+,
( \partial^\nu B) \Big]_+
+\mathrm{h.c.} \Big)\,,
\nonumber\\
{\mathcal L}^{(T)}&=&
\frac{g^{(T)}_1}{8\!\,f^2}\,\tr\bar B \,( \partial_\mu\Phi)
\,i\,\sigma^{\mu \nu}\,\tr( \partial_\nu \Phi) \, B
\nonumber\\
&+& \frac{g^{(T)}_D}{16\!\,f^2}\,
\tr \bar B \,i\,\sigma^{\mu \nu}\,\Big[
\Big[(\partial_\mu \Phi) ,( \partial_\nu \Phi)  \Big]_-, B \Big]_+
\nonumber\\
&+&\frac{g^{(T)}_F}{16\!\,f^2}\,
\tr \bar B \,i\,\sigma^{\mu \nu}\,\Big[
\Big[( \partial_\mu \Phi) ,( \partial_\nu\Phi) \Big]_- ,B\Big]_-
\,.
\label{two-body}
\end{eqnarray}
It is clear that if the heavy-baryon expansion is applied to (\ref{two-body})
the quasi-local 4-point interactions can be mapped onto corresponding terms
of  the heavy-baryon formalism presented for example in \cite{CH-Lee}. Inherent in
the relativistic scheme is the presence of redundant interaction terms which requires
that a systematic regrouping of the interaction terms is performed. This will be
discussed below in more detail when introducing the quasi-local counter terms at
chiral order $Q^3$.

The large-$N_c$ counting rules are applied in order to estimate the
relative importance of the  quasi-local $Q^2$-terms in (\ref{two-body}).
Terms which involve a single-flavor trace are enhanced as compared to the
double-flavor trace terms. This is because a flavor trace in an interaction term
is necessarily accompanied by a corresponding color trace if visualized in terms of quark
and gluon lines. A color trace signals a quark loop and therefore provides the announced
$1/N_c$ suppression factor \cite{Hooft,Witten}. The counting rules are nevertheless
subtle, because a certain combination of double trace expressions can be rewritten in
terms of a single-flavor trace term \cite{Fearing}
\begin{eqnarray}
&&\tr \left( \bar B \, B \right)
\,\tr \Big(\Phi \, \Phi  \Big)
+2\,\tr \left( \bar B \, \Phi \right)
\,\tr \Big(\Phi \, B  \Big)
\nonumber\\
=&&\tr \Big[\bar B, \Phi \Big]_-\,\Big[B, \Phi
\Big]_-
+\frac{3}{2}\,\tr \bar B \Big[
\Big[\Phi , \Phi  \Big]_+, B \Big]_+ \;.
\label{trace-id}
\end{eqnarray}
Thus, one expects for example that both parameters $g_{0}^{(S)}$ and $g_{1}^{(S)}$ may be large
separately but the combination $2\,g_0^{(S)}-g_1^{(S)}$ should be small. A more detailed
operator analysis  leads to
\begin{eqnarray}
&&\langle {\mathcal B}' | {\mathcal L}^{(4)}_2 | {\mathcal B} \rangle  = \frac{1}{16\,f^2}\,
\langle {\mathcal B}' | \,O_{ab}(g_1,g_2)\, | {\mathcal B} \rangle
\,\tr [(\partial_\mu \Phi),\lambda^{(a)}]_-\,[(\partial^\mu \Phi),\lambda^{(b)}]_-
\nonumber\\
&&  \qquad \qquad +\frac{1}{16\,f^2}\,
\langle {\mathcal B}' | \,O_{ab}(g_3,g_4)\, | {\mathcal B} \rangle
\,\tr [(\partial_0 \Phi),\lambda^{(a)}]_-\,[(\partial_0 \Phi),\lambda^{(b)}]_-
\nonumber\\
&& \qquad \qquad  + \frac{1}{16\,f^2}\,
\langle {\mathcal B}' | \,O^{(ij)}_{ab}(g_5,g_6)\, | {\mathcal B} \rangle
\,\tr [(\nabla_i \Phi),\lambda^{(a)}]_-\,[(\nabla_j \Phi),\lambda^{(b)}]_- \;,
\nonumber\\ \nonumber\\
&& O_{ab}(g,h) = g\,d_{abc}\,T^{(c)} + h\,[T_a,T_b]_+
+{\mathcal O}\left(\frac{1}{N_c} \right) \;,
\nonumber\\
&& O_{ab}^{(ij)}(g,h) = i\,\epsilon^{ijk }\,i\,f_{abc}\left(
g\,G_k^{(c)} + h\,J_{k}\,T^{(c)} \right)
+{\mathcal O}\left(\frac{1}{N_c} \right) \;.
\label{Q^2-large-Nc}
\end{eqnarray}
It was checked that other forms for the coupling of the operators $O_{ab}$ to the
meson fields do not lead to new structures. The matching of the coupling constants
$g_{1,..,6}$ onto those of (\ref{two-body}) is straight forward. Identifying the
leading terms in the non-relativistic expansion, we obtain
\begin{eqnarray}
&& g_0^{(S)}= \frac{1}{2}\,g_1^{(S)} = \frac{2}{3}\,g_D^{(S)}
= -2\,g_2\,, \qquad  \; g_F^{(S)}= -3\,g_1 \,,
\nonumber\\
&& g_0^{(V)}= \frac{1}{2}\,g_1^{(V)}=\frac{2}{3}\,g_D^{(V)}
=-2\,\frac{g_4}{M}\,, \qquad
g_F^{(V)}= -3\,\frac{g_3}{M} \,,
\nonumber\\
&& g_1^{(T)}= 0 \,, \qquad
g_F^{(T)}= -g_5-\frac{3}{2}\,g_6 \,,  \qquad g_D^{(T)}= -\frac{3}{2}\,g_5 \;,
\label{Q^2-large-Nc-result}
\end{eqnarray}
where $M$ is the large-$N_c$ value of the baryon octet mass.
Thus, to chiral order $Q^2$ there are only six leading large-$N_c$
coupling constants.

Continue with the quasi-local counter terms to chiral order $Q^3$. It is
useful to discuss first a set of redundant interaction terms:
\begin{eqnarray}
{\mathcal L}^{(R)}&=&\frac{h^{(1)}_0}{8 f^2}\,
\tr(\partial^\mu\bar B)\,
(\partial_\nu B)\,
\tr( \partial_\mu \Phi) \, ( \partial^\nu\Phi)
\nonumber\\
&+&\frac{h^{(1)}_1}{16 f^2}\,
\tr(\partial^\mu\bar B)\,  ( \partial_\nu\Phi)\,
\tr(\partial_\mu\Phi) \, (\partial^\nu B)
\nonumber\\
&+&\frac{h^{(1)}_1}{16 f^2}\,
\tr(\partial^\mu\bar B)\,( \partial_\mu \,\Phi)\,
\tr(\partial^\nu\Phi) \, (\partial_\nu B) \,,
\nonumber\\
&+&\frac{h^{(1)}_F}{16 f^2}\,
\tr (\partial^\mu\bar B)\, \Big[
\Big[( \partial_\mu\Phi) , (\partial^\nu \Phi) \Big]_+ ,
(\partial_\nu B) \Big]_-
\nonumber\\
&+& \frac{h^{(1)}_D}{16 f^2}\,
\tr (\partial^\mu\bar B)\,  \Big[
\Big[( \partial_\mu\Phi) , (\partial^\nu \Phi)\Big]_+,
(\partial_\nu B )\Big]_+ \,.
\label{redundant}
\end{eqnarray}
Performing the non-relativistic expansion of (\ref{redundant}) one finds that the
leading moment is of chiral order $Q^2$. Formally the terms in (\ref{redundant}) are
transformed into terms of subleading order $Q^3$ by subtracting  ${\mathcal L}^{(V)}$ of
(\ref{two-body}) with $g^{(V)} =  \m0_{[8]} \,h^{(1)}$. Bearing this in mind the
terms of (\ref{redundant}) define particular interaction vertices of chiral order $Q^3$.
Note that by analogy with (\ref{Q^2-large-Nc}) and (\ref{Q^2-large-Nc-result})
we expect the coupling constants
$h_F^{(1)}$ and $h_D^{(1)}$ with $h_1^{(1)}=2\,h_0^{(1)}= 4\,h_D^{(1)}/3$
to be leading in the large-$N_c$ limit.
A complete collection of counter terms of chiral order $Q^3$ is presented in
\cite{Lutz:Kolomeitsev}. Including the four terms of (\ref{redundant}) we find ten independent
interaction terms which all contribute exclusively to the
s- and p-wave channels. Here the two additional terms with $h^{(2)}_{F}$ and
$h^{(3)}_{F}$ which are leading in the large-$N_c$ expansion are presented:
\begin{eqnarray}
{\mathcal L}^{(4)}_3 &=& {\mathcal L}^{(R)}- {\mathcal L}^{(V)} [g^{(V)}\! = \m0_{[8]}\,h^{(1)}]
\nonumber\\
&+& \frac{h^{(2)}_{F}}{32\!\,f^2}\,
\tr \bar B \,i\,\gamma^\mu\,\Big[ \big[ ( \partial_\mu\Phi)
, (\partial_\nu \Phi)\big]_- , ( \partial^\nu B) \Big]_- +{\rm h.c}
\nonumber\\
&+&\frac{h^{(3)}_{F}}{16\!\,f^2}\,
\tr \bar B \,i\,\gamma^\alpha\,\Big[ \big[ ( \partial_\alpha \,\partial_\mu\Phi)
,( \partial^\mu \Phi) \big]_- ,  B \Big]_- \;.
\label{local-q-3}
\end{eqnarray}
The interaction vertices in (\ref{local-q-3}) can be mapped onto corresponding
static matrix elements of the large-$N_c$ operator analysis
\begin{eqnarray}
&&\langle {\mathcal B}' | {\mathcal L}^{(4)}_3 | {\mathcal B} \rangle  = \frac{h_2}{16\,f^2}\,
\langle {\mathcal B}' | \,i\,f_{abc}\,T^{(c)}\, | {\mathcal B} \rangle
\,\partial_\mu \,\Big(
\tr [(\partial_0 \Phi),\lambda^{(a)}]_-\,[(\partial^\mu \Phi),\lambda^{(b)}]_- \Big)
\nonumber\\
&& \qquad \qquad \;\; +\frac{h_3}{16\,f^2}\,
\langle {\mathcal B}' | \,i\,f_{abc}\,T^{(c)}\, | {\mathcal B} \rangle
\, \tr [(\partial_0 \,\partial_\mu \Phi),\lambda^{(a)}]_-\,[(\partial^\mu \Phi),\lambda^{(b)}]_- \,,
\label{}
\end{eqnarray}
where $h^{(2)}_{F} \sim h_2$ and $h_{F}^{(3)} \sim h_3$. To summarize the
result for the quasi-local chiral interaction vertices of order $Q^3$: at leading order
the  $1/N_c$ expansion leads to four relevant parameters
$h^{(1)}_{F,D}$ and $h^{(2)}_{F}$ and $h^{(3)}_{F}$,
only. Also one should stress that the $SU(3)$ structure of the
$Q^3$ terms as they contribute to the s- and p-wave channels differ from
the $SU(3)$ structure of the $Q^2$ terms. For instance the $g^{(S)}$
coupling constants contribute to the p-wave channels with four independent
$SU(3)$ tensors. In contrast, to order $Q^3$ the parameters $h^{(2)}_{F}$ and
$h^{(3)}_{F}$, which are in fact the only parameters contribution to the
p-wave channels to this order, contribute with a different and independent
$SU(3)$ tensor. This is to be compared with the static $SU(3)$ prediction
that leads to six independent tensors:
\begin{equation}
8\otimes 8= 1
\oplus 8_S\oplus 8_A \oplus 10\oplus \overline{10}\oplus 27 \;.
\label{}
\end{equation}
Part of the predictive power of the chiral Lagrangian results, because
chiral $SU(3)$ symmetry selects certain subsets of all $SU(3)$ symmetric
tensors to a given chiral order.

\vskip1.5cm \section{Explicit chiral symmetry breaking}

There remain the interaction terms proportional to $\chi_\pm $
which break the chiral $SU(3)$ symmetry explicitly. Here all relevant terms of
chiral order $Q^2$ \cite{Gasser,Kaiser} and $Q^3$ \cite{Mueller} are collected.
It is convenient to visualize the symmetry-breaking
fields $\chi_\pm$ of (\ref{def-chi}) in their expanded forms:
\begin{eqnarray}
\! \!\! \chi_+ = \chi_0 -\frac{1}{8\,f^2}
\Big[ \Phi, \Big[ \Phi ,\chi_0 \Big]_+\Big]_+ \!+{\mathcal O} \left(\Phi^4 \right) \,,
\;\,
\chi_- = \frac{i}{2\,f}\, \Big[ \Phi ,\chi_0 \Big]_+
\!+{\mathcal O} \left(\Phi^3 \right) \,.
\label{chi-exp}
\end{eqnarray}
Begin with the 2-point interaction vertices which all result exclusively from chiral
interaction terms linear in $\chi_+$. They read:
\begin{eqnarray}
{\mathcal L }_{\chi}^{(2)}&=& -\frac{1}{4}\,\tr \Phi\,\Big[ \chi_0, \Phi \Big]_+
+2\,\tr \bar B \left( b_D\,\Big[ \chi_0 , B \Big]_+ +b_F\,\Big[ \chi_0 , B \Big]_-
+b_0 \, B \,\tr \chi_0 \right)
\nonumber\\
&+&2\,d_D\,\tr \Big(\bar \Delta_\mu\cdot \Delta^\mu  \Big)  \,\chi_0
+2\,d_0 \,\tr \left(\bar \Delta_\mu\cdot  \Delta^\mu\right) \,\tr \chi_0
\nonumber\\
&+&\tr \bar B \left(i\,\partialslash-\m0_{[8]}\right) \,
\Big( \zeta_0\,B \, \tr  \chi_0
+  \zeta_D\,[B, \chi_0 ]_+ + \zeta_F\, [B, \chi_0 ]_- \Big) \;,
\nonumber\\ \nonumber\\
\chi_0 &=&\frac{1}{3} \left( m_\pi^2+2\,m_K^2 \right)\,1
+\frac{2}{\sqrt{3}}\,\left(m_\pi^2-m_K^2\right) \lambda_8 \;,
\label{chi-sb}
\end{eqnarray}
where we normalized $\chi_0$ to give the pseudo-scalar mesons their isospin averaged
masses. The first term in (\ref{chi-sb}) leads to the finite masses of the pseudo-scalar
mesons. Note that at chiral order $Q^2$ one has $m_\eta^2 = 4\,(m_K^2-m_\pi^2)/3$.
The parameters $b_D$, $b_F$, and $d_D$ are determined at leading order by the
baryon octet and decuplet mass splitting
\begin{eqnarray}
&&m_{[8]}^{(\Sigma )}-m_{[8]}^{(\Lambda )}=
{\textstyle{16\over 3}}\,b_D\,(m_K^2-m_\pi^2)\,, \quad
m_{[8]}^{(\Xi )}-m_{[10]}^{(N )} =-8\,b_F\,(m_K^2-m_\pi^2)\,,
\nonumber \\
&&m_{[10]}^{(\Sigma )}-m_{[10]}^{(\Delta )}=m_{[8]}^{(\Xi )}-m_{[10]}^{(\Sigma )}
=m_{[10]}^{(\Omega )}-m_{[10]}^{(\Xi )}
=-\textstyle{4\over 3}\, d_D\, (m_K^2-m_\pi^2)\,.
\label{mass-splitting}
\end{eqnarray}
The empirical baryon masses lead to the estimates $b_D \simeq 0.06$~GeV$^{-1}$,
$b_F \simeq -0.21$~GeV$^{-1}$, and $d_D\simeq -0.49$~GeV$^{-1}$.
For completeness we recall the leading large-$N_c$ operators for
the baryon mass splitting (see e.g. \cite{Jenkins}):
\begin{eqnarray}
&&\langle {\mathcal B}' |{\mathcal L }_{\chi}^{(2)} | {\mathcal B} \rangle
=\langle {\mathcal B}' |\,
b_1\,T^{(8)} + b_2\,[J^{(i)}, G^{(8)}_i]_+ \,| {\mathcal B} \rangle
+{\mathcal O}\left(\frac{1}{N_c^{2}}\right)\,,
\nonumber\\
&&b_D = -\frac{\sqrt{3}}{16}\,\frac{3\,b_2}{m_K^2-m_\pi^2} \,, \qquad
b_F = -\frac{\sqrt{3}}{16}\,\frac{2\,b_1+b_2}{m_K^2-m_\pi^2}\,,
\nonumber\\
&& d_D = -\frac{3\,\sqrt{3}}{8}\,\frac{b_1+2\,b_2}{m_K^2-m_\pi^2} \,,
\label{}
\end{eqnarray}
where we matched the symmetry-breaking parts with $\lambda_8$.
One observes that the empirical values for $b_D+b_F$ and $d_D$ are remarkably consistent with
the large-$N_c$ sum rule $b_D+b_F\simeq {\textstyle{1\over 3}}\,d_D$. The parameters $b_0$
and $d_0$ are more difficult to access. They determine the deviation of the octet and
decuplet baryon masses from their chiral $SU(3)$ limit values $\m0_{[8]}$ and $\m0_{[10]}$:
\begin{eqnarray}
&& m_{[8]}^{(N)} = \m0_{[8]}-2\,m_\pi^2\,(b_0+2\,b_F)-4\,m_K^2\,(b_0+b_D-b_F) \;,
\nonumber\\
&& m_{[10]}^{(\Delta )}  = \m0_{[10]}-2\,m_\pi^2 (d_0 +d_D)-4\,m_K^2 \,d_0 \;,
\label{piN-sig-term}
\end{eqnarray}
where terms of chiral order $Q^3$ are neglected. The size of the parameter $b_0$ is
commonly encoded into the pion-nucleon sigma term
\begin{equation}
\sigma_{\pi N}= -2\,m_\pi^2\,(b_D+b_F+2\,b_0) +{\mathcal O}\left(Q^3\right)\,.
\label{spin:naive}
\end{equation}
Note that the former standard value $\sigma_{\pi N}=(45\pm 8)$~MeV of
\cite{piN-sigterm} is currently under debate \cite{pin-news}.

The parameters $\zeta_0,\zeta_D$ and $\zeta_F$ are required to cancel a divergent term in the
baryon wave-function renormalization as it follows from the one loop self-energy correction or
equivalently the unitarization of the s-channel baryon exchange term. It will be demonstrated
explicitly that within our approximation they will not have any observable effect.
They lead to a renormalization of the three-point vertices only, which can be accounted for by
a redefinition of the parameters in (\ref{chi-sb-3}). Thus, one may simply drop these
interaction terms.

The predictive power of the chiral Lagrangian lies in part in the strong correlation of
vertices of different degrees as implied by the non-linear fields $U_\mu $ and $\chi_\pm $.
A powerful example is given by the two-point vertices introduced in (\ref{chi-sb}). Since
they result from chiral interaction terms linear in the $\chi_+$-field (see (\ref{chi-exp})),
they induce particular meson-octet baryon-octet interaction vertices:
\begin{eqnarray}
{\mathcal L }_{\chi}^{(4)}&=&
\frac{i}{16\,f^2}\,\tr\bar B\,\gamma^\mu \Big[\Big[ \Phi ,
(\partial_\mu \,\Phi) \Big]_-,\zeta_0\,B \, \tr  \chi_0
+  \zeta_D\,[B, \chi_0 ]_+ + \zeta_F\, [B, \chi_0 ]_-  \Big]_-
\nonumber\\
&+&\frac{i}{16\,f^2}\,\tr \Big[\zeta_0\,\bar B \, \tr  \chi_0
+  \zeta_D\,[\bar B, \chi_0 ]_+ + \zeta_F\, [\bar B, \chi_0 ]_- ,
\Big[ \Phi ,(\partial_\mu \,\Phi) \Big]_- \Big]_- \,\gamma^\mu\,B
\nonumber\\
&-&\frac{1}{4\,f^2}\,\tr \bar B \left(b_D\,
\Big[ \Big[ \Phi, \Big[ \Phi ,\chi_0 \Big]_+\Big]_+ , B \Big]_+
+b_F\,\Big[ \Big[ \Phi, \Big[ \Phi ,\chi_0 \Big]_+\Big]_+ , B \Big]_-
\right)
\nonumber\\
&-&\frac{b_0}{4\,f^2}\,\tr\bar{B}\, B \,\tr \Big[ \Phi, \Big[ \Phi ,\chi_0 \Big]_+\Big]_+
\;.
\label{chi-sb-4}
\end{eqnarray}
To chiral order $Q^3$ there are no further four-point interaction terms with explicit
chiral symmetry breaking.

Next turn to the three-point vertices with explicit chiral symmetry breaking. Here
the chiral Lagrangian permits two types of interaction terms written as
${\mathcal L }_{\chi }^{(3)}={\mathcal L }_{\chi,\, +}^{(3)}+{\mathcal L }_{\chi, -}^{(3)}$.
In ${\mathcal L }_{\chi, +}^{(3)}$ we collect 16 axial-vector terms, which
result form chiral interaction terms linear in the $\chi_+$ field (see (\ref{chi-exp})),
with a priori unknown coupling constants $F_{0,..,9}$ and $C_{0,...,5}$,
\begin{eqnarray}
{\mathcal L }_{\chi, +}^{(3)}&=&
\frac{1}{4\,f} \,\tr  \bar B
\,\gamma_5\,\gamma^\mu \,\Big(
\Big[\left(\partial_\mu\,\Phi\right),F^{}_0\,\Big[\chi_0, B \Big]_+
+F^{}_1\,\Big[\chi_0, B \Big]_-\Big]_+  \Big)+{\rm h.c.}
\nonumber\\
&+&\frac{1}{4\,f} \,\tr  \bar B
\,\gamma_5\,\gamma^\mu \,\Big(
\Big[\left(\partial_\mu\,\Phi\right),F^{}_2\,\Big[\chi_0, B \Big]_+
+F^{}_3\,\Big[\chi_0, B \Big]_-\Big]_-  \Big)+{\rm h.c.}
\nonumber\\
&+&\frac{1}{2\,f} \,\tr  \bar B
\,\gamma_5\,\gamma^\mu \,\Big( F_4\,\Big[
\Big[\chi_0, \left(\partial_\mu\,\Phi\right)\Big]_+ , B \Big]_+
+F^{}_5\,\Big[\Big[\chi_0, \left(\partial_\mu\,\Phi\right)\Big]_+ , B \Big]_-  \Big)
\nonumber\\
&+&\frac{1}{4\,f} \,\tr  \bar B
\,\gamma_5\,\gamma^\mu \,\Big(
F^{}_{6}\,B \,\tr \Big( \chi_0 \left(\partial_\mu\,\Phi\right)  \Big)
+F^{}_{7} \left(\partial_\mu\,\Phi\right)
\tr \Big(\chi_0 \, B  \Big) \Big) +{\rm h.c.}
\nonumber\\
&+& \frac{1}{2\,f} \,\tr  \bar B
\,\gamma_5\,\gamma^\mu \,\Big(
F_8\,\Big[\left(\partial_\mu\,\Phi\right),B\Big]_+ \,\tr \chi_0
+F_9 \,\Big[\left(\partial_\mu\,\Phi\right),B\Big]_-\,\tr \chi_0 \Big)
\nonumber\\
&-& \frac{1}{2\,f}\,
\tr \left\{C_0 \,\Big( \bar \Delta_\mu \cdot
\Big[ \chi_0 ,(\partial_\mu \,\Phi )\Big]_+
+C_1 \,\Big( \bar \Delta_\mu \cdot
\Big[ \chi_0 ,(\partial_\mu \,\Phi )\Big]_- \Big)\,B +\rm{h.c.}\right\}
\nonumber\\
&-&\frac{1}{2\,f}\,
\tr \left\{
\Big( \bar \Delta_\mu \cdot (\partial_\mu \,\Phi )  \Big)\,
\Big( C_2\,\Big[\chi_0,B\Big]_++C_3\,\Big[\chi_0,B\Big]_- +\rm{h.c.}\right\}
\nonumber\\
&-&\frac{C_4}{2\,f}\,
\tr \left\{ \Big( \bar \Delta_\mu \cdot \chi_0
\Big)\,\big[(\partial_\mu \,\Phi ),B \big]_- +\rm{h.c.}\right\}
\nonumber\\
&-&\frac{C_5}{2\,f}\,
\tr \left\{ \Big( \bar \Delta_\mu \cdot
(\partial_\mu \,\Phi )\Big)\,
B +\mathrm{h.c.}\right\} \tr \, \chi_0  \;.
\label{chi-sb-3}
\end{eqnarray}
Similarly in ${\mathcal L }_{\chi, -}^{(3)}$  the remaining terms
which result from chiral interaction terms linear in $\chi_-$ are collected.
There are three pseudo-scalar interaction terms with $\bar F_{4,5,6}$ and four
additional terms parameterized by $\delta F_{4,5,6}$ and $\delta C_0$
\begin{eqnarray}
{\mathcal L }_{\chi, -}^{(3)}&=& \frac{1}{2\,f} \,\tr  \bar B
 \Big[
2\,i\,\gamma_5\,\m0_{[8]}\,
\bar F_4\,\Big[\chi_0,\Phi\Big]_+
+\gamma_5\,\gamma^\mu \, \delta F_4\,
\Big[\chi_0, \left(\partial_\mu\,\Phi\right)\Big]_+
, B \Big]_+
\nonumber\\
&+&\frac{1}{2\,f} \,\tr  \bar B
 \Big[
2\,i\,\gamma_5\,\m0_{[8]}\,
\bar F_5\,\Big[\chi_0,\Phi\Big]_+
+\gamma_5\,\gamma^\mu \, \delta F_5\,
\Big[\chi_0, \left(\partial_\mu\,\Phi\right)\Big]_+
, B \Big]_-
\nonumber\\
&+&\frac{1}{2\,f} \,\tr  \bar B\,
 \Big(
2\,i\,\gamma_5\,\m0_{[8]}\,
\bar F_6\,\tr \big(\chi_0\,\Phi\big)
+\gamma_5\,\gamma^\mu \, \delta F_6\,\tr
\big( \chi_0\, \left(\partial_\mu \,\Phi\right)\big) \Big)\, B
\nonumber\\
&-& \frac{\delta C_0}{2\,f}\,
\tr  \,\Big( \bar \Delta_\mu \cdot
\Big[ \chi_0 ,(\partial_\mu \,\Phi )\Big]_+ \,B +\rm{h.c.} \Big) \,.
\label{chi-sb-3:p}
\end{eqnarray}
While the parameters $F_i$ and $C_i$ contribute to matrix elements of
the $SU(3)$ axial-vector current $A^\mu_a$, none of the terms in (\ref{chi-sb-3:p})
contribute. This follows once the external axial-vector current is
restored. In (\ref{chi-sb-3}) this is achieved with the replacement
$\partial^\mu\,\Phi_a \to \partial^\mu\,\Phi_a +2\,f\, A^\mu_a$ (see (\ref{def-fields})).
Though it is obvious that the pseudo-scalar terms in (\ref{chi-sb-3:p}) proportional to
$\bar F_i$ do not contribute to the axial-vector current, it is less immediate that the terms
proportional to $\delta F_i$ and $\delta C_i$ also do not contribute. Moreover, the latter
terms appear redundant, because terms with identical structure at the 3-point level
are already listed in (\ref{chi-sb-3:p}). Here one needs to realize that the terms proportional
to $\delta F_i$ and $\delta C_i$ result from chiral interaction terms linear
in $[{\mathcal D}_\mu ,\chi_-]_- $ while the terms proportional to $F_i, C_i$
result from chiral interaction terms linear in $U_\mu $.

The pseudo-scalar parameters $\bar F_i$ and also $\delta F_{i}$ lead to a
tree-level Goldberger-Treiman discrepancy. For instance, we have
\begin{eqnarray}
f \,g_{\pi N N} -g_A \,m_N = 2\,m_N\,m_\pi^2 \,
(\bar F_4+\delta F_4+\bar F_5+\delta F_5 ) + {\mathcal O} \left( Q^3\right)  \,,
\label{GTD}
\end{eqnarray}
where we introduced the pion-nucleon coupling constant $g_{\pi N N}$ and the axial-vector
coupling constant of the nucleon $g_A$. The corresponding generalized Goldberger-Treiman
discrepancies for the remaining axial-vector coupling constants of the baryon octet states
follow easily from the replacement rule $F_i \to F_i+\bar F_i+\delta F_i $ for $i=4,5,6$
(see also \cite{Goity:Lewis}). It is to be
emphasized that (\ref{GTD}) must not be confronted directly with the Goldberger-Treiman
discrepancy as discussed in \cite{GTD,GTD:pi,Goity:Lewis}, because it necessarily involves
the $SU(3)$ parameter $f $ rather than $f_\pi \simeq 92$ MeV or $f_K \simeq 113$ MeV.

The effect of the axial-vector interaction terms in (\ref{chi-sb-3}) is twofold. First, they
lead to renormalized values of the $F_{[8]}, D_{[8]}$ and $C_{[10]}$ parameters in (\ref{lag-Q}).
Secondly they induce interesting $SU(3)$ symmetry-breaking effects which are proportional
to $(m_K^2-m_\pi^2)\,\lambda_8$. Note that the renormalization of the $F_{[8]},D_{[8]}$ and $C_{[10]}$ parameters
requires care, because it is necessary to discriminate between the
renormalization of the axial-vector current and the renormalization of the meson-baryon coupling
constants. The parameters $F_R, D_R$ and $C_R$  are introduced as they enter
matrix elements of the axial-vector current:
\begin{eqnarray}
&& F_R =F_{[8]}+ (m_\pi^2+2\,m_K^2)\left(F_9
+{\textstyle{2\over 3}}\,\big( F_2+F_5\big) \right) \;, \qquad
\nonumber\\
&& D_R= D_{[8]}+ (m_\pi^2+2\,m_K^2)
\left(F_8 +{\textstyle{2\over 3}}\,\big( F_0+F_4\big) \right)\;,
\nonumber\\
&& C_R=C_{[10]} +(m_\pi^2+2\,m_K^2)\,\left(C_5 +{\textstyle{2\over 3}}\,\big( C_0+C_2\big)\right)\,.
\label{ren-FDC}
\end{eqnarray}
and  the renormalized parameters $F_{A,P}, D_{A,P}$ and
$ C_A$ as they are relevant for the meson-baryon 3-point vertices:
\begin{eqnarray}
&& F_A= F_R + {\textstyle{2\over 3}}\,(m_\pi^2 +2\,m_K^2)\,\delta F_5\;, \quad
D_A= D_R + {\textstyle{2\over 3}}\,(m_\pi^2 +2\,m_K^2)\,\delta F_4\;, \quad
\nonumber\\
&& F_P= {\textstyle{2\over 3}}\,(m_\pi^2 +2\,m_K^2)\,\bar F_5\;, \quad
D_P= {\textstyle{2\over 3}}\,(m_\pi^2 +2\,m_K^2)\,\bar F_4\;, \quad
\nonumber\\
&& C_A= C_R + {\textstyle{2\over 3}}\,(m_\pi^2 +2\,m_K^2)\,\delta C_0\;.
\label{ren-FDC-had}
\end{eqnarray}
The index $A$ or $P$ indicates whether the meson couples to the baryon via an axial-vector
vertex (A) or a pseudo-scalar vertex (P).
It is clear that the effects of (\ref{ren-FDC}) and (\ref{ren-FDC-had}) break the chiral
symmetry but do not break the $SU(3)$ symmetry. In this work the renormalized
parameters $F_R, D_R$ and $C_R$ are used. Note that
one can always choose the parameters $F_8, F_9$ and $C_5$ to obtain
$F_R=F_{[8]}$, $D_R=D_{[8]}$ and $C_R=C_{[10]}$. In order to distinguish the renormalized values
from their  bare values one needs to determine the parameters $F_8, F_9$ and $C_5$  by
investigating higher-point Green functions. This is beyond the scope of this work.

The number of parameters inciting $SU(3)$ symmetry-breaking effects can be reduced
significantly by the large-$N_c$ analysis. Recall the five leading operators presented
in \cite{Dashen}:
\begin{eqnarray}
&&\langle {\mathcal B}' | \,{\mathcal L}_\chi^{(3)}\,| {\mathcal B} \rangle
=\frac{1}{f}\,\langle {\mathcal B}'|  \,O_i^{(a)}(\tilde c) |{\mathcal B}\rangle \,
\tr\,\lambda_a\,\nabla^{(i)}\, \Phi
+ 4\,\langle {\mathcal B}'|  \,O_i^{(a)}(c) |{\mathcal B}\rangle \,
A^{(i)}_a \;,
\nonumber\\ \nonumber\\
&& O_i^{(a)}(c) = c_1\,\big( d^{8a}_{\;\;\;b}\,G^{(b)}_{i}
+{\textstyle {1\over 3}}\,\delta^{a8} \,J_{i}\big)
+c_2\,\big( d^{8a}_{\;\;\;b}\,J_{i}\,T^{(b)}+\delta^{a8} \,J_{i}\big)
\nonumber\\
&& \qquad \qquad \quad +\;c_3\,[G^{(a)}_i, T^{(8)}]_+ +c_4 \, [T^{(a)},G^{(8)}_{i}]_+
 +c_5 \, [J^2,[T^{(8)},G^{(a)}_i]_-]_- \;.
\label{ansatz-3}
\end{eqnarray}
It is important to observe that the parameters $c_i$ and $\tilde c_i$ are a priori independent.
They are  correlated by the chiral $SU(3)$ symmetry only. With (\ref{matrix-el}) it is
straightforward to map the
interaction terms (\ref{ansatz-3}), which all break the $SU(3)$ symmetry linearly,
onto the chiral vertices of (\ref{chi-sb-3}) and (\ref{chi-sb-3:p}). This procedure
relates the parameters $c_i $ and $\tilde c_i$. One finds that the matching is possible
for all operators leaving only ten independent parameters $c_{i}$, $\tilde c_{1,2}$,
$\bar c_{1,2}$ and $a$ rather than the twenty-three $F_i$, $C_i$ and $\delta F_i, \delta C_i$ and
$\bar F_i$ parameters in (\ref{chi-sb-3},\ref{chi-sb-3:p}). Simple algebra leads to
$c_i=\tilde c_i$ for $i=3,4,5$ and
\begin{eqnarray}
&& \!\!\!\!F_1 = -\frac{\sqrt{3}}{2}\,\frac{c_3}{m_K^2-m_\pi^2}\;,\quad
F_2 = -\frac{\sqrt{3}}{2}\,\frac{c_4}{m_K^2-m_\pi^2} \,,\quad
F_3 =-\frac{1}{\sqrt{3}}\,\frac{c_3+c_4}{m_K^2-m_\pi^2} \;,
\nonumber\\
&& \!\!\!\!F_4 = -\frac{\sqrt{3}}{4}\,\frac{c_1}{m_K^2-m_\pi^2} \;,\quad
F_5 = -\frac{\sqrt{3}}{4}\,\frac{{\textstyle{2\over 3}}\,c_1+c_2}{m_K^2-m_\pi^2} \;,\quad
F_6=-\frac{\sqrt{3}}{2}\,\frac{{\textstyle{2\over 3}}\,c_1+c_2}{m_K^2-m_\pi^2} \;,
\nonumber\\
&& \!\!\!\! C_0 = -\frac{\sqrt{3}}{2}\,\frac{c_1}{m_K^2-m_\pi^2} \,, \quad
C_1 = -\frac{\sqrt{3}}{2}\,\frac{c_3-c_4+3\,c_5}{m_K^2-m_\pi^2} \;,
\nonumber\\
&& \!\!\!\! C_3 =-\sqrt{3}\,\frac{c_3}{m_K^2-m_\pi^2}\,, \quad
C_4 =-\sqrt{3}\,\frac{c_4}{m_K^2-m_\pi^2}\,, \quad
 F_{0,7}=C_{2,5}=0 \,,
\label{large-Nc-FDC}
\end{eqnarray}
and
\begin{eqnarray}
&& \bar F_4= -\frac{\sqrt{3}}{4}\,\frac{\bar c_1}{m_K^2-m_\pi^2} \;,\quad
\delta F_4 = -\frac{\sqrt{3}}{4}\,\frac{\delta c_1}{m_K^2-m_\pi^2} \;,
\nonumber\\
&& \bar F_5 = -\frac{\sqrt{3}}{4}\,\frac{{\textstyle{2\over 3}}\,\bar c_1+\bar c_2+a}{m_K^2-m_\pi^2}
\,, \quad \delta F_5 = -\frac{\sqrt{3}}{4}\,\frac{{\textstyle{2\over 3}}\,\delta c_1+\delta c_2-a}{m_K^2-m_\pi^2}
\,,
\nonumber\\
&& \bar F_6 =-\frac{\sqrt{3}}{2}\,\frac{{\textstyle{2\over 3}}\,\bar c_1+\bar c_2+a}{m_K^2-m_\pi^2}  \,,
\quad \delta F_6 =-\frac{\sqrt{3}}{2}\,\frac{{\textstyle{2\over 3}}\,\delta c_1+\delta c_2-a}{m_K^2-m_\pi^2}   \,,
\nonumber\\
&& \delta C_0=-\frac{\sqrt{3}}{2}\,\frac{ \tilde c_1- c_1}{m_K^2-m_\pi^2} \;, \qquad
\delta c_i = \tilde c_i -c_i -\bar c_i \;.
\label{large-Nc-FDC:delta}
\end{eqnarray}
In (\ref{large-Nc-FDC:delta})  the pseudo-scalar parameters
$\bar F_{4,5,6}$ are expressed in terms of the more convenient dimensionless
parameters $\bar c_{1,2}$ and $a$. Here we insist that an expansion analogous to
(\ref{ansatz-3}) holds also for the pseudo-scalar vertices in (\ref{chi-sb-3:p}).
\tabcolsep=1.4mm
\renewcommand{\arraystretch}{1.5}
\begin{table}[t]\begin{center}
\begin{tabular}{|c|c||c|c|c|c|c|c|c|}
\hline
& $g_A $ (Exp.) & $ F_R$ & $ D_R$ & $c_1$ & $c_2 $ & $c_3$ & $c_4 $  & $c_5 $\\
\hline \hline
$ n \to p\,e^-\,\bar \nu_e $ & $1.267 \pm 0.004$& 1 & 1 & $\frac{5}{3\,\sqrt{3}}$ &  $\frac{1}{\sqrt{3}}$& $\frac{5}{\sqrt{3}}$&$ \frac{1}{\sqrt{3}}$ & $\;\;0\;\;$\\  \hline

$ \Sigma^- \to \Lambda \,e^-\,\bar \nu_e  $ & $0.601 \pm 0.015$ & 0 & $\sqrt{\frac{2}{3}}$ & $\frac{\sqrt{2}}{3}$ &  0& 0& 0& 0\\  \hline

$ \Lambda \to p \,e^-\,\bar \nu_e $ & $-0.889 \pm 0.015$ & $-\sqrt{\frac{3}{2}}$ & $-\frac{1}{\sqrt{6}}$ & $\frac{1}{2\,\sqrt{2}}$ & $\frac{1}{2\,\sqrt{2}}$ & $-\frac{3}{2\,\sqrt{2}}$& $\frac{1}{2\,\sqrt{2}}$& 0\\  \hline

$ \Sigma^- \to n \,e^-\,\bar \nu_e $ & $0.342 \pm 0.015$ & $-1$ & 1 & $-\frac{1}{6\,\sqrt{3}}$ & $\frac{1}{2\,\sqrt{3}}$  & $\frac{1}{2\,\sqrt{3}}$& $-\frac{\sqrt{3}}{2}$& 0\\  \hline

$ \Xi^- \to \Lambda \,e^-\,\bar \nu_e $ & $0.306 \pm 0.061 $ & $\sqrt{\frac{3}{2}}$ & $-\frac{1}{\sqrt{6}}$  & $-\frac{1}{6\,\sqrt{2}}$ & $-\frac{1}{2\,\sqrt{2}}$ & $-\frac{1}{2\,\sqrt{2}}$& $-\frac{5}{2\,\sqrt{2}}$& 0\\  \hline

$ \Xi^- \to \Sigma^0 \,e^-\,\bar \nu_e $ & $0.929 \pm 0.112 $ & $\frac{1}{\sqrt{2}}$ & $\frac{1}{\sqrt{2}}$ & $-\frac{5}{6\,\sqrt{6}}$ & $-\frac{1}{2\,\sqrt{6}}$ & $-\frac{5}{2\,\sqrt{6}}$& $-\frac{1}{2\,\sqrt{6}}$& 0\\  \hline


\end{tabular}
\end{center}
\caption[TTTTTT]{Axial-vector coupling constants for the weak decay processes of the baryon octet states. The empirical values
for $g_A$ are taken from \cite{Dai}. Here we do not consider  $SU(3)$ symmetry-breaking effects of the
vector current. The last seven columns give the coefficients of the axial-vector
coupling constants $g_A$ decomposed into the $F_R$, $D_R$ and $c_i$ parameters.}
\label{weak-decay:tab}
\end{table}

The analysis of \cite{Dai}, which considers constraints from the weak decay
processes of the baryon octet states and the strong decay widths of the decuplet states, obtains
$c_2 \simeq -0.15$ and $c_3\simeq 0.09$ but finds  values of $c_1$ and $c_4$ which are compatible
with zero\footnote{The parameters of \cite{Dai} are obtained with $c_i \to -\sqrt{3}\,c_i/2$ and
$a= D_R+(c_1+3\,c_3)/\sqrt{3}$ and $b=F_R-2\,a/3+(2\,c_1/3+c_2+2\,c_3+c_4)/\sqrt{3}$.
The analysis of \cite{Dai} does not determine the parameter $c_5$.}. In Tab. (\ref{weak-decay:tab}) we
reproduce the axial-vector coupling constants as given in \cite{Dai} relevant for the various
baryon octet weak-decay processes. Besides $C_A+2\,\tilde c_1/\sqrt{3}$, the empirical strong-decay widths
of the decuplet states constrain the parameters $c_{3}$ and $c_4$ only, as is
evident from the expressions for the decuplet widths,
\begin{eqnarray}
&& \Gamma_{[10]}^{(\Delta )} = \frac{m_N+E_N}{2\,\pi \,f^2}\,\frac{p_{\pi N}^3}{12\,m_{[10]}^{(\Delta )}}
\,\Big( C_A +{\textstyle{2\over \sqrt{3}}}\,(\tilde c_1+3\,c_3)\Big)^2 \,,
\nonumber\\
&& \Gamma_{[10]}^{(\Sigma )} = \frac{m_\Lambda+E_\Lambda}{2\,\pi \,f^2}\,
\frac{p_{\pi \Lambda}^3}{24\,m_{[10]}^{(\Sigma )}}
\,\Big( C_A +{\textstyle{2\over \sqrt{3}}}\,\tilde c_1 \Big)^2
\nonumber\\
&& \qquad +\frac{m_\Sigma+E_\Sigma}{3\,\pi \,f^2}\,
\frac{p_{\pi \Sigma}^3}{24\,m_{[10]}^{(\Sigma )}}
\,\Big( C_A +{\textstyle{2\over \sqrt{3}}}\,(\tilde c_1+6\,c_4)\Big)^2 \,,
\nonumber\\
&& \Gamma_{[10]}^{(\Xi )} = \frac{m_\Xi+E_\Xi}{2\,\pi \,f^2}\,\frac{p_{\pi \Xi}^3}{24\,m_{[10]}^{(\Xi )}}
\,\Big( C_A +{\textstyle{2\over \sqrt{3}}}\,(\tilde c_1-3\,c_3+3\,c_4)\Big)^2 \,,
\label{decuplet-decay}
\end{eqnarray}
where for example $m_\Delta =\sqrt{m_N^2+p_{\pi N}^2}+\sqrt{m_\pi^2+p_{\pi N}^2}$ and
$E_N= \sqrt{m_N^2+p_{\pi N}^2}$. For instance, the values
$ C_A+2\,\tilde c_1/\sqrt{3} \simeq 1.7$,
$c_3 \simeq 0.09$ and $c_4 \simeq 0.0$ together with
$f\simeq f_\pi \simeq 93$ MeV lead to isospin averaged partial
decay widths of the decuplet states which are compatible with the present day empirical estimates. It is
clear that the six data points for the baryon octet decays can be reproduced by a suitable adjustment of the
six parameters $F_R$, $D_R$ and $c_{1,2,3,4}$. The non-trivial issue is to what extent the
explicit $SU(3)$ symmetry-breaking pattern in the axial-vector coupling constants is consistent
with the symmetry-breaking pattern in the meson-baryon coupling constants. Here a possible
strong energy dependence of the decuplet self-energies
may invalidate the use of the simple expressions (\ref{decuplet-decay}). A more
direct comparison with the meson-baryon scattering data may be required.

An implicit assumption one relies on if Tab. \ref{weak-decay:tab}
is applied should be mentioned. In a strict chiral expansion the $Q^2$ effects
included in that table
are incomplete. There are various one-loop diagrams which are not considered but
carry chiral order $Q^2$ also. However, in a combined chiral and $1/N_c$ expansion it is
natural to neglect such loop effects, because they are suppressed by $1/N_c$.
This is immediate with the large-$N_c$ scaling rules $m_\pi \sim N_c^0$ and
$f \sim \sqrt{N_c}$ \cite{Hooft,Witten} together with the fact that the one-loop
effects are proportional to $m_{K,\pi}^2/(4\pi\,f^2) $ \cite{Bijnens:ga}. On the other hand,
it is evident from (\ref{3-point-vertex}) and (\ref{ansatz-3}) that the
$SU(3)$ symmetry-breaking contributions are not necessarily suppressed by $1/N_c$.
This approach differs from previous calculations
\cite{Bijnens:ga,Jenkins:ga1,Jenkins:ga2,Luty-2} where emphasis was put on the one-loop
corrections of the axial-vector current rather than the quasi-local counter terms which were
not considered. It is clear that part of the one-loop effects, in particular their
renormalization scale dependence, can be absorbed into the counter terms
considered in this work.

The large-$N_c$ counting issue will be taken up again when
discussing the approximate scattering kernel and also when presenting
the final set of parameters, obtained from a fit to the data set.

\cleardoublepage

\chapter{Covariant meson--baryon scattering}
\label{k3}
\markboth{\small CHAPTER \ref{k3}.~~~Covariant meson--baryon scattering}{}

In this chapter the basic concepts required for a relativistic
coupled-channel effective field theory of meson--baryon scattering are introduced.
The formalism is developed first for the case of elastic $\pi N$ scattering for
simplicity. The next chapter is then devoted to the inclusion of inelastic
channels which leads to the coupled-channel approach. The developments rely on an
'old' idea present in the literature for many decades. One aims at reducing the
complexity of the relativistic Bethe-Salpeter equation by a suitable reduction scheme
constrained to preserve the relativistic unitarity cuts \cite{Sugar,Gross,Tjon}. A famous
example is the Blankenbecler-Sugar scheme \cite{Sugar} which reduces the Bethe-Salpeter
equation to a 3-dimensional integral equation. For a beautiful variant developed for
pion-nucleon scattering, see the work by Gross and Surya \cite{Gross}. In this work
a scheme is derived which is more suitable for the relativistic chiral Lagrangian. It is not
attempted to establish a numerical solution of the four dimensional Bethe-Salpeter equation
based on phenomenological form factors and interaction kernels \cite{Afnan}.
The merit of the chiral Lagrangian is that a major part of the complexity is already
eliminated by having reduced non-local interactions to 'quasi' local interactions involving
a finite number of derivatives only. The scheme is therefore constructed to be
particularly transparent and efficient for the typical case of 'quasi' local
interaction terms. A further
important aspect to be addressed in this section concerns the independence of the
on-shell scattering amplitude on the choice of chiral coordinates or the choice of
interpolating fields. If one solves the Bethe-Salpeter equation with an interaction
kernel evaluated in perturbation theory the resulting on-shell scattering amplitude
depends on the choice of interpolating fields even though physical quantities should
be independent on that choice \cite{Lahiff}. It will be demonstrated that this problem
can be avoided by introducing an appropriate on-shell reduction scheme for the
Bethe-Salpeter equation. In the course of developing the approach a modified
subtraction scheme within dimensional regularization is proposed. The latter complies
manifestly with the chiral counting rule (\ref{q-rule}).

Consider the on-shell pion-nucleon scattering amplitude
\begin{eqnarray}
\langle \pi^{i}(\bar q)\,N(\bar p)|\,T\,| \pi^{j}(q)\,N(p) \rangle
&=&(2\pi)^4\,\delta^4(p+q-\bar p-\bar q)\,
\nonumber\\
&& \! \times\,\bar u(\bar p)\,
T^{ij}_{\pi N \rightarrow \pi N}(\bar q,\bar p ; q,p)\,u(p)
\label{on-shell-scattering}
\end{eqnarray}
where  $\delta^4(...)$ guarantees energy-momentum conservation and $u(p)$ is the
nucleon isospin-doublet spinor. In quantum field theory the scattering amplitude
$T_{\pi N \rightarrow \pi N}$ is obtained by solving the Bethe-Salpeter matrix
equation
\begin{eqnarray}
T(\bar k ,k ;w ) &=& K(\bar k ,k ;w )
+\int \frac{d^4l}{(2\pi)^4}\,K(\bar k , l;w )\, G(l;w)\,T(l,k;w )\;,
\nonumber\\
G(l;w)&=&-i\,S_N({\textstyle
{1\over 2}}\,w+l)\,D_\pi({\textstyle {1\over 2}}\,w-l)\,,
\label{BS-eq}
\end{eqnarray}
defined in terms of the Bethe-Salpeter scattering kernel $K(\bar k,k;w
)$, the nucleon propagator $S_N(p)=1/(\pslash-m_N+i\,\epsilon)$ and the
pion propagator $D_\pi(q)=1/(q^2-m_\pi^2+i\,\epsilon)$. Self energy corrections
in the propagators are suppressed and therefore not considered in this work.
In (\ref{BS-eq}) convenient kinematical variables are used:
\begin{eqnarray}
w = p+q = \bar p+\bar q\,,
\quad k= {\textstyle{1\over 2}}\,(p-q)\,,\quad
\bar k ={\textstyle{1\over 2}}\,(\bar p-\bar q)\,,
\label{def-moment}
\end{eqnarray}
where $q,\,p,\, \bar q$ and $\bar p$ are the initial and final pion and nucleon 4-momenta.
Here there is no need to specify the form of the scattering kernel $K$ since the
discussion is generic and does not depend on the particular form of the interaction.
The Bethe-Salpeter equation (\ref{BS-eq}) implements Lorentz invariance
and unitarity for the two-body scattering process. It involves the off-shell
continuation of the  on-shell scattering amplitude introduced in
(\ref{on-shell-scattering}). Only the on-shell limit
with $ \bar p^2, p^2\rightarrow m_N^2 $ and $\bar
q^2,q^2\rightarrow m_\pi^2 $ carries direct physical information.
In quantum field theory the off-shell form of the
scattering amplitude reflects the particular choice of the pion and
nucleon interpolating fields chosen in the Lagrangian density and
therefore can be altered basically at will by a redefinition
of the fields~\cite{off-shell,Fearing}.

It is convenient to decompose the interaction kernel and the resulting
scattering amplitude in isospin invariant components
\begin{eqnarray}
K_{ij}(\bar k ,k ;w ) =\sum_I\,K_I(\bar k ,k ;w )\,P^{(I)}_{ij}
\,,\,\,
T_{ij}(\bar k ,k ;w ) &=& \sum_I\,T_I(\bar k ,k ;w )\,P^{(I)}_{ij}
\label{isospin-decom}
\end{eqnarray}
with the isospin projection matrices $P^{(I)}_{ij}$.
For pion-nucleon scattering one has:
\begin{eqnarray}
P^{\left(\frac{1}{2}\right)}_{ij} &=& \frac{1}{3}\,\sigma_i\,\sigma_j \;,\;\;\;
P^{\left(\frac{3}{2}\right)}_{ij} = \delta_{ij}-\frac{1}{3}\,\sigma_i\,\sigma_j
\;,\;\;\;
\sum_k\,P^{\left(I\right)}_{ik}\,P^{\left(I'\right)}_{kj}=
\delta_{I\,I'}\,P^{\left(I\right)}_{ij}\;.
\label{iso:proj}
\end{eqnarray}
The ansatz (\ref{isospin-decom}) decouples the Bethe-Salpeter equation
into the two isospin channels $I=1/2$ and $I=3/2$.

\vskip1.5cm \section{On-shell reduction of the Bethe-Salpeter equation}

It is important to construct a systematic reduction scheme for the Beth-Salpeter
equation as to arrive at a formulation in which the on-shell scattering amplitudes are
independent on the choice of chiral coordinates or the choice of interpolating fields.
The interaction kernel is decomposed into an 'on-shell irreducible' part $\bar K $
and 'on-shell reducible' terms $K_L$ and $K_R$ \cite{nn-lutz} which vanish
if evaluated with on-shell kinematics either in the incoming or outgoing channel
respectively
\begin{eqnarray}
&&K=\bar K+K_L+K_R+K_{LR} \;,
\nonumber\\
&&\bar u_N(\bar p)\,K_L \Big|_{\mathrm{on-shell}} = 0 = K_R
\,u_N(p)\Big|_{\mathrm{on-shell}} \;.
\label{k-decomp}
\end{eqnarray}
The term $K_{LR}$ disappears if evaluated with either incoming or outgoing
on-shell kinematics. The notion of an on-shell irreducible kernel $\bar K$
is not unique per se and needs further specifications. The precise definition of a
particular choice of on-shell irreducibility will be provided  when constructing covariant
partial-wave projectors. In this subsection the generic consequences of decomposing
the interaction kernel according to (\ref{k-decomp}) are studied. With this decomposition
of the interaction kernel the scattering amplitude can be written as follows
\begin{eqnarray}
T &=& \bar T
-\Big(K_L+K_{LR}\Big)\!\cdot\! \Big( 1-G\!\cdot\! K\Big)^{-1}\!\cdot
\!G\! \cdot\!
\Big(K_R+K_{LR}\Big)-K_{LR}
\nonumber\\
&+&\Big(K_L+K_{LR}\Big)\! \cdot\! \Big( 1-G\!\cdot\! K\Big)^{-1}
+\Big( 1-K\!\cdot\! G\Big)^{-1}\!\cdot\!\Big(K_R+K_{LR} \Big)\,,
\nonumber\\
\bar T &=& \Big(1-V\!\cdot\! G \Big)^{-1} \!\cdot\! V \;,
\label{t-eff}
\end{eqnarray}
where we use operator notation with, e.g., $T=K+ K\!\cdot \!G\!\cdot \!T $ representing
the Bethe-Salpeter equation (\ref{BS-eq}). The effective interaction
$V$ in (\ref{t-eff}) is given by
\begin{eqnarray}
V&=&\Big( \bar K+K_R\!\cdot\! G\!\cdot\! X \Big)\!\cdot\!
\Big(1-G\!\cdot\! K_L-G\!\cdot\! K_{LR}\!\cdot\! G\!\cdot\! X  \Big)^{-1} \;,
\nonumber\\
X &=& \Big( 1-(K_R+K_{LR})\!\cdot\! G\Big)^{-1}\!
\cdot\! \Big( \bar K+K_L\Big)
\label{v-eff}
\end{eqnarray}
without any approximations. It should be emphasized that the interaction kernels $V$ and
$K$ are equivalent on-shell by construction. This follows from (\ref{t-eff}) and
(\ref{k-decomp}), which predict the equivalence of $T$ and $\bar T$ for on-shell kinematics
\begin{eqnarray}
\bar{u}_N(\bar p)\, T \,u_N(p)\Big|_{\mathrm{on-shell}}
 \equiv \bar{u}_N(\bar
 p)\,\bar T \,u_N(p)\Big|_{\mathrm{on-shell}} .
\nonumber
\end{eqnarray}

To what extent does the formalism (\ref{k-decomp}-\ref{v-eff}) help in
constructing an approximation scheme to the Bethe-Salpeter equation which does not
show an unwanted dependence on the choice of fields? In order to streamline the
following discussion it is written
\begin{eqnarray}
O= \sum\, O_n \;, \qquad  {\rm for}\;\quad  O=K,V,T, ... \;,
\label{}
\end{eqnarray}
where the moments $O_n$ are defined with respect to some perturbative expansion.
At leading order, say $n=1$, it is evident that given a particular notion of on-shell
irreducibility the moment $\bar K_1 =\bar T_1$ does not depend on the choice of fields.
The conclusion relies on the equivalence theorem which implies that on-shell
quantities do not depend on the choice of fields if evaluated systematically at given
order in perturbation theory. The $\bar O$-notation is used for any object,
$O=K,V,T,...$, as introduced in (\ref{k-decomp}). If one
truncates $V\to V_1=\bar K_1$ the resulting on-shell equivalent
amplitude $\hat T = (1- V_1 \!\cdot\!G)^{-1}\!\cdot\!V_1 $ is invariant under field
redefinitions provided that the two-particle propagator $G$ is invariant
also\footnote{This assumption can be justified upon an appropriate renormalization of
the two-particle propagator $G$ (see (\ref{ren-v})). One may simply identify $G_R$
with the bare two-particle propagator written in terms of physical mass parameters.}.
The extension of this argument to subleading orders is subtle. For instance there is
no reason to expect the invariance of
\begin{eqnarray}
\hat T = (1- \sum_{i=1}^n\,V_i
\!\cdot\!G)^{-1}\!\cdot\!\sum_{j=1}^n\,V_j \;,
\label{}
\end{eqnarray}
at order $n$. Moreover it is evident that subleading moments of $\bar K_n$ may
depend on the choice of fields simply because
they are not observable quantities. On the other hand, given any notion of on-shell
irreducibility the moments $\bar T_n $ are invariant as a direct consequence of the
equivalence theorem. It is then straightforward to identify the uniquely induced
effective interaction kernel $V_{\rm eff}$,
\begin{eqnarray}
V_{\rm eff} =  \Big( 1+\bar T\!\cdot\!G \Big)^{-1}\!\cdot\!\bar T \,, \qquad \bar T
= \overline{\left( 1-K\!\cdot\!G \right)^{-1}\!\cdot\!K} = \left( 1-V_{\rm eff
}\!\cdot\!G \right)^{-1}\!\cdot\!V_{\rm eff} \,,\label{def-gen-Veff}
\end{eqnarray}
which is invariant under field redefinitions by construction. The moments $V_{{\rm
eff},n}$ follow by expanding the result (\ref{def-gen-Veff}). That may be inconvenient
for a given notion of on-shell irreducibility but is in principle straight forward. To
conclude it is possible to construct the invariant effective interaction $V_{\rm eff}$
once one agrees on a particular notion of on-shell irreducibility. Of course there are
infinite many ways to do so, but this is welcome since we may exploit that freedom as
to simplify the process of solving the Bethe-Salpeter equation.

In this work the guide will be the question how the result
(\ref{def-gen-Veff}) relates to the more explicit and convenient
formalism (\ref{k-decomp}-\ref{v-eff}). In general there
appears no transparent relation of $V_{\rm eff}$ and $V$. Only
for a particular choice of on-shell irreducibility there is a
simple relation. This difficulty reflects a subtle property of the
effective invariant interaction, namely $V_{\rm eff} \neq \bar
V_{\rm eff}$. Only if one succeeds in introducing a specific form
of on-shell irreducibility that is invariant under unitary
iteration,
\begin{eqnarray}
\bar K \!\cdot\!G\!\cdot\!\bar K = \overline{\bar K \!\cdot\!G \!\cdot\!\bar K} \,,
\label{def-OSR}
\end{eqnarray}
for any interaction kernel $K$, one obtains $V_{\rm eff}=\bar V_{\rm eff}$. Given
(\ref{def-OSR}) the effective interaction kernel $V_{\rm eff}[K]$ can now be
constructed in perturbation theory applying the convenient formalism
(\ref{k-decomp}-\ref{v-eff}). Here $V[K]$ as given in (\ref{v-eff})
is considered as a functional in $K$. Within this notation one obtains
\begin{eqnarray}
\Big(V_{\rm eff}[K] \Big)_{n+1} = \Big(\overline{V^{(n)}[K]} \Big)_{n+1}\,,\quad
V^{(n+1)} [K] = V [V^{(n)}[K]]\;,\quad V^{(1)}[K] = V[K] \,. \label{def-barV}
\end{eqnarray}
To summarize the crucial property of the on-shell reduction scheme
(\ref{k-decomp}-\ref{v-eff},\ref{def-barV}). Given that scheme using
different chiral coordinates or fields does lead to the same on-shell scattering
amplitude $\bar T$ provided that the effective interaction $ V_{\rm eff}$ is evaluated
systematically to a given order in perturbation theory. It is left
to establish a notion of on-shell irreducibility that is consistent with
(\ref{def-OSR}). This will be accomplished when constructing covariant
partial-wave projectors.

As an explicit simple example for the application of the formalism
(\ref{t-eff},\ref{v-eff}) consider the s-channel nucleon pole
diagram as a particular contribution to the interaction
kernel $K(\bar k,k;w)$ in (\ref{BS-eq}). In the isospin $1/2$ channel
its contribution  evaluated with the pseudo-vector pion-nucleon vertex
reads
\begin{eqnarray}
K(\bar k, k;w ) &=& -\frac{3\,g_A^2}{4\,f^2}\,
\gamma_5\,\left( {\textstyle{1\over 2}}\,\wslash-\barkslash  \right)\,
\frac{1}{\wslash-m_N-\Delta m_N(w)}\,\gamma_5\,
\left( {\textstyle{1\over 2}}\,\wslash- \kslash \right)\;,
\label{nucleon-pole}
\end{eqnarray}
where  a wave-function and mass renormalization
$\Delta m_N(w)$ is included for later convenience\footnote{The corresponding counter terms
in the chiral Lagrangian (\ref{chi-sb}) are linear combinations of $b_0,b_D$ and $b_F$
and $\zeta_0$, $\zeta_D$ and $\zeta_F$.}. The various components of the kernel
according to (\ref{k-decomp}) are derived readily
\begin{eqnarray}
\bar K&=&-\frac{3\,g_A^2}{4\,f^2}\,
\frac{\big( \wslash-m_N \big)^2}{\wslash+\bar m_N}\,,\quad
K_{LR}=\frac{3\,g_A^2}{4\,f^2}\,
\Big( {\barpslash}-m_N \Big)\,
\frac{1}{\wslash+\bar m_N}\,\Big( \pslash- m_N\Big)\;,
\nonumber\\
K_L&=&\frac{3\,g_A^2}{4\,f^2}\,
\Big(\barpslash-m_N \Big)\,
\frac{\wslash- m_N}{\wslash+\bar m_N}\,,\quad
K_R=\frac{3\,g_A^2}{4\,f^2}\,
\frac{\wslash-m_N}{\wslash+\bar m_N}\,\Big( \pslash- m_N\Big)
\,,
\label{nucleon-pole-off}
\end{eqnarray}
where $\bar m_N = m_N+\gamma_5\,\Delta m_N(w)\,\gamma_5$.
The solution of the Bethe-Salpeter equation is derived in two steps. First
solve for the auxiliary object $X$ in (\ref{v-eff})
\begin{eqnarray}
X &=& \frac{3\,g_A^2}{4\,f^2}\,
\Big( \barpslash-\wslash\Big)\,
\frac{1}{\wslash+\bar m_N}\,\Big( \wslash- m_N\Big)
\nonumber\\
&+&\frac{3\,g_A^2}{4\,f^2}\,
\Big( \barpslash+\wslash-2\,m_N\Big)\,
\frac{3\,g_A^2\,I_\pi^{(l)}}
{4\,f^2\,(\wslash+\bar m_N)+3\,g_A^2\,I_\pi}\,
\frac{ \wslash- m_N}{\wslash+\bar m_N} \;,
\label{x-example}
\end{eqnarray}
where one encounters the pionic tadpole integrals:
\begin{eqnarray}
I_\pi &=&i\,
\int  \frac{d^d\,l}{(2\,\pi)^d }
\frac{\mu^{4-d}}{l^2-m_\pi^2+i\,\epsilon }
\,,\quad
I^{(l)}_\pi =i\,
\int  \frac{d^d\,l}{(2\,\pi)^d }
\frac{\mu^{4-d}\,\lslash }{l^2-m_\pi^2+i\,\epsilon }\,,
\label{def-tadpole}
\end{eqnarray}
properly regularized for space-time dimension $d$ in terms of the renormalization scale $\mu $.
Since $I_\pi^{(l)}=0$ and $K_R\cdot  G\cdot
\!(\bar K+K_L)\equiv 0$ for our example the expression (\ref{x-example}) reduces to
$X=\bar K+K_L $. The effective potential $V(w)$
and the on-shell equivalent scattering amplitude $\bar T $ follow
\begin{eqnarray}
V(w)&=& -\frac{3\,g_A^2}{4\,f^2}\,\frac{ (\wslash- m_N)^2}{\wslash+\bar m_N}
\left(1+
\frac{3\,g_A^2\,I_\pi}{4\,f^2}\,\frac{\wslash- m_N}{\wslash+\bar m_N}
\right)^{-1} \;,
\nonumber\\
\bar T(w) &=& \frac{1}{1-V(w)\,J_{\pi N}(w)}\,V(w) \;.
\label{pin-example}
\end{eqnarray}
The divergent loop function $J_{\pi N}$ in (\ref{pin-example}) defined via
$\bar{K}\!\cdot\!G\!\cdot\!\bar{K}= \bar K\, J_{\pi N}\,\bar K$ may be decomposed
into scalar master-loop functions $I_{\pi N}(\sqrt{s}\,)$ and $I_{N},I_\pi$ with
\begin{eqnarray}
J_{\pi N}(w)&=& \left(m_N +
\frac{w^2+m_N^2-m_\pi^2}{2\,w^2}\,
\wslash\right)\,I_{\pi N}(\sqrt{s}\,)
+\frac{I_N-I_{\pi}}{2\,w^2}\,\wslash \;,
\nonumber\\
I_{\pi N}(\sqrt{s}\,)&=&-i\,\int
\frac{d^dl}{(2\pi)^d}\,\frac{\mu^{4-d}}{l^2-m_\pi^2}\,\frac{1}{(w-l)^2-m_N^2}
\;,
\nonumber\\
I_N &=&i\,
\int  \frac{d^d\,l}{(2\,\pi)^d }
\frac{\mu^{4-d}}{l^2-m_N^2+i\,\epsilon }
\label{jpin-def}
\end{eqnarray}
where $s=w^2$. The result (\ref{pin-example}) gives an explicit example of the powerful
formula (\ref{t-eff}). The Bethe-Salpeter equation may be solved in two steps. Once the
effective potential $V$ is evaluated the scattering amplitude $\bar T$ is given in
terms of the loop function $J_{\pi N}$ which is independent on the form of the interaction.
In chapter 3.3 the result (\ref{pin-example}) will be generalized by constructing a complete
set of covariant projectors. The latter will define a particular notion of on-shell
irreducibility. Before discussing the result (\ref{pin-example}) in more detail an appropriate
regularization and renormalization scheme for the relativistic loop
functions in (\ref{jpin-def}) is developed.

\vskip1.5cm \section{Renormalization program}

An important requisite of the chiral Lagrangian is a consistent regularization and renormalization
scheme for its loop diagrams. The regularization scheme should respect all symmetries built
into the theory but should also comply with the power counting rule (\ref{q-rule}).
The standard $MS$ or $\overline{MS}$ subtraction scheme of dimensional regularization appears inconvenient, because
it contradicts standard chiral power counting rules if applied to relativistic Feynman
diagrams \cite{gss,kambor}. A modified subtraction scheme for relativistic diagrams,
properly regularized in space-time dimensions $d$, which complies with the chiral counting rule
(\ref{q-rule}) manifestly, will be suggested.

The discussion centers around the one-loop functions $I_\pi$, $I_N$ and
$I_{\pi N}(\sqrt{s}\,)$ introduced in (\ref{jpin-def}). One encounters some freedom how to
regularize and renormalize those master-loop functions, which are typical
representatives for all one-loop diagrams \cite{nn-lutz}. Recall their well-known properties
at $d=4$. The loop function $I_{\pi N}(\sqrt{s}\,)$ is rendered finite by one
subtraction, for example at $\sqrt{s}=0$,
\begin{eqnarray}
\!\!\!\!I_{\pi N}(\sqrt{s}\,)&=&\frac{1}{16\,\pi^2}
\left( \frac{p_{\pi N}}{\sqrt{s}}\,
\left( \ln \left(1-\frac{s-2\,p_{\pi N}\,\sqrt{s}}{m_\pi^2+m_N^2} \right)
-\ln \left(1-\frac{s+2\,p_{\pi N}\sqrt{s}}{m_\pi^2+m_N^2} \right)\right)
\right.
\nonumber\\
&+&\left.
\left(\frac{1}{2}\,\frac{m_\pi^2+m_N^2}{m_\pi^2-m_N^2}
-\frac{m_\pi^2-m_N^2}{2\,s}
\right)
\,\ln \left( \frac{m_\pi^2}{m_N^2}\right) +1 \right)+I_{\pi N}(0)\;,
\nonumber\\
p_{\pi N}^2 &=&
\frac{s}{4}-\frac{m_N^2+m_\pi^2}{2}+\frac{(m_N^2-m_\pi^2)^2}{4\,s}  \;.
\label{ipin-analytic}
\end{eqnarray}
One finds $I_{\pi N}(m_N)-I_{\pi N}(0)= (4\pi)^{-2}+{\mathcal O}(m_\pi/m_N)\sim Q^0$
in conflict with the expected minimal chiral power $Q$.
On the other hand the leading chiral power of the subtracted loop function
complies with the prediction of the standard chiral power counting rule (\ref{q-rule}) with
$I_{\pi N}(\sqrt{s}\,)-I_{\pi N}(\mu_S ) \sim Q$ rather than the anomalous
power $Q^0$ provided that $\mu_S \sim m_N$ holds. The anomalous contribution
is eaten up by the subtraction constant $I_{\pi N}(\mu_S)$. This can be seen by expanding
the loop function
\begin{eqnarray}
I_{\pi N}(\sqrt{s}\,)&=&i\,\frac{\sqrt{\phi_{\pi N}}}{8\,\pi\,m_N}
+\frac{\sqrt{\phi_{\pi N} }}{16\,\pi^2\,m_N}\,
\ln \left( \frac{\sqrt{s}-m_N+\sqrt{\phi_{\pi N} }}
{\sqrt{s}-m_N-\sqrt{\phi_{\pi N} }} \right)
\nonumber\\
&+&\frac{\sqrt{m^2_\pi-\mu_N^2}}{8\,\pi\,m_N}
-\frac{\sqrt{\mu_N^2-m_\pi^2}}{16\,\pi^2\,m_N}\,
\ln \left( \frac{\mu_N+\sqrt{\mu_N^2-m_\pi^2}}
{\mu_N-\sqrt{\mu_N^2-m_\pi^2}} \right)
\nonumber\\
&-&\frac{\sqrt{s}-\mu_S}{16\,\pi^2\,m_N}\,\ln \left( \frac{m_\pi^2}{m_N^2}\right)
+{\mathcal O}\left( \frac{(\mu_S-m_N)^2}{m_N^2},Q^2 \right)+I_{\pi N}(\mu_S)\;,
\label{ipin-expand}
\end{eqnarray}
in powers of $\sqrt{s}-m_N\sim Q$ and $\mu_N=\mu_S -m_N $. Here
$\phi_{\pi N}= (\sqrt{s}-m_N)^2-m_\pi^2$ is an approximate phase-space factor.

From the simple example of $I_{\pi N}(\sqrt{s}\,)$ in (\ref{ipin-analytic}) it becomes clear
that a manifest realization of the chiral counting rule (\ref{q-rule}) is closely linked
to the subtraction scheme implicit in any regularization scheme. A priori it is unclear
in which way the subtraction constants of various loop functions are related by the
pertinent symmetries\footnote{This aspect was not addressed satisfactorily
in \cite{nn-lutz,Gegelia}.}. Here dimensional regularization has proven to be an extremely
powerful tool to regularize and to subtract loop functions in accordance with all
symmetries. Therefore, it is useful to recall the expressions for
the master-loop function $I_N$, and $I_\pi$ and $I_{\pi N}(m_N)$ as they follow in
dimensional regularization:
\begin{eqnarray}
&& I_N = m_N^2\,\frac{\Gamma (1-d/2)}{(4\pi)^2}
\left(\frac{m_N^2}{4\,\pi \,\mu^2} \right)^{(d-4)/2}
\nonumber\\
&& \quad \;\,= \frac{m_N^2}{(4\,\pi)^2}
\left( -\frac{2}{4-d}+\gamma-1 -\ln (4 \pi)+\ln \left( \frac{m_N^2}{\mu^2}\right)
+ {\mathcal O}\left(4-d \right)\right) \;,
\label{n-tadpole}
\end{eqnarray}
where $d$ is the dimension of space-time and $\gamma $ the Euler constant. The analogous
result for the pionic tadpole follows by replacing the nucleon mass $m_N$ in (\ref{n-tadpole})
by the pion mass $m_\pi$. The merit of dimensional regularization is that one is free
to subtract all poles at $d=4$ including any specified finite term without violating any of the
pertinent symmetries. In the $\overline{MS }$-scheme the pole $1/(4-d)$ is subtracted
including the finite part $\gamma- \ln (4 \pi)$. This leads to
\begin{eqnarray}
&&I_{N,\overline{MS}}= \frac{m_N^2}{(4 \pi)^2}
\left( -1+\ln \left(\frac{m_N^2}{\mu^2}\right) \right)\,,\quad  \;
I_{\pi,\overline{MS}} =\frac{m_\pi^2}{(4\pi)^2}
\left(-1+\ln \left(\frac{m_\pi^2}{\mu^2}\right) \right)\,,
\nonumber\\
&& I_{\pi N,\overline{MS}}\,(m_N)= -\frac{1}{(4\pi)^2} \,
\ln \left(\frac{m_N^2}{\mu^2} \right)
-\frac{m_\pi}{16\,\pi \,m_N}+
{\mathcal O}\left( \frac{m_\pi^2}{m_N^2}\right) \,.
\label{master-dim}
\end{eqnarray}
The result (\ref{master-dim}) confirms the expected chiral power for the pionic
tadpole $I_\pi \sim Q^2$. However, a striking disagreement with the chiral counting rule
(\ref{q-rule}) is found for the $\overline {MS}$-subtracted loop functions
$I_{\pi N}\sim Q^0$ and $I_N \sim Q^0$.
Recall that for the subtracted loop function $I_{\pi N}(\sqrt{s}\,)-I_{\pi N}(m_N)\sim Q $
the expected minimal chiral power is manifest (see (\ref{ipin-expand})).
It is instructive to trace the source of the anomalous chiral powers. By means of the
identities
\begin{eqnarray}
&& I_{\pi N}(m_N) = \frac{I_\pi-I_N}{m_N^2-m_\pi^2} + I_{\pi N}(m_N)-I_{\pi N}(0)  \;,
\nonumber\\
&&  I_{\pi N}(m_N)-I_{\pi N}(0)=
\frac{1}{(4\pi)^2}-\frac{m_\pi}{16\,\pi\,m_N}  \left(1-\frac{m_\pi^2}{8\,m_N^2} \right)
\nonumber\\
&& \qquad \qquad \qquad +\frac{1}{(4 \pi)^2}\left( 1
-\frac{3}{2}\,\ln \left( \frac{m_\pi^2}{m_N^2}\right) \right)
\frac{m_\pi^2}{m_N^2}
+{\mathcal O}\left( \frac{m_\pi^4}{m_N^4}, d-4 \right) \,,
\label{ipin-dim}
\end{eqnarray}
it appears that once the subtraction scheme has been specified for the tadpole terms
$I_\pi$ and $I_N$ the required subtractions for the remaining master-loop functions are
unique. In (\ref{ipin-dim}) the algebraic consistency identity is applied\footnote{Note
that (\ref{algebra-1}) leads to a well-behaved loop function
$J_{\pi N}(w)$ at $w=0$ (see (\ref{jpin-def})).}:
\begin{eqnarray}
I_\pi-I_N= \Big(m_N^2-m_\pi^2\Big)\,I_{\pi N}(0) \,,
\label{algebra-1}
\end{eqnarray}
which holds for any value of the space-time dimension d, and expanded the finite expression
$I_{\pi N}(m_N)-I_{\pi N}(0)$ in powers of $m_\pi/m_N$ at $d=4$. The result (\ref{ipin-dim})
seems to show that one either violates the desired chiral power for the
nucleonic tadpole, $I_N$, or the one for the loop function $I_{\pi N}(m_N)$.
One may for example subtract the pole at $d=4$ including the finite constant
$$
\gamma-1 -\ln (4 \pi)+\ln \left( \frac{m_N^2}{\mu^2}\right) \,.
$$
That leads to a vanishing
nucleonic tadpole $I_N \to 0$, which would be consistent with the expectation $I_N\sim Q^3$,
but we find $I_{\pi N}(m_N) \to 1/(4\pi)^2 +{\mathcal O}\left(m_\pi \right)$, in disagreement
with the expectation $I_{\pi N}(m_N)\sim Q$. This problem can be solved if one succeeds in
defining a subtraction scheme which acts differently on $I_{\pi N}(0)$ and $I_{\pi N}(m_N)$.
It is stressed that this is legitimate, because $I_{\pi N}(0)$  probes our effective theory
outside its applicability domain. Mathematically this can be  achieved most economically and
consistently by subtracting a pole in $I_{\pi N}(\sqrt{s}\,)$ at $d=3$ which arises in
the limit $m_\pi /m_N \to 0$.  To be explicit recall the expression for
$I_{\pi N}(\sqrt{s}\,)$ at arbitrary space-time dimension $d$ (see eg. \cite{Becher}):
\begin{eqnarray}
&& I_{\pi N}(\sqrt{s}\,) = \left(\frac{m_N}{\mu}\right)^{d-4}
\frac{\Gamma(2-d/2)}{(4\pi)^{d/2}}\,\int_0^1 \,d z \; C^{d/2-2} \;,
\nonumber\\
&& \qquad  C= z^2-
\frac{s-m_N^2-m_\pi^2}{m^2_N}\,z \,(1-z)+\frac{m_\pi^2}{m_N^2}\,(1-z)^2-i\,\epsilon  \,.
\label{explicit-rep}
\end{eqnarray}
The explicit representation (\ref{explicit-rep}) demonstrates that at $d=3$ the loop
function $I_{\pi N}(\sqrt{s}\,)$ is finite at $\sqrt{s}=0$
but infinite at $\sqrt{s}=m_N$ if one applies the limit $m_\pi /m_N \to 0$. One finds:
\begin{eqnarray}
&& I_{\pi N}(m_N\,) = \frac{1}{8\,\pi}\,\frac{\mu}{m_N}\,\frac{1}{d-3}
-\frac{1}{16\,\pi}\,\frac{\mu}{m_N}
\left(\ln (4\,\pi)+ \ln \left(\frac{\mu^2}{m_N^2}\right)+\frac{\Gamma'(1/2)}{\sqrt{\pi}}\right)
\nonumber\\
&&\qquad \qquad \,+\,{\mathcal O}\left(  \frac{m_\pi}{m_N},d-3\right) \;.
\label{}
\end{eqnarray}
A minimal chiral subtraction scheme is now introduced. It may be viewed as
a simplified variant of the scheme of Becher and Leutwyler in \cite{Becher}. As usual one
first needs to evaluate the contributions to an observable quantity at arbitrary space-time
dimension $d$. The result shows poles at $d=4$ and $d=3$ if considered in the non-relativistic
limit with $m_\pi/m_N \to 0$. Our modified subtraction scheme is  defined by the
replacement rules:
\begin{eqnarray}
&& \frac{1}{d-3} \to -\frac{m_N}{2\pi\,\mu} \;, \qquad
\frac{2}{d-4} \to  \gamma-1 -\ln (4 \pi)+\ln \left( \frac{m_N^2}{\mu^2}\right) \,,
\label{def-sub}
\end{eqnarray}
where it is understood that poles at $d=3$ are isolated in the non-relativistic
limit with $m_\pi/m_N \to 0$. The $d=4$ poles are isolated with the ratio $m_\pi/m_N$ at its
physical value. The limit $d\to 4$ is applied after the pole terms are replaced according
to (\ref{def-sub}). It is to be emphasized that there are no infrared singularities in the residuum
of the $1/(d-3)$-pole terms. In particular we observe that the anomalous subtraction implied in
(\ref{def-sub}) does not lead to a potentially troublesome pion-mass dependence of the
counter terms\footnote{In the scheme of
Becher and Leutwyler  \cite{Becher} a pion-mass dependent subtraction scheme for the scalar
one-loop functions is suggested. To be specific the master-loop function $I_{\pi N}(\sqrt{s})$
is subtracted by a regular polynomial in $s, m_N$ and $m_\pi$ of infinite order. In our scheme
we subtract only a constant which agrees with the non-relativistic limit
of the suggested polynomial of Becher and Leutwyler. Our subtraction constant does not
depend on the pion mass.}. To make contact with a non-relativistic scheme one needs to
expand the loop function in powers of $p/m_N $ where $p$ represents any external three
momentum.

The results for the loop functions $I_N$, $I_\pi$ and $I_{\pi N}(m_N)$
as implied by the subtraction prescription (\ref{def-sub}) are:
\begin{eqnarray}
&& \bar I_N = 0 \,, \qquad
\bar I_\pi = \frac{m_\pi^2}{(4\pi)^2}\,\ln \left( \frac{m_\pi^2}{m_N^2}\right)\;,
\nonumber\\
&& \bar I_{\pi N}(m_N) = -\frac{m_\pi}{16\,\pi\,m_N}  \left(1-\frac{m_\pi^2}{8\,m_N^2} \right)
\nonumber\\
&& \qquad \qquad \; +\frac{1}{(4 \pi)^2}\left( 1
-\frac{1}{2}\,\ln \left( \frac{m_\pi^2}{m_N^2}\right) \right)
\frac{m_\pi^2}{m_N^2}
+{\mathcal O}\left( \frac{m_\pi^4}{m_N^4} \right) \,,
\label{result-bar}
\end{eqnarray}
where the '$\bar{\phantom{X}}$' signals a  subtracted loop functions\footnote{
The renormalized loop function $J_{\pi N}(w)$ is no longer well-behaved at $w=0$. This was
expected and does not cause any harm, because the point $\sqrt{s}=0$ is far outside the
applicability domain of our effective field theory.}. The result shows
that now the loop functions behave according to their expected minimal chiral power (\ref{q-rule}).
Note that the one-loop expressions (\ref{result-bar}) do no longer depend on the
renormalization scale $\mu$ introduced in dimensional regularization\footnote{One
may make contact with the non-relativistic so-called PDS-scheme of \cite{KSW} by
slightly modifying the replacement rule for the pole at $d=3$. With
$$\frac{1}{d-3} \to 1-\frac{m_N}{2\pi\,\mu} \;, \qquad
I_{\pi N}(m_N) \to  \frac{1}{16\,\pi}\,\frac{2\,\mu-m_\pi}{m_N}
+{\mathcal O}\left( \frac{m_\pi^2}{m_N^2}\right)\,,$$
power counting is manifest if one counts $\mu \sim Q$. This is completely
analogous to the PDS-scheme}. This should not be too surprising, because the renormalization
prescription (\ref{def-sub}) has a non-trivial effect on the counter terms of the chiral
Lagrangian.  Note that the prescription (\ref{def-sub}) defines also a unique subtraction
for higher loop functions in the same way the $\overline {MS}$-scheme does. The
renormalization scale dependence will be explicit at the two-loop level reflecting the
presence of so-called overall divergencies.

It was checked that all scalar one-loop functions, subtracted according to (\ref{def-sub}),
comply with their expected minimal chiral power (\ref{q-rule}). Typically, loop functions
which are finite at $d=4$ are not affected by the subtraction prescription (\ref{def-sub}).
Also, loop functions involving pion propagators only do not show singularities at $d=3$ and
therefore can be  related easily to the corresponding loop functions of the
$\overline{MS}$-scheme. Though, in the nucleon sector it may be tedious to relate the standard
$\overline{MS}$-scheme to the scheme introduced here, in particular
when multi-loop diagrams are considered, it is asserted that  a well defined
prescription for regularizing all divergent loop integrals is proposed. A prescription
which is far more convenient than the $\overline {MS}$-scheme, because it complies manifestly
with the chiral counting rule (\ref{q-rule})\footnote{It is unclear how to
set up a prescription in the presence of two massive fields with respective masses
$m_1$ and $m_2$ and
$m_1 \gg m_2$. For the decuplet baryons one may count $m_{[8]}-m_{[10]} \sim Q $ as suggested
by large-$N_c$ counting arguments and therefore start with a common mass for the baryon
octet and decuplet states. The presence of the $B_\mu^*$ field in the chiral Lagrangian does
cause a problem. Since we include the $B_\mu^*$ field only at tree-level in the interaction
kernel of pion-nucleon and kaon-nucleon scattering, we do not further investigate the possible
problems in this work. Formally one may avoid such problems all together if one counts
$m_{[8]}-m_{[9]} \sim Q$ even though this assignment may not be effective.}.

If one were to perform calculations within standard chiral perturbation theory in terms of the
relativistic chiral Lagrangian everything is all set for any computation. However,
as advocated before a non-perturbative chiral theory is asked for. That requires a
more elaborate renormalization program, because it is necessary to  discriminate carefully
reducible and irreducible diagrams and sum the reducible
diagrams to infinite order. The idea is to take over the renormalization program of standard chiral
perturbatioon theory to the interaction kernel. In order to apply the standard perturbative
renormalization program for the interaction kernel one has to move all divergent parts lying
in reducible diagrams into the interaction kernel. That problem is solved in part by constructing
an on-shell equivalent interaction kernel according to (\ref{k-decomp}-\ref{v-eff}).
It is evident that the 'moving' of divergencies needs to be controlled  by an additional
renormalization condition. Any such condition imposed should be constructed so as to respect
crossing symmetry approximatively. While standard chiral perturbation theory leads directly to
cross symmetric amplitudes at least approximatively, it is not automatically so in a resummation
scheme.

Before introducing the general scheme the above issues are examined explicitly in terms of
the example worked out in detail in the previous section. Taking the s-channel nucleon pole
term as the driving term in the Bethe-Salpeter equation the explicit result
(\ref{pin-example}) was derived. First its implicit nucleon mass renormalization is discussed.
The result (\ref{pin-example}) shows a pole at the physical nucleon mass with $\wslash =-m_N$,
only if the mass-counter term $\Delta m_N$ in (\ref{nucleon-pole}) is identified as follows
\begin{eqnarray}
\Delta m_N =  \frac{3\,g_A^2}{4\,f^2}\,2\,m_N\,\Big (m_\pi^2\,I_{\pi N}(m_N)-I_N \Big)\,.
\label{mass-ren}
\end{eqnarray}
In the $\overline{MS}$-scheme the divergent parts of $\Delta  m_N$, or more precisely
the renormalization scale dependent parts, may be absorbed into the nucleon bare mass
\cite{gss}. Renormalization within the chiral minimal subtraction scheme
is achieved by the replacements
$I_{\pi N}\to \bar I_{\pi N} $, $I_\pi \to \bar I_\pi$  and
$I_N \to \bar I_N= 0$. The result (\ref{mass-ren}) together
with (\ref{result-bar}) then reproduces the well known result \cite{gss,kambor}
$$
\Delta m_N = -\frac{3\,g_A^2\,m_\pi^3}{32\,\pi\,f^2}+\cdots \;,
$$
commonly derived in terms of the one-loop nucleon self
energy $\Sigma_N(p)$. In order to offer a more direct comparison of (\ref{pin-example})
with the  nucleon self-energy $\Sigma_N(p)$ recall the one-loop result
\begin{eqnarray}
&&\!\!\!\!\!\! \Sigma_N (p) = \frac{3\,g_A^2}{4\,f^2}\, \Bigg(
 m_N \,\Big( m_\pi^2\,I_{\pi N}(\sqrt{p^2}\,) -I_N\Big)
-\pslash \,\Bigg(\frac{p^2-m_N^2}{2\,p^2} \, I_\pi
+\frac{p^2+m_N^2}{2\,p^2} \,I_N
\nonumber\\
&& \qquad \qquad \qquad
+\left( \frac{(p^2-m_N^2)^2}{2\,p^2}- m_\pi^2\,\frac{p^2+m_N^2}{2\,p^2}
\right) I_{\pi N}(\sqrt{p^2}\,)
\Bigg) \Bigg)\;,
\label{n-self}
\end{eqnarray}
in terms of the convenient master-loop function $I_{\pi N}(\sqrt{p^2})$ and the
tadpole terms $I_N$ and $I_\pi$. It is emphasized that the
expression (\ref{n-self}) is valid for arbitrary space-time dimension $d$. The wave-function
renormalization, ${\mathcal Z}_N$, of the nucleon can be read off (\ref{n-self}) leading  to
\begin{eqnarray}
{\mathcal Z}^{-1}_N &=& 1-\frac{\partial \,\Sigma_N }{\partial \,\pslash}\Bigg|_{\pslash \,=m_N}
\!\!= 1 -\frac{3\,g_A^2}{4\,f^2}
\left( 2\,m_\pi^2\,m_N \,\frac{\partial \,I_{\pi N}(\sqrt{p^2})}{\partial \,\sqrt{p^2}}
 \Bigg|_{p^2=m_N^2}
\!\!-I_\pi
\right)
\nonumber\\
&=& 1- \frac{3\,g_A^2}{4\,f^2}\,\frac{m_\pi^2 }{(4\pi)^2}
\left( -4 -3\,\ln \left(\frac{m_\pi^2}{m_N^2}\right)
+3\,\pi\,\frac{m_\pi}{m_N}
+{\mathcal O}\left( \frac{m^2_\pi}{m^2_N} \right)
\right) \,,
\label{def-zn}
\end{eqnarray}
where in the last line of (\ref{def-zn}) relies on the minimal chiral
subtraction scheme (\ref{def-sub}).
The result (\ref{def-zn}) agrees with the expressions obtained previously in
\cite{gss,kambor,Becher}. Note that in (\ref{def-zn}) the contribution of
the counter terms $\zeta_0, \zeta_D$ and $\zeta_F$ are suppressed.

It is illuminating to discuss the role played by the pionic tadpole
contribution, $I_\pi$, from $V(w)$ in (\ref{pin-example}) and from $J_{\pi N}(w)$
in (\ref{jpin-def}). In the mass renormalization (\ref{mass-ren}) both tadpole contributions
cancel identically. If one dropped the pionic tadpole term $I_\pi$ in the effective
interaction $V(w)$ one would find a mass renormalization
\begin{eqnarray}
\Delta m_N \sim m_N\,I_\pi/f^2 \sim Q^2 \;,
\label{}
\end{eqnarray}
in conflict with the expected minimal chiral power $\Delta m_N \sim Q^3$ (see (\ref{mass-ren})).
Therefore it is mandatory to 'move' all tadpole contributions from the loop function $J_{\pi N}(w)$
to the effective interaction kernel $V(w)$ if the latter is evaluated
in perturbation theory. This is readily achieved by splitting the meson-baryon propagator
\begin{eqnarray}
G={\mathcal Z}_N\,(G_R+\Delta G) \;,
\label{}
\end{eqnarray}
into two terms $G_R$ and $\Delta G$. The form of
$\Delta G$ is chosen such that the renormalized loop function implied by $G_R$ is free of
reduced tadpole contributions. This procedure renormalizes the effective
interaction kernel $V(w)\to V_R(w)$ as follows
\begin{eqnarray}
\bar T_R= \frac{{\mathcal Z}_N}{1-V\cdot G}\cdot V=\frac{1}{1-V_R\cdot G_R}\cdot V_R \;,
\quad
V_R = \frac{{\mathcal Z}_N}{1-{\mathcal Z}_N\,V\cdot\Delta G}\cdot V \;,
\label{ren-v}
\end{eqnarray}
where the renormalized scattering amplitude
\begin{eqnarray}
\bar T_R = {\mathcal Z}^{\frac{1}{2}}_N \,\bar T\, {\mathcal Z}^{\frac{1}{2}}_N \;,
\label{}
\end{eqnarray}
as implied by the LSZ reduction scheme is introduced. Now the cancellation of the pionic
tadpole contributions is easily implemented by applying
the chiral expansion to $V_R(w)$. The renormalized interaction kernel $V_R(w)$
is real by construction.

The renormalization scheme is still incomplete \footnote{Here
the most general case not necessarily imposing the minimal chiral subtraction
scheme (\ref{def-sub}) is discussed.}. It is necessary to specify how to absorb the
remaining logarithmic divergence in $I_{\pi N}(\sqrt{s}\,)$. The strategy is to move
all divergences from the unitarity loop function $J_{\pi N}(w)$ into the renormalized
potential $V_R(w)$ via (\ref{ren-v}). For the effective potential one may then apply
the standard perturbative renormalization program. The remaining ambiguity is fixed
by the renormalization condition that the effective potential
$V_R(w)$ matches the scattering amplitudes at a subthreshold energy $\sqrt{s}=\mu_S$:
\begin{eqnarray}
\bar T_R (\mu_S) = V_R(\mu_S) \;.
\label{ren-cond}
\end{eqnarray}
The choice for the subtraction point $\mu_S$ is rather well determined
by the crossing symmetry constraint. In fact one is lead almost uniquely
to the convenient point $\mu_S =m_N$. For the case of pion-nucleon
scattering one observes that the renormalized effective interaction $V_R$ is real only if
$m_N-m_\pi<\mu_S < m_N+m_\pi$ holds. The first condition reflects the s-channel
unitarity cut with  $\Im I_{\pi N}(\mu_S) =0 $ only if $\mu_S < m_N+m_\pi$. The
second condition signals the u-channel unitarity cut.
A particular convenient choice for the subtraction point is $\mu_S=m_N$, because it protects
the nucleon pole term contribution. It leads to
\begin{eqnarray}
\bar T_R (w)&=& V_R(w)+{\mathcal O}\Big(\left( \wslash+m_N\right)^0\Big)
\nonumber\\
&=&-\frac{3\,({\mathcal Z}_N\,g_A)^2}{4\,f^2}\,
\frac{4\,m^2_N }{\wslash+m_N}
+{\mathcal O}\Big(\left( \wslash+m_N\right)^0\Big) \,,
\label{pole-protection}
\end{eqnarray}
if the renormalized loop function is subtracted in such a way that
\begin{eqnarray}
I_{\pi N,R}(m_N) =0 \;, \qquad I_{\pi N,R}'(m_N)=0 \,,
\label{}
\end{eqnarray}
hold in $J_{\pi N}(w)$ as specified in (\ref{jpin-def}).
It is insisted on a minimal subtraction for the
scalar loop function $I_{\pi N}(\sqrt{s}\,)$ 'inside' the full loop function $J_{\pi N}(w)$
as suggested by dimensional regularization.
According to (\ref{ipin-analytic}) one subtraction suffices to render $I_{\pi N}(\sqrt{s}\,)$
finite. A direct subtraction for $J_{\pi N}(w)$ would require a subtraction polynomial
that would not be specified by the simple renormalization condition (\ref{ren-cond}) and would
most likely be at odds with chiral symmetry constraints.
It is stressed that the double subtraction in $I_{\pi N}(\sqrt{s}\,)$ is necessary in order
to meet the condition (\ref{ren-cond}). Only with $I_{\pi N,R}'(m_N)=0$
the renormalized effective potential $V_R(w)$ in (\ref{pole-protection}) represents
the s-channel nucleon pole term in terms of the physical coupling constant ${\mathcal Z}_N\,g_A$
properly. For example, it is evident that if $V_R$ is truncated at chiral order $Q^2$
one finds ${\mathcal Z}_N=1$. The one-loop wave-function renormalization (\ref{def-zn}) is needed
if one considers the $Q^3$-terms in $V_R$.

One observes that the subtracted loop function $I_{\pi N}(\sqrt{s}\,)-I_{\pi N}(\mu_S)$
is in fact independent on the subtraction point to order $Q^2$ if one counted
$\mu_S-m_N\sim Q^2$. In this case one derives from (\ref{ipin-expand}) the expression
\begin{eqnarray}
I_{\pi N}(\sqrt{s}\,)-I_{\pi N}(\mu_S)
&=&i\,\frac{\sqrt{\phi_{\pi N}}}{8\,\pi\,m_N}
+\frac{m_\pi}{16\,\pi\,m_N}
-\frac{\sqrt{s}-m_N}{16\,\pi^2\,m_N}\,\ln \left( \frac{m_\pi^2}{m_N^2}\right)
\nonumber\\
&+&\frac{\sqrt{\phi_{\pi N}}}{16\,\pi^2\,m_N}\,
\ln \left( \frac{\sqrt{s}-m_N+\sqrt{\phi_{\pi N}}}
{\sqrt{s}-m_N-\sqrt{\phi_{\pi N}}} \right)
+{\mathcal O}\left( Q^2 \right) \;,
\label{ipin-expand:b}
\end{eqnarray}
where $\phi_{\pi N}= (\sqrt{s}-m_N)^2-m_\pi^2$.  The result (\ref{ipin-expand:b})
suggests that one may set up a systematic expansion scheme in powers of
$\mu_S-m_N \sim Q^2$. The subtraction-scale independence of physical
results then implies that all powers $(\mu_S-m_N)^n$ cancel except for the leading term
with $n=0$. In this sense one may say that the scattering amplitude is independent of the
subtraction point $\mu_S$. Note that such a scheme does not necessarily require a
perturbative expansion of the scattering amplitude. It may be advantageous instead to
restore that minimal $\mu_S$-dependence in the effective potential $V$, leading to a
subtraction scale independent scattering amplitude at given order in $(\mu_S-m_N)$.
Equivalently, it is legitimate to directly insist on the 'physical'
subtraction with $\mu_S=m_N$. There is a further important point to be made: we would reject
a conceivable scheme in which the inverse effective potential $V^{-1}$ is expanded in
chiral powers, even though it would obviously facilitate the construction of that minimal
$\mu_S$-dependence. As examined in detail in \cite{nn-lutz} expanding the inverse effective
potential requires a careful analysis as to determine if the effective potential has a zero within
its applicability domain. If this is the case one must reorganize the expansion scheme.
This is particularly cumbersome in a coupled-channel scenario where one must
ensure that $\det V \neq 0$ holds. Since in the $SU(3)$ limit the Weinberg-Tomozawa interaction
term, which is the first term in the chiral expansion of $V $, leads
to $\det V_{WT}=0$ (see (\ref{WT-k})) one should not pursue this path \footnote{There is a
further strong indication that the expansion of the inverse effective potential indeed
requires a reorganization. The effective p-wave interaction kernel is troublesome, because
the nucleon-pole term together with a smooth background term will lead necessarily to a
non-trivial zero.}.

An important further aspect, the approximate crossing symmetry, needs to be
addressed. It will turn out that this problem is closely related to the
renormalization program required to define the scattering equation. At first,
one may insist on either a strict perturbative
scheme or an approach which performs a simultaneous iteration of the s- and
u-channel in order to meet the crossing symmetry constraint. It is
pointed out, however, that a simultaneous iteration of the s- and
u-channel is not required, if the s-channel iterated and u-channel
iterated amplitudes match at subthreshold energies $\sqrt{s}\simeq
\mu_S \simeq m_N $ to high accuracy. This is a sufficient
condition in the chiral framework, because the overlap of the
applicability domains of the s- and u-channel iterated amplitudes
are restricted to a small matching window at subthreshold energies
in any case. This will be discussed in more detail in chapter 4.3.
Crossing symmetry is not necessarily observed in a
cutoff regularized approach. In the scheme developed here the meson-baryon
scattering process is {\it perturbative} below the s- and
u-channel unitarity thresholds by construction and therefore meets
the crossing symmetry constraints approximatively. {\it
Non-perturbative} effects as implied by the unitarization are then
expected at energies outside the matching window.

Before turning to the coupled channel formalism it is instructive to
study the degree of convergence of standard $\chi$PT. The poor
convergence of the standard $\chi$PT scheme in the $\bar K\,N$
sector will be illustrated.

\vskip1.5cm \section{Weinberg-Tomozawa interaction and convergence of $\chi$PT}

Consider the Weinberg-Tomozawa interaction term $K^{}_{WT}(\bar k,k;w)$ as the driving term
in the Bethe-Salpeter equation. This is an instructive example because it exemplifies the
non-perturbative nature of the strangeness channels and it also serves as a transparent first
test of the renormalization scheme. The on-shell equivalent scattering amplitude
$\bar T_{WT}(w)$ is
\begin{eqnarray}
&& \!\! V_{WT}(w) =\left( 1-\frac{c}{4 f^2}\,I_L\right)^{-1}
\!\! \frac{c}{4 f^2}  \left(2 \;\wslash -2\!\,M
-I_{LR}\,\frac{c}{4 f^2} \right)
\left(1-I_R\,\frac{c}{4 f^2} \right)^{-1},
\nonumber\\
&&\!\! \bar T_{WT}(w) = \frac{1}{1-V_{WT}(w)\,J(w)}\,V_{WT}(w) \;,\quad
K_{WT}(\bar k,k;w) = c^{}_{}\,\frac{\qslash+\barqslash}{4\,f^2}\;,
\label{wt-tadpole}
\end{eqnarray}
where the Bethe-Salpeter scattering equation (\ref{BS-eq}) was solved algebraically
following the construction (\ref{k-decomp},\ref{t-eff}). The result is obtained
by replacing the typical building blocks in (\ref{v-eff}) according to
\begin{eqnarray}
&& K_R\cdot G \to I_R \;, \qquad G\cdot K_L \to I_L \;,\qquad
K_R\cdot G\cdot K_L \to I_{LR}\;,
\nonumber\\
&& I_L=I_R=I_M \;, \quad \;I_{LR}=(\wslash-M)\,I_{M}\;,
\label{}
\end{eqnarray}
with the mesonic tadpole loop $I_M$ introduced in (\ref{def-tadpole}).
The meson-baryon loop function $J(w)$ is
defined for the $\pi N$ system in (\ref{jpin-def}).
The coupling constants $c$ specify the strength of the
Weinberg-Tomozawa interaction in a given meson-baryon channel with isospin $I$.
Note that (\ref{wt-tadpole}) is written in a way such that coupled channel effects
are easily included by identifying the proper matrix structure of its building blocks.
A detailed account of these effects will be presented in subsequent chapters.
It is emphasized that the mesonic tadpole $I_M$ cannot be absorbed systematically in $f$,
in particular when the coupled-channel generalization of (\ref{wt-tadpole}) with its
non-diagonal matrix $c$ is considered. Note that at leading chiral order $Q$ the tadpole
contribution should be dropped in the effective interaction $V$ in any case.

It is instructive to consider the s-wave $\bar K N $ scattering lengths up
to second order in the Weinberg-Tomozawa interaction vertex.
The coefficients
\begin{eqnarray}
c^{(0)}_{\bar K N\to \bar KN}=3\;,\qquad c^{(1)}_{\bar K N\to \bar KN}=1 \;,
\label{}
\end{eqnarray}
lead to
\begin{eqnarray}
4\,\pi \left(1+\frac{m_K}{m_N} \right) a_{\bar  K N}^{(I=0)} &=&
\frac{3}{2}\,\frac{m_K}{f^2} \left(
1+\frac{3\,m^2_K}{16\,\pi^2\,f^2}
\left(\pi-\ln \left(\frac{m_K^2}{m_N^2}\right) \right)\right)+\cdots \;,
\nonumber\\
4\,\pi \left(1+\frac{m_K}{m_N} \right) a_{\bar  K N}^{(I=1)} &=&
\frac{1}{2}\,\frac{m_K}{f^2} \left(
1+\frac{m^2_K}{16\,\pi^2\,f^2}
\left(\pi-\ln \left(\frac{m_K^2}{m_N^2}\right) \right)\right)+\cdots \;,
\label{wt-square-KN}
\end{eqnarray}
where only the s-channel iteration effects following
from (\ref{wt-tadpole}) are included. The loop function $I_{\bar KN}(\sqrt{s})$ was
subtracted at the nucleon mass
and all tadpole contributions are dropped. Though the expressions (\ref{wt-square-KN}) are
incomplete in the $\chi $PT framework (terms of chiral order $Q^2$ and $Q^3$ terms are
neglected) it is highly instructive to investigate the convergence property of a reduced
chiral Lagrangian with the Weinberg-Tomozawa interaction only. According to
(\ref{wt-square-KN}) the relevant expansion parameter $m_K^2/(8\pi f^2) \simeq 1$ is about one
in the $\bar KN$ sector. One observes the enhancement factor $2 \pi$ as compared to irreducible
diagrams which would lead to the typical factor $ m^2_K/(4\pi \,f)^2$.
The perturbative treatment of the Weinberg-Tomozawa
interaction term is therefore unjustified and a change in approximation
scheme is required. In the isospin zero $\bar KN$ system the Weinberg-Tomozawa interaction
if iterated to all orders in the s-channel (\ref{wt-tadpole}) leads to a pole in the
scattering amplitude at subthreshold
energies $\sqrt{s}< m_N+m_K$. This pole is a precursor of
the $\Lambda(1405)$ resonance \cite{dalitz-1,Siegel,Kaiser,Ramos,Hirschegg}.

\begin{figure}[t]
\begin{center}
\includegraphics[width=12cm,clip=true]{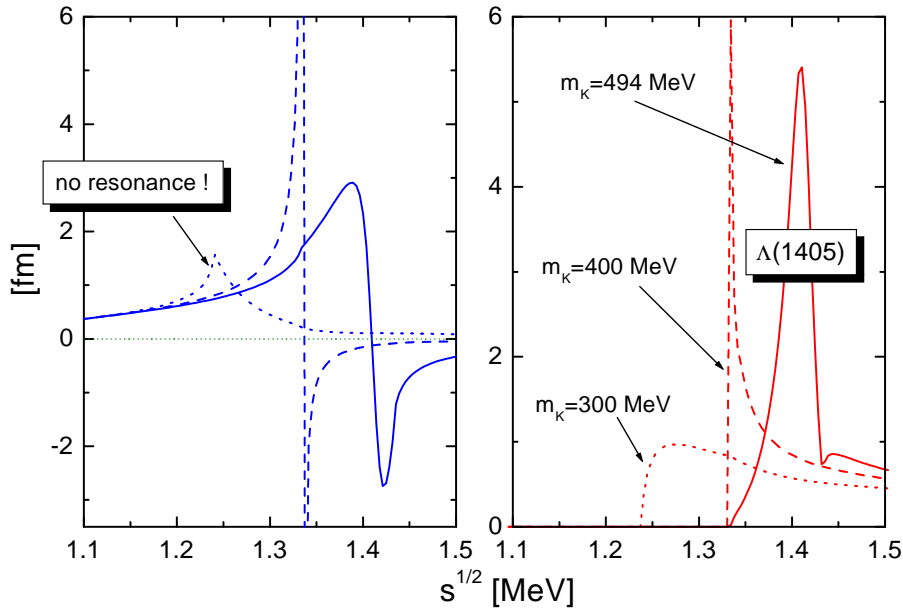}
\end{center}
\caption{Real and imaginary part of the isospin zero
s-wave $K^-$-nucleon scattering amplitude as it follows from the
$SU(3)$ Weinberg-Tomozawa interaction term in a coupled channel
calculation. It is used  $f = 93 $ MeV.  The
subtraction point is identified with the $\Lambda(1116)$ mass.}
\label{fig:wt}
\end{figure}

In Fig.~\ref{fig:wt}, which was published previously in \cite{Hirschegg},
the final result based on the leading interaction term of the
chiral $SU(3)$ Lagrangian density as suggested by Tomozawa and Weinberg is shown.
If taken as input for the multi-channel Bethe-Salpeter equation,
properly furnished with a renormalization scheme leading to a subtraction point close to
the baryon octet mass, a rich structure of the scattering amplitude arises \cite{Hirschegg}.
Details for the coupled-channel generalization of (\ref{wt-tadpole}) are presented in the
subsequent chapters. Fig.~\ref{fig:wt} shows the s-wave solution of the multi-channel
Bethe-Salpeter as a
function of the kaon mass. For physical kaon masses the isospin zero scattering amplitude
exhibits a resonance structure at energies where one would expect the $\Lambda(1405)$
resonance. It is pointed out that the resonance structure disappears as the kaon mass is decreased.
Already at a hypothetical kaon mass of $300$ MeV the $\Lambda(1405)$ resonance is no longer
formed. Fig.~\ref{fig:wt} demonstrates that the chiral $SU(3)$ Lagrangian is
necessarily non-perturbative in the strangeness sector. This confirms the findings of
\cite{Kaiser,Ramos}. In previous works, however, the $\Lambda(1405)$ resonance
is the result of a fine-tuned cutoff parameter which gives rise to a different kaon mass
dependence of the scattering amplitude \cite{Ramos}. In the scheme developed here
the choice of subtraction point close to the baryon octet mass follows necessarily from the
compliance of the expansion scheme with approximate crossing symmetry.
Moreover, the identification of the subtraction point with the $\Lambda$-mass in the
isospin zero channel protects the hyperon exchange s-channel pole contribution and
therefore avoids possible pathologies at subthreshold energies.

In the light of the above arguments it is important to study their possible relevance
for the pion-nucleon sector. The chiral $SU(2)$ Lagrangian has been
successfully applied to pion-nucleon scattering in standard chiral
perturbation theory \cite{Bernard,Meissner,pin-q4}. Here the typical expansion parameter
$m^2_\pi/(8\,\pi\,f^2) \ll 1$ characterizing the unitarization  is sufficiently
small and one would expect good convergence properties. The application of the chiral $SU(3)$
Lagrangian to pion-nucleon scattering on the other hand is not completely worked out so far.
In the SU(3) world the $\pi\, N$ channel couples for example to the $K \,\Sigma $ channel. Thus,
the slow convergence of the unitarization in the $K\, \Sigma$ channel suggests
to expand the interaction kernel rather than the scattering amplitude also in the
strangeness zero channel. This may improve the convergence properties of the chiral
expansion and extend its applicability domain to larger energies. Also,
if the same set of parameters are to be used in the pion-nucleon and kaon-nucleon
sectors the analogous partial resummation of higher order counter terms included
by solving the Bethe-Salpeter equation should be applied.
Such effects are illustrated for the case of the Weinberg-Tomozawa interaction.
With
\begin{eqnarray}
c^{(\frac{3}{2})}_{\pi N\to \pi N}=-1\;,\qquad
c^{(\frac{3}{2})}_{\pi N\to K \Sigma}=-1 \;,
\label{}
\end{eqnarray}
the isospin three half s-wave pion-nucleon
scattering length, $a_{\pi N}^{(\frac{3}{2})}$, receives the typical correction terms
\begin{eqnarray}
&&4\,\pi \left(1+\frac{m_\pi}{m_N} \right) a_{\pi N}^{(\frac{3}{2})} =
-\frac{m_\pi}{2\,f^2}\, \Bigg(
1-\frac{m^2_\pi}{16\,\pi^2\,f^2}
\left(\pi-\ln \left(\frac{m_\pi^2}{m_N^2}\right) \right)
\nonumber\\
&&\qquad \qquad - \frac{(m_\pi+m_K)^2}{32\,\pi^2\,f^2}
\Bigg( -1-\frac{1}{2}\,\ln \left(\frac{m_K^2}{m_\Sigma^2}\right)
\nonumber\\
&& \qquad \qquad
+\pi \left( \frac{3\,m_K}{4\,m_N}-\frac{m_\Sigma-m_N}{2\,m_K} \right)
+{\mathcal O}\left( Q^2\right)\Bigg)\Bigg)
+ {\mathcal O}\left( m_\pi^2\right)\;.
\label{wt-square-piN}
\end{eqnarray}
Again, exclusively the unitary correction terms are considered. Note that the
ratio $(m_\Sigma -m_N)/m_K$  arises in (\ref{wt-square-piN}), because we first expand
in powers of $m_\pi$ and only then expand further with $m_\Sigma-m_N \sim Q^2$ and
$m_K^2 \sim Q^2 $. The correction terms in (\ref{wt-square-piN}) induced by the
kaon-hyperon loop, which is subtracted at the nucleon
mass, exemplify the fact that the parameter $f$ is renormalized by the strangeness sector
and therefore must not be identified with the chiral limit value of $f$ as derived for the
$SU(2)$ chiral Lagrangian. This is evident if one confronts the Weinberg-Tomozawa
theorem of the chiral $SU(2)$ symmetry with (\ref{wt-square-piN}).
The expression (\ref{wt-square-piN})  demonstrates further
that this renormalization of $f$ appears poorly convergent in the kaon mass. Note in
particular the anomalously large term $\pi \,m_K/m_N $. Hence it is advantageous to consider
the partial resummation induced by a unitary coupled channel treatment of pion-nucleon
scattering.

\vskip1.5cm \section{Partial-wave decomposition of the Bethe-Salpeter equation}

The Bethe-Salpeter equation (\ref{BS-eq}) can be solved analytically for
quasi-local interaction terms which typically arise in the chiral Lagrangian.
The scattering equation is decoupled by introducing relativistic projection
operators ${Y}^{(\pm)}_n(\bar q,q;w)$ with good total angular momentum:
\begin{eqnarray}
&&{Y}^{(\pm )}_n(\bar q,q;w)=\frac{1}{2}\,\Bigg(\frac{\wslash}{\sqrt{w^2}}\pm 1\Bigg)\,
\bar Y_{n+1}(\bar q,q;w)
\nonumber\\
&&\;\;\;\;\;\;\;\;\;\;-\frac{1}{2}\,\Bigg(  \barqslash -\frac{w\cdot \bar q}{w^2}\,\wslash \Bigg)
\Bigg(\frac{\wslash}{\sqrt{w^2}} \mp 1\Bigg)\,
\Bigg( \qslash -\frac{w\cdot q}{w^2}\,\wslash \Bigg)
\bar Y_{n}(\bar q,q;w)\;,
\nonumber\\
&&\bar Y_{n}(\bar q,q;w)= \sum_{k=0}^{[(n-1)/2]}\,\frac{(-)^k\,(2\,n-2\,k) !}{2^n\,k !\,(n-k)
!\,(n-2\,k -1) !}\,Y_{\bar q \bar q}^{k}\,Y_{\bar q q}^{n-2\,k-1}\,Y_{q q}^{k}\;,
\nonumber\\
&&Y_{\bar q \bar q}=\frac{(w\cdot \bar q )\,(\bar q\cdot w)}{w^2}
-\bar q \cdot \bar q
\;,\;\;\;
Y_{q  q}=\frac{(w\cdot  q )\,( q\cdot w)}{w^2} -q \cdot  q \;,
\nonumber\\
&&Y_{\bar q q}=\frac{(w\cdot \bar q )\,(q\cdot w)}{w^2} -\bar q \cdot q  \;.
\label{cov-proj}
\end{eqnarray}
For the readers convenience we provide the leading order projectors
${Y}_n^{(\pm)}$ relevant for the $J={\textstyle{1\over 2}}$ and $J={\textstyle{3\over 2}}$
channels explicitly:
\begin{eqnarray}
{Y}_0^{(\pm )}(\bar q,q;w) &=& \frac{1}{2}\,\left( \frac{\wslash}{\sqrt{w^2}}\pm 1 \right)\;,
\nonumber\\
{Y}_1^{(\pm)}(\bar q,q;w) &=& \frac{3}{2}\,\left( \frac{\wslash}{\sqrt{w^2}}\pm 1 \right)
\left(\frac{(\bar q \cdot w )\,(w \cdot q)}{w^2} -\big( \bar q\cdot q\big)\right)
\nonumber\\
&-&\frac{1}{2}\,\Bigg(  \barqslash -\frac{w\cdot \bar q}{w^2}\,\wslash \Bigg)
\Bigg(\frac{\wslash}{\sqrt{w^2}}\mp 1\Bigg)\,
\Bigg( \qslash -\frac{w\cdot q}{w^2}\,\wslash \Bigg)\;.
\label{}
\end{eqnarray}

The objects ${Y}^{(\pm)}_n(\bar q,q;w)$ are constructed to have the following
convenient property: suppose that the interaction kernel $K$ in (\ref{BS-eq}) can be
expressed as linear combinations of the ${Y}^{(\pm)}_n(\bar q,q;w)$ with a set of
coupling functions $V^{(\pm)}(\sqrt{s}\,, n)$, which may depend on the variable $s$,
\begin{eqnarray}
K(\bar k ,k ;w ) &=& \sum_{n=0}^\infty \left(
V^{(+)}(\sqrt{s};n)\,{Y}^{(+)}_n(\bar q,q;w)
+V^{(-)}(\sqrt{s};n)\,{Y}^{(-)}_n(\bar q,q;w) \right) \,
\label{k-sum}
\end{eqnarray}
with $ w = p+q$, $ k= (p-q)/2 $ and $\bar k =(\bar p-\bar q)/2 $.
Then in a given isospin channel the unique solution reads
\begin{eqnarray}
&& T(\bar k ,k ;w ) = \sum_{n=0}^\infty \left(
M^{(+)}(\sqrt{s};n )\,{Y}^{(+)}_n(\bar q,q;w)
+ M^{(-)}(\sqrt{s};n)\,{Y}^{(-)}_n(\bar q,q;w)\right) \;,
\nonumber\\
&& M^{(\pm)}(\sqrt{s};n) = \frac{V^{(\pm )}(\sqrt{s};n)}{1-V^{(\pm)}(\sqrt{s};n)
\,J^{(\pm)}(\sqrt{s};n)} \;,
\label{t-sum}
\end{eqnarray}
with a set of divergent loop functions $J^{(\pm)}_{\pi N }(\sqrt{s};n)$ defined by
\begin{eqnarray}
J^{(\pm)}_{\pi N}(\sqrt{s};n)\,{Y}^{(\pm)}_n(\bar q,q;w)
&&= -i\int \frac{d^4l}{(2\pi)^4}\,{Y}^{(\pm)}_n(\bar q,l;w)\,
S_N(w-l)\,
\nonumber\\
&& \;\;\;\;\;\;\;\;\;\;\;\;\;\;\;\;\;\;\;\;\;
\times D_\pi(l)\,{Y}^{(\pm)}_n(l,q;w)\;.
\label{jpin-n-def}
\end{eqnarray}
It is underlined that the definition of the loop functions in (\ref{jpin-n-def}) is non
trivial, because it assumes that $Y_n^{(\pm )}\cdot G \cdot Y_n^{(\pm )} $ is indeed
proportional to $Y_n^{(\pm)}$. An explicit derivation of this property, which is in fact
closely linked to our renormalization scheme, is given in \cite{Lutz:Kolomeitsev}. Recall
that the loop functions $J^{(\pm)}_{\pi N}(\sqrt{s};n)$, which are badly divergent, have a
finite and well-defined imaginary part
\begin{eqnarray}
\Im\,J^{(\pm)}_{\pi N }(\sqrt{s};n) &=& \frac{p^{2\,n+1}_{\pi N}}{8\,\pi\,\sqrt{s}}
\left(  \frac{\sqrt{s}}{2}+ \frac{m_N^2-m_\pi^2}{2\,\sqrt{s}}\pm m_N \right)\;.
\label{}
\end{eqnarray}
It will be specified how to renormalize the loop functions. In dimensional regularization
the loop functions can be written as linear combinations of scalar one
loop functions
$I_{\pi N}(\sqrt{s}\,)$, $I_\pi$, $I_N $ and $I^{(n)}$,
\begin{eqnarray}
I^{(n)}=i\,\int \frac{d^4l}{(2\pi)^4}\,\Big(l^2 \Big)^n\;,
\label{tadpole:a}
\end{eqnarray}
According to our renormalization procedure we drop $I^{(n)}$, the tadpole
contributions $I_\pi$, $I_N $ and replace
$I_{\pi N}(s)$ by $I_{\pi N}(\sqrt{s}\,)-I_{\pi N}(\mu_S )$. This leads to
\begin{eqnarray}
J^{(\pm)}_{\pi N }(\sqrt{s}; n) &=&
p_{\pi N}^{2\,n}(\sqrt{s}\,)
\left( \frac{\sqrt{s}}{2}+ \frac{m_N^2-m_\pi^2}{2\,\sqrt{s}}\pm m_N \right)
\Delta I_{\pi N}(\sqrt{s}\,)\;,
\nonumber\\
\Delta I_{\pi N}(\sqrt{s}\,) &=& I_{\pi N}(\sqrt{s}\,)-I_{\pi N}(\mu_S ) \;,
\label{result-loop}
\end{eqnarray}
with the master loop function $I_{\pi N}(\sqrt{s}\,)$ and $p_{\pi N}(\sqrt{s}\,)$
given in (\ref{ipin-analytic}). In the center of mass frame $p_{\pi N}$ represents
the relative momentum. It is to be emphasized that the loop functions
$J^{(\pm)}_{\pi N }(\sqrt{s}; n)$ are renormalized in accordance
with (\ref{ren-cond}) and (\ref{ren-v}) where $\mu_S=m_N$\footnote{
Note that consistency with the renormalization condition (\ref{ren-cond}) requires
a further subtraction in the loop function $J^{(-)}_{\pi N}(\sqrt{s},0)$
if the potential $V^{(-)}_{\pi N}(\sqrt{s},0) \sim 1/(s-m_N^2+i\,\epsilon)$ exhibits the
s-channel nucleon pole (see (\ref{pole-protection})).}.
This leads to tadpole-free loop functions and also to $M^{(\pm )}(m_N,n)=V^{(\pm )}(m_N,n)$.
The behavior of the loop functions $J^{(\pm )}_{\pi N}(\sqrt{s},n)$ close to threshold
\begin{eqnarray}
\Im \,J^{(+)}_{\pi N}(\sqrt{s},n) \sim p_{\pi N}^{2\,n+1} \;, \qquad
\Im \,J^{(-)}_{\pi N}(\sqrt{s},n) \sim p_{\pi N}^{2\,n+3} \,,
\label{}
\end{eqnarray}
already tells the angular momentum, $l$, of a given channel with $l=n$ for the $'+'$ and
$l=n+1$ for the $'-'$ channel.

It is instructive to compare the form of the unitarity
loop functions (\ref{result-loop}) with those suggested by Oller and Meissner \cite{Oller-Meissner}
as they arise in the N/D method of Chew and Mandelstam \cite{Chew-Mandelstam}.
Despite the conceptual differences it is useful to compare the form of the
final unitarity loop functions generated by the two schemes. In any approach that is
unitary the partial wave amplitudes, $M^{(\pm)}(\sqrt{s};n)$, may be cast into the form (\ref{t-sum}). Therefore
different schemes may be compared at the level of the unitarity loop function
$J^{(\pm)}(\sqrt{s};n)$ and the effective interaction kernel $V^{(\pm)}(\sqrt{s};n)$.
It is evident that, once the precise form of the unitarity loop functions
$J^{(\pm)}(\sqrt{s},n)$ is specified it is well defined how to evaluate the effective
interaction kernel in perturbation theory. In
\cite{N-D-Oller-Oset,N-D-Meissner-Oller,Oller-Meissner} it was suggested to take
\begin{eqnarray}
\frac{p^{2\,l+1}_{\pi N}}{8\pi \,\sqrt{s}} \,,
\label{}
\end{eqnarray}
as the imaginary part of the unitarity loop function of given angular momentum $l$ and
then reconstruct the real part of
the loop function by means of a subtracted dispersion-integral representation.
That requires the determination of l+1 subtraction constants \cite{N-D-Oller-Oset}, which leads to a flood
of free parameters that are not controlled systematically by
the chiral SU(3) symmetry. This is particularly cumbersome when dealing with many coupled channels all of which
would require a priori independent subtraction coefficients. On the other hand,
in the scheme developed here the application of dimensional regularization strongly constrains
the subtraction procedure leaving one universal subtraction point $\mu_S$ which is fixed by
the requirement that chiral counting rules and approximative crossing symmetry hold.
Here the construction of the covariant projection operators (\ref{cov-proj}) proves particularly useful since
by means of those it is possible to apply dimensional regularization required to control the subtraction parameters
in conformity with symmetry constraints. It was demonstrated that all reduced tadpole
contributions (see e.g. (\ref{jpin-def}))
must be moved into the effective interaction $V^{(\pm)}(\sqrt{s};n)$ as to avoid an anomalous chiral power of the
baryon mass renormalization which otherwise would arise when solving the scattering equation. An independent
argument that lead to the same renormalization requirement was based on the observation that the orthogonality
of covariant projectors with different total angular momentum $J$ follows only if all reduced tadpole contributions are
considered as part of the effective interaction. A further important difference are the prefactors $E_N\pm m_N$
in (\ref{result-loop}), which are required in a covariant scheme but were dropped in \cite{Oller-Meissner}. If those
factors would be restored into the scheme of Oller and Meissner, already the s-wave unitarity loop function required
two subtraction constants.

The Bethe-Salpeter equation (\ref{t-sum}) decouples into reduced scattering amplitudes
$M^{(\pm)}(\sqrt{s},n)$ with well-defined angular momentum. In order to unambiguously
identify the total angular momentum $J$ we recall the partial-wave decomposition of
the on-shell scattering amplitude \cite{Landoldt}. The latter amplitude $T$ is decomposed
into invariant amplitudes $F^{(I)}_\pm(s,t)$ carrying good isospin,
\begin{eqnarray}
T &=& \sum_{I}\,\left(
\frac{1}{2}\,\Bigg(\frac{\wslash}{\sqrt{w^2}}+1
\Bigg)\, F^{(I)}_+(s,t)+ \frac{1}{2}\,\Bigg(\frac{\wslash}{\sqrt{w^2}}-1
\Bigg)\,F^{(I)}_-(s,t)\right) P_{I} \,,
\label{define-fpm}
\end{eqnarray}
where $s= (p+q)^2=w^2$ and $t=(\bar q-q)^2$ and $P_I$ are the isospin
projectors introduced in (\ref{iso:proj}).
The choice of invariant amplitudes is not unique. The particular choice
used in (\ref{define-fpm}) is particularly
convenient to make contact with the covariant projection operators (\ref{cov-proj}).
For different choices, see \cite{Landoldt}.
The amplitudes $F_{\pm}(s,t)$ are decomposed into partial-wave amplitudes
$f^{(l)}_{J=l\pm \frac{1}{2}}(s)$ with\cite{Landoldt}:
\begin{eqnarray}
F_+(s,t) &=& \frac{8\,\pi\,
\sqrt{s}}{E+m_N}\,\sum_{n=1}^\infty
\,\Big( f_{J=n+\frac{1}{2}}^{(n-1)}(\sqrt{s}\,)-f_{J=n- \frac{1}{2}}^{(n+1)}(\sqrt{s}\,)\Big)\,P'_n(\cos \theta)
\;,
\nonumber\\
F_-(s,t) &=&\frac{8\,\pi\,
\sqrt{s}}{E-m_N}\,\sum_{n=1}^\infty
\,\Big( f_{J=n-\frac{1}{2}}^{(n)}(\sqrt{s}\,)-f_{J=n+ \frac{1}{2}}^{(n)}(\sqrt{s}\,)\Big)\,P'_n(\cos \theta)\;,
\nonumber\\
P'_n (\cos \theta)&=& \sum_{k=0}^{[(n-1)/2]}\,\frac{(-)^k\,(2\,n-2\,k) !}{2^n\,k !\,(n-k)
!\,(n-2\,k -1) !}\,\Big(\cos  \theta\Big)^{n-2\,k-1} \,,
\label{t-on-decomp}
\end{eqnarray}
where $[n/2]= (n-1)/2$ for $n$ odd and $[n/2]= n/2$ for $n$
even. $P'_n(z)$ is the derivative
of the Legendre polynomials. In the center of mass frame $E$ represents the nucleon
energy and $\theta$ the scattering angle:
\begin{eqnarray}
&&E=\frac{1}{2}\,\sqrt{s}+\frac{m_N^2-m_\pi^2}{2\,\sqrt{s}} \;,\;\;\;
t=(\bar q-q)^2=-2\,p_{\pi N}^2\,\Big(1-\cos \theta \Big)  \;.
\label{}
\end{eqnarray}
The unitarity condition formulated for the partial-wave amplitudes
$f_J^{(l)}$ leads to their representation in terms of the scattering phase
shifts $\delta^{(l)}_J$
\begin{eqnarray}
&&p_{\pi N}\,f^{(l)}_{J=l\pm \frac{1}{2}}(\sqrt{s}\,) =
\frac{1}{2\,i}\left( e^{2\,i\,\delta^{(l)}_{J=l\pm \frac{1}{2}} (s)}-1 \right)
=\frac{1}{\cot \delta^{(l)}_{J=l\pm \frac{1}{2}} (s)-i}\;.
\label{}
\end{eqnarray}
One can now match the reduced amplitudes
$M^{(\pm)}_n(s)$ of (\ref{t-sum}) and the partial-wave amplitudes
$f^{(l)}_{J=l\pm \frac{1}{2}}(s)$
\begin{eqnarray}
f^{(l)}_{J=l\pm\frac{1}{2}}(\sqrt{s}\,) &=& \frac{p^{2\,J-1}_{\pi N}}{8\,\pi\,\sqrt{s}}
\left( \frac{\sqrt{s}}{2}+\frac{m_N^2-m_\pi^2}{2\,\sqrt{s}} \pm m_N\right)
M^{(\pm )}_{}(\sqrt{s},J-{\textstyle{1\over2}})\;.
\label{match}
\end{eqnarray}
It is useful to consider the basic building block $\bar Y_n(\bar q,q;w)$ of the covariant
projectors ${Y}_n^{(\pm)}(\bar q,q;w)$ in (\ref{cov-proj}) and observe the formal similarity
with $P'_n(\cos \theta )$ in (\ref{t-on-decomp}). In fact in the center of mass frame with
$w_{cm}=(\sqrt{s},0)$ one finds $p_{\pi N}^{2\,n-2}\,P'_n(\cos \theta )=Y_n(\bar q,q;w_{cm})$.
This observation leads to a straightforward proof of (\ref{t-sum})
and (\ref{jpin-n-def}). It is sufficient to prove the loop-orthogonality of the projectors
in the center of mass frame, because they are free of kinematical singularities
\cite{Lutz:Kolomeitsev}. One readily finds that the imaginary part of the unitary products
${Y}^{(\pm)}_n\,G\,{Y}^{(\pm)}_m$ vanish unless both projectors are the same. It follows that
the unitary product of projectors which are expected to be orthogonal can at most be a real
polynomial involving the tadpole functions $I_\pi, I_N$ and $I^{(n)}$. Then the renormalization
procedure as described in chapter 3.2 leads to (\ref{t-sum}) and (\ref{jpin-n-def}). It is
emphasized that the argument relies crucially on the fact that the projectors
are free of kinematical singularities in $q$ and $\bar q$. This implies in particular that the
object $Y_n(\bar q,q;w)$ must not be identified with $p_{\pi N}^{2\,n-2}\,P'_n(\cos \theta )$ as
one may expect naively.

It is necessary to return to the assumption made in (\ref{k-sum}) that the interaction
kernel $K$ can be decomposed in terms of the covariant projectors. Of course this is not possible
for a general interaction kernel $K$. It is pointed out, however, that the set of projectors
(\ref{cov-proj}) can be used to introduce a particular notion of on-shell irreducibility defining
what is meant with $\bar K$ in (\ref{k-decomp}). The on-shell irreducible kernel $\bar K$
is identified by decomposing the  interaction kernel according to
\begin{eqnarray}
&&\bar K ^{(I)}(\bar q, q;w)=
\sum_{n=0}^\infty \left(\bar K^{(I)}_+(\sqrt{s};n)\,{Y}_n^{(+)}(\bar q, q;w)
+\bar K^{(I)}_-(\sqrt{s};n)\,{Y}_n^{(-)}(\bar q, q;w)\right) \;,
\nonumber\\
&&\bar K^{(I )}_\pm(\sqrt{s};n)= \int_{-1}^1 \frac{dz}{2} \,
\frac{K^{(I)}_\pm(s,t)}{\,p^{2\,n}_{\pi N}}\,P_n(z)
\nonumber\\
&&\;\;\;\;\;\;\;\;\;\;\;\;\;\;\;\;+ \int_{-1}^1 \frac{dz}{2} \,
\left( \frac{1}{2}\,\sqrt{s}+\frac{m_N^2-m_\pi^2}{2\,\sqrt{s}}\mp m_N\right)^2
\frac{K^{(I)}_\mp(s,t)}{p^{2\,n+2}_{\pi N}}\,P_{n+1}(z) \;,
\label{bark-def-pin}
\end{eqnarray}
where $t= -2\,p_{\pi N}^2\,(1-x)$ and $K_\pm^{(I)}(s,t)$ follows from the decomposition of the
interaction kernel $K$ into the scalar components $K_\pm^{(I)}(s,t)$,
\begin{eqnarray}
&&K^{(I)}(\bar q, q; w) \Big|_{\rm{on-shell}}=
\frac{1}{2}\,\Bigg(\frac{\wslash}{\sqrt{w^2}}+1
\Bigg)\, K_+^{(I)}(s,t) + \frac{1}{2}\,\Bigg(\frac{\wslash}{\sqrt{w^2}}-1
\Bigg)\,K_-^{(I)}(s,t) \;.
\label{}
\end{eqnarray}
Then $K-\bar K $ is on-shell reducible by construction and therefore can be
decomposed into $K_L,\,K_R $ and $K_{LR}$. The decomposition in (\ref{bark-def-pin})
generalizes the partial wave decomposition of the scattering kernel,
usually defined only in the center of mass frame. Now, with a well defined
notion of on-shell irreducibility, that complies with (\ref{def-OSR}), one can
apply (\ref{def-barV}) to set up a systematic expansion of the on-shell reduced
effective scattering kernel $V_{\rm eff}$. The merit of the formulation presented here
lies in the fact that for a given choice of the chiral Lagrangian it is in principal
straight forward to reconstruct the off-shell part of the scattering amplitude as it
may be required for calculations of higher n-point Green's functions or
current-current correlation functions. If one aims at an
evaluation of the on-shell scattering amplitudes only it is of course equivalent to
evaluate the partial wave amplitudes $M^{(\pm)}(\sqrt{s},n)$ to a given order and
derive from the representation (\ref{t-sum}) the effective interaction
$V^{(\pm)}_{\rm eff}(\sqrt{s},n)$ to that same order. The latter calculational
scheme relies of course on the form of the loop functions as derived above.

\cleardoublepage

\chapter{SU(3) coupled-channel dynamics}
\label{k4}
\markboth{\small CHAPTER \ref{k4}.~~~SU(3) coupled-channel dynamics}{}

The Bethe-Salpeter equation (\ref{BS-eq}) is readily generalized for a coupled
channel system. The chiral  $SU(3)$ Lagrangian with baryon octet and
pseudo-scalar meson octet couples the $\bar K N$ system  to five inelastic channels
$\pi \Sigma $, $\pi \Lambda $, $\eta \Lambda$, $\eta \Sigma$ and $K \Xi $ and the
$\pi N$ system to the three channels $K \Sigma$, $\eta N$ and $K\Lambda$. The strangeness
plus one sector with the $K N$ channel is treated separately in chapter 4.3 when
discussing constraints from crossing symmetry. For simplicity good isospin symmetry is
assumed in the following discussion. How to deal with isospin symmetry breaking
effects is described in \cite{Lutz:Kolomeitsev}. In order to set up a convenient notation
consider for example the two-body meson-baryon interaction terms in (\ref{two-body}) in momentum
space. The Lagrangian density ${\mathcal L}(x)$ in coordinate space
is related to its momentum space representation through
\begin{eqnarray}
\int d^4x\,{\mathcal L}(x) &=& \int \frac{d^4k}{(2\,\pi)^4}\,
\frac{d^4\bar k}{(2\,\pi)^4}\,\frac{d^4w}{(2\,\pi)^4}\,
{\mathcal L}(\bar k ,k ;w)\;,
\nonumber\\
R(q,p)&=& \int d^4x \, d^4y\,e^{-i\,q\,x-i\,p\,y}\,\Phi(x)\,B(y)\,,
\label{lag-momentum}
\end{eqnarray}
where the bilinear field $R(q,p)$ is introduced in terms of
the meson and baryon fields $\Phi(x)$ and $B(y)$. In momentum space
${\mathcal L}(\bar k ,k ;w)$ determines the scattering kernel $K(\bar k,k;w)$ as is
required for the Bethe-Salpeter equation rather directly in the following way
\begin{eqnarray}
{\mathcal L}(\bar k ,k ;w)&=&
\sum_{I=0,\frac{1}{2},1,\frac{3}{2}}\,R^{(I)\,\dagger }(\bar q,\bar p)\,\gamma_0
\,K^{(I)}(\bar k ,k ;w )\,R^{(I)}(q,p) \,,\;\;\;\;
\nonumber\\
R^{(0)}&=& \left(
\begin{array}{c}
\textstyle{1\over\sqrt{2}}\,K^\dagger\,N \\
\textstyle{1\over\sqrt{3}}\,\vec{\pi}_c \; \vec{\Sigma} \\
\eta_c \,\Lambda\\
\textstyle{1\over\sqrt{2}}\,K^t\,i\,\sigma_2\,\Xi
\end{array}
\right) \;,\;\;\;
\vec{R}^{(1)}= \left(
\begin{array}{c}
\textstyle{1\over\sqrt{2}}\,K^\dagger\,\vec{\sigma}\,N \\
\textstyle{1\over i\sqrt{2}}\,\vec{\pi}_c \, \times \vec{\Sigma} \\
\vec{\pi}_c \,\Lambda\\
\eta_c \,\vec{\Sigma} \\
\textstyle{1\over\sqrt{2}}\,K^t\,i\,\sigma_2\,\vec{\sigma}\,\Xi
\end{array}
\right) \;,
\nonumber\\
R^{(\frac{1}{2})} &=&
\left(
\begin{array}{c}
\textstyle{1\over\sqrt{3}}\,\pi_{c} \cdot \sigma\,N \\
\textstyle{1\over\sqrt{3}}\,\Sigma \cdot \sigma  \,K \\
\eta_c \,N \\
K\,\Lambda
\end{array}
\right) \;,\;\;\;\;
R^{(\frac{3}{2})}= \left(
\begin{array}{c}
\pi_c \cdot S\,N \\
\Sigma \cdot S \,K
\end{array}
\right) \;.
\label{r-def}
\end{eqnarray}
In (\ref{r-def}) the pion field and the eta field are decomposes with
$\vec \pi = \vec \pi_c+\vec \pi_c^\dagger $ and $\eta = \eta_c+\eta_c^\dagger$ \footnote{
For a neutral scalar field $\phi(x)=\phi_c(x)+\phi^\dagger_c(x)$ with mass $m$ we write
$$\phi_c(0,\vec x)= \int \frac{d^3 k}{(2\pi)^3} \,\frac{e^{i\,\vec k \cdot \vec x}}{2\,\omega_k}\,a(\vec k)
\;, \qquad \phi^\dagger_c(0,\vec x)= \int \frac{d^3 k}{(2\pi)^3} \,
\frac{e^{-i\,\vec k \cdot \vec x}}{2\,\omega_k}\,a^\dagger(\vec k)$$
where $\omega_k=(m^2+\vec k^2\,)^{\frac{1}{2}}$ and
$[a(k),a^\dagger(k')]_-= (2\pi)^3\,2\,\omega_k\,\delta^3(k-k')$.
In (\ref{r-def})  terms are suppressed which do not contribute to the two-body scattering
process at tree-level. For example terms like $\bar N\,\eta_c \,N\,\eta_c$
or $\bar N \,\eta_c^\dagger\,N\,\eta_c^{\dagger }$ are dropped.}. Also the
isospin decomposition of (\ref{field-decomp}) is applied. The isospin
$1/2$ to $3/2$ transition matrices
$S_i$ in (\ref{r-def}) are normalized by $S^\dagger_i\,S_j=\delta_{ij}-\sigma_i\,\sigma_j/3$.

The merit of the notation (\ref{r-def}) is threefold. First, the phase convention for the
isospin states is specified. Second, it defines the convention for the interaction kernel
$K$ in the Bethe-Salpeter equation. Last it provides also a convenient scheme to  read
off the isospin decomposition for the interaction kernel $K$ directly
from the interaction Lagrangian (see \cite{Lutz:Kolomeitsev}). The coupled channel Bethe-Salpeter matrix equation reads
\begin{eqnarray}
T^{(I)}_{ab}(\bar k ,k ;w ) &=& K^{(I)}_{ab}(\bar k ,k ;w )
+\sum_{c,d}\int\!\! \frac{d^4l}{(2\pi)^4}\,K^{(I)}_{ac}(\bar k , l;w )\,
G^{(I)}_{cd}(l;w)\,T^{(I)}_{db}(l,k;w )\;,
\nonumber\\
G^{(I)}_{cd}(l;w)&=&-i\,D_{\Phi(I,d)}({\textstyle{1\over 2}}\,w-l)\,S_{B(I,d)}(
{\textstyle{1\over 2}}\,w+l)\,\delta_{cd}
\label{BS-coupled}
\end{eqnarray}
where $D_{\Phi(I,d)}(q)$ and $S_{B(I,d)}(p)$ denote the
meson propagator and baryon propagator respectively for a given channel $d$ with
isospin $I$. The matrix structure of the coupled-channel interaction kernel $K_{ab}(\bar k ,k ;w )$ is defined
via (\ref{r-def}) and
\begin{eqnarray}
&&\Phi(0,a)=(\bar K,\pi,\eta , K)_a\;,\;\;\;\;\;\;\ B(0,a)=(N,\Sigma,\Lambda,\Xi)_a\;,
\nonumber\\
&&\Phi(1,a)=(\bar K,\pi,\pi,\eta , K)_a\;,\;\;\ B(1,a)=(N,\Sigma,\Lambda,\Sigma ,\Xi)_a\;,
\nonumber\\
&&\Phi({\textstyle{1\over 2}},a)=(\pi,K, \eta, K)_a\;,\;\;\;\;\;\;\
B({\textstyle{1\over 2}},a)=(N,\Sigma,N, \Lambda )_a\;,
\nonumber\\
&&\Phi({\textstyle{3\over 2}},a)=(\pi,K)_a\;,\;\;\qquad  \quad \;\;
B({\textstyle{3\over 2}},a)=(N,\Sigma )_a\;.
\label{def-channel}
\end{eqnarray}

The on-shell equivalent coupled channel interaction kernel $V_{\rm eff}$
of (\ref{v-eff},\ref{def-barV}) will be constructed. To leading chiral
orders it is legitimate to identify $V^{(I)}_{ab}$ with
$\bar K^{(I)}_{ab}$ of (\ref{k-decomp}), because the loop corrections in (\ref{v-eff}) are
of minimal chiral power $Q^3$ (see (\ref{q-rule})). To chiral order $Q^3$ the interaction
kernel receives additional terms from one loop diagrams involving the on-shell reducible
interaction kernels $K_{L,R}$ as well as from irreducible one-loop diagrams. In the notation
of (\ref{k-decomp}) one finds
\begin{eqnarray}
V = \bar K + K_R \cdot G \cdot \bar K +\bar K\,\cdot G \cdot K_L +
K_R \cdot G \cdot K_L +{\mathcal O}\left(Q^4 \right) \;.
\label{v-idef}
\end{eqnarray}
Typically the $Q^3$ terms induced by $K_{L,R}$ in (\ref{v-idef}) are tadpoles
(see e.g. (\ref{pin-example},\ref{wt-tadpole})). The only non-trivial contribution arise
from the on-shell reducible parts of the u-channel baryon octet terms. However, by construction
those contributions have the same form as the irreducible $Q^3$ loop-correction terms of
$\bar K$. In particular they do not show the typical enhancement factor of $2 \pi$ associated
with the s-channel unitarity cuts. In the large-$N_c$ limit all loop correction terms to
$V$ are necessarily suppressed by $1/N_c$. This follows, because any hadronic loop
function if visualized in terms of quark-gluon diagrams involves at least one quark-loop,
which in turn is $1/N_c$ suppressed \cite{Hooft,Witten}. Thus, it is legitimate to take
$V = \bar K$ in this work.

The baryon octet and decuplet exchange contributions ask for special attention.
Individually the baryon exchange diagrams are of forbidden order $N_c$.
Only the complete large-$N_c$ baryon ground state multiplet with
$J=({\textstyle {1 \over 2}}, ...,{\textstyle {N_c\over 2}} )$ leads to an exact cancellation
and a scattering amplitude of order $N_c^0$ \cite{Manohar}. However this cancellation persists
only in the limit of degenerate baryon octet $\m0_{[8]}$ and decuplet mass $\m0_{[10]}$.
With $m_\pi < \m0_{[10]}-\m0_{[8]}$ the cancellation is incomplete and thus leads to an
enhanced sensitivity of the scattering amplitude on the physical baryon-exchange contributions.
Therefore one should sum the $1/N_c$ suppressed contributions of the form
\begin{eqnarray}
\left(\frac{\m0_{[10]}-\m0_{[8]}}{m_\pi} \right)^n \sim \frac{1}{N_c^n}\;.
\label{}
\end{eqnarray}
This is taken into account by evaluating the
baryon-exchange contributions to subleading chiral orders but avoid the expansion in either
$(\m0_{[10]}-\m0_{[8]})/m_\pi $ or $m_\pi/(\m0_{[10]}-\m0_{[8]})$. We also include the $SU(3)$
symmetry-breaking counter terms of the 3-point meson-baryon vertices and the quasi-local
two-body counter terms of chiral order $Q^3$ which are leading in the $1/N_c$ expansion.
Note that the quasi-local two-body counter terms of large chiral order
are not necessarily suppressed by $1/N_c$ relatively to the terms of
low chiral order. This is plausible, because for example a t-channel
vector-meson exchange, which has a definite large-$N_c$ scaling behavior, leads to
contributions in all partial waves. Thus, quasi-local counter terms with different partial-wave
characteristics may have identical large-$N_c$ scaling behavior even though they
carry different chiral powers.  In chapter 3.3 it was observed that reducible diagrams are enhanced
close to their unitarity threshold. The typical enhancement factor of $2 \pi$ per unitarity
cut, measured relatively to irreducible diagrams (see (\ref{wt-square-KN})), is larger than
the number of colors $N_c=3$ of our world. Therefore it is justified to perform the
partial $1/N_c$ resummation of all reducible diagrams implied by solving the on-shell
reduced Bethe-Salpeter equation (\ref{t-eff}).

By analogy with (\ref{t-sum}) the coupled-channel scattering amplitudes
$T^{(I)}_{ab} $ are decomposed into their on-shell equivalent
partial-wave amplitudes $M^{(I,\pm)}_{ab} $
\begin{eqnarray}
\bar T^{(I)}_{ab}(\bar k,k; w)&=& \sum_{n=0}^\infty\,M^{(I,+)}_{ab}(\sqrt{s};n)\,{Y}_n^{(+)}(\bar q, q;w)
\nonumber\\
&+&\sum_{n=0}^\infty\,M^{(I,-)}_{ab}(\sqrt{s};n)\,{Y}_n^{(-)}(\bar q, q;w) \;,
\label{tttt}
\end{eqnarray}
where $k=\! {\textstyle{1\over 2}}\,w -q$ and
$\bar k=\! {\textstyle{1\over 2}}\,w -\bar q$ and $s= w_\mu\, w^\mu$. The covariant
projectors ${Y}_n^{(\pm )}(\bar q, q;w)$ were introduced in (\ref{cov-proj}).
Equations for the differential cross sections as expressed in
terms of the partial-wave amplitudes $M_{ab}^{(\pm)}$ can be found in \cite{Lutz:Kolomeitsev}.
The form of the scattering amplitude (\ref{tttt}) follows, because
the effective interaction kernel $V^{(I)}_{ab}$ of (\ref{v-eff}) is decomposed
accordingly
\begin{eqnarray}
V^{(I)}_{ab}(\bar k,k;w)&=&
\sum_{n=0}^\infty\,V^{(I,+)}_{ab}(\sqrt{s};n)\,{Y}_n^{(+)}(\bar q, q;w)
\nonumber\\
&+&\sum_{n=0}^\infty\,V^{(I,-)}_{ab}(\sqrt{s};n)\,{Y}_n^{(-)}(\bar q, q;w) \;.
\label{not-def}
\end{eqnarray}
The coupled-channel Bethe-Salpeter equation (\ref{BS-coupled}) reduces to
a convenient matrix equation for the effective interaction kernel $V^{(I)}_{ab}$
and the invariant amplitudes $M^{(I,\pm)}_{ab} $
\begin{eqnarray}
M^{(I,\pm)}_{ab}(\sqrt{s};n) &=&  V^{(I,\pm )}_{ab}(\sqrt{s};n)
\nonumber\\
&+& \sum_{c,d}
V^{(I,\pm)}_{ac}(\sqrt{s};n)\,J^{(I,\pm )}_{cd}(\sqrt{s};n)\,M^{(I,\pm)}_{db}(\sqrt{s};n) \;,
\label{}
\end{eqnarray}
which is readily solved with:
\begin{eqnarray}
M^{(I,\pm)}_{ab}(\sqrt{s};n)&=& \Bigg[\left( 1- V^{(I,\pm )}_{}(\sqrt{s};n)\,J^{(I,\pm )}_{}(\sqrt{s};n)\right)^{-1}
V^{(I,\pm )}_{}(\sqrt{s};n)\Bigg]_{ab} \,.
\label{}
\end{eqnarray}
It remains to specify the coupled-channel loop matrix function
$J^{(I,\pm )}_{ab} \!\! \sim \delta_{a b }$, which is diagonal in the coupled
channel space. The diagonal elements read
\begin{eqnarray}
&& J^{(I,\pm )}_{aa}(\sqrt{s};n)=
\left( \frac{\sqrt{s}}{2}+ \frac{m_{B(I,a)}^2
-m_{\Phi(I,a)}^2}{2\,\sqrt{s}}\pm m_{B(I,a)} \right)
\Delta I^{(k)}_{\Phi(I,a)\,B(I,a)}(\sqrt{s}\, )
\nonumber\\
&& \qquad \qquad \qquad \! \times
\left(\frac{s}{4}-\frac{m_{B(I,a)}^2+m_{\Phi(I,a)}^2}{2}
+\frac{\big(m_{B(I,a)}^2-m_{\Phi(I,a)}^2\big)^2}{4\,s} \right)^n
\;,
\nonumber\\
&& \Delta I^{(k)}_{\Phi(I,a)\,B(I,a)}(\sqrt{s}\,) = I_{\Phi(I,a)\,B(I,a)}
(\sqrt{s}\,)
\nonumber\\
&& \qquad \qquad \qquad \quad -
\sum_{l=0}^k\,\frac{1}{l} \left( \sqrt{s}-\mu_S\right)^l
\left(\frac{\partial }{\partial \,\sqrt{s}} \right)^l\,
\Bigg|_{\sqrt{s}= \mu_S}\,I_{\Phi(I,a)\,B(I,a)}(\sqrt{s}\,)\;.
\label{result-loop:ab}
\end{eqnarray}
The index $a$ labels a specific channel consisting of a  meson-baryon pair of given
isospin ($\Phi(I,a)$, $B(I,a)$). In particular $m_{\Phi(I,a)}$ and $m_{B(I,a)}$ denote
the empirical isospin averaged meson and baryon octet mass respectively.
The scalar master-loop integral, $I_{\Phi(I,a)\,B(I,a)}(\sqrt{s}\,)$,
was introduced explicitly in (\ref{ipin-analytic}) for the pion-nucleon channel.
Note that we do not use the expanded form of (\ref{ipin-expand}).
The subtraction point is identified with $\mu_S\!=\!m_\Lambda$ and
$\mu_S\!=\!m_\Sigma$ in the isospin zero and isospin one channel respectively
so as to protect the hyperon s-channel pole structures. Similarly we use
$\mu_S=m_N$ in the pion-nucleon sectors. In the p-wave loop functions $J^{(-)}(\sqrt{s},0)$ and
$J^{(+)}(\sqrt{s},1)$  a double subtraction of the internal master-loop
function with $k=1$ in (\ref{result-loop:ab}) is performed
whenever a large-$N_c$ baryon ground state manifests itself with a
s-channel pole contribution in the associated partial-wave scattering amplitude.
In all remaining channels it is used $k=0$ in (\ref{result-loop:ab}).
This leads to consistency with the renormalization condition
(\ref{ren-cond}), in particular in the large-$N_c$ limit with $m_{[8]}=m_{[10]}$.

The on-shell irreducible interaction kernel $\bar K^{(I)}_{ab}$
in (\ref{v-idef}) needs to be identified. The result (\ref{bark-def-pin}) is generalized to
the case of inelastic scattering. This leads to the matrix structure of the interaction
kernel $K_{ab}$ defined in (\ref{r-def}). In a given partial wave the effective interaction kernel
$\bar K^{(I,\pm )}_{ab}(\sqrt{s};n)$ reads:
\begin{eqnarray}
&&\bar K ^{(I)}_{ab}=
\sum_{n=0}^\infty \left(\bar K^{(I,+)}_{ab}(\sqrt{s};n)\,{Y}_n^{(+)}(\bar q, q;w)
+\bar K^{(I,-)}_{ab}(\sqrt{s};n)\,{Y}_n^{(-)}(\bar q, q;w)\right) \;,
\nonumber\\
&&\bar K^{(I,\pm)}_{ab}(\sqrt{s};n)= \int_{-1}^1 \frac{dz}{2} \,
\frac{K^{(I,\pm)}_{ab}(s,t^{(I)}_{ab})}{\big( p^{(I)}_{a}\,
p^{(I)}_{b}\big)^{n}}\,P_n(z)
\nonumber\\
&&\;\;\;\;\;\;\;\;\;\;\;+\int_{-1}^1 \frac{dz}{2} \,
\,\Big(E^{(I)}_a\mp m_{B(I,a)} \Big) \,\Big(E^{(I)}_b \mp m_{B(I,b)} \Big)
\frac{K^{(I,\mp)}_{ab}(s,t^{(I)}_{ab})}{\big( p^{(I)}_{a}\,
p^{(I)}_{b}\big)^{n+1}}\,P_{n+1}(z)\;,
\label{bark-def}
\end{eqnarray}
where $P_n(z)$ are the Legendre polynomials and
\begin{eqnarray}
&&t^{(I)}_{ab}= m_{\Phi(I,a)}^2+m_{\Phi(I,b)}^2 -2\,\omega^{(I)}_a\,\omega^{(I)}_b
+2\,p^{(I)}_a\,p^{(I)}_b\,z \;,\;\;\;
E^{(I)}_a=\sqrt{s}-\omega^{(I)}_a\;,
\nonumber\\
&&\omega^{(I)}_a=\frac{s+m^2_{\Phi(I,a)}-m^2_{B(I,a)}}{2\,\sqrt{s}}\;,\;\;\;
\Big(p^{(I)}_{a}\Big)^2 =\Big(\omega^{(I)}_a\Big)^2-m_{\Phi(I,a)}^2\;.
\label{}
\end{eqnarray}
The construction of the on-shell irreducible interaction kernel requires the identification of the
invariant amplitudes $K_{ab}^{(I,\pm )}(s,t) $ in a given channel $ab$:
\begin{eqnarray}
&&K^{(I)}_{ab} \Big|_{\rm{on-sh.}}
= \frac{1}{2}\,\Bigg(\frac{\wslash}{\sqrt{w^2}}+1
\Bigg)\, K_{ab}^{(I,+)}(s,t) + \frac{1}{2}\,\Bigg(\frac{\wslash}{\sqrt{w^2}}-1
\Bigg)\,K_{ab}^{(I,-)}(s,t) \;.
\label{}
\end{eqnarray}

It is underlined that the approach developed here
deviates from the common chiral expansion scheme as implied by
the heavy-fermion representation of the chiral Lagrangian. A strict chiral expansion of the
unitarity loop function $I_{\Phi(I,a)\,B(I,a)}(\sqrt{s}\,)$ does not reproduce the correct
s-channel unitarity cut. One must perform an infinite summation of interaction terms in
the heavy-fermion chiral Lagrangian to recover the correct threshold behavior.
This is achieved more conveniently  by working directly with the manifest relativistic scheme,
where it is natural to write down the loop functions in terms of the physical masses.
In this work results are systematically expressed in terms of physical parameters avoiding the
use of bare parameters like $\m0_{[8]}$ whenever possible.

\vskip1.5cm \section{Construction of the effective interaction kernel}

After having specified the basic ingredients of the chiral coupled channel approach
it is necessary to collect all terms of the chiral Lagrangian as they
contribute to the interaction kernel $K^{(I)}_{ab}(\bar k ,k ;w )$ of (\ref{BS-coupled})
to chiral order $Q^3$:
\begin{eqnarray}
K^{(I)}(\bar k,k;w) &=& K^{(I)}_{WT}(\bar k,k;w)+K^{(I)}_{s-[8]}(\bar k,k;w)
+K^{(I)}_{u-[8]}(\bar k,k;w)
\nonumber\\
&+&K^{(I)}_{s-[10]}(\bar k,k;w)
+K^{(I)}_{u-[10]}(\bar k,k;w)+K^{(I)}_{s-[9]}(\bar k,k;w)
\nonumber\\
&+&K^{(I)}_{u-[9]}(\bar k,k;w)
+K^{(I)}_{[8][8]}(\bar k,k;w)+K^{(I)}_{\chi}(\bar k,k;w) \,.
\label{k-all}
\end{eqnarray}
The contributions to the interaction kernel will be  written in a form
which facilitates the derivation of $K^{(I,\pm)}_{ab}(s,t)$ in (\ref{bark-def}).
It is then straightforward to derive the on-shell irreducible interaction kernel $V$.

To start the discussion of the various terms in (\ref{k-all}) consider the Weinberg-Tomozawa
interaction term $K^{(I)}_{WT}$ in (\ref{lag-Q}). To chiral order
$Q^3 $ the effects of the baryon wave-function
renormalization factors ${\mathcal Z}$ and of further counter terms
introduced in (\ref{chi-sb-4}) must be considered. All together one finds
\begin{eqnarray}
\Big[\,K^{(I)}_{W\,T\,}(\bar k,k;w)\Big]_{ab}&=&
\Big[C_{WT}^{(I)}\Big]_{ab}\,
\Big(1+{\textstyle{1\over 2}}\,\Delta \zeta_{B(I,a)}
+{\textstyle{1\over 2}}\,\Delta \zeta_{B(I,b)} \Big)
\,\frac{\barqslash+\qslash}{4\,f^2}\;.
\label{WT-k}
\end{eqnarray}
The dimensionless coefficient matrix
$C_{WT}^{(I)}$ of the Weinberg-Tomozawa interaction is given in Tab. \ref{tabkm-1} for the
strangeness minus one channels and in \cite{Lutz:Kolomeitsev} for the strangeness zero channels.
The  constants $\Delta \zeta_{B(c)}$ in (\ref{WT-k}) receive
contributions from the baryon octet wave-function renormalization factors ${\mathcal Z}$ and
the parameters $\zeta_0,\zeta_D$ and $\zeta_F$ of (\ref{chi-sb-4}):
\begin{eqnarray}
&&\Delta \zeta = \zeta +{\mathcal Z}-1 +\cdots \,,
\nonumber\\
&&\zeta_N = \big( \zeta_0+2\,\zeta_F  \big)\, m_\pi^2
+ \big( 2\,\zeta_0 +2\,\zeta_D-2\,\zeta_F\big) \,m_K^2 \;,
\nonumber\\
&& \zeta_\Lambda =\big( \zeta_0-{\textstyle{2\over 3}}\,\zeta_D  \big)\, m_\pi^2
+ \big( 2\,\zeta_0 +{\textstyle{8\over 3}}\,\zeta_D\big) \,m_K^2 \;,
\nonumber\\
&& \zeta_\Sigma =\big( \zeta_0+2\,\zeta_D  \big)\, m_\pi^2 + 2\,\zeta_0  \,m_K^2 \;,
\nonumber\\
&& \zeta_\Xi = \big( \zeta_0-2\,\zeta_F  \big)\, m_\pi^2
+ \big( 2\,\zeta_0 +2\,\zeta_D+2\,\zeta_F\big) \,m_K^2 \;.
\label{zeta-par}
\end{eqnarray}
The dots in (\ref{zeta-par}) represent corrections terms of order
$Q^3$ and further contributions from irreducible one-loop diagrams not considered here.
It is pointed out that the coupling constants $\zeta_0$, $\zeta_D$ and $\zeta_F$, which
appear to renormalize the strength of the Weinberg-Tomozawa interaction term, also
contribute to the baryon wave-function factor ${\mathcal Z}$ as is
evident from (\ref{chi-sb}). In fact, it will be demonstrated that the explicit and
implicit dependence via the wave-function renormalization factors ${\mathcal Z}$ cancel
identically to leading order. Generalizing (\ref{def-zn})  the
wave-function renormalization constants for the $SU(3)$ baryon octet fields are derived
\begin{eqnarray}
&& {\mathcal Z}^{-1}_{B(c)}-1 = \zeta_{B(c)}
-\frac{1}{4\,f^2}\,\sum_{a} \,\xi_{B(I,a)}\,m_{\Phi(I,a)}^2
\left( G^{(B(c))}_{\Phi (I,a)\, B(I,a)} \right)^2 \,,
\label{zeta-result}\\
&& \xi_{B(I,a)}=  2\,m_{B(c)}\,
\frac{\partial \,I_{\Phi (I,a)\, B(I,a)}(\sqrt{s}\,)}{\partial \sqrt{s}}
\Bigg|_{\sqrt{s}=m_{B(c)}} -\frac{I_{\Phi (I,a)}}{m_{\Phi(I,a)}^2 }
+{\mathcal O}\left( Q^4\right)\,,
\nonumber
\end{eqnarray}
where the sum in (\ref{zeta-result}) includes all $SU(3)$ channels as listed in (\ref{r-def}).
The result (\ref{zeta-result}) is  expressed in terms of the dimensionless
coupling constants $G_{\Phi B}^{(B)}$. For completeness  all
required 3-point coupling coefficients are collected:
\begin{eqnarray}
&&G_{\pi N}^{(N)} = \sqrt{3}\,\Big( F+D \Big) \;, \quad
G_{K \,\Sigma}^{(N)} = -\sqrt{3}\,\Big( F-D \Big) \;,\quad
G_{\eta \,N}^{(N)} = {\textstyle{1\over\sqrt{3}}}\,\Big( 3\,F-D \Big)\;,
\nonumber\\
&& G_{K \,\Lambda}^{(N)} = -{\textstyle{1\over\sqrt{3}}}\,\Big( 3\,F+D \Big) \;,\quad
G_{\bar K N}^{(\Lambda)} = \sqrt{2}\,G_{K \,\Lambda}^{(N)}
\;, \quad G_{\pi \Sigma}^{(\Lambda)} = 2\,D \;,
\nonumber\\
&& G_{\eta \,\Lambda}^{(\Lambda)} = -{\textstyle{2\over\sqrt{3}}}\,D \;,\quad
G_{K \Xi}^{(\Lambda)} = -\sqrt{{\textstyle{2\over 3}}}\,\Big( 3\,F-D\Big)  \;,
\quad  G_{\bar K N}^{(\Sigma)} =
\sqrt{{\textstyle{2\over 3}}}\,G_{K \Sigma}^{(N)}\;,
\nonumber\\
&& G_{\pi \Sigma}^{(\Sigma)} = -\sqrt{8}\,F \;, \quad
G_{\pi \Lambda}^{(\Sigma)} =
{\textstyle{1\over\sqrt{3}}}\,G_{\pi \Sigma}^{(\Lambda)} \;,\quad
G_{\eta \Sigma }^{(\Sigma)} = {\textstyle{2\over\sqrt{3}}}\,D\;,
\nonumber\\
&&G_{K\Xi}^{(\Sigma)} = \sqrt{2}\,\Big( F+D\Big) \;,\quad
G_{\bar K \Lambda}^{(\Xi)} = -{\textstyle{1\over\sqrt{2}}}\,
G_{K \Xi}^{(\Lambda)}\;, \quad
G_{\bar K \Sigma}^{(\Xi)} = -\sqrt{{\textstyle{3\over2}}}\,
G_{K \Xi}^{(\Sigma)}\;,
\nonumber\\
&&G_{\eta \,\Xi}^{(\Xi)} = -{\textstyle{1\over\sqrt{3}}}\,\Big( 3\,F+D \Big)\;,\quad
G_{\pi \,\Xi}^{(\Xi)} = \sqrt{3}\,\Big( F-D \Big) \;,
\label{G-explicit}
\end{eqnarray}
to leading chiral order $Q^0$. The loop function $I_{\Phi (I,a)\, B(I,a)}(\sqrt{s}\,)$ and
the mesonic tadpole term $I_{\Phi(I,a)}$ are the obvious generalizations of
$I_{\pi N}(\sqrt{s}\,)$ and $I_\pi$ introduced in (\ref{ipin-analytic}).
It is emphasized that the contribution $\zeta_{B(c)}$ in (\ref{zeta-result}) is identical
to the corresponding contribution in $ \Delta \zeta_{B(c)}$.
Therefore, if all one-loop effects are dropped, one finds the result
\begin{eqnarray}
{\mathcal Z}^{-1}=1 +\zeta \;, \qquad
\Delta \zeta =0 +{\mathcal O} \left(Q^3, \frac{1}{N_c}\,Q^2 \right) \;.
\label{delta-zeta}
\end{eqnarray}
Given the renormalization condition (\ref{ren-cond}), which requires the
baryon s-channel pole contribution to be represented by the renormalized effective
potential $V_R$, it follows that (\ref{delta-zeta}) holds here.
This argument relies on the requirement that the loop correction to ${\mathcal Z}$
as given in (\ref{zeta-result}) must be considered as part of the renormalized
effective potential $V_R$. Consequently such loop corrections should be dropped as
they are suppressed by the factor $Q^3/N_c$. It is evident that the one-loop contribution
to the baryon ${\mathcal Z}$-factor is naturally moved into $V_R$ by that double subtraction
with $k=1$ in (\ref{result-loop:ab}), as explained previously.

\tabcolsep=1.45mm
\renewcommand{\arraystretch}{1.4}
\begin{table}[p]
\begin{tabular}{|r||c||c|c|c||c|c||c|c|c|c||c|c|c|c|}
 \hline
& $ C_{WT}^{(0)} $  &  $ C_{N_{[8]}}^{(0)}$  & $C_{\Lambda_{[8]}}^{(0)}$  &
$C_{\Sigma_{[8]}}^{(0)}$
& $C_{\Delta_{[10]}}^{(0)}$ & $C_{\Sigma_{[10]} }^{(0)}$
    &$\widetilde{C}_{N_{[8]}}^{(0)}$ & $\widetilde{C}_{\Lambda_{[8]} }^{(0)}$ &
    $\widetilde{C}_{\Sigma_{[8]}}^{(0)}$ & $\widetilde {C}_{\Xi_{[8]} }^{(0)}$   &
$\widetilde {C}_{\Delta_{[10]} }^{(0)}$ &
    $\widetilde {C}_{\Sigma_{[10]}}^{(0)}$& $\widetilde C_{\Xi_{[10]}}^{(0)}$
\\ \hline \hline
$11$& $3$ & 0& $1$  & $0$ & 0 & 0  &0  &0  & 0 &0 &0 &0 & 0\\ \hline
$12$& $\sqrt{\frac{3}{2}}$ & 0& $1$  & $0$ & 0 & 0  &$\sqrt{\frac{2}{3}}$  &0  & 0 &0 &$2\sqrt{\frac{2}{3}}$ &0 & 0\\ \hline
$13$& $\frac{3}{\sqrt{2}}$ & 0& $1$  & $0$  & 0& 0  & $\sqrt{2}$  &0  & 0 &0 &0 &0 & 0\\ \hline
$14$& $0$ & 0& $1$  & $0$  & 0 & 0  &0  & $\frac{1}{2}$  & -$\frac{3}{2}$ &0 &0 &-$\frac{3}{2}$ & 0\\ \hline \hline

$22$& $4$ & 0& $1$  & $0$  & 0 & 0  &0  &$\frac{1}{3}$  & -1 &0 &0 &-1 & 0\\ \hline
$23$& $0$& 0 & $1$  & $0$  & 0& 0  &0  &0  & $\sqrt{3}$ &0 &0 &$\sqrt{3}$ & 0\\ \hline
$24$& -$\sqrt{\frac{3}{2}}$ & 0 & $1$  & $0$ & 0 & 0  &0  &0  & 0 &-$\sqrt{\frac{2}{3}}$ &0 &0 & -$\sqrt{\frac{2}{3}}$\\ \hline  \hline

$33$& $0$ & 0 & $1$  & $0$  & 0& 0  &0  &1  & 0 &0 &0 &0 & 0\\ \hline
$34$& -$\frac{3}{\sqrt{2}}$ & 0 & $1$  & $0$ & 0 & 0  &0  &0  & 0 &-$\sqrt{2}$ &0 &0 & -$\sqrt{2}$\\ \hline \hline

$44$& $3$& 0  & $1$  & $0$  & 0 & 0  &0  &0  & 0 &0 &0 &0 & 0\\ \hline \hline

\hline
& $ C_{WT}^{(1)} $  &  $ C_{N_{[8]}}^{(1)}$  & $C_{\Lambda_{[8]}}^{(1)}$  &
$C_{\Sigma_{[8]}}^{(1)}$
& $C_{\Delta_{[10]}}^{(1)}$ & $C_{\Sigma_{[10]} }^{(1)}$
    &$\widetilde{C}_{N_{[8]}}^{(1)}$ & $\widetilde{C}_{\Lambda_{[8]} }^{(1)}$ &
    $\widetilde{C}_{\Sigma_{[8]}}^{(1)}$ & $\widetilde {C}_{\Xi_{[8]} }^{(1)}$   &
$\widetilde {C}_{\Delta_{[10]} }^{(1)}$ &
    $\widetilde {C}_{\Sigma_{[10]}}^{(1)}$& $\widetilde C_{\Xi_{[10]}}^{(1)}$
\\ \hline \hline
$11$& 1 & 0& 0 & 1 & 0 & 1  &0  &0  & 0 &0 &0 &0 & 0\\ \hline
$12$& 1 & 0& 0 & 1 & 0 & 1  &-$\frac{2}{3}$  &0  & 0 &0 &$\frac{2}{3}$ &0 & 0\\ \hline
$13$& $\sqrt{\frac{3}{2}}$ & 0 & $0$  & $1$ & 0 & 1  &$\sqrt{\frac{2}{3}}$  &0  & 0 &0 &0 &0 & 0\\ \hline
$14$& $\sqrt{\frac{3}{2}}$ & 0 & $0$  & $1$ & 0 & 1  &$\sqrt{\frac{2}{3}}$  &0  & 0 &0 &0 &0 & 0\\ \hline
$15$& $0$ & 0 & $0$  & $1$  & 0 & 1  &0  &-$\frac{1}{2}$  & -$\frac{1}{2}$ &0 &0 &-$\frac{1}{2}$ & 0\\ \hline \hline

$22$& $2$ & 0 & $0$  & $1$ & 0 & 1  &0  &-$\frac{1}{3}$  & $\frac{1}{2}$ &0 &0 &$\frac{1}{2}$ & 0\\ \hline
$23$& $0$ & 0 & $0$  & $1$ & 0 & 1  &0  &0  & -1 &0 &0 &-1 & 0\\ \hline
$24$& $0$ & 0 & $0$  & $1$ & 0 & 1  &0  &0  & 1 &0 &0 &1 & 0\\ \hline
$25$& $-1$ & 0 & $0$  & $1$ & 0 & 1  &0  &0  & 0 &$\frac{2}{3}$ &0 &0 & $\frac{2}{3}$\\ \hline  \hline

$33$& $0$ & 0& $0$  & $1$ & 0 & 1  &0  &0  & 1 &0 &0 &1 & 0\\ \hline
$34$& $0$ & 0& $0$  & $1$  & 0& 1  &0  &$\frac{1}{\sqrt{3}}$ & 0 &0 &0 &0 & 0\\ \hline
$35$& $\sqrt{\frac{3}{2}}$ & 0 & $0$  & $1$& 0  & 1  &0  &0  & 0 &-$\sqrt{\frac{2}{3}}$ &0 &0 & -$\sqrt{\frac{2}{3}}$\\ \hline \hline

$44$& $0$ & 0 & $0$  & $1$ & 0 & 1  &0  &0  & 1 &0 &0 &1 & 0\\ \hline
$45$& $\sqrt{\frac{3}{2}}$ & 0 & $0$  & $1$ & 0 & 1  &0  &0  & 0 &-$\sqrt{\frac{2}{3}}$ &0 &0 & -$\sqrt{\frac{2}{3}}$\\ \hline  \hline

$55$& $1$ & 0 & $0$  & $1$ & 0 & 1  &0  &0  & 0 &0 &0 &0 & 0\\ \hline \hline

& $ C_{WT}^{(2)} $  &  $ C_{N_{[8]}}^{(2)}$  & $C_{\Lambda_{[8]}}^{(2)}$  &
$C_{\Sigma_{[8]}}^{(2)}$
& $C_{\Delta_{[10]}}^{(2)}$ & $C_{\Sigma_{[10]} }^{(2)}$
    &$\widetilde{C}_{N_{[8]}}^{(2)}$ & $\widetilde{C}_{\Lambda_{[8]} }^{(2)}$ &
    $\widetilde{C}_{\Sigma_{[8]}}^{(2)}$ & $\widetilde {C}_{\Xi_{[8]} }^{(2)}$   &
$\widetilde {C}_{\Delta_{[10]} }^{(2)}$ &
    $\widetilde {C}_{\Sigma_{[10]}}^{(2)}$& $\widetilde C_{\Xi_{[10]}}^{(2)}$
\\ \hline \hline
$11$& -2 & 0& 0 & 0 & 0 & 0  &0  &$\frac{1}{3}$ & $\frac{1}{2}$ &0 &0 &$\frac{1}{2}$ & 0\\ \hline

\end{tabular}
\vspace*{2mm} \caption{Weinberg-Tomozawa interaction strengths and baryon exchange
coefficients of the strangeness minus one channels as defined in
(\ref{WT-k}) and (\ref{wave-ren}).}
\label{tabkm-1}
\end{table}

It is illuminating to explore whether the parameters
$\zeta_0, \zeta_D $ and $\zeta_F $ can be dialed to give ${\mathcal Z}=1$. In
the $SU(3)$ limit with degenerate meson and also degenerate baryon masses
one finds a degenerate wave-function renormalization factor ${\mathcal Z}$
\begin{eqnarray}
&&{\mathcal Z}^{-1}-1=3\,\zeta_0+2\,\zeta_D
- \frac{m_\pi^2}{(4\pi\,f)^2}
\left( \frac{5}{3}\,D^2+3\,F^2\right)
\Bigg( -4 -3\,\ln \left(\frac{m_\pi^2}{m_N^2}\right)
\nonumber\\
&& \qquad \qquad \qquad \qquad \qquad \qquad \qquad
+3\,\pi\,\frac{m_\pi}{m_N}
+{\mathcal O}\left( \frac{m^2_\pi}{m^2_N} \right)
\Bigg) \;,
\label{su3-limit:z}
\end{eqnarray}
where the minimal chiral subtraction prescription (\ref{def-sub}) was applied.
The result (\ref{su3-limit:z}) agrees identically with the $SU(2)$-result (\ref{def-zn})
if one formally replaces  $g^2_A $ by $4\, F^2+ {\textstyle{20\over 9}}\,D^2 $
in (\ref{def-zn}). It is clear from (\ref{su3-limit:z}) that in the $SU(3)$ limit the
counter term $3\,\zeta_0+2\,\zeta_D$ may be dialed so as to impose ${\mathcal Z}=1$. This
is no longer possible once the explicit $SU(3)$ symmetry-breaking effects are included.
Note however that consistency of the perturbative renormalization procedure suggests that
for hypothetical $SU(3)$-degenerate $\xi$-factors in (\ref{zeta-result}) it should be
possible to dial $\zeta_0$, $\zeta_D$ and $\zeta_F$ so as to find ${\mathcal Z}_B =1$
for all baryon octet fields. This is expected, because for example in the
$\overline {MS}$-scheme
one finds a $SU(3)$-symmetric renormalization scale\footnote{Note that in the
minimal chiral subtraction scheme the counter terms $\zeta_{0,D,F}$ are already renormalization
scale independent. The consistency of this procedure follows from the symmetry conserving
property of dimensional regularization.} dependence of $\xi_{B(c)}$
in (\ref{zeta-result}) with $\xi_{B(c)}  \sim  \ln \mu^2$. Indeed in this case the choice
\begin{eqnarray}
&&\zeta_0 = \frac{\xi}{4\,f^2}\left(\frac{26}{9}\,D^2+2\,F^2\right)\, , \qquad
\zeta_D = \frac{\xi}{4\,f^2}\left(-D^2+3\,F^2\right)\,, \quad
\nonumber\\
&& \zeta_F = \frac{\xi}{4\,f^2}\,\frac{10}{3}\,D\,F \,,
\label{}
\end{eqnarray}
would lead to ${\mathcal Z}=1$ for all baryon octet wave functions.
The $SU(3)$ symmetry-breaking effects in the baryon wave-function renormalization factors
will be discussed again below. They affect the meson-baryon 3-point vertices
at subleading order.

The s- and u-channel exchange diagrams of the baryon octet
contribute as follows
\begin{eqnarray}
\Big[K^{(I)}_{s-[8]}(\bar k,k;w)\Big]_{ab} &=& -\sum_{c\,=\,1}^3\,
\Big(\barqslash -R^{(I,c)}_{L,ab}\Big)\,
\frac{\Big[C^{(I,c)}_{[8]}\Big]_{ab}}
{4\,f^2\,\big(\wslash+m^{(c)}_{[8]}\big)}\,\Big( \qslash -R^{(I,c)}_{R,ab}\Big)\;,
\nonumber\\
\Big[K^{(I)}_{u-[8]}(\bar k,k;w)\Big]_{ab}  &=&
\sum_{c\,=\,1}^4\,
\frac{\Big[\widetilde C^{(I,c)}_{[8]}\Big]_{ab} }{4\,f^2 }\,\Bigg(
\wslash+m^{(c)}_{\,[8]}+\tilde R^{(I,c)}_{L,ab}+ \tilde R^{(I,c)}_{R,ab}
\nonumber\\
&-& \Big(\barpslash+m^{(c)}_{\,[8]}+ \tilde R^{(I,c)}_{L,ab} \Big)\,
\frac{1}{\barwslash+m^{(c)}_{\,[8]} }
\, \Big(\pslash+m^{(c)}_{\,[8]}+ \tilde R^{(I,c)}_{R,ab}\Big)\Bigg)\;,
\label{wave-ren}
\end{eqnarray}
where $\widetilde w_\mu=p_\mu-\bar q_\mu$.
The index $c$ in (\ref{wave-ren}) labels the baryon octet exchange with
$c\to (N,\Lambda, \Sigma ,\Xi)$. In particular $m^{(c)}_{[8]}$ denotes
the physical baryon octet masses and $R^{(I,c)}$ and  $\tilde R^{(I,c)}$ characterize
the ratios of pseudo-vector to pseudo-scalar terms in the meson-baryon vertices.
The dimensionless matrices $C^{}_{[8]}$ and $\widetilde C^{}_{[8]}$
give the renormalized strengths of the s-channel and
u-channel baryon exchanges respectively. They are expressed most conveniently
in terms of s-channel $C^{}_{B(c)}$ and u-channel $ \tilde C^{}_{B(c)}$ coefficient matrices
\begin{eqnarray}
&& \Big[C^{(I,c)}_{[8]}\Big]_{ab} = \Big[C^{(I)}_{B(c)}\Big]_{ab}\,
\bar A_{\Phi (I,a) B(I,a)}^{(B(c))}\,\bar A_{\Phi (I,b) B(I,b)}^{(B(c))} \;,
\nonumber\\
&& \Big[\widetilde C^{(I,c)}_{[8]}\Big]_{ab}=
\Big[\widetilde C^{(I)}_{B(c)}\Big]_{ab}\,
\bar A_{\Phi (I,b) B(I,a)}^{(B(c))}\,\bar A_{\Phi (I,a) B(I,b)}^{(B(c))}
 \;,
\label{def-not}
\end{eqnarray}
and the renormalized three-point coupling constants  $\bar A$.
According to the LSZ-scheme the bare coupling constants, $A$, are related to the renormalized
coupling constants, $\bar A$, by means of the  baryon octet ${\mathcal Z}$-factors:
\begin{eqnarray}
\bar A_{\Phi (I,a) B(I,b)}^{(B(c))}= {\mathcal Z}^{\frac{1}{2}}_{B(c)}\,A_{\Phi (I,a) B(I,b)}^{(B(c))}
\,{\mathcal Z}_{B(I,b)}^{\frac{1}{2}}\,.
\label{g-ren:z}
\end{eqnarray}
At leading order $Q^0$ the bare coupling constants $A = G(F,D)+{\mathcal O}\left(Q^2 \right)$ are
specified in (\ref{G-explicit}). The chiral correction terms to the coupling constants $A$ and
$R $ will be discussed in great detail later.
In the convention applied here the s-channel coefficients $C_{B(c)}$ are either one
or zero and the u-channel coefficients $\tilde C^{}_{B(c)}$ represent
appropriate Fierz factors resulting from the interchange of initial and final meson states.
To illustrate the notation explicitly the coefficients are listed in Tab. \ref{tabkm-1} for the
strangeness minus one channels. For the strangeness zero channel see \cite{Lutz:Kolomeitsev}.
It is stressed that the expressions for the s- and u-channel exchange contributions
(\ref{wave-ren}) depend on the result (\ref{delta-zeta}), because there would otherwise be a
factor ${\mathcal Z}^{-1}/(1+\zeta)$ in front of both contributions. That would necessarily
lead to an asymmetry in the treatment of s- versus u-channel, simply because the
s-channel contribution would be further renormalized by the unitarization.
In contrast, in the present scheme the proper balance of s- and
u-channel exchange contributions in the scattering amplitude as is required for the
realization of the large-$N_c$ cancellation mechanism \cite{Manohar} is observed.

The $SU(3)$ symmetry-breaking effects in the meson-baryon coupling constants are
particularly interesting. The 3-point vertices have pseudo-vector and pseudo-scalar
components determined by $F_i+\delta F_i$ and $\bar F_i$ of (\ref{chi-sb-3},\ref{chi-sb-3:p})
respectively. Collect first the symmetry-breaking terms in the axial-vector coupling
constants $A$ introduced in (\ref{g-ren:z}). They are of chiral order $Q^2$ but lead to
particular $Q^3$-correction terms in the scattering amplitudes. One finds
\begin{eqnarray}
&& A = G(F_A,D_A) -\frac{2}{\sqrt{3}}\,(m_K^2-m_\pi^2)\,\Delta A \;,\quad
\tilde F_i = F_i + \delta F_i \;,
\nonumber\\
&&\Delta A_{\pi N}^{(N)} =
3\,(F_1+F_3)-F_0-F_2+2\,(\tilde F_4+\tilde F_5) \;,
\nonumber\\
&&\Delta A_{K \Sigma}^{(N)} =
{\textstyle{3\over 2}}\,(F_1-F_3) +{\textstyle{1\over 2}}\,(F_0-F_2)-\tilde F_4+\tilde F_5 \;,
\nonumber\\
&&\Delta A_{\eta \,N}^{(N)} =
-F_1+3\,F_3+{\textstyle{1\over3}}\,F_0-F_2
+{\textstyle{2\over3}}\,\tilde F_4-2\,\tilde F_5 +2\,(\tilde F_6+{\textstyle{4\over3}}\,\tilde F_4) \;,
\nonumber\\
&&\Delta A_{K \Lambda}^{(N)} =
-{\textstyle{1\over 2}}\,F_1-{\textstyle{3\over 2}}\,F_3+{\textstyle{1\over 2}}\,F_0+
{\textstyle{3\over2}}\,F_2 +{\textstyle{1\over3}}\,\tilde F_4+\tilde F_5
+(F_7+{\textstyle{4\over 3}}\,F_0)\;,
\nonumber\\
&& \Delta A_{\bar K N}^{(\Lambda)} = \sqrt{2}\,\Delta A_{K \,\Lambda}^{(N)} \,,\quad
\nonumber\\
&& \Delta A_{\pi \Sigma}^{(\Lambda)} =
 {\textstyle{4\over \sqrt{3}}}\,\tilde F_4 +\sqrt{3}\,(F_7+{\textstyle{4\over3}}\,F_0)  \;,
\nonumber\\
&&\Delta A_{\eta \,\Lambda}^{(\Lambda)} =
{\textstyle{4\over 3}}\,(F_0 +\tilde F_4)+2\,(\tilde F_6+{\textstyle{4\over3}}\,\tilde F_4)
+2\,(F_7+{\textstyle{4\over3}}\,F_0) \;,
\nonumber\\
&& \Delta A_{K \Xi}^{(\Lambda)} =
-{\textstyle{1\over \sqrt{2}}}\,(F_0+F_1)
+{\textstyle{3\over \sqrt{2}}}\,(F_2+F_3)
-{\textstyle{\sqrt{2}\over 3}}\,(\tilde F_4-3\,\tilde F_5)
-\sqrt{2}\,(F_7 +{\textstyle{4\over3}}\,F_0) \;,
\nonumber\\
&& \Delta A_{\bar K N}^{(\Sigma)} =
\sqrt{{\textstyle{2\over 3}}}\,\Delta A_{K \Sigma}^{(N)} \;, \quad
\Delta A_{\pi \Sigma}^{(\Sigma)} = -4\,\sqrt{{\textstyle{2\over 3}}}\,(F_2+\tilde F_5) \;, \quad
\Delta A_{\pi \Lambda}^{(\Sigma)} =
{\textstyle{1\over\sqrt{3}}}\,\Delta A_{\pi \Sigma}^{(\Lambda)} \;,
\nonumber\\
&& \Delta A_{\eta \Sigma }^{(\Sigma)} =  {\textstyle{4\over 3}}\,(F_0-\tilde F_4)
+2\,(\tilde F_6+{\textstyle{4\over3}}\,\tilde F_4) \;,
\nonumber\\
&& \Delta A_{K \Xi}^{(\Sigma)} =  -\sqrt{{\textstyle{3\over 2}}}\,(F_1+F_3)+
\sqrt{{\textstyle{1\over 6}}}\,(F_0+F_2)
-\sqrt{{\textstyle{2\over 3}}}\,(\tilde F_4+\tilde F_5)\;,
\nonumber\\
&& \Delta A_{\bar K \Lambda}^{(\Xi)} = -{\textstyle{1\over\sqrt{2}}}\,
\Delta A_{K \Xi}^{(\Lambda)} \;, \quad
\Delta A_{\bar K \Sigma}^{(\Xi)} = -\sqrt{{\textstyle{3\over2}}}\,
\Delta A_{K \Xi}^{(\Sigma)} \;,
\nonumber\\
&& \Delta A_{\eta \,\Xi}^{(\Xi)} =
F_1+3\,F_3+{\textstyle{1\over3}}\,F_0+F_2
+{\textstyle{2\over3}}\,\tilde F_4+2\,\tilde F_5+2\,(\tilde F_6+{\textstyle{4\over3}}\,\tilde F_4) \;,
\nonumber\\
&& \Delta A_{\pi \Xi}^{(\Xi)} = F_0+3\,F_1-F_2-3\,F_3+2\,(\tilde F_5-\tilde F_4) \;,
\label{G-explicit-new}
\end{eqnarray}
where the renormalized coupling constants $D_A$ and $F_A$
are given in (\ref{ren-FDC-had}) and $G(F,D)$ in (\ref{G-explicit}).
It is emphasized that to order $Q^2$ the effect of the ${\mathcal Z}$-factors in (\ref{g-ren:z})
can be accounted for by the following redefinition of the $F_R,D_R$ and
$F_i$ parameters in (\ref{G-explicit-new})
\begin{eqnarray}
&& F_R \to F_R +\, \Big( 2\,m_K^2+m_\pi^2\Big)\,
\Big(\zeta_0+{\textstyle{2\over 3}}\,\zeta_D\Big) \,F_R \;, \qquad
\nonumber\\
&& D_R \to D_R + \Big( 2\,m_K^2+m_\pi^2\Big)\,
\Big( \zeta_0+{\textstyle{2\over 3}}\,\zeta_D\Big) \,D_R \;,
\nonumber\\
&& F_0 \to F_0 +\zeta_D\,D_R\,, \qquad F_1 \to F_1+\zeta_F \,F_R
\,, \qquad F_2 \to F_2 +\zeta_D \,F_R\,,
\nonumber\\
&& F_3 \to F_3 + \zeta_F \,F_R\,,\qquad \,F_7 \to F_7 -{\textstyle{4\over 3}}\,\zeta_D\,\,D_R\,.
\label{zeta-redundant}
\end{eqnarray}
As a consequence the parameters $\zeta_0$, $\zeta_D$ and $\zeta_F$ are redundant in
the approximation applied here and can therefore be dropped. It is legitimate to identify the
bare coupling constants $A$ with the renormalized coupling constants $\bar A$.
The same renormalization holds for the matrix elements of the axial-vector current.

Recall the large-$N_c$ result (\ref{large-Nc-FDC}). The many parameters $F_i$ and
$\delta F_i$ in (\ref{G-explicit-new}) are expressed in terms of the seven parameters
$c_{i}$ and $\delta c_i= \tilde c_i-c_i-\bar c_i$. We give explicitly
the axial-vector coupling constants, which are most relevant for the pion-nucleon and
kaon-nucleon scattering processes, in terms of those large-$N_c$ parameters
\begin{eqnarray}
&& A_{\pi N}^{(N)} = \sqrt{3}\,\Big( F_A+D_A \Big) +
{\textstyle{5\over 3}}\, (c_1+\delta c_1) + c_2+\delta c_2 -a+ 5\, c_3 + c_4\;,
\nonumber\\
&& A_{\bar K N}^{(\Lambda)} = -\sqrt{{\textstyle{2\over3}}}\,\Big( 3\,F_A+D_A \Big)
+{\textstyle{1\over\sqrt{2}}}\,\Big(c_1+\delta c_1+c_2+\delta c_2-a-3\,c_3+c_4 \Big)\,,
\nonumber\\
&& A_{\bar K N}^{(\Sigma)} = -\sqrt{2 }\,\Big( F_A-D_A \Big)
-{\textstyle{1\over\sqrt{6}}}\,\Big({\textstyle{1\over 3}}\,(c_1+\delta c_1)-c_2-\delta c_2+a-c_3+3\,c_4 \Big)\,,
\nonumber\\
&& A_{\pi \Sigma}^{(\Lambda)} = 2\,D_A +{\textstyle{2\over\sqrt{3}}}\,(c_1+\delta c_1)\;,
\label{c1234}
\end{eqnarray}
where
\begin{eqnarray}
F_A = F_R
-\frac{\beta}{\sqrt{3}}\,\Big({\textstyle{2\over 3}}\,\delta c_1 +\delta c_2 -a\Big) \,,
\quad \; \;D_A = D_R - \frac{\beta}{\sqrt{3}}\,\delta c_1 \,, \quad
\beta = \frac{m_K^2+m_\pi^2/2}{m_K^2-m_\pi^2}  \,.
\label{c1234-b}
\end{eqnarray}
The result (\ref{c1234}) shows that the axial-vector coupling
constants $A_{\pi N}^{(N)}$, $A_{\bar K N}^{(\Lambda)}$, $A_{\bar K N}^{(\Sigma)}$ and
$A_{\pi \Sigma}^{(\Lambda)}$, can be fine tuned with $c_i $ in (\ref{large-Nc-FDC}) to be
off their $SU(3)$ limit values. Moreover, the SU(3) symmetric
contribution proportional to $ F_A$ or $ D_A$ deviate from their corresponding
$F_R$ and $D_R$ values relevant for matrix elements of the axial-vector current, once
non-zero values for $\delta c_{1,2}$ are established. However,
the values of $F_R, D_R$ and $c_i$ are strongly constrained by the weak decay widths of the
baryon-octet states (see Tab. \ref{weak-decay:tab}). Thus, the SU(3) symmetry-breaking pattern
in the axial-vector current and the one in the meson-baryon axial-vector coupling constants
are closely linked.

It is left to specify the pseudo-scalar part, $P$, of the meson-baryon vertices introduced
in (\ref{chi-sb-3:p}). In (\ref{wave-ren}) their effect was encoded
into the $R$ and $\tilde R $ parameters with
$R, \tilde R \sim \bar F_{}\sim \bar c_{},a_{}$. Applying the large-$N_c$ result
of (\ref{large-Nc-FDC:delta}) we obtain
\begin{eqnarray}
&& R_{L,ab}^{(I,c)} \,A_{\Phi (I,a)\, B(I,a)}^{(B(c))} =
\Big(m_{B(I,a)}+m_{B(c)} \Big)\,P_{\Phi (I,a)\, B(I,a)}^{(B(c))}
\;,
\nonumber\\
&& R_{R,ab}^{(I,c)} \,A_{\Phi (I,b)\, B(I,b)}^{(B(c))} =
\Big(m_{B(I,b)}+m_{B(c)} \Big)\,P_{\Phi (I,b)\, B(I,b)}^{(B(c))}
\;,
\nonumber\\
&& \tilde R_{L,ab}^{(I,c)} \,A_{\Phi (I,b)\, B(I,a)}^{(B(c))} =
\Big(m_{B(I,a)}+m_{B(c)} \Big)\,P_{\Phi (I,b)\, B(I,a)}^{(B(c))}
\;,
\nonumber\\
&& \tilde R_{R,ab}^{(I,c)} \,A_{\Phi (I,a)\, B(I,b)}^{(B(c))} =
\Big(m_{B(I,b)}+m_{B(c)} \Big)\,P_{\Phi (I,a)\, B(I,b)}^{(B(c))}
\;,
\nonumber\\ \nonumber\\
&&P_{\pi N}^{(N)} = -\Big({\textstyle{5\over3}}\, \bar c_1 +\bar c_2+a\Big)\, \Big(\beta -1\Big)\;, \quad
P_{K \Sigma}^{(N)} = -\Big( {\textstyle{1\over3}}\,\bar c_1 -\bar c_2-a \Big)\,
\Big( \beta +{\textstyle{1\over2}} \Big) \;,
\nonumber\\
&&P_{\eta \,N}^{(N)} = {\textstyle{7\over3}}\, \bar c_1 +\bar c_2+a
- \beta \,\Big( {\textstyle{1\over3}}\,\bar c_1 +\bar c_2+a \Big)\;, \qquad
\nonumber\\
&& P_{K \Lambda}^{(N)} = \Big( \bar c_1 +\bar c_2+a\Big)\,\Big(\beta +{\textstyle{1\over2}}\Big)\;,
\qquad
P_{\bar K N}^{(\Lambda)} = \sqrt{2}\,P_{K \,\Lambda}^{(N)} \,,\qquad
\nonumber\\
&&P_{\pi \Sigma}^{(\Lambda)} = - {\textstyle{2\over \sqrt{3}}}\,\bar c_1\,\Big( \beta -1 \Big)\;,
\quad P_{\eta \,\Lambda}^{(\Lambda)} = {\textstyle{2\over3}}\,\bar c_1\,\Big( 5+\beta \Big) +2\,\bar c_2+2\,a \;, \qquad
\nonumber\\
&&  P_{K \Xi}^{(\Lambda)} = {\textstyle{1\over \sqrt{2}}}\,\Big( {\textstyle{1\over3}}\,\bar c_1+\bar c_2+a \Big)\,
\Big(2\, \beta +1 \Big)\;, \quad
P_{\bar K N}^{(\Sigma)} =
\sqrt{{\textstyle{2\over 3}}}\,P_{K \Sigma}^{(N)} \;, \quad \!\!
\nonumber\\
&& P_{\pi \Sigma}^{(\Sigma)} = \sqrt{{\textstyle{2\over3}}}\,\Big( {\textstyle{4\over 3}}\,\bar c_1 +2\,\bar c_2+2\,a \Big)\,
 \Big(\beta -1\Big) \;, \quad
P_{\pi \Lambda}^{(\Sigma)} =
{\textstyle{1\over\sqrt{3}}}\,P_{\pi \Sigma}^{(\Lambda)} \;,
\nonumber\\
&& P_{\eta \Sigma }^{(\Sigma)} = 2\,\bar c_1 \,\Big(1-{\textstyle{1\over 3}}\,\beta \Big)+2\,\bar c_2+2\,a \;,\qquad
\nonumber\\
&&  P_{K \Xi}^{(\Sigma)} = -{\textstyle{1\over \sqrt{6}}}\,\Big( {\textstyle{5\over 3}}\,\bar c_1 +\bar c_2+a\Big) \,
\Big(2\,\beta +1\Big)  \;,\quad
P_{\bar K \Lambda}^{(\Xi)} = -{\textstyle{1\over\sqrt{2}}}\,
P_{K \Xi}^{(\Lambda)} \;, \quad
\nonumber\\
&& P_{\bar K \Sigma}^{(\Xi)} = -\sqrt{{\textstyle{3\over2}}}\,
P_{K \Xi}^{(\Sigma)} \;,\quad
P_{\eta \,\Xi}^{(\Xi)} = {\textstyle{11\over3}}\,\bar c_1 +3\,\bar c_2 + 3\,a
+\beta \,\Big(  \bar c_1 +\bar c_2+a\Big)
\;,\quad \!\!
\nonumber\\
&&  P_{\pi \Xi}^{(\Xi)} = \Big({\textstyle{1\over3}}\,\bar c_1 -\bar c_2-a\Big)\,\Big( \beta -1 \Big)\;,
\label{P-result}
\end{eqnarray}
where $\beta \simeq 1.12$ was introduced in (\ref{c1234-b}). The pseudo-scalar
meson-baryon vertices show $SU(3)$ symmetric contribution linear in
$\beta \,\bar c_i$ and symmetry-breaking terms proportional to $\bar c_i$.
It is emphasized that, even though the physical  meson-baryon coupling constants, $G=A+P$, are
the sum of their axial-vector and pseudo-scalar components, it is important to carefully
discriminate both types of vertices, because they give rise to quite different behaviors
for the partial-wave amplitudes off the baryon-octet pole. Such effects are expected to be
particularly important in the strangeness sectors since there (\ref{P-result}) leads to
$P^{(\Lambda, \Sigma )}_{K N} \sim m_K^2$ to be compared to $P_{\pi N}^{(N)} \sim m_\pi^2$.

Next study the decuplet exchange terms $K^{(I)}_{s-[10] }$ and $K^{(I)}_{u-[10] }$
in (\ref{k-all}). Their respective interaction kernels are written in a form which
facilitates the identification of their on-shell irreducible parts
\begin{eqnarray}
K^{(I)}_{s-[10] }(\bar k,k;w) &=&\sum_{c\,=\,1}^2\,
\frac{C^{(I,c)}_{[10]}}{4\,f^2 }\,\Bigg(
\frac{\bar q\cdot q}{\wslash-m_{[10]}^{(c)}}\,
-\frac{(\bar q\cdot w)\,(w\cdot q)}
{(m_{[10]}^{(c)})^2\,\big(\wslash-m_{[10]}^{(c)}\big) }
\nonumber\\
&+&\frac{1}{3}\left(\barqslash +\frac{\bar q \cdot w}{m_{[10]}^{(c)}}\right)
\frac{1}{\wslash+m_{[10]}^{(c)}}
\left(\qslash +\frac{w\cdot q}{m_{[10]}^{(c)}}\right)  -\frac{Z_{[10]}}{3\,m_{[10]}^{(c)}}\,\barqslash \,\qslash
\nonumber\\
&-&\frac{Z_{[10]}^2}{6\,(m_{[10]}^{(c)})^2}\,\barqslash \,\Big(\wslash-2\,m_{[10]}^{(c)}\Big)\,\qslash
+Z_{[10]}\,\barqslash \,\frac{w\cdot q}{3\,(m_{[10]}^{(c)})^2}
\nonumber\\
&+&Z_{[10]} \,\frac{\bar q \cdot w}{3\,(m_{[10]}^{(c)})^2}\,\qslash \Bigg)\;,
\nonumber\\
K^{(I)}_{u-[10] }(\bar k,k;w) &=&\sum_{c\,=\,1}^3\,
\frac{\widetilde C^{(I,c)}_{[10]}}{4\,f^2 }\,\Bigg(
\frac{\bar q\cdot q}{\barwslash-m_{[10]}^{(c)}}
-\frac{(\bar q\cdot \widetilde w)\,(\widetilde w\cdot q)}
{(m_{[10]}^{(c)})^2\,\big(\barwslash-m_{[10]}^{(c)}\big)}
\nonumber\\
&+&\frac{1}{3}\left(\barpslash +m_{[10]}^{(c)}+\frac{q \cdot \widetilde w}{m_{[10]}^{(c)}}\right)
\frac{1}{\barwslash+m_{[10]}^{(c)}}
\left(\pslash +m_{[10]}^{(c)}+\frac{\widetilde w\cdot \bar q}{m_{[10]}^{(c)}}\right)
\nonumber\\
&-&\frac{\widetilde w\cdot (q+\bar q)}{3\,m_{[10]}^{(c)}}
-\frac{1}{3}\,\Big(\wslash+m_{[10]}^{(c)}\Big)
\nonumber\\
&+&\frac{Z_{[10]}}{3\,(m_{[10]}^{(c)})^2}\,\Big(
\barqslash \,\big(\widetilde w\cdot q\big)
+\big(\bar q \cdot \widetilde w\big) \,\qslash
+m_{[10]}^{(c)}\,\big(
\barqslash\,\qslash-2\,(\bar q\cdot q) \big) \Big)
\nonumber\\
&-&\frac{Z_{[10]}^2}{6\,(m_{[10]}^{(c)})^2}\,
\Big( \barpslash \;\barwslash\,\pslash -\widetilde w^2\,\wslash
+2\,m_{[10]}^{(c)}\,\big(\barqslash\,\qslash -2\,q\cdot q\big) \Big)\Bigg) \;.
\label{k-nonlocal}
\end{eqnarray}
The index $c$ in (\ref{k-nonlocal}) labels the decuplet exchange with
$c\to (\Delta_{[10]} , \Sigma_{[10]}, \Xi_{[10]})$. The dimensionless matrices $C^{(I,c)}_{[10]}$
and  $\widetilde C^{(I,c)}_{[10]}$ characterize the strength of the s-channel and
u-channel resonance exchanges respectively. Here we apply the notation introduced in (\ref{def-not})
also for the resonance exchange contributions. In particular the decuplet coefficients
$C^{(I,c)}_{[10]}$ and $\tilde C^{(I,c)}_{[10]}$ are determined by
\begin{eqnarray}
&& A = G( C_A) -\frac{2}{\sqrt{3}}\,(m_K^2-m_\pi^2)\,\Delta A \;,\qquad
\tilde C_0 = C_0 +\delta C_0 \;,
\nonumber\\
&&\Delta A_{\pi N}^{(\Delta_{[10]} )} = \sqrt{2}\,
\big( -{\textstyle{1\over \sqrt{3}}}\,C_2+\sqrt{3}\,C_3+{\textstyle{2\over \sqrt{3}}}\,\tilde C_0
\big) \;, \quad
\nonumber\\
&&\Delta A_{K \,\Sigma}^{(\Delta_{[10]} )} = -\sqrt{2}\,
\big({\textstyle{2\over \sqrt{3}}}\,C_2
- {\textstyle{1\over \sqrt{3}}}\,\tilde C_0+\sqrt{3}\,C_1\big) \;,\quad
\nonumber\\
&& \Delta A_{\bar K N}^{(\Sigma_{[10]})} =\sqrt{{\textstyle{2\over 3}}}\,
\big(-{\textstyle{1\over \sqrt{3}}}\,C_2+\sqrt{3}\,C_3
- {\textstyle{1\over \sqrt{3}}}\,\tilde C_0-\sqrt{3}\,C_1 \big)-\sqrt{2}\,C_4  \;, \quad
\nonumber\\
&&\Delta A_{\pi \Sigma}^{(\Sigma_{[10]})} = -\sqrt{{\textstyle{2\over 3}}}\,
\big({\textstyle{2\over \sqrt{3}}}\,C_2
+ {\textstyle{2\over \sqrt{3}}}\,\tilde C_0\big)-2\,\sqrt{2}\,C_4 \;, \qquad
\nonumber\\
&& \Delta A_{\pi \Lambda}^{(\Sigma_{[10]})} = {\textstyle{2\over \sqrt{3}}}\,(C_2-\tilde C_0) \;,  \qquad
\Delta A_{\eta \Sigma}^{(\Sigma_{[10]})} = {\textstyle{2\over \sqrt{3}}}\,C_2
- {\textstyle{2\over \sqrt{3}}}\,\tilde C_0\;,\quad
\nonumber\\
&& \Delta A_{K \Xi}^{(\Sigma_{[10]})} = -\sqrt{{\textstyle{2\over 3}}}\,
\big( -{\textstyle{1\over \sqrt{3}}}\,C_2-\sqrt{3}\,C_3
- {\textstyle{1\over \sqrt{3}}}\,\tilde C_0 +\sqrt{3}\, C_1\big)+\sqrt{2}\,C_4\;,
\nonumber\\
&&\Delta A_{\bar K \Lambda}^{(\Xi_{[10]})} =
-{\textstyle{1\over \sqrt{3}}}\, \tilde C_0 -\sqrt{3}\,C_1
-{\textstyle{1\over \sqrt{3}}}\,(3\,C_4+2\,C_2) \;, \quad
\nonumber\\
&& \Delta A_{\bar K \,\Sigma}^{(\Xi_{[10]})} = {\textstyle{2\over \sqrt{3}}}\,C_2
- {\textstyle{1\over \sqrt{3}}}\,\tilde C_0- \sqrt{3}\, C_1 +\sqrt{3}\,C_4  \;,\quad
\nonumber\\
&&\Delta A_{\eta \,\Xi}^{(\Xi_{[10]})} = {\textstyle{1\over \sqrt{3}}}\,C_2+\sqrt{3}\,C_3
+ {\textstyle{2\over \sqrt{3}}}\,\tilde C_0 +\sqrt{3}\,C_4 \;,\quad
\nonumber\\
&&\Delta A_{\pi \,\Xi}^{(\Xi_{[10]})} = {\textstyle{1\over \sqrt{3}}}\,C_2+\sqrt{3}\,C_3
- {\textstyle{2\over \sqrt{3}}}\,\tilde C_0 -\sqrt{3}\,C_4  \;,
\label{}
\end{eqnarray}
where the $SU(3)$ symmetric contributions $G(C)$ are
\begin{eqnarray}
&&G_{\pi N}^{(\Delta_{[10]} )} = \sqrt{2}\,C \;, \quad G_{K \,\Sigma}^{(\Delta_{[10]} )} = -\sqrt{2}\,C \;,
\quad
G_{\bar K N}^{(\Sigma_{[10]})} =\sqrt{{\textstyle{2\over 3}}}\,C  \;,
\nonumber\\
&& G_{\pi \Sigma}^{(\Sigma_{[10]})} = -\sqrt{{\textstyle{2\over 3}}}\,C \;,\quad
G_{\pi \Lambda}^{(\Sigma_{[10]})} = -C \;,  \quad G_{\eta \Sigma}^{(\Sigma_{[10]})} = C\;,
\quad G_{K \Xi}^{(\Sigma_{[10]})} = -\sqrt{{\textstyle{2\over 3}}}\,C\;,
\nonumber\\
&&G_{\bar K \Lambda}^{(\Xi_{[10]})} = C \;, \quad
G_{\bar K \,\Sigma}^{(\Xi_{[10]})} = C \;,\quad
G_{\eta \,\Xi}^{(\Xi_{[10]})} = -C\;,\quad
G_{\pi \,\Xi}^{(\Xi_{[10]})} = -C \;.
\label{}
\end{eqnarray}
Recall that to leading order in the $1/N_c$ expansion the five symmetry-breaking
parameters $C_i$ introduced in (\ref{chi-sb-3} are all given in terms of the
$c_i$ parameters (\ref{large-Nc-FDC}). The parameter $\delta C_0 \sim \tilde c_1-c_1$
and also implicitly $\tilde C_R $ (see (\ref{ren-FDC-had})) probe the $\tilde c_1$ parameter
introduced in (\ref{ansatz-3}). All $c_i$ parameters but $c_5$ are
determined to a large part by the weak decay widths of the baryon octet states. The
$\tilde c_1$ parameter is constrained strongly by the $SU(3)$ symmetry-breaking pattern of
the meson-baryon-octet coupling constants (\ref{ansatz-3}).

The baryon octet resonance contributions $K^{(I)}_{s-[9] }$ and $K^{(I)}_{u-[9] }$
require special attention, because the way to incorporate systematically these resonances
in a chiral SU(3) scheme is not clear. We present first the s-channel and u-channel
contributions as they follow from (\ref{lag-Q}) at tree-level
\begin{eqnarray}
K^{(I)}_{s-[9] }(\bar k,k;w) &=&\sum_{c\,=\,0}^4\,
\frac{C^{(I,c)}_{[9]}}{4\,f^2 }\,\Bigg(
-\frac{\bar q\cdot q}{\wslash+m_{[9]}^{(c)}}\,
+\frac{(\bar q\cdot w)\,(w\cdot q)}
{(m_{[9]}^{(c)})^2\,\big(\wslash+m_{[9]}^{(c)}\big) }
\nonumber\\
&-&\frac{1}{3}\left(\barqslash -\frac{\bar q \cdot w}{m_{[9]}^{(c)}}\right)
\frac{1}{\wslash-m_{[9]}^{(c)}}
\left(\qslash -\frac{w\cdot q}{m_{[10]}^{(c)}}\right)  \Bigg)\;,
\nonumber\\
K^{(I)}_{u-[9] }(\bar k,k;w) &=&\sum_{c\,=\,0}^4\,
\frac{\widetilde C^{(I,c)}_{[9]}}{4\,f^2 }\,\Bigg(
-\frac{\bar q\cdot q}{\barwslash+m_{[9]}^{(c)}}
+\frac{(\bar q\cdot \widetilde w)\,(\widetilde w\cdot q)}
{(m_{[9]}^{(c)})^2\,\big(\barwslash+m_{[9]}^{(c)}\big)}
\nonumber\\
&-&\frac{1}{3}\left(\barpslash -m_{[9]}^{(c)}-\frac{q \cdot \widetilde w}{m_{[9]}^{(c)}}\right)
\frac{1}{\barwslash-m_{[9]}^{(c)}}
\left(\pslash -m_{[9]}^{(c)}-\frac{\widetilde w\cdot \bar q}{m_{[9]}^{(c)}}\right)
\nonumber\\
&-&\frac{\widetilde w\cdot (q+\bar q)}{3\,m_{[9]}^{(c)}}
+\frac{1}{3}\,\Big(\wslash-m_{[9]}^{(c)}\Big)\Bigg)
\nonumber\\
&-&\frac{Z_{[9]}}{3\,(m_{[9]}^{(c)})^2}\,\Big(
\barqslash \,\big(\widetilde w\cdot q\big)
+\big(\bar q \cdot \widetilde w\big) \,\qslash
-m_{[9]}^{(c)}\,\big(
\barqslash\,\qslash-2\,(\bar q\cdot q) \big) \Big)
\nonumber\\
&+&\frac{(Z_{[9]})^2}{6\,(m_{[9]}^{(c)})^2}\,
\Big( \barpslash \;\barwslash\,\pslash -\widetilde w^2\,\wslash
-2\,m_{[9]}^{(c)}\,\big(\barqslash\,\qslash -2\,q\cdot q\big) \Big)\Bigg)
\label{8stern}
\end{eqnarray}
where the index $c\to (\Lambda(1520),N(1520), \Lambda (1690), \Sigma(1680) ,\Xi(1820))$ extends over
the nonet resonance states.  The coefficient matrices $C^{(I,c)}_{[9]}$ and
$\widetilde C^{(I,c)}_{[9]}$ are constructed by analogy with those of the octet and decuplet contributions
(see (\ref{G-explicit},\ref{def-not})). In particular one has $C_{B^*(c)}^{(I)}=C_{B(c)}^{(I)}$ and
$\tilde C_{B^*(c)}^{(I)}=\tilde C_{B(c)}^{(I)}$. The coupling constants
$A^{(B^*)}_{\Phi B}=G^{(B)}_{\Phi B}(F_{[9]},D_{[9]})$  are given in
terms of the $F_{[9]}$ and $D_{[9]}$ parameters introduced in (\ref{lag-Q}) for all
contributions except those of the $\Lambda (1520)$ and $\Lambda (1690)$ resonances
\begin{eqnarray}
&& G_{\bar K N}^{(\Lambda (1520))} = \sqrt{{\textstyle{2\over 3}}}\,\Big( 3\,F_{[9]}+D_{[9]} \Big)\,\sin \vartheta
+\sqrt{3}\,C_{[9]}\,\cos \vartheta
 \;,\quad
\nonumber\\
&& G_{\pi \Sigma}^{(\Lambda(1520))} = -2\,D_{[9]} \,\sin \vartheta
+{\textstyle{3\over \sqrt{2}}}\,C_{[9]}\,\cos \vartheta\;,\quad
\nonumber\\
&& G_{\eta \,\Lambda}^{(\Lambda(1520))} = {\textstyle{2\over\sqrt{3}}}\,D_{[9]}\,\sin \vartheta
+\sqrt{{\textstyle{3\over 2}}}\,C_{[9]}\,\cos \vartheta\;,\quad
\nonumber\\
&& G_{K \Xi}^{(\Lambda(1520))} = \sqrt{{\textstyle{2\over 3}}}\,\Big( 3\,F_{[9]}-D_{[9]}\Big)\,\sin \vartheta
-\sqrt{3}\,C_{[9]}\,\cos \vartheta\;,
\label{G-9-explicit-1}
\end{eqnarray}
and
\begin{eqnarray}
&& G_{\bar K N}^{(\Lambda (1690))} = -\sqrt{{\textstyle{2\over 3}}}\,\Big( 3\,F_{[9]}+D_{[9]} \Big)\,\cos \vartheta
+\sqrt{3}\,C_{[9]}\,\sin \vartheta
 \;,\quad
\nonumber\\
&& G_{\pi \Sigma}^{(\Lambda(1690))} = 2\,D_{[9]} \,\cos \vartheta
+{\textstyle{3\over \sqrt{2}}}\,C_{[9]}\,\sin \vartheta\;,\quad
\nonumber\\
&& G_{\eta \,\Lambda}^{(\Lambda(1690))} = -{\textstyle{2\over\sqrt{3}}}\,D_{[9]}\,\cos \vartheta
+\sqrt{{\textstyle{3\over 2}}}\,C_{[9]}\,\sin \vartheta\;,\quad
\nonumber\\
&& G_{K \Xi}^{(\Lambda(1690))} = -\sqrt{{\textstyle{2\over 3}}}\,\Big( 3\,F_{[9]}-D_{[9]}\Big)\,\cos \vartheta
-\sqrt{3}\,C_{[9]}\,\sin \vartheta\;.
\label{G-9-explicit-2}
\end{eqnarray}

The baryon nonet field $B^*_\mu $ asks for special considerations
in a chiral SU(3) scheme as there is no straightforward systematic
approximation strategy. Given that the baryon octet and decuplet
states are degenerate in the large-$N_c$ limit, it is natural to
impose $m_{[10]} -m_{[8]} \sim Q$. In contrast with that there is
no fundamental reason to insist on $m_{N(1520)}-m_N \sim Q$, for
example. But, only with $m_{[9]}-m_{[8]} \sim Q $ is it feasible
to establish consistent power counting rules needed for the
systematic evaluation of the chiral Lagrangian. The
presence of the baryon nonet resonance states in the large-$N_c$
limit of QCD is far from obvious. Referring to the detailed
exposition in chapter 2.1, our opinion differs here from the one
expressed in \cite{Carone-1,Carone-2,Schat-Goity-Scoccola} where
the d-wave baryon octet resonance states are considered as part of
an excited large-$N_c$ ${\bf 70}$-plet. It was argued that reducible
diagrams summed by the Bethe-Salpeter equation are typically
enhanced by a factor of $2 \pi$ relatively to irreducible
diagrams. Therefore there are a priori two extreme possibilities: the
baryon resonances are a consequence of important coupled channel
dynamics or they are already present in the interaction kernel.
As was motivated in detail in chapter 2.1 we favor the latter case.

The description of resonances has a subtle consequence for the
treatment of the u-channel baryon resonance exchange contributions.
If a resonance is formed primarily by the coupled channel dynamics
one should not include an explicit bare u-channel resonance
contribution in the interaction kernel. The then necessarily
strong energy dependence of the resonance self-energy would
invalidate the use of (\ref{8stern}), because for physical
energies $\sqrt{s}> m_N+m_\pi$ the resonance self-energy is probed
far off the resonance pole. This argument has non-trivial
implications for the chiral SU(3) dynamics. Naively one may object
that the effect of the u-channel baryon resonance exchange
contribution in (\ref{8stern}) can be absorbed to good accuracy
into chiral two-body interaction terms in any case. However, while
this is true in  a chiral $SU(2)$ scheme this is no longer
possible in a chiral $SU(3)$ approach. This follows because chiral
symmetry selects a specific subset of all possible
$SU(3)$-symmetric two-body interaction terms to given chiral
order. In particular one finds that the effect of the $Z_{[9]}$
parameter in (\ref{8stern}) can not be absorbed into the chiral
two-body parameters $g^{(S)}, g^{(V)}$ or $g^{(T)}$ of
(\ref{two-body}). For that reason we discriminate two possible
scenarios. In scenario I we conjecture that the baryon nonet
resonance states are primarily generated by the coupled channel
dynamics of the vector-meson baryon-octet channels. Therefore in
this scenario the u-channel resonance exchange contribution of
(\ref{8stern}) is neglected and only its s-channel contribution is
included as a reminiscence of the neglected vector meson channels.
This is analogous the treatment of the $\Lambda
(1405)$ resonance, which is generated dynamically in the chiral
$SU(3)$ scheme (see Fig. \ref{fig:wt}). Here a s-channel pole term
is generated by the coupled channel dynamics but the associated
u-channel term is effectively left out as a much suppressed
contribution. In scenario II we explicitly include the s- and the
u-channel resonance exchange contributions as given in
(\ref{8stern}), thereby assuming that the resonance was preformed
already in the large-$N_c$ limit of QCD and only slightly affected
by the coupled channel dynamics. According to
the discussion of chapter 2.1 one may expect that the data set
should favor scenario I): if one member of the {\bf 70}-plet is dynamically
generated the remaining members should be of dynamic origin too.
Since the $\Lambda(1405)$ appears to be the result of coupled channel dynamics
the same should hold for the baryon nonet states, in particular the $\Lambda(1520)$.

Detailed analyses of the data set clearly favor scenario I). It appears impossible
to obtain any reasonable fit within scenario II). The inclusion of the u-channel resonance
exchange contributions destroys the subtle balance of chiral s-wave range
terms. From a physical point of view this mechanism is easily understood.
Once the coupling constants of the baryon nonet states to the
meson bayron final state are estimated in terms of the resonance decay width,
the u-channel diagrams predict quite large s-wave effective range terms that
can not be reabsorbed into the quasi-local counter terms of the chiral Lagrangian.
Therefore an attempt to fit the antikaon-nucleon scattering data in scenario II)
is unable to generate the s-wave $\Lambda(1405)$ resonance and support the proper
resonance structure for the nonet states at the same time.
Thus, all results presented in this work will be based on
scenario I). In this scenario the background parameter
$Z_{[9]}$ drops out completely (see (\ref{v-result-1})).

\tabcolsep=1.8mm
\renewcommand{\arraystretch}{1.4}
\begin{table}[p]
\vspace{-1.2cm}
\begin{tabular}{|r||c||c|c||c|c|c||c|c|c|c||c|c|c|}
\multicolumn{14}{c}{}\\ \multicolumn{14}{c}{}\\
 \hline
    & $C_{\pi,0}^{(0)}$ &  $C_{\pi,D}^{(0)}$ &  $C_{\pi,F}^{(0)}$
    & $C_{K,0}^{(0)}$   & ${C}_{K,D}^{(0)}$  & ${C}_{K,F}^{(0)}$
    & ${C}_{0}^{(0)}$   &  ${C}_{1}^{(0)}$  &
    ${C}_{D}^{(0)}$ & ${C}_{F}^{(0)}$
    & $\bar C_{1}^{(0)}$ & $\bar C_{D}^{(0)}$ & $\bar C_{F}^{(0)}$\\
\hline \hline
$11$& 0 & 0 & 0
    & -4 &-6 & -2
    & 2 & 2 & 3& 1
    & 2 &1& 3   \\ \hline
$12$& 0& -$\sqrt{\frac32}$ &  $\sqrt{\frac32}$
    & 0& -$\sqrt{\frac32}$ & $\sqrt{\frac32}$
    & 0 & $\sqrt 6$ & $\sqrt{\frac32}$ &-$\sqrt{\frac32}$
    & $\sqrt 6$ & -$\sqrt{\frac32}$ &$\sqrt{\frac32}$      \\ \hline
$13$& 0& $\frac{1}{\sqrt 2}$ & $\frac{3}{\sqrt 2}$
    & 0& -$\frac{5}{3\,\sqrt 2}$  & -$\frac{5}{\sqrt 2}$
    & 0  & $\sqrt 2$ & $\frac{1}{3\sqrt 2}$ & $\frac{1}{\sqrt 2}$
    & $\sqrt 2$ &$\frac{1}{\sqrt 2}$&$\frac{3}{\sqrt 2}$   \\  \hline
$14$& 0 & 0  & 0
    & 0 & 0  & 0
    & 0  & -3& 0 & 0
    & -1& 0 & 0   \\ \hline\hline
$22$& -4 & -4 & 0
    & 0  & 0 & 0
    & 2  & 4& 2 & 0
    & 2 & 0& 4    \\ \hline
$23$& 0 &  -$\frac{4}{\sqrt 3}$ & 0
    &  0  & 0  & 0
    & 0  & $\sqrt 3$& $\frac{2}{\sqrt 3}$ &  0
    & $\sqrt 3$ &  0 &  0  \\ \hline
$24$& 0&  $\sqrt{\frac32}$ & $\sqrt{\frac32}$
    & 0&  $\sqrt{\frac32}$ & $\sqrt{\frac32}$
    & 0 & -$\sqrt 6$& -$\sqrt{\frac32}$ &  -$\sqrt{\frac32}$
    & -$\sqrt 6$ &-$\sqrt{\frac32}$&  -$\sqrt{\frac32}$   \\ \hline\hline
$33$& $\frac43$ & $\frac{28}{9}$ & 0
    & -$\frac{16}{3}$ & -$\frac{64}{9}$& 0
    & 2 & 2 & 2 & 0
    & 0 & 0  & 0 \\ \hline
$34$& 0&  -$\frac{1}{\sqrt 2}$ & $\frac{3}{\sqrt{2}}$
    & 0& $\frac{5}{3\sqrt 2}$  & -$\frac{5}{\sqrt 2}$
    & 0 & -$\sqrt 2$ & -$\frac{1}{3\sqrt 2}$ & $\frac{1}{\sqrt 2}$
    & -$\sqrt 2$ &$\frac{1}{\sqrt 2}$ & -$\frac{3}{\sqrt 2}$    \\ \hline\hline
$44$& 0 & 0 & 0
    & -4 &-6 & 2
    & 2 & 2 & 3&-1
    & 2 & -1 &3  \\ \hline \hline

  & $C_{\pi,0}^{(1)}$    &  $C_{\pi,D}^{(1)}$      &  $C_{\pi,F}^{(1)}$
  & $C_{K,0}^{(1)}$  &${C}_{K,D}^{(1)}$ & ${C}_{K,F}^{(1)}$
  & ${C}_{0}^{(1)}$ & ${C}_{1}^{(1)}$& ${C}_{D}^{(1)}$ & ${C}_{F}^{(1)}$
  & $\bar C_{1}^{(1)}$ & $\bar C_{D}^{(1)}$ & $\bar C_{F}^{(1)}$\\
\hline \hline
$11$&0 & 0 & 0
    & -4 & -2& 2
    & 2  & 0  & 1& -1
    & 0& -1 &  1   \\ \hline
$12$& 0& -1 & 1
    & 0& -1 & 1
    & 0 & 0 & 1& -1
    & 0 &-1& 1    \\ \hline
$13$& 0 &  $\frac{1}{\sqrt 6}$ & $\sqrt{\frac32}$
    & 0 & $\frac{1}{\sqrt 6}$& $\sqrt{\frac32}$
    & 0 & 0 & -$\frac{1}{\sqrt 6}$
    &-$\sqrt{\frac32}$  & 0 &$\frac{1}{\sqrt 6}$& $\sqrt{\frac32}$   \\  \hline
$14$& 0 & -$\sqrt{\frac32}$&  $\sqrt{\frac32}$
    & 0 & $\frac{5}{\sqrt 6}$ &-$\frac{5}{\sqrt 6}$
    & 0 & 0 & -$\frac{1}{\sqrt 6}$   & $\frac{1}{\sqrt 6}$
    &0 & -$\sqrt{\frac32}$& $\sqrt{\frac32}$    \\  \hline
$15$& 0  &  0 & 0
    & 0  &  0 & 0
    & 0 & 1 & 0  &  0
    & -1 & 0  &  0  \\ \hline\hline
$22$& -4&-4 & 0
    & 0& 0  & 0
    & 2  & -1 & 2 & 0
    & 1 & 0  & 2  \\ \hline
$23$& 0  & 0 & 0
    & 0  & 0 & 0
    & 0  & 0 & 0  & 0
    & 0  & -$2\sqrt{\frac23}$& 0   \\ \hline
$24$& 0& 0  &  $4\sqrt{\frac23}$
    & 0  & 0  & 0
    & 0 & 0 & 0   & -$2\sqrt{\frac23}$
    & 0  & 0 &0 \\ \hline
$25$& 0 & 1& 1
    & 0 & 1 & 1
    &0  & 0 & -1  & -1
    &0 & -1  & -1  \\ \hline\hline
$33$& -4 & -$\frac43$ & 0
    & 0    & 0   & 0
    & 2 & 0 & $\frac23$& 0
    &0 & 0  & 0  \\ \hline
$34$& 0 & -$\frac43$  & 0
    & 0   &0  & 0
    & 0 & 1 & $\frac23$ & 0
    &-1 & 0   & 0  \\ \hline
$35$& 0 & $\frac{1}{\sqrt 6}$  &  -$\sqrt{\frac32}$
    & 0 & $\frac{1}{\sqrt 6}$ &-$\sqrt{\frac32}$
    & 0 & 0 &-$\frac{1}{\sqrt 6}$& $\sqrt{\frac32}$
    & 0 & -$\frac{1}{\sqrt 6}$ & $\sqrt{\frac32}$    \\ \hline\hline
$44$& $\frac43$   & -$\frac43$  &  0
    & -$\frac{16}{3}$ & 0  & 0
    & 2 & 0 & $\frac23$ & 0
    & 0 & 0 & 0  \\ \hline
$45$& 0  & -$\sqrt{\frac32}$  &  -$\sqrt{\frac32}$
    & 0  & $\frac{5}{\sqrt 6}$  &  $\frac{5}{\sqrt 6}$
    & 0 & 0 & -$\frac{1}{\sqrt 6}$  & -$\frac{1}{\sqrt 6}$
    & 0 & $\sqrt{\frac32}$ & $\sqrt{\frac32}$   \\ \hline\hline
$55$& 0 & 0  & 0
    & -4 & -2  & -2
    & 2 & 0 & 1   & 1
    & 0 & 1   & 1  \\ \hline \hline

  & $C_{\pi,0}^{(2)}$    &  $C_{\pi,D}^{(2)}$      &  $C_{\pi,F}^{(2)}$
  & $C_{K,0}^{(2)}$  &${C}_{K,D}^{(2)}$ & ${C}_{K,F}^{(2)}$
  & ${C}_{0}^{(2)}$ & ${C}_{1}^{(2)}$& ${C}_{D}^{(2)}$ & ${C}_{F}^{(2)}$
  & $\bar C_{1}^{(2)}$ & $\bar C_{D}^{(2)}$ & $\bar C_{F}^{(2)}$\\
\hline \hline
$11$ &-4 & -4 & 0  & 0 & 0& 0 & 2  & 1  & 2 & 0
    & -1& 0 &  -2   \\ \hline
\end{tabular}
\vspace*{2mm} \caption{Coefficients of quasi-local interaction terms in the strangeness minus one
channels as defined in (\ref{def-local-1}).}
\label{tabkm-2}
\end{table}

Consider the quasi-local two-body interaction terms $K_{[8][8]}^{(I)}$ and
$K_{\chi}^{(I)}$ in (\ref{k-all}).
It is convenient to represent the strength
of an interaction term in a given channel $(I,a)\to (I,b)$ by means of the dimensionless
coupling coefficients $\big[C^{(I)}_{..}\big]_{ab}$. The $SU(3)$ structure of the interaction terms
${\mathcal L}^{(S)}$ and ${\mathcal L}^{(V)}$ in (\ref{two-body}) are characterized by the
coefficients $C^{(I)}_{0}$, $C^{(I)}_{1}$, $C^{(I)}_{D}$, $C^{(I)}_{F}$ and
${\mathcal L}^{(T)}$  by $\bar C^{(I)}_{1}$, $\bar C^{(I)}_{D}$ and
$\bar C^{(I)}_{F}$. Similarly the interaction terms (\ref{chi-sb}) which break chiral symmetry
explicitly are written in terms of the coefficients
$C^{(I)}_{\pi,0}$, $C^{(I)}_{\pi,D}$, $C^{(I)}_{\pi,F}$ and $C^{(I)}_{K,0}$,
$C^{(I)}_{K,D}$, $C^{(I)}_{K,F}$. We have
\begin{eqnarray}
K^{(I)}_{[8][8]}(\bar k, k; w)&=&  \frac{\bar q \cdot q }{4 f^2} \,
\Big(g^{(S)}_0\,C^{(I)}_0+g^{(S)}_1\,C^{(I)}_1 + g^{(S)}_D\,C^{(I)}_D
+g^{(S)}_F\,C^{(I)}_F\Big)
\nonumber\\
&+&\frac{1}{16 f^2}\,\Big( \qslash\,\big( p+\bar p\big) \cdot \bar q
+\barqslash \,\big(p+ \bar p\big) \cdot  q\Big)\,
\Big( g^{(V)}_0\,C^{(I)}_0+g^{(V)}_1\,C^{(I)}_1 \Big)
\nonumber\\
&+&\frac{1}{16 f^2}\,\Big( \qslash\,\big( p+\bar p\big) \cdot \bar q
+\barqslash \,\big(p+ \bar p\big) \cdot  q\Big)
\,\Big( g^{(V)}_D\,C^{(I)}_D +g^{(V)}_F\,C^{(I)}_F \Big)
\nonumber\\
&+&\frac{i}{4 f^2}\,\Big( \bar q^\mu\,\sigma_{\mu \nu}\,q^\nu \Big)
\,\Big( g^{(T)}_1\,\bar C^{(I)}_{1}+g^{(T)}_{D}\,\bar C^{(I)}_{D} +g^{(T)}_F\,\bar C^{(I)}_{F}\Big)\;,
\nonumber\\
K^{(I)}_{\chi}(\bar k, k; w)&=&
\frac{b_0}{f^2}\,\Big(
m_\pi^2\,C^{(I)}_{\pi,0} +m_K^2\,C^{(I)}_{K,0} \Big)
+\frac{b_D}{f^2}\,\Big(
m_\pi^2\,C^{(I)}_{\pi,D} +m_K^2\,C^{(I)}_{K,D} \Big)
\nonumber\\
&+&\frac{b_F}{f^2}\,\Big(
m_\pi^2\,C^{(I)}_{\pi,F} +m_K^2\,C^{(I)}_{K,F} \Big)\;,
\label{def-local-1}
\end{eqnarray}
where the coefficients $C^{(I)}_{..,ab}$ are listed in Tab. \ref{tabkm-2} for the
strangeness minus one channels but shown in \cite{Lutz:Kolomeitsev} for the strangeness
zero channels. A complete listing of the $Q^3$ quasi-local two-body interaction terms
can be found in \cite{Lutz:Kolomeitsev}.

\vskip1.5cm \section{Chiral expansion and covariance}

It remains to perform a chiral expansion of the interaction kernel
(\ref{k-all}). This is necessary since here the relativistic
chiral Lagrangian is applied. In contrast, if the heavy-baryon representation
of the chiral Lagrangian is used part of the chiral expansion is already
implemented at the level of the Lagrangian. A novel chiral expansion strategy for
the interaction kernel will be proposed that complies manifestly with covariance.

For this purpose it is instructive to cast the meson energy $\omega $ and
the baryon energy $E$
introduced in (\ref{bark-def}) into the following form
\begin{eqnarray}
&&\omega^{(I)}_a(\sqrt{s}) =\sqrt{s}-m^{(I)}_{B(I,a)}
-\frac{\phi^{(I)}_a(\sqrt{s})}{2\,\sqrt{s}}\,,
\;\;\;E_a^{(I)}(\sqrt{s}) = m_{B(I,a)}+\frac{\phi^{(I)}_a(\sqrt{s})}{2\,\sqrt{s}}\;,
\nonumber\\
&&\phi^{(I)}_{a}(\sqrt{s})= \Big(\sqrt{s}-m_{B(I,a)}\Big)^2-m_{\Phi(I,a)}^2 \,,
\label{rel-power}
\end{eqnarray}
expressed in terms of the approximate phase-space factor $\phi $.
It is sufficient to consider the above kinematical variables since it is assumed that
the partial wave decomposition of the scattering kernel was performed previously.
In the standard counting scheme the meson energy $\omega $ is assigned the formal power
$Q$. On the other hand, the baryon energy $E = m+ {\mathcal O}(Q^2)$ is interpreted
as a leading term, the baryon mass (more precisely its chiral limit value) and a series of
correction terms. As a consequence also the Mandelstam variable,
$\sqrt{s} = \omega + E$, is split into an infinite hierarchy of terms characterized by
increasing chiral orders. Here it is proposed to count
\begin{eqnarray}
\sqrt{s} \sim Q^0 \;,
\label{}
\end{eqnarray}
instead, a manifest covariant assignment. Then the representation (\ref{rel-power}) suggests
that it may be advantageous to
write the meson energy $\omega $ as a sum of two terms, the first of order $Q$ and the
second term of order $Q^2$. This is achieved once the formal powers are assigned as follows
\begin{eqnarray}
\sqrt{s} \sim Q^0\;, \qquad \phi^{(I)}_a/\sqrt{s} \sim Q^2 \;, \qquad
m_{\Phi(I,a)}^2 \sim Q^2 \;.
\label{}
\end{eqnarray}
A unique decomposition of the meson energy $ \omega $ into the leading term
$ \sqrt{s}- m \sim Q$ and the subleading term $-\phi/(2 \sqrt{s})\sim Q^2$
is implied. It is stressed that the assignment leads to $m$ as the leading chiral
moment of the baryon energy $E$.

The implications of the novel covariant power counting assignment are first exemplified
for the case of the quasi-local two-body interaction terms.
It is straight forward to derive the effective interaction kernel relevant for the
s- and p-wave channels
\begin{eqnarray}
V^{(I,+)}_{[8][8]}(\sqrt{s};0)&=& \frac{1}{4 f^2}\,
\Big(g^{(S)}_0+\sqrt{s}\,g_0^{(V)}\Big)\,
\Big( \sqrt{s}-M^{(I)}\Big)\,C_0^{(I)}\,\Big( \sqrt{s}-M^{(I)}\Big)
\nonumber\\
&+& \frac{1}{4 f^2}\,\Big(g^{(S)}_1+\sqrt{s}\,g_1^{(V)}\Big)\,
\Big( \sqrt{s}-M^{(I)}\Big)\,C_1^{(I)}\,\Big( \sqrt{s}-M^{(I)}\Big)
\nonumber\\
&+& \frac{1}{4 f^2}\, \Big(g^{(S)}_D+\sqrt{s}\,g_D^{(V)}\Big)\,
\Big( \sqrt{s}-M^{(I)}\Big)\,C_D^{(I)}\,\Big( \sqrt{s}-M^{(I)}\Big)
\nonumber\\
&+& \frac{1}{4 f^2}\,\Big(g^{(S)}_F+\sqrt{s}\,g_F^{(V)}\Big)\,
\Big( \sqrt{s}-M^{(I)}\Big)\,C_F^{(I)}\,\Big( \sqrt{s}-M^{(I)}\Big)
\nonumber\\
&+&{\mathcal O} \left(Q^3\right)\;,
\nonumber\\
V^{(I,-)}_{[8][8]}(\sqrt{s};0)&=&
-\frac{1}{3 f^2} \,M^{(I)}\,
\Big(g^{(S)}_0\,C_0^{(I)}+g^{(S)}_1\,C_1^{(I)}\Big)\, M^{(I)}
\nonumber\\
&-&\frac{1}{3 f^2} \,M^{(I)}\,
\Big(g^{(S)}_D\,C_D^{(I)} +g^{(S)}_F\,C_F^{(I)}\Big)\, M^{(I)}
\nonumber\\
&+& \frac{1}{4 f^2} \left(\sqrt{s}+M^{(I)}\right)
\Big(g^{(T)}_1\,\bar C_{1}^{(I)}+g^{(T)}_D\,\bar C_{D}^{(I)}\Big)
\left(\sqrt{s}+M^{(I)}\right)
\nonumber\\
&+& \frac{1}{4 f^2} \left(\sqrt{s}+M^{(I)}\right)
g^{(T)}_F\,\bar C_{F}^{(I)} \left(\sqrt{s}+M^{(I)}\right)
\nonumber\\
&-&\frac{1}{3 f^2}\,M^{(I)}\,
\Big(g^{(T)}_1\,\bar C_{1}^{(I)}+g^{(T)}_D\,\bar C_{D}^{(I)}+g^{(T)}_F\,\bar C_{F}^{(I)} \Big)\,
M^{(I)}
+ {\mathcal O} \left(Q\right)\;,
\nonumber\\
V^{(I,+)}_{[8][8]}(\sqrt{s};1)&=& -\frac{1}{12 f^2}\,
\Big(g^{(S)}_0\,C_0^{(I)}+g^{(S)}_1\,C_1^{(I)}+g^{(S)}_D\,C_D^{(I)}
+g^{(S)}_F\,C_F^{(I)}\Big)
\nonumber\\
&-&\frac{1}{12 f^2}\,
\Big(g^{(T)}_1\,\bar C_{1}^{(I)}+g^{(T)}_D\,\bar C_{D}^{(I)}+g^{(T)}_T\,\bar C_{F}^{(I)}\Big)
+ {\mathcal O} \left(Q \right)\;.
\nonumber\\
V^{(I,+)}_{\chi}(\sqrt{s},0)&=& \frac{b_0}{f^2}\,\Big(
m_\pi^2\,C^{(I)}_{\pi,0} +m_K^2\,C^{(I)}_{K ,0} \Big)
+\frac{b_D}{f^2}\,\Big(
m_\pi^2\,C^{(I)}_{\pi,D} +m_K^2\,C^{(I)}_{K ,D} \Big)
\nonumber\\
&+&\frac{b_F}{f^2}\,\Big(
m_\pi^2\,C^{(I)}_{\pi,F} +m_K^2\,C^{(I)}_{K ,F}\Big)
+{\mathcal O}\left(Q^3\right)\;,
\label{local-v}
\end{eqnarray}
where the diagonal baryon mass matrix $M^{(I)}_{ab}=\delta_{ab}\,m_{B(I,a)}$ was introduced.
The notation in (\ref{local-v}) implies a matrix multiplication of the mass
matrix $M^{(I)}$ by the coefficient matrices $C^{(I)}$ in the '$ab$' channel-space.
Recall that the d-wave interaction kernel $V^{(I,-)}_{[8][8]}(\sqrt{s},1)$
does not receive any contribution from quasi-local counter
terms to chiral order $Q^3$. Similarly the chiral symmetry-breaking interaction kernel
$V_\chi$ can only contribute to s-wave scattering to this order.
It is pointed out that the result (\ref{local-v}) is uniquely determined by expanding the meson
energy according to (\ref{rel-power}). In particular we keep in $V^{(I,+)}_{\Phi B}(\sqrt{s},0)$
the $\sqrt{s}$ factor in front of $g^{(V)}$. Any further expansion in $\sqrt{s}-\Lambda $ with
some scale $\Lambda \simeq m_N$ is refrained from. The relativistic chiral Lagrangian supplied
with (\ref{rel-power}) leads to well-defined kinematical factors included in (\ref{local-v}).
These kinematical factors, which are natural ingredients of the relativistic chiral
Lagrangian, can be generated also in the heavy-baryon formalism by a proper regrouping of
interaction terms. The $Q^3$-terms induced by the interaction kernel (\ref{def-local-1}) are
shown in \cite{Lutz:Kolomeitsev} as part of a complete collection of relevant $Q^3$-terms.

Next consider the Weinberg-Tomozawa term and the baryon octet and decuplet s-channel
pole contributions
\begin{eqnarray}
V^{(I,\pm)}_{WT}(\sqrt{s};0)&=&\frac{1}{2\,f^2}\,\left( \sqrt{s}\,C^{(I)}_{WT}
\mp \frac{1}{2}\,\Big[M^{(I)}\,,C^{(I)}_{WT}\Big]_+ \right) \;,
\nonumber\\
V^{(I,\pm)}_{s-[8]}(\sqrt{s};0)
&=& -\sum_{c=1}^3\,\Big(\sqrt{s}\mp M^{(I,c)}_5 \Big)\,
\frac{C^{(I,c)}_{[8]}}{4\,f^2\,\big(\sqrt{s}\pm  m^{(c)}_{[8]}\big) }\,
\Big(\sqrt{s}\mp M^{(I,c)}_5 \Big)\,,
\nonumber\\
V^{(I,+)}_{s-[10]}(\sqrt{s};0)
&=& -\frac{2}{3}\,\sum_{c=1}^2\,\frac{\sqrt{s}+m^{(c)}_{[10]}}{(m^{(c)}_{[10]})^2}\,
\Big(\sqrt{s}- M^{(I)} \Big)\,\frac{C^{(I,c)}_{[10]}}{4\,f^2 }\,\Big(\sqrt{s}- M^{(I)} \Big)
\nonumber\\
&+&\sum_{c=1}^2\,Z_{[10]}\,\frac{2\,\sqrt{s}-m^{(c)}_{[10]}}{3\,(m^{(c)}_{[10]})^2}\,
\Big(\sqrt{s}- M^{(I)} \Big)\,\frac{C^{(I,c)}_{[10]}}{4\,f^2 }\,\Big(\sqrt{s}- M^{(I)} \Big)
\nonumber\\
&+&\sum_{c=1}^2\,Z_{[10]}^2\,\frac{2\,m^{(c)}_{[10]}-\sqrt{s}}{6\,(m^{(c)}_{[10]})^2}\,
\Big(\sqrt{s}- M^{(I)} \Big)\,\frac{C^{(I,c)}_{[10]}}{4\,f^2 }\,\Big(\sqrt{s}- M^{(I)} \Big)
\nonumber\\
&+&{\mathcal O}\left(Q^3 \right)\;,
\nonumber\\
V^{(I,-)}_{s-[10]}(\sqrt{s};0)
&=&\sum_{c=1}^2\, \frac{Z_{[10]}}{3\,m^{(c)}_{[10]}}\,
\Big(\sqrt{s}+ M^{(I)} \Big)\,\frac{C^{(I,c)}_{[10]}}{4\,f^2 }\,\Big(\sqrt{s}+ M^{(I)} \Big)
\nonumber\\
&-&\sum_{c=1}^2\,Z_{[10]}^2\,\frac{\sqrt{s}+2\,m^{(c)}_{[10]}}{6\,(m^{(c)}_{[10]})^2}\,
\Big(\sqrt{s}+ M^{(I)} \Big)\,\frac{C^{(I,c)}_{[10]}}{4\,f^2 }\,\Big(\sqrt{s}+ M^{(I)} \Big)
\nonumber\\
&+&{\mathcal O}\left(Q \right)\;,
\nonumber\\
V^{(I,+)}_{s-[10]}(\sqrt{s};+1)
&=& -\frac{1}{3}\,\sum_{c=1}^2\,
\frac{C^{(I,c)}_{[10]}}{4\,f^2\,(\sqrt{s}-m^{(c)}_{[10]})}
+{\mathcal O}\left(Q \right)\;,
\nonumber\\
V^{(I,-)}_{s-[9]}(\sqrt{s};-1)
&=& -\frac{1}{3}\,\sum_{c=1}^2\,
\frac{C^{(I,c)}_{[9]}}{4\,f^2\,(\sqrt{s}-m^{(c)}_{[9]})}
+{\mathcal O}\left(Q^0 \right)\;
\label{v-result-1}
\end{eqnarray}
where terms of order $Q^3$ are suppressed. Furthermore observe the short hand notation
\begin{eqnarray}
\Big[M^{(I,c)}_5 \Big]_{ab} = \delta_{ab}\, \Big( m_{B(I,a)}+R^{(I,c)}_{L,aa}\Big) \;,
\label{}
\end{eqnarray}
used in the s-channel exchange contribution. The Weinberg-Tomozawa interaction term
$V_{WT}$ contributes to the s-wave and p-wave interaction kernels with $J=\frac{1}{2}$
to chiral order $Q$ and $Q^2$ respectively but not in the $J=\frac{3}{2}$ channels. Similarly
the baryon octet s-channel exchange $V_{s-[8]}$ contributes only in the $J=\frac{1}{2}$
channels and the baryon decuplet s-channel exchange $V_{s-[10]}$ to all considered
channels but the d-wave channel. The $Q^3$-terms not shown in (\ref{v-result-1}) are given
in \cite{Lutz:Kolomeitsev}. In (\ref{v-result-1}) the baryon-nonet resonance contribution
is expanded applying in particular the questionable formal rule $\sqrt{s}-m_{[9]}\sim Q$.
To order $Q^3$ one then finds that only the  d-wave interaction kernels are affected.
In fact, the vanishing of all contributions except the one in the d-wave channel does not
depend on the assumption $\sqrt{s}-m_{[9]}\sim Q$. It merely reflects the phase space
properties of the resonance field. This strategy preserves the correct pole contribution in
$V^{(I,-)}_{s-[9]}(\sqrt{s};-1)$ but discards smooth background terms in all partial-wave
interaction kernels. This is consistent with the discussion of chapter 4.1 which
implies that those background terms are not controlled in any case. It is emphasized that
the physical mass matrix $M^{(I)}$ is used in the interaction kernel rather than its
chiral $SU(3)$ limit. Since the mass matrix follows from the on-shell reduction of the
interaction kernel it necessarily involves the physical mass matrix $M^{(I)}$. Similarly
$M^{(I)}_5$ is kept unexpanded since only this leads to the proper meson-baryon coupling
strengths. This procedure is analogous to keeping the physical masses in the unitarity
loop functions $J_{MB}(w)$ in (\ref{jpin-n-def}).

Proceed with the baryon octet and baryon decuplet
u-channel contributions. After performing their proper angular average
as implied by the partial-wave projection in (\ref{bark-def}) their
contributions to the scattering kernels are written in terms of  the matrix valued functions
${ h}^{(I)}_{n \pm}(\sqrt{s},m),{q}^{(I)}_{n \pm }(\sqrt{s},m)$ and
${ p}^{(I)}_{n \pm }(\sqrt{s},m)$ as
\begin{eqnarray}
\Big[V^{(I,\pm )}_{u-[8]}(\sqrt{s};n)\Big]_{ab} &=& \sum_{c=1}^4\,
\frac{1}{4\,f^2 }\,\Big[\widetilde C^{(I,c)}_{[8]}\Big]_{ab}\,
\Big[{ h}^{(I)}_{n \pm }(\sqrt{s},m^{(c)}_{[8]})\Big]_{ab} \;,
\nonumber\\
\Big[V^{(I,\pm )}_{u-[10]}(\sqrt{s};n)\Big]_{ab} &=&\sum_{c=1}^3\,
\frac{1}{4\,f^2 }\,\Big[\widetilde C^{(I,c)}_{[10]}\Big]_{ab}\,
\Big[{ p}^{(I)}_{n \pm }(\sqrt{s},m_{[10]}^{(c)})\Big]_{ab}\; ,
\nonumber\\
\Big[V^{(I,\pm )}_{u-[9]}(\sqrt{s};n)\Big]_{ab} &=& \sum_{c=1}^4\,
\frac{1}{4\,f^2 }\,\Big[\widetilde C^{(I,c)}_{[9]}\Big]_{ab}\,
\Big[{q}^{(I)}_{n \pm }(\sqrt{s},m^{(c)}_{[9]})\Big]_{ab} \;.
\label{u-result}
\end{eqnarray}
The functions ${h}^{(I)}_{n \pm}(\sqrt{s},m), { q}^{(I)}_{n \pm }(\sqrt{s},m)$ and
${p}^{(I)}_{n \pm}(\sqrt{s},m)$ can be expanded in
chiral powers once a formal powers to the typical building blocks
\begin{eqnarray}
\mu_{\pm,ab}^{(I)}({s},m)=  m_{B(I,a)}+m_{B(I,b)}-\sqrt{s}\pm m \;,
\label{mupm-scale}
\end{eqnarray}
is assigned. It is counted $\mu_- \sim Q$ and $\mu_+ \sim Q^0$ but it is refrained from any
further expansion. Then the baryon octet functions ${h}^{(I)}_{n \pm}(\sqrt{s},m)$
to order $Q^3$ read
\begin{eqnarray}
&&\Big[{h}_{0+}^{(I)} (\sqrt{s},m)\Big]_{ab}=\sqrt{s}+m +\tilde R^{(I,c)}_{L,ab}
+\tilde R^{(I,c)}_{R,ab}-
\frac{m+M_{ab}^{(L)}}{\mu_{+,ab}^{(I)}({s},m)}\,
\Bigg(\sqrt{s}+
\nonumber\\
&& \qquad\qquad
+\frac{\phi^{(I)}_{a}({s})\,(\sqrt{s}-m_{B(I,b)})}
{\mu_{+,ab}^{(I)}({s},m)\,\mu_{-,ab}^{(I)}({s},m)}
+\frac{(\sqrt s-m_{B(I,a)})\,\phi^{(I)}_{b}({s})}
{\mu_{+,ab}^{(I)}({s},m)\,\mu_{-,ab}^{(I)}({s},m)}
\nonumber\\
&& \qquad\qquad
+\frac{1}{3}\,\frac{\phi^{(I)}_{a}({s})\,(4\,\sqrt s -\mu_{+,ab}^{(I)}({s},m))\,\phi^{(I)}_{b}({s}) }
{\big( \mu_{+,ab}^{(I)}({s},m)\big)^2\,\big(\mu_{-,ab}^{(I)}({s},m)\big)^2}
\Bigg)\,\frac{m+M_{ab}^{(R)}}{\sqrt{s}}+ {\mathcal O}\left( Q^3\right)\,,
\nonumber\\
&&\Big[{h}_{0-}^{(I)} (\sqrt{s},m)\Big]_{ab}=
\frac{m+M_{ab}^{(L)}}{\mu_{-,ab}^{(I)}({s},m)}\,
\Bigg(2\,\sqrt{s}+\mu_{-,ab}^{(I)}({s},m)
\nonumber\\
&&\qquad\qquad
-\frac{8}{3}\,\frac{m_{B(I,a)}\,m_{B(I,b)}}
{ \mu_{+,ab}^{(I)}({s},m)} \Bigg)\,
\frac{m+M_{ab}^{(R)}}{\mu_{+,ab}^{(I)}({s},m)}
+ {\mathcal O}\left( Q\right)\,,
\nonumber\\
&&\Big[{h}_{1+}^{(I)} (\sqrt{s},m)\Big]_{ab}=-\frac23\,
\frac{(m+M_{ab}^{(L)})\,(m+M_{ab}^{(R)}) }{\mu_{-,ab}^{(I)}({s},m)\,
\big(\mu_{+,ab}^{(I)}({s},m)\big)^2}
+{\mathcal O}\left( Q\right)
\nonumber\\
\nonumber\\
&&\Big[{h}_{1-}^{(I)} (\sqrt{s},m)\Big]_{ab}=
-\frac{32}{15}\,
\frac{(m+M_{ab}^{(L)})\,m_{B(I,a)}\,m_{B(I,b)}\,(m+M_{ab}^{(R)}) }{\big(\mu_{-,ab}^{(I)}({s},m)\big)^2\,
\big(\mu_{+,ab}^{(I)}({s},m)\big)^3}
\nonumber\\
&&\qquad\qquad
+\frac43\,
\frac{(m+M_{ab}^{(L)})\,\sqrt{s}\,(m+M_{ab}^{(R)}) }{\big(\mu_{-,ab}^{(I)}({s},m)\big)^2\,
\big(\mu_{+,ab}^{(I)}({s},m)\big)^2}
+{\mathcal O}\left( Q^{-1}\right)
\label{u-approx-1}\;,
\end{eqnarray}
where  $M_{ab}^{(L)}=m_{B(I,a)}+\tilde R^{(I,c)}_{L,ab}$ and
$M_{ab}^{(R)}=m_{B(I,b)}+\tilde R^{(I,c)}_{R,ab}$ were introduced.
The terms of order $Q^3$ can be found in \cite{Lutz:Kolomeitsev} where also the
analogous expressions for the decuplet and baryon-octet resonance
exchanges are presented. Two important points related to the expansion in
(\ref{u-approx-1}) should be emphasized. First, it leads to a separable interaction
kernel. Thus, the induced
Bethe-Salpeter equation is solved conveniently by the covariant projector method of
chapter 3.4. Secondly such an expansion is only meaningful in conjunction with the
renormalization procedure outlined in chapter 3.2. The expansion leads necessarily to
further divergences which require careful attention.

This section is closed with a more detailed discussion of the u-channel exchange. Its
non-local nature necessarily leads to singularities in the partial-wave scattering
amplitudes at subthreshold energies. For instance, the expressions (\ref{u-approx-1})
as they stand turn useless at energies $\sqrt{s}\simeq m_{B(I,a)}+m_{B(I,b)}-m$ due to
unphysical multiple poles at $\mu_-=0$. One needs to address this problem, because
the subthreshold amplitudes are an important input for the many-body treatment of
the nuclear meson dynamics. It is stressed that a singular behavior in the vicinity
of $\mu_-\sim 0$ is a rather general property of any u-channel exchange contribution. It is
not an artifact induced by the chiral expansion. The partial-wave decomposition
of a u-channel exchange contribution represents the pole term only for sufficiently
large or small $\sqrt{s}$. To be explicit consider the u-channel nucleon pole term
contribution of elastic $\pi N$ scattering
\begin{eqnarray}
\sum_{n=0}^\infty \,\int_{-1}^1\frac{d x}{2}\,
\frac{P_n(x)}{\mu_{\pi N}^{(+)}\,\mu_{\pi N}^{(-)}-2\,\phi_{\pi N}\,x
+{\mathcal O}\left( Q^3\right)}\; ,
\label{pin-example:b}
\end{eqnarray}
where $\mu^{(\pm)}_{\pi N}=2\,m_N -\sqrt{s}\pm m_N$.
Upon inspecting the branch points induced by the angular average one concludes that the partial
wave decomposition (\ref{pin-example:b}) is valid only if
$\sqrt{s} >\Lambda_+$ or $\sqrt{s}<\Lambda_-$ with
$\Lambda_\pm=m_N\pm m^2_\pi/m_N+{\mathcal O}(Q^3)$.
For any value in between, $\Lambda_-<\sqrt{s} <\Lambda_+$, the partial-wave decomposition
is not converging. Therefore the following prescription for the
$u$-channel contributions is adopted. The unphysical pole terms in
(\ref{u-approx-1}) are systematically replaced by
\begin{eqnarray}
&& m\,\Lambda^{(\pm)}_{ab} (m)= m\,(m_{B(I,a)}+m_{B(I,b)})-m^2
\nonumber\\
&& \qquad \qquad \pm \left(\Big((m-m_{B(I,b)})^2 - m_{\Phi(I,a)}^2\Big)\,
\Big((m-m_{B(I,a)})^2-m_{\Phi(I,b)}^2\Big)\right)^{1/2} \;,
\nonumber\\
&& \left(\mu^{-1}_{-,ab}(\sqrt{s},m)\right)^n
 \rightarrow
\left(\mu^{-1}_{-,ab}(\Lambda_{ab}^{(-)},m)\right)^n
\,\frac{\sqrt{s}-\Lambda_{ab}^{(+)}}{\Lambda_{ab}^{(-)}-\Lambda_{ab}^{(+)}}
\nonumber\\
&&\qquad \qquad \qquad \quad \;+\,
\left(\mu^{-1}_{-,ab}(\Lambda_{ab}^{(+)},m)\right)^n
\,\frac{\sqrt{s}-\Lambda_{ab}^{(-)}}{\Lambda_{ab}^{(+)}-\Lambda_{ab}^{(-)}} \;,
\label{prescription}
\end{eqnarray}
for  $\Lambda_{ab}^{(-)} < \sqrt{s}< \Lambda_{ab}^{(+)}$
but kept unchanged for $\sqrt{s}>\Lambda^{(+)}_{ab}$ or $\sqrt{s}<\Lambda^{(-)}_{ab}$ in
a given channel (a,b). The prescription (\ref{prescription}) properly generalizes the
$SU(2)$ sector result $\Lambda_\pm\simeq m_N\pm m^2_\pi/m_N $ to the $SU(3)$ sector.
We underline  that (\ref{prescription}) leads to regular expressions for
$h_{\pm n}(\sqrt{s},m)$ but leaves the u-channel pole contributions unchanged above
threshold. The reader may ask why such a prescription is imposed at all. Outside the
convergence radius of the partial-wave expansion the amplitudes do not make much sense in any
case. The prescription is nevertheless required due to a coupled-channel effect.
Consider for example elastic kaon-nucleon scattering in the strangeness -1 channel.
Since there are no u-channel exchange contributions in this channel the physical
partial-wave amplitudes  do not show any induced singularity structures. In the
$\pi \Sigma \to \pi \Sigma $ channel, on the other hand, the u-channel hyperon exchange does
contribute and therefore leads through the coupling of the $\bar K N$ and $\pi \Sigma $ channels
to a singularity also in the $\bar K N$ amplitude. Such induced singularities are
an immediate consequence of the approximate treatment of the u-channel exchange contribution
and must be absent in an exact treatment. As a measure for the quality of the prescription
the accuracy to which the resulting subthreshold forward kaon-nucleon scattering
amplitudes satisfy a dispersion integral representation is proposed. This issue will be taken up
in the result chapter.

\vskip1.5cm \section{Crossing symmetry and the $S=1$ channels}

This section addresses an issue of central importance for coupled channel
analyses. To what extent do the coupled channel scattering amplitudes
as constructed in the previous sections comply with the crossing symmetry constraint
of quantum field theory. The discussion necessarily involves the $K^+$-nucleon
scattering process which is related to the $K^-$-nucleon scattering
process by crossing symmetry.

The scattering amplitudes $T^{(I)}_{K N \rightarrow K N}$ follow from the
antikaon-nucleon scattering amplitudes $T^{(I)}_{\bar K N \rightarrow \bar K N}$
via the transformation
\begin{eqnarray}
&&T^{(0)}_{K N \rightarrow K N}(\bar q, q; w)
=-\frac{1}{2}\, T^{(0)}_{\bar K N \rightarrow \bar K N}(-q, -\bar q; w-\bar q-q)
\nonumber\\
&&\quad \quad \quad \quad \quad
+\frac{3}{2}\,T^{(1)}_{\bar K N \rightarrow \bar K N}(-q, -\bar q; w-\bar q-q) \;,
\nonumber\\
&&T^{(1)}_{K N \rightarrow K N}(\bar q, q; w)
=+\frac{1}{2}\, T^{(0)}_{\bar K N \rightarrow \bar K N}(-q, -\bar q; w-\bar q-q)
\nonumber\\
&&\quad \quad \quad \quad \quad
+\frac{1}{2}\,T^{(1)}_{\bar K N \rightarrow \bar K N}(- q, -\bar q; w-\bar q-q) \;.
\label{cross-symmetry}
\end{eqnarray}

Consider the two important remarks. First, if the solution of the coupled channel
Bethe-Salpeter equation of the antikaon-nucleon system of the previous chapters is
used to construct the kaon-nucleon scattering amplitudes via (\ref{cross-symmetry})
one finds real partial-wave amplitudes in conflict with unitarity. Second, in any
case the antikaon-nucleon scattering amplitudes must not be applied far below the antikaon-nucleon
threshold. The first point is obvious because in the kaon-nucleon channel only two-particle
irreducible diagrams are summed. The reducible diagrams in the $K\,N$ sector
correspond to irreducible contributions in the $\bar K\, N$ sector and vice versa.
Since the interaction kernel of the $\bar K \,N$ sector is evaluated perturbatively it is clear
that the scattering amplitude does not include that infinite sum of reducible diagrams
required for unitarity in the crossed channel.
The second point follows, because the chiral $SU(3)$ Lagrangian
is an effective field theory where the heavy-meson exchange contributions are integrated out.
It is important to identify the applicability domain correctly. Inspecting the
singularity structure induced by the light t-channel vector meson exchange contributions one
observes that they, besides restricting the applicability domain with
\begin{eqnarray}
\sqrt{s} < \sqrt{m_N^2+m_\rho^2/4}+\sqrt{m_K^2+m_\rho^2/4}\simeq 1640\; {\rm MeV}\;,
\label{}
\end{eqnarray}
from above, induce cut-structures in between the kaon- and antikaon-nucleon thresholds.
This will be discussed in more detail below.
Though close to the kaon and antikaon-nucleon thresholds the tree-level contributions
of the vector meson exchange are successfully represented by quasi-local interaction vertices,
basically the Weinberg-Tomozawa interaction term, it is not justified to
extrapolate a loop evaluated with the effective antikaon-nucleon vertex down to the
kaon-nucleon threshold. Thus, one should not identify the antikaon-nucleon scattering
amplitudes as obtained in this work with the Bethe-Salpeter kernel of the kaon-nucleon
system. This would lead to a manifestly crossing symmetric approach,
a so-called 'parquet' approximation, if set up in a self consistent manner.

The fatal drawback of a parquet type of approach within the present chiral
framework is reiterated: the antikaon-nucleon amplitudes would be probed far
below the $\bar K N$ threshold at $\sqrt{s}\simeq m_N-m_K$ outside their validity
domain. A clear signal for the
unreliability of the antikaon-nucleon scattering amplitudes at $\sqrt{s}\simeq m_N-m_K$ is the
presence of unphysical pole structures which typically arise at $\sqrt{s}< 700$ MeV. For
example, the Fig. \ref{fig:wt} of chapter 3.3  would show unphysical poles if
extended for $\sqrt{s}< 1$ GeV.

\tabcolsep=1.3mm
\renewcommand{\arraystretch}{1.5}
\begin{table}[t]
\begin{tabular}{|r||c||c|c|c||c|c||c|c|c|c||c|c|c|c|}
 \hline
& $ C_{WT}^{(I)} $  &  $ C_{N_{[8]}}^{(I)}$  & $C_{\Lambda_{[8]}}^{(I)}$  &
$C_{\Sigma_{[8]}}^{(I)}$
& $C_{\Delta_{[10]}}^{(I)}$ & $C_{\Sigma_{[10]} }^{(I)}$
    &$\widetilde{C}_{N_{[8]}}^{(I)}$ & $\widetilde{C}_{\Lambda_{[8]} }^{(I)}$ &
    $\widetilde{C}_{\Sigma_{[8]}}^{(I)}$ & $\widetilde {C}_{\Xi_{[8]} }^{(I)}$   &
$\widetilde {C}_{\Delta_{[10]} }^{(I)}$ &
    $\widetilde {C}_{\Sigma_{[10]}}^{(I)}$& $\widetilde C_{\Xi_{[10]}}^{(I)}$
\\ \hline \hline
$I=0$& 0 & 0 & 0 & 0 & 0 & 0
     & 0 & -$\frac{1}{2}$ & $\frac{3}{2}$ & 0 & 0 & $\frac{3}{2}$ & 0\\ \hline \hline

$I=1$& -2 & 0 & 0 & 0 & 0 & 0
     & 0  & $\frac{1}{2}$ & $\frac{1}{2}$ & 0 & 0 & $\frac{1}{2}$ & 0\\ \hline
\end{tabular}
\vspace*{2mm} \caption{Weinberg-Tomozawa interaction strengths and baryon exchange coefficients in the
strangeness plus one channels as defined in (\ref{k-nonlocal}).}
\label{tabkp-1}
\end{table}

For the above reasons  the antikaon-nucleon interaction kernel is evaluated in chiral
perturbation theory. Equivalently the interaction kernels,
$V_{K N}$, could be derived from the antikaon-nucleon interaction kernels
$V_{\bar K N}$ by applying the crossing identities (\ref{cross-symmetry}).
The interaction kernels $V_{K N}(\sqrt{s};n)$ in the
$K^+$-nucleon channel are given by (\ref{local-v},\ref{v-result-1},\ref{u-result}) with the
required coefficients listed in Tab. \ref{tabkp-1} and Tab. \ref{tabkp-2}. By analogy with
the treatment of the antikaon-nucleon scattering process the kaon-nucleon interaction
evaluated to chiral order $Q^3$ is taken as input in the Bethe-Salpeter equation.
Again only those $Q^3$-correction terms are considered which are leading in the $1/N_c$
expansion. The partial-wave scattering
amplitudes $M_{KN}^{(I,\pm )}(\sqrt{s};n)$ then follow
\begin{eqnarray}
M_{KN}^{(I,\pm )}(\sqrt{s};n) &=& \frac{V_{KN}^{(I,\pm )}(\sqrt{s};n)}
{1-V_{KN}^{(I,\pm)}(\sqrt{s};n)\,J^{(\pm)}_{KN}(\sqrt{s}; n)} \;
\label{M:kplus}
\end{eqnarray}
where the loop functions $J^{(\pm)}_{KN}(\sqrt{s};n)$ are given in (\ref{result-loop:ab}) with
$m_{B(I,a)}=m_N$ and $m_{\Phi(I,a)}=m_K$. The subtraction point $\mu^{(I)}$ is identified
with the averaged hyperon mass $\mu^{(I)}=(m_\Lambda+m_\Sigma)/2$.

It is pointed out that in the scheme developed here one arrives at a
crossing symmetric amplitude by matching the amplitudes  $M^{(I,\pm)}_{KN}(\sqrt{s},n)$
and $M^{(I,\pm)}_{\bar KN}(\sqrt{s},n)$ at subthreshold energies. The matching interval
must be chosen so that both the kaon and antikaon amplitudes are still within their
validity domains. A complication arises due to the light vector meson exchange contributions
in the t-channel, which was already identified above to restrict the validity domain of the
present chiral approach. To be explicit consider the t-channel $\omega $-exchange. It leads
to a branch point at $\sqrt{s}= \Lambda_\omega $ with
\begin{eqnarray}
\Lambda_\omega =(m_N^2-m_\omega^2/4)^{1/2}+(m_K^2-m_\omega^2/4)^{1/2} \simeq  1138 \;{\rm MeV}\,.
\label{}
\end{eqnarray}
Consequently, one expects the partial-wave kaon and antikaon amplitudes to be reliable
for $\sqrt{s} > \Lambda_\omega $ only. That implies, however, that in the crossed channel
the amplitude is needed for
\begin{eqnarray}
\sqrt{s}>(2\,m_N^2-\Lambda^2_\omega +2\,m_K^2)^{1/2}\simeq  978 \;{\rm MeV }\;
 < \Lambda_\omega \;.
\label{}
\end{eqnarray}

One may naively conclude that crossing symmetry appears outside the scope of the present
approach because the matching window is closed.  The minimal critical point
$\Lambda_{\rm opt}$ needed to open the matching window is
\begin{eqnarray}
\Lambda_{\rm opt.}\simeq \sqrt{m_N^2+m_K^2}\simeq  1061 {\rm MeV}\;.
\label{opt-match}
\end{eqnarray}
It is determined by the condition $s=u$ at $\cos \theta =1$. One concludes that matching
the kaon and antikaon amplitudes requires that both amplitudes are within their applicability
domain at $\sqrt{s}= \Lambda_{\rm opt.}$. The point $\sqrt{s}=\Lambda_{\rm opt.}$ is optimal,
because it identifies the minimal reliability domain required for the matching of the
subthreshold amplitudes. The complication implied by the t-channel vector
meson exchanges could be circumvented by reconstructing that troublesome branch point
explicitly; after all it is determined by a tree-level diagram.
On the other hand, it is clear that one may avoid this complication altogether if one
considers the forward scattering amplitudes only. For the latter amplitudes the branch cut at
$\sqrt{s}=\Lambda_\omega $ cancels identically and consequently the forward scattering
amplitudes should be reliable for energies smaller than $ \Lambda_\omega$ also.
For that reason, the matching will be demonstrated for the forward scattering
amplitudes only. The approximate forward scattering amplitudes
$T^{(I)}_{KN}(s)$ are reconstructed in terms the partial-wave amplitudes considered
in this work
\begin{eqnarray}
T^{(I)}_{KN}(s) &=&
\frac{1}{2\,\sqrt{s}}\left(\frac{s+m_N^2-m_K^2}{2\,m_N}+\sqrt{s} \right)
M^{(I,+)}_{KN}(\sqrt{s};0)
\nonumber\\
&+&\frac{1}{2\,\sqrt{s}}\left(\frac{s+m_N^2-m_K^2}{2\,m_N} -\sqrt{s}\right)
M^{(I,-)}_{KN}(\sqrt{s};0)
\nonumber\\
&+& \frac{1}{\sqrt{s}}\left(\frac{s+m_N^2-m_K^2}{2\,m_N}+\sqrt{s} \right)
\,p^2_{KN}\,M^{(I,+)}_{KN}(\sqrt{s};1)
\nonumber\\
&+& \frac{1}{\sqrt{s}}\left(\frac{s+m_N^2-m_K^2}{2\,m_N}-\sqrt{s} \right)
\,p^2_{KN}\,M^{(I,-)}_{KN}(\sqrt{s};1) +\cdots \,,
\label{forward-amplitude}
\end{eqnarray}
where $\sqrt{s}=\sqrt{m_N^2+p_{KN}^2}+\sqrt{m_K^2+p_{KN}^2}$. The analogous
expression holds for $T^{(I)}_{\bar KN}(s)$.  The crossing identities for
the forward scattering amplitudes
\begin{eqnarray}
T^{(0)}_{KN}(s) &=& -\frac{1}{2}\,T^{(0)}_{\bar K N} (2\,m_N^2+2\,m_K^2-s)
+\frac{3}{2}\,T^{(1)}_{\bar K N} (2\,m_N^2+2\,m_K^2-s) \;,
\nonumber\\
T^{(1)}_{KN}(s) &=& +\frac{1}{2}\,T^{(0)}_{\bar K N} (2\,m_N^2+2\,m_K^2-s)
+\frac{1}{2}\,T^{(1)}_{\bar K N} (2\,m_N^2+2\,m_K^2-s)\;
\label{forward-crossing}
\end{eqnarray}
are expected to hold approximatively within some matching window centered around
$\sqrt{s}\simeq \Lambda_{\rm opt}$.
A further complication arises due to the approximate treatment of the u-channel exchange
contribution in the $K \,N$-channel. Since the optimal matching point
$\Lambda_{\rm opt.}$ is not too far away from the hyperon poles with
$\Lambda_{\rm opt.} \sim  m_\Lambda, m_\Sigma$ it is advantageous to investigate
approximate crossing symmetry for the hyperon-pole term subtracted scattering
amplitudes.

\tabcolsep=1.75mm
\renewcommand{\arraystretch}{1.4}
\begin{table}[t]
\vspace{-1.2cm}
\begin{tabular}{|r||c||c|c||c|c|c||c|c|c|c||c|c|c|}
\multicolumn{14}{c}{}\\ \multicolumn{14}{c}{}\\
 \hline
    & $C_{\pi,0}^{(I)}$ &  $C_{\pi,D}^{(I)}$ &  $C_{\pi,F}^{(I)}$
    & $C_{K,0}^{(I)}$   & ${C}_{K,D}^{(I)}$  & ${C}_{K,F}^{(I)}$
    & ${C}_{0}^{(I)}$   &  ${C}_{1}^{(I)}$  &
    ${C}_{D}^{(I)}$ & ${C}_{F}^{(I)}$
    & $\bar C_{1}^{(I)}$ & $\bar C_{D}^{(I)}$ & $\bar C_{F}^{(I)}$\\
\hline \hline
$I=0$& 0 & 0 & 0
    & -4 & 0 & 4
    & 2 & -1 & 0& -2
    & 1 & 2& 0   \\ \hline  \hline
$I=1$&0 & 0 & 0
    & -4 & -4& 0
    & 2  & 1  & 2& 0
    & -1& 0 &  -2   \\ \hline
\end{tabular}
\caption{Coefficients of quasi-local interaction terms in the strangeness plus one
channels as defined in (\ref{def-local-1}).}
\label{tabkp-2}
\end{table}

It should be stressed that the expected approximate crossing symmetry is
closely linked to the renormalization condition (\ref{ren-cond}). Since all
loop functions $J^{(\pm )}(\sqrt{s},n)$
vanish close to $\sqrt{s}= (m_\Lambda+m_\Sigma)/2$ by construction, the
kaon and antikaon amplitudes turn perturbative sufficiently close to the optimal matching point
$\Lambda_{\rm opt.} $. Therefore the approximate crossing symmetry of the scheme
developed here follows directly from the crossing symmetry of the interaction kernel.
In the result chapter the final $K^\pm N$ amplitudes are confronted with the
expected approximate crossing symmetry.

Finally, for pion-nucleon scattering the situation is rather different. Given
the rather small pion mass, the t-channel vector meson exchange contributions do not induce
singularities in between the $\pi^+$ and $\pi^-$-nucleon thresholds but restrict the
applicability domain of the chiral Lagrangian to
\begin{eqnarray}
\sqrt{s} < \sqrt{m_N^2+m_\rho^2/4}+\sqrt{m_\pi^2+m_\rho^2/4}\simeq 1420\;{\rm MeV}\;.
\label{}
\end{eqnarray}
The approximate crossing symmetry follows directly from the perturbative character of
the pion-nucleon sector.

\cleardoublepage

\chapter{Goldstone bosons in cold nuclear matter}
\label{k5}
\markboth{\small CHAPTER \ref{k5}.~~~Goldstone bosons in cold nuclear matter}{}

In this chapter a novel and covariant many-body framework to evaluate the Goldstone boson
propagation in cold nuclear matter is introduced. While the
formalism is quite general and applicable for pions and etas as well, the
focus here is on the properties of antikaons in nuclear matter. The self consistent
approach is based on the chiral SU(3) dynamics derived in the previous chapters.
Besides deriving a realistic antikaon spectral function quantitative results on the
in-medium structure of the s-wave $\Lambda(1405)$, p-wave $\Sigma (1385)$ and the
d-wave $\Lambda(1520)$ resonances are obtained.

The starting point of any microscopic evaluation of meson propagation in nuclear matter
should be a well established theory of the meson's scattering process off nucleons, the
constituents of nuclear matter. According to the low-density theorem (\ref{LDT-energy}) the
meson-nucleon scattering amplitudes determine the self energy of a meson at small
density \cite{dover,njl-lutz}. Once the appropriate scattering amplitudes are available
a difficult task remains, namely to explore what small density really means
in (\ref{LDT-energy}). For the case of antikaons it was demonstrated in \cite{ml-sp}
that the density expansion stops converging at roughly $\rho \simeq 0.02$ fm$^{-3}$, about
one tens of the nuclear saturation density $\rho_0 \simeq 0.17$ fm$^{-3}$. In contrast to
that the density expansion is expected to predict solid results for the kaon self energy
\cite{ml-sp} up to about $2 \,\rho_0$ provided that sufficiently many terms in the expansion
are kept \cite{ml-sp}. This is nicely illustrated upon studying the leading and subleading
terms in the density expansion of the mass change $\Delta m_K^2$ of a kaon at rest.
All what is required are the isospin zero and isospin one s-wave kaon-nucleon scattering
lengths:
\begin{eqnarray}
\Delta m_{K}^2 &=& -\pi \left( 1+\frac{m_K}{m_N} \right)
\left( a^{(I=0)}_{KN}+3\,a^{(I=1)}_{KN} \right)
\,\rho
\nonumber\\
&+& \alpha \left( \left(a_{KN}^{(I=0)}\right)^2 +3\,\left(a_{KN}^{(I=1)}\right)^2 \right)
 k_F^4 +{\cal O}\left( k_F^5 \right) \,,
\label{}
\end{eqnarray}
where
\begin{eqnarray}
\alpha &=&\frac{1-x^2+x^2\,\log
\left(x^2\right)}{\pi^2\,(1-x)^2}
\simeq 0.166 \,, \qquad x= \frac{m_K}{m_N}\,,\qquad \rho = \frac{2\,k_F^3}{3\,\pi^2 }\,.
\label{lowdensity}
\end{eqnarray}
Using typical values for the kaon-nucleon scattering lengths,
the correction term of order $k_F^4$ is indeed small. At nuclear
saturation density with $k_F \simeq 265 $ MeV it increases the
repulsive mass shift of the kaon from $28$ MeV to $ 35 $ MeV by about $20
\%$. Thus, the density expansion is useful in this case. Any
microscopic model consistent with the low-energy kaon-nucleon
scattering data is bound to give similar results for the
$K^+$ and $K^0$ propagation in nuclear matter at densities $\rho
\simeq \rho_0 $ sufficiently small to maintain the density expansion rapidly
convergent. Applying the low density expansion (\ref{lowdensity}) for the antikaon
a leading repulsive mass shift of $23$ MeV with a width of about $147 $ MeV arises at saturation
density. The correction term results in a total repulsive mass shift of $55 $
MeV and a width of about $195 $ MeV. At nuclear saturation the density expansion
for the antikaon-mode is poorly convergent if at all. Furthermore,
the leading terms appear to contradict kaonic atom data \cite{Gal}
which suggest sizable attraction at small density.

The physical reason for the breakdown of the
density expansion for antikaons is the presence of the $\Lambda (1405)$ resonance just below the antikaon-nucleon
threshold.
In order to obtain reliable results for the antikaon spectral function at saturation density $\rho_0$ or above
an infinite resummation scheme is asked for. Since the breakdown of the density expansion is closely linked to
a rapid in-medium change of the hyperon resonance properties it is sufficient to evaluate the antikaon and
hyperon resonance spectral functions in a self consistent manner. That leads to the required infinite
resummation scheme of the density expansion. Schematically this effect can be illustrated in
terms of an elementary $\Lambda (1405 )$ field dressed by an antikaon-nucleon loop.
The repulsive $\Lambda (1405 )$ mass shift due to the Pauli
blocking  of the nucleon is then given by:
\begin{eqnarray}
\Delta\,m_\Lambda =\frac{g_{\Lambda NK}^2}{\pi^2}\,\frac{m_N}{m_\Lambda} \,
\left(1-\frac{\mu_\Lambda}{k_F}\,\arctan \left( \frac{k_F}{\mu_\Lambda} \right)
\right)k_F
\label{pauli-lambda}
\end{eqnarray}
with the 'small' scale
\begin{eqnarray}
\mu_\Lambda^2 &=&\frac{m_N}{m_\Lambda }\left(m_K^2-\Big(m_\Lambda-m_N\Big)^2 \right)
\simeq \left( 144\, {\rm MeV } \right)^2
\label{scale}
\end{eqnarray}
and the $\Lambda(1405)  $ kaon nucleon coupling constant
$g_{\Lambda NK}$. The result (\ref{pauli-lambda}) clearly demonstrates that a density
expansion of the resonance properties must break down at $k_F \simeq \mu_\Lambda $ where one
hits the convergence radius of the arctan. Since the scale $\mu_\Lambda $ is rather small and
also sensitively dependent on a mass shift of the antikaon a self consistent scheme is
indispensable to study the physics at nuclear saturation density quantitatively.
The relevant Fermi momentum
$k_F \simeq 270$ MeV is much larger than the small scale $\mu_\Lambda \simeq 140$ MeV and therefore a
low-density expansion is inappropriate. It is evident that analogous arguments hold for the p-wave
$\Sigma (1385)$ resonance which therefore should be part of a self consistent computation.

Self consistent evaluations of the antikaon spectral function in nuclear matter are so far typically
based on s-wave dynamics only \cite{ml-sp,ramossp,Cieply}. Although these calculations agree qualitatively
they are based on different realizations of the coupled channel s-wave interactions and lead to
significant differences in the antikaon self energy important when describing kaonic atom data
\cite{Florkowski,Cieply}. Moreover, since a strong momentum
dependence of the antikaon spectral function is predicted, it is important to generalize these calculations and include
higher partial waves \cite{Hirschegg}. The first work that considers higher partial waves in
this context but relying on a quasi-particle approximation for the antikaon spectral function
was by Tolos et al. \cite{Tolos}. A meson-exchange model of the J\"ulich group was
applied \cite{Juelich:2}. Unfortunately, the latter model has so far been confronted with total
cross section data only and therefore it remains
unclear whether it describes the dynamics of higher partial waves correctly. More seriously, as was
already emphasized in
\cite{Juelich:2} that model leads to unrealistic scattering amplitudes at subthreshold energies, an energy domain
of upmost importance for the nuclear antikaon dynamics.

\vskip1.5cm \section{Self consistent and covariant nuclear dynamics}

Here a covariant framework in which self consistency is implemented in terms of the vacuum
meson-nucleon scattering amplitudes is constructed. The scheme is capable to treat
consistently the spectral distribution of the meson and also
the in-medium mixing of partial wave amplitudes. In order to make this chapter self contained
all requisites needed for the formulation are briefly recalled.

Of central importance is the free-space on-shell antikaon-nucleon scattering amplitude:
\begin{eqnarray}
\langle \bar K^{j}(\bar q)\,N(\bar p)|\,T\,| \bar K^{i}(q)\,N(p) \rangle
&=&(2\pi)^4\,\delta^4(p+q-\bar q-\bar p)\,
\nonumber\\
&& \! \!\! \! \times \,\bar u(\bar p)\,
T^{ij}_{\bar K N \rightarrow \bar K N}(\bar q,\bar p ; q,p)\,u(p)\,,
\label{on-shell-scattering}
\end{eqnarray}
where $u(p)$ is the
nucleon isospin-doublet spinor. The isospin doublet notation is used also for the antikaons with
$\bar K =(K^-, \bar K^0)$. The vacuum scattering amplitude is decomposed into its isospin
zero and isospin one channels with:
\begin{eqnarray}
&&T^{i j}_{\bar K N \to \bar K N}(\bar q,\bar p \,; q,p)
= T^{(0)}_{\bar K N\to \bar K N}(\bar k,k;w)\,P^{ij}_{(I=0)}+
T^{(1)}_{\bar K N\to \bar K N }(\bar k,k;w)\,P^{ij}_{(I=1)}\;,
\nonumber\\
&& P^{ij}_{(I=0)}= \frac{1}{4}\,\Big( \delta^{ij}\,1
+ \big( \vec \tau\,\big)^{ij}\,\vec \tau \,\Big)\, , \quad
P^{ij}_{(I=1)}= \frac{1}{4}\,\Big( 3\,\delta^{ij}\,1-
\big(\vec \tau \,\big)^{ij}\,\vec \tau \,\Big)\;,
\label{}
\end{eqnarray}
where
\begin{eqnarray}
w = p+q = \bar p+\bar q\,,
\quad k= {\textstyle{1\over 2}}\,(p-q)\,,\quad
\bar k ={\textstyle{1\over 2}}\,(\bar p-\bar q)\,.
\label{def-moment}
\end{eqnarray}
In quantum field theory the scattering amplitudes $T^{(I)}_{\bar K N}$ arise
in terms of the Bethe-Salpeter matrix equation,
\begin{eqnarray}
T(\bar k ,k ;w ) &=& K(\bar k ,k ;w )
+\int \frac{d^4l}{(2\pi)^4}\,K(\bar k , l;w )\, G(l;w)\,T(l,k;w )\;,
\nonumber\\
G(l;w)&=&-i\,S_N({\textstyle
{1\over 2}}\,w+l)\,D_{\bar K}({\textstyle {1\over 2}}\,w-l)\,,
\label{BS-eq}
\end{eqnarray}
specified by the Bethe-Salpeter scattering kernel $K(\bar k,k;w)$, the free-space nucleon
propagator $S_N(p)=1/(\pslash-m_N+i\,\epsilon)$ and kaon propagator
$D_K(q)=1/(q^2-m_K^2+i\,\epsilon)$. Following previous considerations
self energy corrections in the nucleon and kaon propagators that
arise in free-space are neglected. In a chiral scheme such
effects are of subleading order. The Bethe-Salpeter equation (\ref{BS-eq}) implements
properly  Lorentz invariance and unitarity for the two-body scattering process. The
generalization of (\ref{BS-eq}) to a coupled channel system is straightforward (see (\ref{BS-coupled})).

The antikaon-nucleon scattering process is readily generalized from the
vacuum to the nuclear matter case. It is convenient to introduce the compact notation:
\begin{eqnarray}
&& {\mathcal T} = {\mathcal K} + {\mathcal K} \cdot {\mathcal G} \cdot {\mathcal T}  \;,\quad
{\mathcal T} = {\mathcal T}(\bar k,k; w,u) \;, \quad {\mathcal G} = {\mathcal G} (l;w,u) \,,
\label{hatt}
\end{eqnarray}
where the  in-medium scattering amplitude ${\mathcal T}(\bar k,k;w,u)$ and the two-particle
propagator ${\mathcal G}(l;w,u)$ depend now on the 4-velocity $u_\mu$
characterizing the nuclear matter frame. For nuclear matter moving with a velocity
$\vec v$ we write:
\begin{eqnarray}
u_\mu =\left(\frac{1}{\sqrt{1-\vec v\,^2/c^2}},\frac{\vec v/c}{\sqrt{1-\vec v\,^2/c^2}}\right)
\;, \quad u^2 =1\,.
\label{}
\end{eqnarray}
It is important to realize that (\ref{hatt}) is well defined from a Feynman diagrammatic point
of view even in the case where the in-medium scattering process is not well defined anymore due
to a broad antikaon spectral function. In this work the medium modifications of the interaction
kernel is not considered, i.e. the approximation ${\mathcal K} = K$ is applied.
It is studied exclusively the effect of a modified two-particle propagator ${\mathcal G}$ with
\begin{eqnarray}
&& \Delta S_N (p,u) = 2\,\pi\,i\,\Theta \Big(p\cdot u \Big)\,
\delta(p^2-m_N^2)\,\big( \pslash +m_N \big)\,
\Theta \big(k_F^2+m_N^2-(u\cdot p)^2\big)\,,
\nonumber\\
&&{\mathcal S}_N(p,u) = S_N(p)+ \Delta S_N(p,u)\,, \quad
{\mathcal D}_{\bar K}(q,u)=\frac{1}{q^2-m_K^2-\Pi_{\bar K}(q,u)} \;,
\nonumber\\
&& {\mathcal G}(l;w,u) = -i\,{\mathcal S}_N({\textstyle
{1\over 2}}\,w+l,u)\,{\mathcal D}_{\bar K}({\textstyle {1\over 2}}\,w-l,u)  \;.
\label{hatg}
\end{eqnarray}
For isospin symmetric nuclear matter the nucleon density $\rho $ follows with
\begin{eqnarray}
\rho = -2\,\tr \,\gamma_0\,\int \frac{d^4p}{(2\pi)^4}\,i\,\Delta S_N(p,u)
= \frac{2\,k_F^3}{3\,\pi^2\,\sqrt{1-\vec v\,^2/c^2}}  \;,
\label{rho-u}
\end{eqnarray}
where $\rho= \rho_n+ \rho_p$ and $k_F=k_{F,n}=k_{F,p}$.
In the rest frame of the bulk with $u_\mu=(1,\vec 0\,)$ one recovers with (\ref{rho-u}) the
standard result $\rho = 2\,k_F^3/(3\,\pi^2)$. In this work nucleonic correlation effects
are not considered. Such effects have been estimated to be small
at moderate densities \cite{Waas2} but should nevertheless be subject of a more complete
consideration. The antikaon self energy $\Pi_{\bar K}(q,u)$ is evaluated self
consistently in terms of the in-medium scattering amplitudes
${\mathcal T}_{\bar K N}^{(I)}(\bar k,k;w,u)$:
\begin{eqnarray}
\Pi_{\bar K}(q,u) &=& -2\,\tr \int \frac{d^4p}{(2\pi)^4}\,i\,\Delta S_N(p,u)\,
\bar {\mathcal T}_{\bar K N \to \bar K N}\big({\textstyle{1\over 2}}\,(p-q),
{\textstyle{1\over 2}}\,(p-q);p+q,u \big)\,,
\nonumber\\
\bar {\mathcal T}_{\bar K N\to \bar K N}&=& \frac{1}{4}\,{\mathcal T}_{\bar K N \to \bar K N}^{(I=0)}+
\frac{3}{4}\,{\mathcal T}_{\bar K N \to \bar K N}^{(I=1)} \;.
\label{k-self}
\end{eqnarray}
The generalization of (\ref{k-self}) to asymmetric nuclear matter
with $\rho_p \neq \rho_n$ is straightforward. In this case, the two
isospin channels couple. Hence it is more convenient to apply the particle basis
rather than the isospin basis. The relevant Clebsch Gordan coefficients are:
\begin{eqnarray}
T^{}_{K^-p\to K^- p } \!\!\!\!&=&\!\!\!\!
\frac{1}{2}\left( T_{\bar K N \to \bar K N}^{(0)}+T_{\bar K N\to \bar K N}^{(1)}\right)
\,, \quad \!\!
T^{}_{K^-n\to K^- n } = T_{\bar K N \to \bar K N}^{(1)} = T^{}_{\bar K^0p\to \bar K^0 p }\;,
\nonumber\\
T^{}_{\bar K^0n\to \bar K^0 n } \!\!\!\!&=& \!\!\!\! \frac{1}{2}\left( T_{\bar K N\to \bar K N}^{(0)}+T_{\bar K N\to \bar K N}^{(1)}\right)
\,, \quad \!\!
T^{}_{\bar K^0n\to \bar K^- p } = \frac{1}{2}\left( T_{\bar K N\to \bar K N}^{(0)}-T_{\bar K N\to \bar K N}^{(1)}\right)\,.
\label{}
\end{eqnarray}
In order to solve the self consistent set of equations (\ref{hatt},\ref{hatg},\ref{k-self})
it is convenient to rewrite the scattering amplitude in the following way
\begin{eqnarray}
&&{\mathcal T}= K+K\cdot {\mathcal G}\cdot {\mathcal T}
= T+T\cdot \Delta {\mathcal G} \cdot {\mathcal T}\;,\quad
\Delta {\mathcal G}={\mathcal G}-G\;,
\label{rewrite}
\end{eqnarray}
where $T=K+K\cdot G\cdot T$  is the vacuum scattering amplitude. The idea is to
start with a set of tabulated coupled channel scattering amplitudes $T$ in terms of which
self consistency is implemented. Thus, the vacuum interaction kernel $K$ need not to be
specified.

Coupled channel effects, which are known to be important
for $\bar K N$ scattering, are included by assigning ${\mathcal T}$, ${\mathcal G}$ and $K$
the appropriate matrix structures. Here we use the convention introduced in (\ref{r-def}). For
example the isospin zero loop matrix $\Delta {\mathcal G}^{(I=0)}$ reads:
\begin{eqnarray}
\Delta {\mathcal G}^{(I=0)} =
\left(
\begin{array}{cccc}
\Delta {\mathcal G}_{\bar K N} & 0 & 0 & 0 \\
0& \Delta {\mathcal G}_{\pi \Sigma} & 0 & 0 \\
0 & 0 &\Delta {\mathcal G}_{\pi \Lambda} & 0 \\
0 & 0 & 0 & \Delta {\mathcal G}_{K \Xi }
\end{array}
\right) \;.
\label{}
\end{eqnarray}
This work focuses on the in-medium modification of the $\bar K N$ channel.
Ultimately it would be desirable to include also the effects of the in-medium modified
$\pi \Sigma$, $\pi \Lambda$ and $K \Xi$ channels. Since this requires a realistic
in-medium pion propagator to be determined ultimately also by a self consistent
scheme \cite{Korpa} such effects will be considered in a separate work. Here approximation
$\Delta {\mathcal G}_{\pi \Sigma} =0$ and $\Delta {\mathcal G}_{\pi \Lambda}=0$ is applied.
Since the $K^+$ spectral density is affected only moderately by a nuclear environment
it is assumed $\Delta {\mathcal G}_{K \Xi }=0$ also.  With these assumptions
the coupled channel problem reduces to a single channel problem if
rewritten in terms of the vacuum $\bar K N$ amplitude $T_{\bar K N \to \bar K N}$:
\begin{eqnarray}
&&{\mathcal T}_{\bar K N \to \bar K N}^{(I)}=  T_{\bar K N \to \bar K N}^{(I)}
+T_{\bar KN \to \bar K N}^{(I)}\cdot \Delta {\mathcal G}_{\bar KN}
\cdot {\mathcal T}_{\bar K N \to \bar K N}^{(I)}\;,\quad
\nonumber\\
&& \Delta {\mathcal G}_{\bar K N}={\mathcal G}_{\bar K N}-G_{\bar K N}\;.
\label{rewrite:b}
\end{eqnarray}
It should be pointed out that the self consistent set of equations
(\ref{hatg},\ref{k-self},\ref{rewrite:b})
are now completely determined by the vacuum amplitudes $T_{\bar K N \to \bar K N}^{(I)}$.
Even though the antikaon spectral function is
a functional of the elastic $\bar K N \to \bar K N$ scattering process only,
the inelastic channels $\bar K N \to X$ are nevertheless affected by
$\Delta {\mathcal G}_{\bar K N}$. Simple algebra leads to:
\begin{eqnarray}
&&{\mathcal T}^{(I)}_{\bar K N \to X} = T^{(I)}_{\bar K N \to X} +
{\mathcal T}^{(I)}_{\bar K N \to \bar K N} \cdot \Delta {\mathcal G}_{\bar K N}
\cdot T^{(I)}_{\bar K N \to X} \;,
\nonumber\\
&& {\mathcal T}^{(I)}_{Y \to X} = T^{(I)}_{Y \to X}+
T^{(I)}_{Y \to \bar K N} \cdot \Delta {\mathcal G}_{\bar K N} \cdot
{\mathcal T}^{(I)}_{\bar K N \to X}\;,
\label{inelastic}
\end{eqnarray}
where $X, Y \neq \bar K N$. In (\ref{inelastic}) the effect of
$\Delta {\mathcal G}_{\bar K N} \neq 0$ on the reaction
$X \to Y$ with $X,Y\neq \bar K N$ is demonstrated.

In the following section it is outlined how to solve the self consistent set of
equations (\ref{hatg},\ref{k-self},\ref{rewrite:b}) by introducing an appropriate
projector algebra. The scheme takes into account the spectral distribution of the
meson as well as the medium induced mixing of partial waves usually neglected in
conventional G-matrix calculations (see e.g. \cite{Tolos}).

\vskip1.5cm \section{Covariant projector algebra method in nuclear matter}

It is useful to briefly recall the coupled channel theory of chapter 4.
The scattering amplitudes were obtained in a relativistic chiral $SU(3)$ approach
by decomposing the amplitudes systematically into covariant projectors $Y^{(\pm)}_n$ with
good angular momentum $J=n+{\textstyle{1\over 2}}$
\begin{eqnarray}
T^{(I)}(\bar k,k;w) &=& \sum_{n=0}^\infty\,
Y^{(+)}_n (\bar q, q;w)\,M^{(I,+)} (\sqrt{s}\,, n)
\nonumber\\
&+& \sum_{n=0}^\infty\,
 Y^{(-)}_n (\bar q, q;w)\,M^{(I,-)} (\sqrt{s}\,, n)  \,,
\label{t-vacuum}
\end{eqnarray}
where $w^2=s$ and $k= {\textstyle{1\over 2}}\,(p-q)$ and $ \bar k ={\textstyle{1\over 2}}\,(\bar p-\bar q)$. The
representation (\ref{t-vacuum}) implies a particular off-shell behavior of the scattering amplitude, which was
optimized as to simplify the task of solving the covariant Bethe-Salpeter equation. This is legitimate because
the off-shell structure of the scattering amplitude is not determined by the two-body scattering processes in
any case and moreover reflects the choice of meson and baryon interpolation fields only \cite{off-shell,Fearing}.
The leading projectors relevant for the $J= {\textstyle{1\over 2}}$
and  $J={\textstyle{3\over 2}}$ channels are $ Y_0^{(+)}$ (s-wave), $ Y_0^{(-)}$ (p-wave)
and $ Y_1^{(+)}$ (p-wave), $ Y_1^{(-)}$ (d-wave) where
\begin{eqnarray}
Y_0^{(\pm )}(\bar q,q;w) &=& \frac{1}{2}\,\left( \frac{\wslash}{\sqrt{w^2}}\pm 1 \right)\;,
\nonumber\\
Y_1^{(\pm)}(\bar q,q;w) &=&
\frac{3}{2}\,\left( \frac{\wslash}{\sqrt{w^2}}\pm 1 \right)
\left(\frac{(\bar q \cdot w )\,(w \cdot q)}{w^2} -\big( \bar q\cdot q\big)\right)
\nonumber\\
&- &\frac{1}{2}\,\Bigg(  \barqslash -\frac{w\cdot \bar q}{w^2}\,\wslash \Bigg)
\Bigg(\frac{\wslash}{\sqrt{w^2}}\mp 1\Bigg)\,
\Bigg( \qslash -\frac{w\cdot q}{w^2}\,\wslash \Bigg)\;.
\label{}
\end{eqnarray}
For instance the projector $Y_1^{(-)}$ probes the d-wave $\Lambda (1520)$
resonance and $Y_1^{(+)}$ the p-wave $\Sigma (1385)$ resonance.
For more details on the construction and the properties of these projectors see
\cite{Lutz:Kolomeitsev}.

In the following the self consistent set of equations (\ref{rewrite:b},\ref{hatg},\ref{k-self})
is solved for $n=0,1$ in (\ref{t-vacuum}). The solution of the Bethe-Salpeter equation in
nuclear matter is considerably complicated by the fact that the partial waves, which nicely
decouple in the vacuum, start to mix in the medium. This manifests itself in the presence of
further tensor structures in the scattering amplitude not required in the vacuum. For example
if one starts to iterate the Bethe-Salpeter equation (\ref{rewrite:b}) in terms of the vacuum
amplitudes
$M^{(I,\pm)}(\sqrt{s},\,n)$ incorporating $\Delta {\mathcal G}_{\bar K N}$ as determined by (\ref{k-self})
with ${\mathcal T}_{\bar K N} = T_{\bar K N}$, one finds that ${\mathcal T}_{\bar K N}$ can
no longer be decomposed into the projectors $ Y^{(\pm)}_n (\bar q, q;w)$. The
in-medium solution of the Bethe-Salpeter equation requires a more general ansatz
for the scattering amplitude.

Some tedious algebra shows that the following ansatz solves the self consistent set of
equations:
\begin{eqnarray}
&&{\mathcal T}(\bar k,k;w,u) = {\mathcal T}(w,u)+{\mathcal T}^\nu \,(w,u)\,q_\nu
+\bar q_\mu \,{\bar {\mathcal T}}^\mu (w,u)+ \bar q_\mu\,{\mathcal T}^{\mu \nu}(w,u)\,q_\nu \;,
\nonumber\\ \nonumber\\
&&\bar {\mathcal T}^\mu =\sum_{j=3}^8\,\Big( {\mathcal M}_{[1j]}^{(p)}\,\bar P_{[1j]}^\mu
+ {\mathcal M}_{[2 j]}^{(p)}\,\bar P_{[2j]}^\mu\Big)\,, \;
{\mathcal T}^\mu = \sum_{j=3}^8\,\Big( {\mathcal M}_{[j1]}^{(p)}\,P_{[j1]}^\mu
+ {\mathcal M}_{[j2]}^{(p)}\,P_{[j2]}^\mu\Big) \,,
\nonumber\\
&& {\mathcal T} = \sum_{i,j=1}^2 \,{\mathcal M}^{(p)}_{[ij]}\,P_{[ij]} \,, \quad
{\mathcal T}^{\mu \nu} = \sum_{i,j=3}^8\,{\mathcal M}_{[ij]}^{(p)}\,P_{[ij]}^{\mu \nu}+
\sum_{i,j=1}^2\,{\mathcal M}^{(q)}_{[ij]}\,Q_{[ij]}^{\mu \nu}\,,
\label{ansatz}
\end{eqnarray}
where appropriate projectors $P_{[ij]}(w,u)$ which generalize the
vacuum projectors $ Y_n^{(\pm)}(\bar q,q;w)$ are introduced \cite{Lutz:Korpa}.
The matrix valued  functions
${\mathcal M}_{[ij]}^{(p,q)}(w,u)$ are scalar and therefore depend only  on $s =w^2$ and
$w \cdot u$. The set of projectors is constructed
to satisfy the convenient algebra
\begin{eqnarray}
&&P_{[ik]}\cdot P_{[lj]} =\delta_{kl}\,P_{[ij]} \;, \quad
P^\mu_{[ik]}\;\bar P^\nu_{[lj]}= \delta_{kl}\,P_{[ij]}^{\mu \nu}\,,\quad
\bar P^\mu_{[ik]}\,g_{\mu \nu}\,P^\nu_{[lj]}= \delta_{kl}\,P_{[ij]} \;,
\nonumber\\
&& Q_{[ik]}^{\mu \alpha }\,g_{\alpha \beta}\,P_{[lj]}^{\beta \nu }
= 0 = P_{[ik]}^{\mu \alpha }\,g_{\alpha \beta}\,Q_{[lj]}^{\beta \nu }
\;, \quad
Q_{[ik]}^{\mu \alpha }\,g_{\alpha \beta}\,P_{[lj]}^{\beta }
= 0 = \bar P_{[ik]}^{\alpha }\,g_{\alpha \beta}\,Q_{[lj]}^{\beta \nu }\;,
\nonumber\\
&& Q_{[ik]}^{\mu \alpha }\,g_{\alpha \beta}\,Q_{[lj]}^{\beta \nu }
= \delta_{kl}\,Q_{[ij]}^{\mu \nu} \;,\quad
P_{[ik]}^{\mu \alpha }\,g_{\alpha \beta}\,P_{[lj]}^{\beta \nu }
= \delta_{kl}\,P_{[ij]}^{\mu \nu}\;.
\label{proj-algebra}
\end{eqnarray}
The projector algebra (\ref{proj-algebra}) properly implements the coupling of
the partial waves in the medium. Consider for example the projectors $P_{[ij]}(w,u)$ for
$i,j=1,2$
\begin{eqnarray}
&&P_{[11]}= +Y_0^{(+)} \;, \quad \,
P_{[12]} = -Y_0^{(+)} \,
\frac{\gamma \cdot u}{\sqrt{1-(w\cdot u)^2/w^2}}\,Y_0^{(-)}\,,
\nonumber\\
&& P_{[22]}= -Y_0^{(-)} \,,\quad \,
P_{[21]} = -Y_0^{(-)} \,
\frac{\gamma \cdot u}{\sqrt{1-(w\cdot u)^2/w^2}}\,Y_0^{(+)} \,,
\label{res-proj}
\end{eqnarray}
which demonstrate that the vacuum projectors start to mix in the medium due
to the presence of the matter 4-velocity $u_\mu$. If
we neglected the $J={\textstyle{3\over 2}}$ amplitudes $M^{(\pm )}(\sqrt{s},\,1)$  the
Bethe-Salpeter equation is solved with the restricted set of projectors $P_{[ij]}$ given
in (\ref{res-proj}). Explicit results for the remaining projectors are presented in
\cite{Lutz:Korpa}. In particular it is found
\begin{eqnarray}
&& Y_1^{(+)} =-3\, \bar q_\mu\,
\Big( Q_{[11]}^{\mu \nu}+ P_{[77]}^{\mu \nu}\Big)\, q_\nu \,,\quad
Y_1^{(-)} =+3\, \bar q_\mu\,
\Big( Q_{[22]}^{\mu \nu}+ P_{[88]}^{\mu \nu}\Big)\, q_\nu \;.
\label{}
\end{eqnarray}
From the form of the algebra in (\ref{proj-algebra}) it follows that
P- and Q-space projectors decouple. Whereas the $P_{[ij]}$-projectors
couple to all considered partial waves the $Q_{[ij]}$-projectors
require only the two $J={\textstyle{3\over 2}}$ partial wave amplitudes.
This reflects the fact that the four polarizations of a spin
${\textstyle{3\over 2}}$ fermion are no longer degenerate in a
nuclear environment. By analogy with a spin 1 boson with one
longitudinal and two degenerate transverse modes one expects two
independent modes for a spin ${\textstyle{3\over 2}}$ fermion in
nuclear matter also. Such modes have been discussed for the case
of the $\Delta (1232)$ isobar properties in nuclear matter (see
e.g. \cite{Isobar:Oset}). They reflect two independent terms in
its self energy, a scalar and a tensor, characterizing a
non-relativistic spin 3/2 state in nuclear matter. Like for a vector
meson at rest where longitudinal and transverse modes are
degenerate in nuclear matter the analogous property holds for the
spin 3/2 states.

With the ansatz (\ref{ansatz}) the antikaon-nucleon propagator
$\Delta {\mathcal  G}_{\bar K N}$ translates into the set of loop functions
\begin{eqnarray}
&& \int \frac{d^4l }{(2\pi)^4}\,
\Delta {\mathcal G}(l-{\textstyle{1\over 2}} w;w,u)
=\sum_{i,j=1}^2 \,\Delta J^{(p)}_{[ij]}(w,u)\,P_{[ij]}(w,u) \;,
\nonumber\\
&& \int \frac{d^4l }{(2\pi)^4}\,l^\mu
\,\Delta {\mathcal G}(l-{\textstyle{1\over 2}} w;w,u)
\nonumber\\
&& \qquad \qquad \quad
=\sum_{j=3}^8\,\Big( \Delta J_{[j1]}^{(p)}(w,u)\,P_{[j1]}^\mu (w,u)
+ \Delta J_{[j2]}^{(p)}(w,u)\,P_{[j2]}^\mu(w,u)\Big)
\nonumber\\
&& \qquad \qquad \quad
= \sum_{j=3}^8\,\Big( \Delta J_{[1j]}^{(p)}(w,u)\,\bar P_{[1j]}^\mu(w,u)
+ \Delta J_{[2j]}^{(p)}(w,u)\,\bar P_{[2j]}^\mu(w,u)\Big)\;,
\nonumber\\
&& \int \frac{d^4l }{(2\pi)^4}\,l^\mu \,l^\nu \,
\Delta {\mathcal G}(l-{\textstyle{1\over 2}} w;w,u)
\nonumber\\
&& \qquad \qquad \quad
= \sum_{i,j=3}^8\,\Delta J_{[ij]}^{(p)}(w,u)\,P_{[ij]}^{\mu \nu}(w,u)
+\sum_{i,j=1}^2\,\Delta J^{(q)}_{[ij]}(w,u)\,Q_{[ij]}^{\mu \nu}(w,u)  \;,
\nonumber\\
&& \Delta {\mathcal G}(l-{\textstyle{1\over 2}} w;w,u) =-i\,\Big(
{\mathcal S}_N (l,u)\,{\mathcal D}_K(w -l,u)
-S_N(l)\,D_K(w-l)\Big)  \,,
\label{def-loop-in-medium}
\end{eqnarray}
which can be decomposed in terms of the projectors $P_{[ij]}$ and $Q_{[ij]}$. The reduced loop functions
$\Delta J_{[ij]}(w,u)$ acquire the generic form
\begin{eqnarray}
&&\Delta J_{[ij]}(w,u) = \int \frac{d^4l}{(2 \pi)^4}\,g(l;w,u)\,\Delta J_{[ij]}(l;w,u)\,,
\nonumber\\ \nonumber\\
&& g(l;w,u) = -2\,\pi\,\Theta \Big(l\cdot u \Big)\,\delta(l^2-m_N^2)\,
\frac{\Theta \Big(k_F^2+m_N^2-(u \cdot l)^2 \Big)}{(w-l)^2-m_K^2-\Pi_{\bar K}(w-l,u)}
\nonumber\\
&& \qquad \qquad -\frac{i}{l^2-m_N^2+i\,\epsilon} \,
\frac{1}{(w-l)^2-m_K^2-\Pi_{\bar K}(w-l,u)}
\nonumber\\
&& \qquad \qquad +\frac{i}{l^2-m_N^2+i\,\epsilon} \,\frac{1}{(w-l)^2-m_K^2+i\,\epsilon}\;,
\label{j-exp}
\end{eqnarray}
where the scalar polynomials $\Delta J_{[ij]}(l;w,u)$ involve typically some
power of $l^2,l \cdot w$ or $l \cdot u$. They are listed in \cite{Lutz:Korpa}. Note
that the reduced loop functions $\Delta J_{[ij]}(w,u)$ are scalar and therefore
depend only on $w^2$ and $w \cdot u$. Hence the loop functions can be evaluated in any
convenient frame without loss of information. In practice the loop integration is performed
in the rest frame of nuclear matter with $u_\mu=(1,\vec 0)$.

Choosing the center of mass frame of the antikaon-nucleon system with
$w_\mu = (\sqrt{s},\vec 0\,)$ instead does not necessarily
lead to any further simplification as suggested in \cite{Ramos} since in this frame a nonzero
bulk velocity $\vec v \neq 0 $ is required. After performing the energy and azimuthal angle
integration in (\ref{j-exp}) one is left with a two dimensional integration which must
be evaluated numerically. Note that the scalar self energy $\Pi_{\bar K}(l-w,u)$ depends on
the two invariants $(l-w)^2$ and $(l-w)\cdot u$ only. Consequently in the nuclear matter rest
frame the first entry involves the angle $\vec l \cdot \vec w$ as compared to the
frame with $w_\mu= (\sqrt{s},\vec 0\,)$ which leads the angle $\vec l \cdot \vec v$
in second entry. This confirms the expected conservation of complexity. A
strong 'vector' potential in the antikaon self energy, as suggested by phenomenology,
actually means a strong dependence on $(l-w)\cdot u $ in (\ref{j-exp}). Therefore it
appears unjustified to neglect that dependence as was done
in \cite{ramossp,Schaffner}.

With (\ref{ansatz},\ref{proj-algebra},\ref{def-loop-in-medium}) the in-medium Bethe-Salpeter
equation (\ref{hatt}) reduces to a simple matrix equation
\begin{eqnarray}
&& {\mathcal M}^{(p)}_{[ij]}(w,u) = M^{(p)}_{[ij]}(\sqrt{s}\,)
+ \sum_{l,k\,=1}^8\,M^{(p)}_{[ik]}(\sqrt{s}\,) \,\Delta J^{(p)}_{[kl]}(w,u) \,
{\mathcal M}^{(p)}_{[lj]}(w,u) \;,
\nonumber\\
&& {\mathcal M}^{(q)}_{[ij]}(w,u) = M^{(q)}_{[ij]}(\sqrt{s}\,)
+ \sum_{l,k\,=1}^2\,M^{(q)}_{[ik]}(\sqrt{s}\,) \,\Delta J^{(q)}_{[kl]}(w,u) \,
{\mathcal M}^{(q)}_{[lj]}(w,u) \;.
\label{m:eq}
\end{eqnarray}
The isospin index $I$ is suppressed and it is implied $s= w_0^2-\vec w\,^2$.
All nonzero vacuum matrix elements are identified with
\begin{eqnarray}
&& M^{(q)}_{[11]}(\sqrt{s}\,)  = -3\,M^{(+)}(\sqrt{s},\,1)  \, , \quad \!\!
M^{(q)}_{[22]}(\sqrt{s}\,)  = 3\, M^{(-)}(\sqrt{s},\,1)   \,,\quad\!\!
\nonumber\\
&& M^{(p)}_{[11]}(\sqrt{s}\,)  = M^{(+)}(\sqrt{s},\,0)  \, ,\quad \;\;\;\,
 M^{(p)}_{[22]}(\sqrt{s}\,)  = - M^{(-)}(\sqrt{s},\,0)  \,, \quad \!\!
\nonumber\\
&& M^{(p)}_{[77]}(\sqrt{s}\,)  = -3\,M^{(+)}(\sqrt{s},\,1)  \, , \quad \! \!
M^{(p)}_{[88]}(\sqrt{s}\,)  = 3\, M^{(-)}(\sqrt{s},\,1)  \,.
\label{}
\end{eqnarray}
Note that in fact all amplitudes ${\mathcal M}^{(p)}_{[ij]}(w,u)$ with either
$i\in \{3,4,5,6\}$ or $j\in \{3,4,5,6\}$ are zero. That is a direct consequence of the
analogous property for the amplitudes $M_{[ij]}$ in free space.
In principle the vacuum scattering amplitudes could
have contributions proportional to $P_{33}, P_{44}, P_{55}$ or $P_{66}$ also. The reason why
such terms are not considered here is that they are not determined by the on-shell antikaon-nucleon
scattering process. The size of such terms merely reflects the choice of the interpolating
fields \cite{off-shell,Fearing}. They acquire physical significance only, once the vacuum
3-body scattering processes are addressed systematically. It should be stressed that
a detailed analysis of the loop matrices $\Delta J(w,u)$ reveals that for $u_\mu =(1,0)$
and $\vec w=0$ all off diagonal elements vanish identically \cite{Lutz:Korpa}. This
property guarantees that all partial wave amplitudes decouple in this limit.

It is left to derive expressions for the antikaon self energy more explicit than (\ref{k-self}). The required
formulae can be found in \cite{Lutz:Korpa}. The antikaon self energy follows
\begin{eqnarray}
\Pi_{\bar K}(q,u) &=& -\sum_{i,j=1}^8\,\int_0^{k_F} \frac{d^3 p}{(2\pi)^3} \,
\frac{2}{E_p}\,c^{(p)}_{[ij]}(q;w,u)\,\bar {\mathcal M}^{(p)}_{[ij]}(w,u)
\nonumber\\
&-& \sum_{i,j=1}^2\,\int_0^{k_F} \frac{d^3 p}{(2\pi)^3} \,
\frac{2}{E_p}\,c^{(q)}_{[ij]}(q;w,u)\,\bar {\mathcal M}^{(q)}_{[ij]}(w,u) \,,
\label{kaon-final}
\end{eqnarray}
where $u_\mu=(1, \vec 0\,), q_\mu=(\omega, \vec q\,)$ and
$w_\mu = (\omega+E_p, \vec q+\vec p\, )$ with $E_p=(m_N^2+\vec p\,^2)^{1/2}$. The scalar
coefficient functions $c^{(p,q)}_{[ij]}(q;w,u)$ are listed in \cite{Lutz:Korpa}. In (\ref{kaon-final})
we introduced the isospin averaged amplitudes
\begin{eqnarray}
\bar {\mathcal M}_{[ij]}(w,u) = \frac{1}{4}\,{\mathcal M}^{(I=0)}_{[ij]}(w,u)
+\frac{3}{4}\,{\mathcal M}^{(I=1)}_{[ij]}(w,u) \,,
\label{}
\end{eqnarray}
appropriate for symmetric nuclear matter. The generalization to the asymmetric case
is straight forward. With (\ref{kaon-final}), (\ref{m:eq}) and (\ref{j-exp}) the final
self consistent set of equations are derived. They are solved numerically by iteration. First
one determines the leading antikaon self energy $\Pi_{\bar K}(\omega ,\vec q\,)$
by (\ref{kaon-final}) with ${\mathcal M}_{[ij]} =M_{[ij]}$. That leads via the loop
functions (\ref{j-exp}) and the in-medium Bethe-Salpeter equation (\ref{m:eq}) to medium
modified scattering amplitudes ${\mathcal M}_{[ij]}(w_0,\vec w\,) $. The latter are used
to determine the antikaon self energy of the next iteration. This procedure typically
converges after 3 to 4 iterations. Note that the manifest covariant form of the self
energy and scattering amplitudes are recovered with
$\Pi_{\bar K}(q^2,\omega)= \Pi_{\bar K}(q^2,q \cdot u)$
and ${\mathcal M}_{[ij]}(w^2,w_0)={\mathcal M}_{[ij]}(w^2,w \cdot u)$ in a straight forward
manner if considered as functions of $q^2,\omega $ and $w^2, w_0$ respectively .

\cleardoublepage

\chapter{Results}
\label{k6}
\markboth{\small CHAPTER \ref{k6}.~~~Results}{}

In this final chapter results for the chiral $SU(3)$ analysis of the low-energy
meson-baryon scattering data and for the in-medium meson-baryon
scattering processes are presented. The acronym '$\chi $-BS(3)' is used
for chiral Bethe-Salpeter dynamics of the flavor $SU(3)$ symmetry as developed in the
previous chapters. The chapter falls into two different parts. First the microscopic
scattering processes are discussed where the strangeness sectors are emphasized, i.e.
the kaon and antikaon nucleon scattering processes. In the second part results for the
self consistent solution of the antikaon and hyperon resonance propagation properties
in cold nuclear matter are discussed.

The main assumptions and crucial arguments the $\chi $-BS(3) approach
is based on are briefly summarized. The number of colors ($N_c$) in QCD is
considered as a large parameter. A systematic expansion of the scattering kernel in
powers of $1/N_c$ is performed. The coupled-channel Bethe-Salpeter kernel is
evaluated in a combined chiral and $1/N_c$ expansion including terms of chiral
order $Q^3$. Contributions of s- and u-channel baryon octet and decuplet
states are included explicitly but only the s-channel contributions of the d-wave
$J^P={\textstyle{3\over 2}}^-$ baryon nonet resonance states. Therewith
the s-channel baryon nonet contributions to the interaction kernel are considered
as a reminiscence of further inelastic channels not included in the present scheme
like for example the $K\,\Delta_\mu $ or $K_\mu \,N$ channel. All baryon
resonances, with the important exception of those resonances which belong to the
large-$N_c$ baryon ground states, are expected to be generated by coupled-channel
dynamics. This conjecture is based on the observation that unitary (reducible) loop
diagrams are typically enhanced by a factor of $2 \pi $ close to threshold relatively
to irreducible diagrams. That factor invalidates the perturbative evaluation  of the
scattering amplitudes and leads necessarily to a non-perturbative scheme with reducible
diagrams summed to all orders. Indeed the $\chi $-BS(3) approach demonstrates precursor
effects of the N(1440), the $\Lambda (1600)$, $\Lambda (1890)$, $\Sigma (1750)$ and
the $\Sigma (1660)$ baryon resonances once agrement with the low-energy data set is
achieved \cite{Lutz:Kolomeitsev}.

The scattering amplitudes for the meson-baryon scattering processes are obtained from
the solution of the coupled-channel Bethe-Salpeter scattering equation where a systematic
on-shell reduction guarantees that the scattering amplitudes do not depend on the particular
choice of the chiral Lagrangian. The method leads to results that are similar to
those of the N/D method introduced by Chew and Mandelstam \cite{Chew-Mandelstam}.
However, there are important differences linked to the fact that we constructed covariant
projection operators that permit the solution of the Bethe-Salpeter equation in dimensional
regularization. That novel technique constrains the many unknown subtraction constants
required in the N/D method. A remaining ambiguity in the renormalization condition
of the Bethe-Salpeter equation is fixed basically by the requirement that the final scattering
amplitudes are crossing symmetric approximatively. First a complete collection
of the parameters as they are adjusted to the data set is presented. Particular emphasis
is put on the SU(3) symmetry breaking effects in the meson-baryon and axial vector
coupling constants of the baryon octet states. In the subsequent sections we report
on details of the fit strategy and confront our results with the empirical data in the
strangeness sectors. Details on the results in the strangeness zero sector can be
found in \cite{Lutz:Kolomeitsev}, where a reasonable description of the elastic
s- and p-wave pion-nucleon scattering phase shifts is documented. It follows a detailed
analysis of the obtained scattering amplitudes, demonstrating their good analyticity properties
as well as their approximate compliance with crossing symmetry.
The result chapter will be closed with a presentation and discussion of the in-medium properties
of antikaons and hyperons \cite{Lutz:Korpa}. It will be argued that the $\Sigma(1385)$ resonance
should play an important role in the microscopic understanding of the subthreshold
production of antikaons in heavy-ion reactions at SIS energies.

\vskip1.5cm \section{Parameters}

It should be emphasized that any fit to a data set, based on a model calculation
that is less accurate in some channels than experimental errors,
involving many different reactions with different statistical and systematical errors
is subject to some bias. In order to obtain a uniform fit a weighting procedure of the various
channels must be devised that cannot be unique. Ultimately one should judge the quality of
the fit by direct comparison with the data set rather than a total chisquare. In this analysis
a data point is included typically if $p_{\rm lab} < 350$ MeV. At higher energies one would
expect that further partial wave amplitudes and inelastic channels not considered in this
work become important. There exist low-energy elastic and inelastic $K^-p$
cross section data including in part angular distributions and polarizations. Also the low-energy
differential $K^+p \to K^+ p$ cross sections are included in the global fit. The empirical
constraints set by the $K^+$-deuteron scattering data above $p_{\rm lab} > 350$ MeV are
considered by requiring a reasonable matching to the single energy $S_{01}$ and $P_{03}$
phase shifts of \cite{Hashimoto}. This resolves an ambiguity in the parameter set.
Finally, the single-energy pion-nucleon phase shifts of \cite{pion-phases} are fitted.
A uniform quality of the data description is aimed at. Therefore the
$\chi^2$ per data point is formed in each sector. The latter are added up to the
total $\chi^2$ which is minimized by the search algorithms
of Minuit \cite{Minuit}. In cases where the empirical error bars are much smaller than the
accuracy to which we expect the $\chi$-BS(3) to work to the given order, we artificially increase
those error bars in our global fit. In Tab. \ref{q1param:tab}-\ref{q3param:tab}
the parameter set of the best fit to the data are presented.
Note that part of the parameters are
predetermined to a large extent and therefore fine tuned in a small interval only.
It is found that the set of parameters is well determined by the available
scattering data and weak decay widths of the baryon octet states.

\tabcolsep=1.5mm
\renewcommand{\arraystretch}{1.5}
\begin{table}[h]\begin{center}
\begin{tabular}{|c|c|c|c|c|}
\hline
$f$ [MeV] & $ C_R$ & $F_R$ & $D_R$
\\
\hline \hline
90.04  & 1.734 & 0.418 & 0.748

\\  \hline

\end{tabular}
\end{center}
\caption{Leading chiral parameters which contribute to meson-baryon
scattering to order $Q$.}
\label{q1param:tab}
\end{table}

A qualitative understanding of the typical strength in the various channels can
be obtained already at leading chiral order $Q$. In particular the
$\Lambda(1405)$ resonance is formed as a result of the coupled-channel
dynamics defined by the Weinberg-Tomozawa interaction vertices (see Fig. \ref{fig:wt}).
There are four parameters relevant to that order $f$, $C_R$, $F_R$ and $D_R$.
Their respective values as given in Tab. \ref{q1param:tab} are the result
of our global fit to the data set including all parameters of the $\chi$-BS(3) approach. At leading
chiral order the parameter $f$ determines the weak pion- and kaon-decay processes and at the same
time the strength of the Weinberg-Tomozawa interaction vertices. At subleading order $Q^2$ the
Weinberg-Tomozawa terms and the weak-decay constants of the pseudo-scalar meson octet receive independent
correction terms. The result $f\simeq  90$ MeV is sufficiently close to the
empirical decay parameters $f_\pi \simeq 92.4$ MeV and
$f_K \simeq 113.0$ MeV to expect that the $Q^2$ correction terms lead indeed
to values rather close to the empirical decay constants. The value for $f$ is consistent
with the estimate of \cite{GL85} which leads to $f_\pi/f = 1.07 \pm 0.12$.
The baryon octet and decuplet s- and u-channel exchange contributions to the interaction kernels are determined
by the $F_R, D_R$ and $C_R$ parameters at leading order. Note that $F_R$ and $D_R$ predict
the baryon octet weak-decay processes and $C_R$ the strong decay widths of the baryon decuplet states
to this order also.

A quantitative description of the data set requires the inclusion
of higher order terms. Initially it was tried to establish a consistent
picture of the existing low-energy meson-baryon scattering data
based on a truncation of the interaction kernels to chiral order
$Q^2$. This attempt failed due to the insufficient quality of the
kaon-nucleon scattering data at low energies. In particular some of the
inelastic $K^-$-proton differential cross sections are strongly influenced by
the d-wave $\Lambda(1520)$ resonance at energies where the data points
start to show smaller error bars. Hence, on the one hand,
one must include an effective baryon-nonet resonance field and, on the other
hand, perform minimally a chiral $Q^3$ analysis to extend the
applicability domain to somewhat higher energies.
Since the effect of the d-wave resonances is only necessary in the strangeness minus one sector,
they are only considered in that channel. The resonance parameters will be presented
when discussing the strangeness minus one sector.

\tabcolsep=1.5mm
\renewcommand{\arraystretch}{1.5}
\begin{table}[t]\begin{center}
\begin{tabular}{|c|c||c|c||c|c||c|c|}
\hline
$g_F^{(V)}$[GeV$^{-2}$] & 0.293 & $g_F^{(S)}$[GeV$^{-1}$] & -0.198 &
$g_F^{(T)}$[GeV$^{-1}$] & 1.106 & $Z_{[10]}$  &  0.719
\\
\hline
$g_D^{(V)}$[GeV$^{-2}$] & 1.240 & $g_D^{(S)}$[GeV$^{-1}$] & -0.853 &
$g_D^{(T)}$[GeV$^{-1}$] & 1.607 & - & -
\\
\hline
\end{tabular}
\end{center}
\caption{Chiral $Q^2$-parameters resulting from a fit to low-energy
meson-baryon scattering data. Further parameters to this order are determined by the
large-$N_c$ sum rules.}
\label{q2param:tab}
\end{table}

At subleading order $Q^2$ the chiral $SU(3)$
Lagrangian predicts the relevance of 12 basically unknown parameters,
$g^{(S)}, g^{(V)}$, $g^{(T)}$ and $Z_{[10]} $, which all need to be adjusted to the
empirical scattering data.
It is important to realize that chiral symmetry is largely predictive in the $SU(3)$ sector
in the  sense that it reduces the number of parameters  beyond
the static $SU(3)$ symmetry. For example one should compare the six tensors which
result from decomposing $8\otimes 8= 1
\oplus 8_S\oplus 8_A \oplus 10\oplus \overline{10}\oplus 27$ into its
irreducible components with the subset of SU(3) structures selected
by chiral symmetry in a given partial wave. Thus, static $SU(3)$
symmetry alone would predict 18 independent terms for the s-wave
and two p-wave channels rather than the 11 chiral $Q^2$ background
parameters, $g^{(S)}, g^{(V)}$ and $g^{(T)}$. In this work the number of parameters was
further reduced significantly by insisting on the large-$N_c$ sum rules
\begin{eqnarray}
g_1^{(S)}=2\,g_0^{(S)}= 4\,g_D^{(S)}/3 \,, \qquad
g_1^{(V)}=2\,g_0^{(V)}= 4\,g_D^{(V)}/3 \,, \qquad g_1^{(T)}=0 \,,
\nonumber
\label{q2-largenc}
\end{eqnarray}
for the symmetry conserving quasi-local two body interaction terms (see (\ref{Q^2-large-Nc-result})).
In Tab. \ref{q2param:tab} we collect the values of all free parameters as they result from our
best global fit. All parameters are found to have natural size. This is an important result, because
only then is the application of the chiral power counting rule (\ref{q-rule}) justified.
We point out that the large-$N_c$ sum rules derived in chapter 2 implicitly assume that other inelastic channels
like $K \,\Delta_\mu $ or $K_\mu\,N $ are not too important. The effect of such
channels can be absorbed to some extent into the quasi-local counter terms, however possibly at the
prize that their large-$N_c$ sum rules are violated. It is therefore a highly non-trivial result
that we obtain a successful fit imposing (\ref{q2-largenc}). Note that the only previous analysis \cite{Kaiser},
which truncated the interaction kernel at chiral order $Q^2$ but did not include p-waves, found values for the s-wave
range parameters largely inconsistent with the large-$N_c$ sum rules. This may be due in part to the use of channel
dependent cutoff parameters and the fact that that analysis missed octet and decuplet exchange contributions,
which are important for the s-wave interaction kernel already to chiral order $Q^2$.

The parameters $b_0, b_D$ and $b_F$ to this order characterize the
explicit chiral symmetry-breaking effects of QCD via the finite current
quark masses.  The parameters $b_D$ and $b_F$ are well estimated
from the baryon octet mass splitting (see (\ref{mass-splitting})) whereas
$b_0$ must be extracted directly from the meson-baryon scattering data.
It drives the size of the pion-nucleon sigma term for which conflicting values are
still being discussed in the literature \cite{pin-news}. The values
\begin{eqnarray}
b_0 = -0.346\, {\rm GeV}^{-1} \,, \quad  b_D = 0.061\, {\rm GeV}^{-1} \,, \quad  b_F =-0.195 \,{\rm GeV}^{-1} \,,
\label{b-result}
\end{eqnarray}
are rather close to values expected from the baryon octet mass splitting (\ref{mass-splitting}).
The pion-nucleon sigma term $\sigma_{\pi N}$ if evaluated to chiral order $Q^2$
(see (\ref{spin:naive})) would be $\sigma_{\pi N} \simeq 32$ MeV. That value should not be
compared directly with $\sigma_{\pi N}$ as extracted usually from pion-nucleon scattering data
at the Cheng-Dashen point. The required subthreshold extrapolation involves further poorly
convergent expansions \cite{pin-news}. Here we do not attempt to add
anything new to this ongoing debate. Consider
the analogous symmetry-breaking parameters $d_0$ and $d_D$
for the baryon decuplet states. Like for the baryon octet states the isospin averaged empirical
values for the baryon masses are used in any u-channel exchange contribution. That implies
$m^{(\Delta )}_{[10]} = 1232.0$
MeV, $m^{(\Sigma )}_{[10]} = 1384.5$ MeV  and $m^{(\Xi)}_{10} = 1533.5$ MeV in the
decuplet exchange expressions. In the s-channel decuplet expressions the slightly
different values $m^{(\Delta )}_{10} = 1223.2$ MeV and $m^{(\Sigma )}_{10} = 1374.4$ MeV are
used in order to compensate for a small mass renormalization induced by the unitarization.
Those values are rather consistent with $d_D \simeq  -0.49$ GeV$^{-1}$ (see (\ref{mass-splitting})).
All values used are quite compatible with the large-$N_c$ sum rule
$$b_D+b_F = d_D/3 \,.$$ The parameter $d_0$ is not determined by the present analysis.
Its determination required the study of the meson baryon-decuplet
scattering processes.

\tabcolsep=1.1mm
\begin{table}[t]\begin{center}
\begin{tabular}{|c|c||c|c||c|c||c|c||c|c||}
\hline
$h_F^{(1)}$[GeV$^{-3}$]  & -0.129 &
$h_D^{(1)}$[GeV$^{-3}$]  & -0.548 &
$h_{F}^{(2)}$[GeV$^{-2}$]  & 0.174 &
$h_{F}^{(3)}$[GeV$^{-2}$] & -0.221
\\
\hline
\end{tabular}
\end{center}
\caption{Chiral $Q^3$-parameters resulting from a fit to low-energy
meson-baryon scattering data. Further parameters to this order are determined by the
large-$N_c$ sum rules.}
\label{q3param:tab}
\end{table}

To chiral order $Q^3$ the number of parameters increases
significantly unless further constraints from QCD are imposed.
Recall for example that \cite{q3-meissner} presents a large
collection of already 102 chiral $Q^3$ interaction terms.  A
systematic expansion of the interaction kernel in powers of $1/N_c$
leads to a much reduced parameter set. For example the $1/N_c$
expansion leads to only four further parameters $h^{(1)}_{F}$,
$h^{(1)}_{D}$, $h^{(2)}_{F}$ and $h^{(3)}_{F}$ describing the refined
symmetry-conserving two-body interaction vertices. This is to be compared with
the ten parameters established in \cite{Lutz:Kolomeitsev}, which were found to be relevant to order
$Q^3$ if large-$N_c$ sum rules are not imposed. In the global fit it is insisted on the
large-$N_c$ sum rules
$$
h_1^{(1)}=2\,h_0^{(1)}= 4\,h_D^{(1)}/3  \;, \quad h^{(2)}_1 = h^{(2)}_D=0
\;, \quad h^{(3)}_1 = h^{(3)}_D=0 \,.
$$
Note that to order $Q^3$ there are no symmetry-breaking 2-body interaction vertices. To that
order the only
symmetry-breaking effects result from the refined 3-point vertices. Here a particularly rich picture emerges.
To order $Q^3$ we  established 23 parameters describing symmetry-breaking effects in the 3-point meson-baryon
vertices.  For instance, to that order the baryon-octet states may couple to the pseudo-scalar mesons
also via pseudo-scalar vertices rather than only via the leading axial-vector vertices. Out of those
23 parameters 16 contribute at the same time to matrix elements of the axial-vector current. Thus,
in order to control the symmetry breaking effects, it is mandatory to include constraints from the weak decay widths
of the baryon octet states also. A detailed analysis of the 3-point vertices in the $1/N_c$ expansion of QCD
reveals that in fact only ten parameters $c_{1,2,3,4,5}$, $\delta c_{1,2}$ and $\bar c_{1,2}$ and $a$, rather
than the 23 parameters, are needed to leading order in that expansion. Since the leading parameters $F_R, D_R$
together with the symmetry-breaking parameters $c_i$ describe at the same time the
weak decay widths of the baryon octet and decuplet ground states (see Tab. \ref{weak-decay:tab},\ref{weak-decay:tabb}),
the number of free parameters does not increase significantly at the $Q^3$ level if
the large-$N_c$ limit is applied.

\tabcolsep=1.5mm
\renewcommand{\arraystretch}{1.5}
\begin{table}[t]\begin{center}
\begin{tabular}{|c|c||c|c||c|c||c|c||c|c||c|c||}
\hline
$c_1 $ & -0.0707 &
$c_2 $ & -0.0443 &
$c_3 $ & 0.0624  &
$c_4 $ & 0.0119  &
$c_5 $ & -0.0434 \\
\hline
$\bar c_1   $ & 0.0754  &
$\bar  c_2  $ & 0.1533  &
$\delta c_1 $ & 0.0328  &
$\delta c_2 $ & -0.0043 &
$ a $ & -0.2099  \\
\hline
\end{tabular}
\end{center}
\caption{Chiral $Q^3$-parameters, which break the $SU(3)$ symmetry explicitly, resulting
from a fit to low-energy meson-baryon scattering data. }
\label{q3param:tab}
\end{table}

The parameter reduction achieved in this work by insisting on chiral and large-$N_c$
sum rules is significant. In this respect it is instructive to recall that for instance the
analysis by Kim \cite{kim}, rather close in spirit to modern effective field theories, required
already 44 parameters in the strangeness minus one sector only. As was pointed out in
\cite{Hurtado} that analysis, even though troubled with severe shortcomings \cite{piN:Hurtado},
was the only one so far which included s- and p-waves and still reproduced the
most relevant features of the subthreshold $\bar K N$ amplitudes. Thus, a
combined chiral and large-$N_c$ analysis leads to a scheme with a reasonably small number
of parameters at the $Q^3$ level.

\vskip1.5cm \section{Axial-vector and meson-baryon coupling constants}

The result of the global fit for the axial-vector coupling constants of the baryon octet
states are presented in Tab. \ref{weak-decay:tabb}. The six data points, which
strongly constrain the parameters $F_R, D_R$ and $c_{1,2,3,4}$, are well reproduced.
The recent measurement of the decay process
$\Xi^0 \to \Sigma^+ \,e^-\,\bar \nu_e$ by the KTeV experiment does not provide a
further stringent constraint so far \cite{KTeV:1,KTeV:2}. The axial-vector coupling constant
of that decay
$g_A(\Xi^0 \to \Sigma^+ \,e^-\,\bar \nu_e) = \sqrt{2}\,g_A (\Xi^- \to \Sigma^0 \,e^-\,\bar \nu_e )$
is related to the decay process $\Xi^- \to \Sigma^0 \,e^-\,\bar \nu_e$ included in Tab. \ref{weak-decay:tabb}
by isospin symmetry. The value given in \cite{KTeV:2} is $g_A(\Xi^0 \to \Sigma^+ \,e^-\,\bar \nu_e) = 1.23 \pm 0.44 $.
As emphasized in \cite{KTeV:2} it would be important to reduce the uncertainties by more data taking.
The result of \cite{Dai} which favors values for $g_A (\Xi^- \to \Sigma^0 \,e^-\,\bar \nu_e )$
and $g_A (\Xi^- \to \Lambda \,e^-\,\bar \nu_e )$ which are somewhat smaller than the central
values given in Tab. \ref{weak-decay:tabb} is confirmed. This is a non-trivial result
because in the present approach the parameters $c_{1,2,3,4}$
are constrained not only by the weak decay processes of the baryon octet states but also by
the meson-baryon scattering data.

\tabcolsep=1.5mm
\renewcommand{\arraystretch}{1.5}
\begin{table}[t]\begin{center}
\begin{tabular}{|c|c||c|c|c|c|c|c|c|}
\hline
& $g_A $ (Exp.) & $\chi$-BS(3) & SU(3) \\
\hline \hline

$ n \to p\,e^-\,\bar \nu_e $ & $1.267 \pm 0.004$& 1.26 & 1.26
\\  \hline

$ \Sigma^- \to \Lambda \,e^-\,\bar \nu_e  $ & $0.601 \pm 0.015$
& 0.58 & 0.65
\\  \hline

$ \Lambda \to p \,e^-\,\bar \nu_e $ & $-0.889 \pm 0.015$ &
-0.92 & -0.90
\\  \hline

$ \Sigma^- \to n \,e^-\,\bar \nu_e $ & $0.342 \pm 0.015$ & 0.33 & 0.32
\\  \hline

$ \Xi^- \to \Lambda \,e^-\,\bar \nu_e $ & $0.306 \pm 0.061 $ &
0.19  & 0.25
\\  \hline

$ \Xi^- \to \Sigma^0 \,e^-\,\bar \nu_e $ & $0.929 \pm 0.112 $ &
0.79 & 0.89
\\  \hline


\end{tabular}
\end{center}
\caption{Axial-vector coupling constants for the weak decay processes of the baryon octet
states. The empirical values
for $g_A$ are taken from \cite{Dai}. Here we do not consider
small $SU(3)$ symmetry-breaking effects of the  vector current. The column labelled by SU(3)
shows the axial-vector coupling constants as they follow from
$F_R=0.47$ and $D_R=0.79$ and $c_{i}=0$.}
\label{weak-decay:tabb}
\end{table}

SU(3) symmetry-breaking effects can be studied also in terms of
the meson-baryon coupling constants. To subleading order the Goldstone bosons
couple to the baryon octet states via axial-vector but also via suppressed pseudo-scalar
vertices (see (\ref{chi-sb-3:p})).
In Tab. \ref{meson-baryon:tab} results are collected for all the axial-vector meson-baryon coupling
constants, $A^{(B)}_{\Phi B}$, and their respective pseudo-scalar parts $P^{(B)}_{\Phi B}$.
The $SU(3)$ symmetric part of the axial-vector vertices is characterized by parameters $F_A$ and $D_A$
\begin{eqnarray}
F_A &=& F_R -\frac{\beta}{\sqrt{3}}\,\Big({\textstyle{2\over 3}}\,\delta c_1 +\delta c_2 -a\Big) =
0.270\,,
\quad \!
\nonumber\\
D_A &=& D_R - \frac{\beta}{\sqrt{3}}\,\delta c_1 = 0.726\,,
\label{FDA:def}
\end{eqnarray}
where $\beta \simeq 1.12$. To subleading order the parameters $F_A$ and $D_A$ differ from
the corresponding parameters $F_R = 0.418$ and $D_R =0.748$ relevant for matrix elements
of the axial-vector current (see Tab. \ref{weak-decay:tab}) by a sizeable amount. The
SU(3) symmetry-breaking effects in the axial-vector coupling constants $A$
are determined by the parameters $c_i$, which are already tightly constrained by the weak decay
widths of the baryon octet states, and $\delta c_{1,2}$ and $a$. Similarly
the four parameters $\bar c_{1,2}$ and $a$ characterize the pseudo-scalar meson-baryon 3-point vertices.
Their symmetric contributions are determined by $F_P$ and $D_P$
\begin{eqnarray}
F_P =  -\frac{\beta}{\sqrt{3}}\,\Big({\textstyle{2\over 3}}\,\bar c_1 +\bar c_2 +a\Big) = 0.004\,,
\quad \!
D_P =  - \frac{\beta}{\sqrt{3}}\,\bar c_1 = -0.049\,.
\label{FDP:def}
\end{eqnarray}
Note that in the on-shell coupling constants $G=A+P$ the parameter
$a$ drops out. In that sense that parameter should be viewed as representing an effective
quasi-local 2-body interaction term, which breaks the $SU(3)$ symmetry explicitly.

\tabcolsep=1.7mm
\renewcommand{\arraystretch}{1.5}
\begin{table}[t]\begin{center}
\begin{tabular}{|c|c|c|c|c|c|c|c|c|c|c|c|c|}
\hline
& $A^{(N)}_{\pi N} $  & $A^{(N)}_{\eta N} $  & $A^{(\Lambda)}_{\bar
K N} $  & $A^{(\Lambda)}_{\pi \Sigma } $  & $A^{(\Lambda)}_{\eta
\Lambda} $ & $A^{(\Lambda)}_{K
\Xi} $  & $A^{(\Sigma)}_{\bar  K N} $ & $A^{(\Sigma)}_{\pi \Sigma} $  &
$A^{(\Sigma)}_{\eta \Sigma} $  & $A^{(\Sigma)}_{K \Xi} $  &
$A^{(\Xi)}_{\pi \Xi} $  & $A^{(\Xi)}_{\eta \Xi} $
\\
\hline \hline
$\chi$-BS(3) & 2.15  & 0.20  &-1.29   & 1.41   & -0.64  & 0.12   & 0.73  &
-1.02  & 1.09  & 1.24   & -0.59  & -0.32
 \\  \hline

SU(3) & 1.73 & 0.05 & -1.25 & 1.45 & -0.84 & -0.07 &0.65  & -0.76 & 0.84 & 1.41 &-0.79 &-0.89
 \\  \hline \hline

& $P^{(N)}_{\pi N} $  & $P^{(N)}_{\eta N} $  &
$P^{(\Lambda)}_{\bar K N} $  & $P^{(\Lambda)}_{\pi \Sigma
} $  & $P^{(\Lambda)}_{\eta
\Lambda} $ & $P^{(\Lambda)}_{K
\Xi} $  & $P^{(\Sigma)}_{\bar  K N} $ & $P^{(\Sigma)}_{\pi \Sigma} $  &
$P^{(\Sigma)}_{\eta \Sigma} $  & $P^{(\Sigma)}_{K \Xi} $
& $P^{(\Xi)}_{\pi \Xi} $  & $P^{(\Xi)}_{\eta \Xi} $
\\
\hline \hline

$\chi$-BS(3) & -0.01 & 0.16 & 0.04 & -0.01 & 0.20 & -0.07 & -0.11 &-0.00  &-0.02  & -0.09 & 0.01& 0.13
 \\  \hline
\end{tabular}
\end{center}
\caption{Axial-vector (A) and pseudo-scalar (P) meson-baryon coupling constants for the baryon octet states.
The row labelled by SU(3) gives results excluding SU(3) symmetry-breaking
effects with $F_A =0.270$ and $D_A= 0.726$. The total strength of the on-shell meson-baryon vertex
is determined by $G=A+P$.}
\label{meson-baryon:tab}
\end{table}

The meson-baryon coupling constants are strongly constrained by the
axial-vector coupling constants. In particular a sum rule derived first
by Dashen and Weinstein is reproduced \cite{Dashen:Weinstein}
\begin{eqnarray}
G_{\pi N}^{(N)}-\sqrt{3}\,g_A (n \to p\,e^-\,\bar \nu_e )
&=& -\frac{m_\pi^2}{\sqrt{6}\,m_K^2}\,\Big( G_{\bar K N }^{(\Sigma)}
-g_A(\Sigma^-\!\to n \,e^-\,\bar \nu_e) \Big)
\nonumber\\
&-& \frac{m_\pi^2}{\sqrt{2}\,m_K^2}\,\Big( G_{\bar K N}^{(\Lambda)}
-2\,g_A (\Lambda \to p \,e^-\,\bar \nu_e ) \Big) \;.
\label{DW-sum}
\end{eqnarray}
We observe that, given the expected range of values for $g_{\pi NN}$, $g_{\bar K N \Lambda }$ and
$g_{\bar K  N \Sigma }$ together with the empirical axial-vector coupling constants,
the Dashen-Weinstein relation strongly favors a small $f$ parameter value close to $f_\pi$.

In Tab. \ref{meson-baryon:tab:2}  we confront our results with a representative
selection of published meson-baryon coupling constants. For the clarity of this comparison
we recall here the connection with our convention
\begin{eqnarray}
&& g_{\pi NN} = \frac{m_N}{\sqrt{3}\,f}\,G^{(N)}_{\pi N} \,, \! \quad
g_{\bar K N \Lambda } = \frac{m_N+m_\Lambda}{\sqrt{8}\,f}\,G^{(\Lambda)}_{\bar KN} \,, \!\quad
g_{\pi  \Lambda \Sigma }= \frac{m_\Lambda+m_\Sigma }{\sqrt{12}\,f }\,G^{(\Lambda)}_{\pi \Sigma }\, ,
\nonumber\\
&& g_{\bar K N \Sigma }= \frac{m_N+m_\Sigma}{\sqrt{8}\,f}\,G^{(\Sigma)}_{\bar K N} \,,\! \quad
g_{\pi \Sigma \Sigma} = \frac{m_\Sigma }{\sqrt{2}\,f}\,G^{(\Sigma)}_{\pi \Sigma } \,.
\label{trans:tab}
\end{eqnarray}
Further values from previous analyses can be found
in \cite{Dumbrajs}. Note also the interesting recent results within the QCD sum rule approach
\cite{Kim:Doi:Oka:Lee,Doi:Kim:Oka} and also \cite{Buchmann:Henley}. We do not confront
our values with those of \cite{Kim:Doi:Oka:Lee,Doi:Kim:Oka} and \cite{Buchmann:Henley}
because the $SU(3)$ symmetry-breaking effects are not yet fully under control in these
works. The analysis \cite{Stoks} is based on nucleon-nucleon and hyperon-nucleon
scattering data where $SU(3)$ symmetry breaking effects in the meson-baryon coupling
constants are parameterized according to the model of \cite{p3-model}.
The values given in \cite{Juelich:2,Juelich:1,Juelich:3} do not
allow for $SU(3)$ symmetry-breaking effects and moreover rely on $SU(6)$
quark-model relations. Particularly striking are the extreme $SU(3)$ symmetry-breaking
effects claimed in \cite{keil}. The parameters result from a K-matrix fit to the
phase shifts of $K^-N$ scattering as given in \cite{gopal}. We do not confirm these results.
Note also the recent analysis \cite{Loiseau:Wycech} which deduces the value
$g_{\pi \Lambda \Sigma } = 12.9 \pm 1.2$ from hyperonic atom data, a value somewhat larger than
our result of $10.4$. For the most recent and accurate pion-nucleon coupling constant
$g_{\pi NN} = 13.34 \pm 0.09$ we refer to \cite{gpinn:best}. We do not compete
with the high precision and elaborate analyses of this work.

\tabcolsep=1.5mm
\renewcommand{\arraystretch}{1.5}
\begin{table}[t]\begin{center}
\begin{tabular}{|c||c|c|c|c|c|c|c|c|}
\hline
& $\chi $-BS(3) &  \cite{Stoks} & \cite{Juelich:2,Juelich:1,Juelich:3} & \cite{keil}
& \cite{A.D.Martin} & \cite{Dalitz:g}     \\
\hline \hline
$|g_{\pi NN }|$ & 12.9  & 13.0  &  13.5  &  13.1 &  - &  -
 \\  \hline

$|g_{\bar K N \Lambda}|$ &10.1  & 13.5 & 14.0 &  10.1 & 13.2 & 16.1
 \\  \hline

$|g_{\pi \Sigma\, \Lambda}|$ & 10.4 & 11.9 & 9.3 & 6.4 & - & -
 \\  \hline

$|g_{\bar K N \Sigma }|$ & 5.2 & 4.1 & 2.7 & 2.4 & 6.4 & 3.5
 \\  \hline

$|g_{\pi \Sigma \,\Sigma}|$ & 9.6 & 11.8 & 10.8 & 0.7 & - & -
 \\  \hline

\end{tabular}
\end{center}
\caption{On-shell meson-baryon coupling constants for the baryon octet states (see (\ref{trans:tab})). We
give the central values only because reliable error analyses are not available in most cases.}
\label{meson-baryon:tab:2}
\end{table}

\tabcolsep=2.0mm
\renewcommand{\arraystretch}{1.5}
\begin{table}[b]\begin{center}
\begin{tabular}{|c|c|c|c|c|c|c|c|c|c|c|c|}
\hline
& $A^{(\Delta )}_{\,\pi\, N} $  & $A^{(\Delta )}_{\,K\, \Sigma} $  & $A^{(\Sigma )}_{\,\bar
K \,N} $  & $A^{(\Sigma)}_{\,\pi \,\Sigma } $  & $A^{(\Sigma )}_{\,\pi \,\Lambda} $
& $A^{(\Sigma )}_{\,\eta \,\Sigma } $
& $A^{(\Sigma )}_{\,K \,\Xi} $ & $A^{(\Xi )}_{\,\bar K \,\Lambda} $  &
$A^{(\Xi )}_{\,\bar K \,\Sigma} $  & $A^{(\Xi )}_{\,\eta \,\Xi} $  &
$A^{(\Xi )}_{\,\pi \,\Xi} $
\\
\hline \hline
$\chi$-BS(3) & 2.62 & -2.03  & 1.54  & -1.40  & -1.64  &  1.55  &-0.96   & 1.67 & 1.75  &-1.29  &-1.46
 \\  \hline

SU(3) & 2.25 & -2.25 & 1.30 & -1.30 & -1.59 & 1.59 & -1.30  & 1.59 & 1.59 & -1.59 &-1.59
 \\  \hline
\end{tabular}
\end{center}
\caption{Meson-baryon coupling constants for the baryon decuplet states. The row
labelled by SU(3) gives results obtained with $C_A =1.593$ excluding all symmetry-breaking
effects.}
\label{meson-baryon-decuplet:tab}
\end{table}

In Tab. \ref{meson-baryon-decuplet:tab}  results for the meson-baron coupling constants
of the decuplet states are collected. Again only moderate $SU(3)$ symmetry-breaking effects
are found in the coupling constants. This is demonstrated by comparing the two rows of
Tab. \ref{meson-baryon-decuplet:tab}. The SU(3) symmetric part is determined by the
parameter $C_A$
\begin{eqnarray}
C_A = C_R -2\,\frac{\beta}{\sqrt{3}}\,\Big(\bar c_1 +\delta c_1 \Big) = 1.593 \,.
\label{}
\end{eqnarray}
Note that the coupling constants as given in (\ref{meson-baryon-decuplet:tab}) should not
be used in the simple expressions (\ref{decuplet-decay}) for the decuplet widths.
For example with $A_{\pi N}^{(\Delta )} \simeq 2.62 $ one would estimate $\Gamma_\Delta \simeq $
102 MeV not too close to the empirical value of $\Gamma_\Delta \simeq $ 120 MeV \cite{fpi:exp}.
As demonstrated in \cite{Lutz:Kolomeitsev} nevertheless the $P_{33}$ phase shift of the
pion-nucleon scattering process, which probes the $\Delta $ resonance width, is
reproduced accurately. This reflects an important energy dependence in
the decuplet self energy.

To summarize the main findings of this section. All established parameters
prove the $SU(3)$ flavor symmetry to be an extremely useful and accurate tool.
Explicit symmetry breaking effects are quantitatively important but sufficiently small to
permit an evaluation within the $\chi$-BS(3) approach. This confirms a beautiful analysis
by Hamilton and Oades \cite{Hamilton:Oades} who strongly supported the $SU(3)$ flavor symmetry
by a discrepancy analysis of antikaon-nucleon scattering data.

\vskip1.3cm \section{Kaon- and antikaon-nucleon scattering}

\renewcommand{\arraystretch}{1.5}
\tabcolsep=1.9mm
\begin{table}[b]\begin{center}
\begin{tabular}{|c|c|c||c|c|c|c|c|}
\hline
& $a^{(K N)}_{S_{01}}$ [fm] & $a^{(K N)}_{S_{21}}$ [fm]  &
$a^{(K N)}_{P_{01}}$ $[m_\pi^{-3}]$ & $a^{(K N)}_{P_{21}}$ $[m_\pi^{-3}]$ &
$a^{(K N)}_{P_{03}}$ $[m_\pi^{-3}]$ & $a^{(K N)}_{P_{23}}$ $[m_\pi^{-3}]$  \\
\hline \hline
$\chi$-BS(3)  & 0.06 & -0.30 & 0.033 & -0.017 & -0.003 & 0.012 \\
\hline
\protect\cite{Hyslop} & 0.0 & -0.33 & 0.028 & -0.056 & -0.046 & 0.025 \\
\hline
\protect\cite{BR:Martin} & -0.04 & -0.32 & 0.030 & -0.011 & -0.007 & 0.007 \\
\hline
\end{tabular}
\end{center}
\caption{$K^+$-nucleon threshold parameters. The values of the $\chi$-BS(3) analysis are given in the first row.
The last two rows recall the threshold parameters as given in \cite{Hyslop} and \cite{BR:Martin}.}
\label{tab-r-kp}
\end{table}

Since it is impossible to give here a comprehensive
discussion of the many works dealing with kaon-nucleon scattering we refer to the review
article by Dover and Walker \cite{Dover} which is still up-to-date in many respects.
The data situation can be summarized as follows: there exist precise low-energy differential cross
sections for $K^+ p$ scattering but no scattering data for the $K^+$-deuteron scattering
process at low energies. Thus, all low-energy results in the isospin zero
channel necessarily follow from model-dependent extrapolations. The
available differential cross section are included in the global fit. They are nicely
reproduced as shown in \cite{Lutz:Kolomeitsev}.

\begin{figure}[t]
\begin{center}
\includegraphics[width=14cm,clip=true]{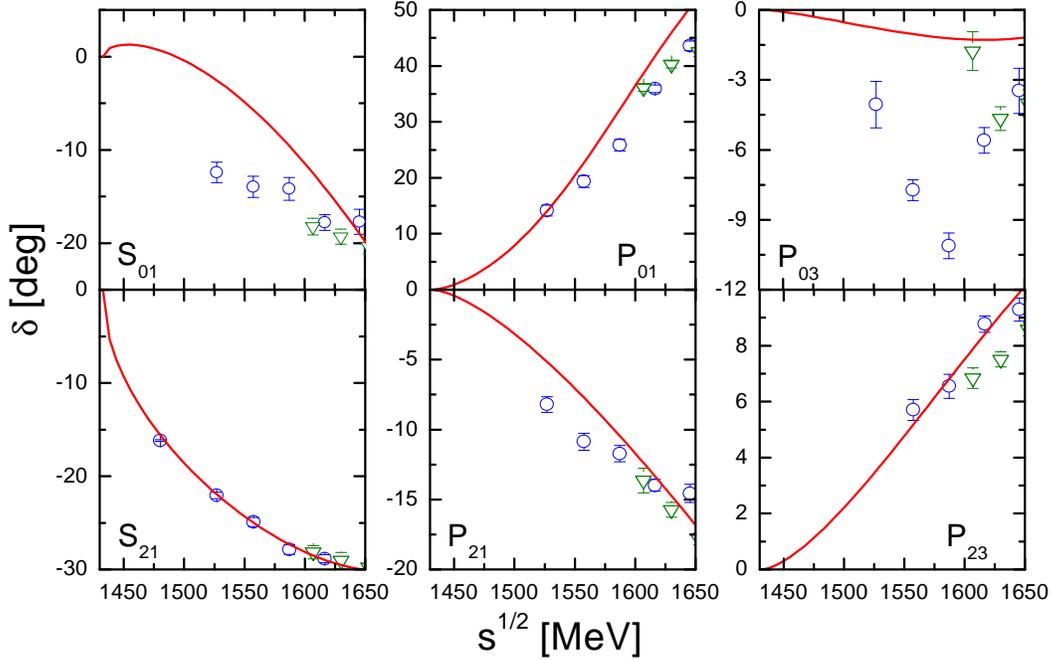}
\end{center}
\caption{S- and p-wave $K^+$-nucleon phase shifts. The solid lines represent the results of the
$\chi$-BS(3) approach. The open circles are from the Hyslop analysis \cite{Hyslop} and the
open triangles from the Hashimoto analysis \cite{Hashimoto}.}
\label{fig:K+phases}
\end{figure}

It is instructive to consider the threshold amplitudes in detail. In the $\chi$-BS(3) approach
the threshold parameters are determined by the threshold values of the effective interaction kernel
$V^{(\pm)}_{KN}(m_N\!+\!m_K; n)$
and the partial-wave loop function $J^{(\pm)}_{KN}(m_N\!+\!m_K; n)$ (see (\ref{result-loop:ab},\ref{M:kplus})).
Since all loop functions vanish at threshold but the one for the s-wave channel, all p-wave scattering
volumes remain unchanged by the unitarization and are directly given by the
threshold values of the appropriate effective interaction kernel $V_{K N}$. Explicit
expressions for the scattering volumes to leading order can be found in \cite{Lutz:Kolomeitsev}.
In contrast the s-wave scattering lengths are renormalized strongly by the loop function
$J^{(+)}_{KN}(m_N\!+\!m_K; 0) \neq 0$.
At leading order the s-wave scattering lengths are
\begin{eqnarray}
4\,\pi\left( 1+\frac{m_K}{m_N}\right) a^{(KN)}_{S_{21}} &=&
-m_K \left( f^2+\frac{m^2_K}{8\,\pi}
\left(1-\frac{1}{\pi}\,\ln \frac{m_K^2}{m_N^2} \right)\right)^{-1} \;,
\nonumber\\
4\,\pi\left( 1+\frac{m_K}{m_N}\right) a^{(KN)}_{S_{01}} &=&0 \;,
\label{}
\end{eqnarray}
and lead to $a^{(KN)}_{S_{21}} \simeq -0.22$ fm and $a^{(KN)}_{S01} =0$ fm
close to our final values to subleading orders as given in Tab. \ref{tab-r-kp}.
In that table we collected  typical results for the p-wave scattering volumes also.
The large differences in the isospin zero channel reflect the fact that
this channel is not constrained by scattering data directly \cite{Dover}.
We find that some of our p-wave scattering volumes, also shown in Tab. \ref{tab-r-kp}, differ
significantly from the values obtained by previous analyses. Such discrepancies may  be explained
in part by important cancellation mechanisms among the u-channel baryon octet and decuplet contributions
(see \cite{Lutz:Kolomeitsev}). An accurate description of the scattering volumes requires a precise input for the meson-baryon
3-point vertices. Since the $\chi$-BS(3) approach describes the 3-point vertices in accordance
with all chiral constraints and large-$N_c$ sum rule of QCD the values for
the scattering volumes presented here should be rather reliable.

In Fig.~\ref{fig:K+phases} results for the s- and p-wave $K^+$-nucleon
phase shifts are confronted with the most recent analyses by Hyslop et al. \cite{Hyslop}
and Hashimoto \cite{Hashimoto}. We find that our partial-wave phase shifts
are reasonably close to the single energy phase shifts of \cite{Hyslop}
and \cite{Hashimoto} except the $P_{03}$ phase for which we obtain much
smaller strength. Note however, that at higher energies we smoothly reach
the single energy phase shifts of Hashimoto \cite{Hashimoto}. A possible
ambiguity in that phase shift is already suggested by the conflicting scattering
volumes found in that channel by earlier works (see Tab. \ref{tab-r-kp}). The
isospin one channel, on the other hand, seems well-established even though the
data set does not include polarization measurements close to threshold, which are
needed to unambiguously determine the p-wave scattering volumes.

In contrast to the kaon-nucleon scattering process the antikaon-nucleon scattering
process shows a large variety of interesting phenomena. Inelastic channels are
already open at threshold leading to a rich coupled channel
dynamics. Also the $\bar K N$ state couples to many of the
observed hyperon resonances for which competing dynamical
scenarios are conceivable. In this work the available data set is fitted directly
rather than any partial wave analysis. Comparing for instance the
energy-dependent analyses \cite{gopal} and \cite{garnjost} one
finds large uncertainties in the s- and p-waves in particular at
low energies. This reflects on the one hand a model dependence of
the analysis and on the other hand an insufficient data set. A
partial wave analysis of elastic and inelastic antikaon-nucleon
scattering data without further constraints from theory is
inconclusive at present \cite{Dover,Gensini}. For a detailed
overview of former theoretical analyses, we refer to the review
article by Dover and Walker \cite{Dover}.

As motivated above  the d-wave baryon resonance nonet
field is considered here. An update of the analysis \cite{Plane} leads to the
estimates $F_{[9]} \simeq 1.8$, $D_{[9]} \simeq 0.84$ and $C_{[9]}
\simeq 2.5$ for the resonance parameters. The singlet-octet mixing angle
$\vartheta \simeq 28^\circ $ confirms the finding of \cite{Tripp:2} that the
$\Lambda(1520)$ resonance is predominantly a flavor singlet state.
The values for $F_{[9]}$ and $D_{[9]} $
describe the decay widths and branching ratios of the $\Sigma (1670)$ and $\Xi (1820)$
reasonably well within their large empirical uncertainties. Less emphasis is put on the
properties of the $N(1520)$ resonance since that resonance is strongly influenced by the
$\pi \Delta_\mu $ channel not considered here. In the global fit $F_{[9]}$ and
$D_{[9]}$ were fixed as given above but  the mixing angle $\vartheta =27.74^\circ$
and $C_{[9]} =2.509 $ were fine-tuned. To account for further small inelastic
three-body channels the 'bare' $\Lambda(1520)$ resonance was given an energy independent
decay width of $\Gamma_{\Lambda(1520)}^{(3-{\rm body})}\simeq $ 1.4 MeV .
The total cross sections are included in the fit for $p_{{\rm lab.}}< $ 500 MeV. For the
bare masses of the d-wave resonances  the values $m_{\Lambda (1520)} \simeq 1528.2$ MeV,
$m_{\Lambda (1690)}\simeq 1705.3$ MeV and $m_{\Sigma (1680)} \simeq 1690.7 $ MeV were used.

\begin{figure}[h]
\begin{center}
\includegraphics[width=12.0cm,clip=true]{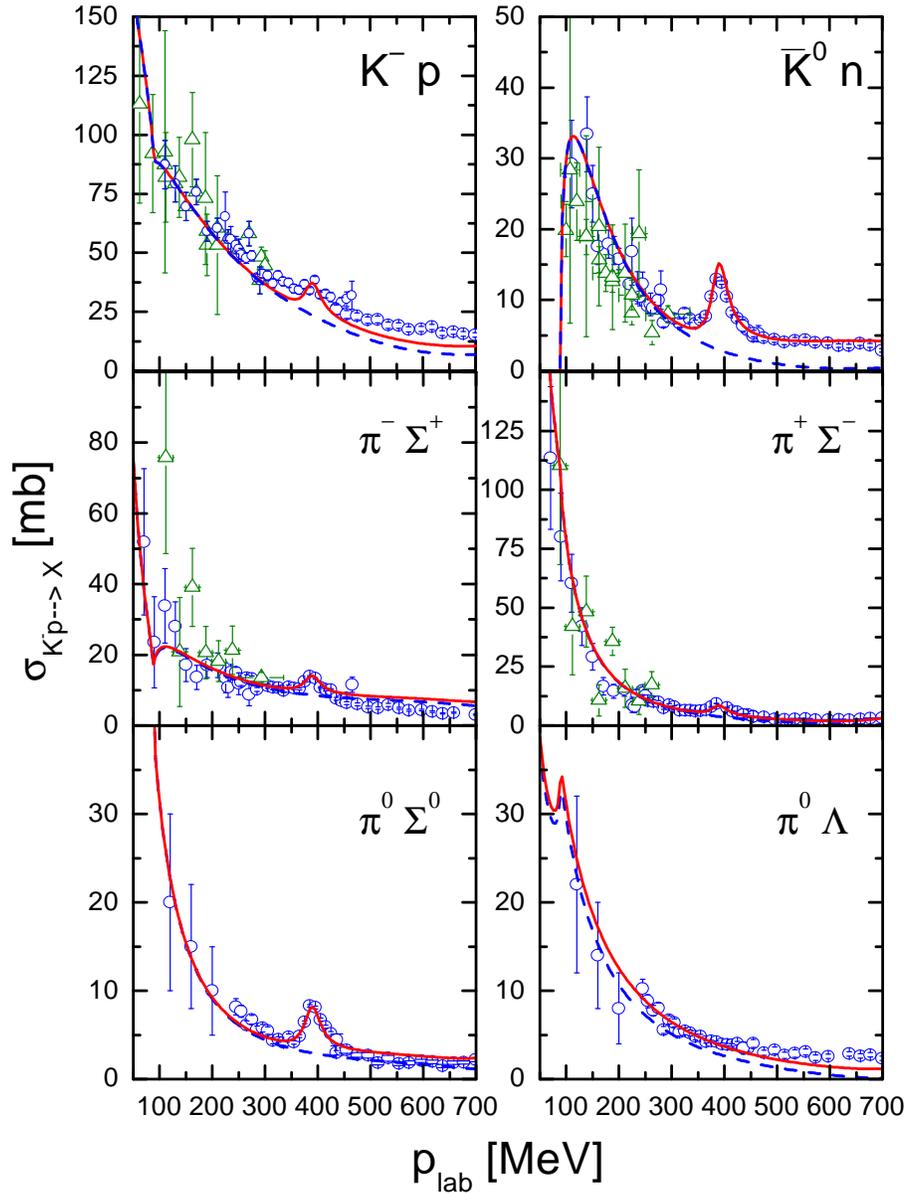}
\end{center}
\caption{$K^-$-proton elastic and inelastic cross sections. The data
are taken from \cite{mast-pio,sakit,Evans,Ciborowski,mast-ko,bangerter-piS,Armenteros,old-scat}. The solid lines
show the results of the $\chi$-BS(3) theory including
all effects of s-, p- and d-waves. The dashed lines represent the s-wave contributions only. We fitted
the data points given by open circles~\cite{mast-pio,sakit,Evans,Ciborowski,mast-ko,bangerter-piS,Armenteros}.
Further data points represented by open triangles~\cite{old-scat} were not considered in the global fit.}
\label{fig:totcross}
\end{figure}

In Fig.~\ref{fig:totcross} the result of the fit for the elastic and inelastic
$K^-p$ cross sections are shown. The data set is nicely reproduced including
the rather precise data points for laboratory momenta 250 MeV$<p_{\rm lab}<$ 500 MeV.
In Fig.~\ref{fig:totcross} the s-wave contribution to the total cross section is shown
with a dashed line. Important p-wave contributions are found at low energies only in
the $\Lambda \pi^0$ production cross section. Note that the $\Lambda \pi^0$ channel
carries isospin one and therefore provides valuable constraint on the poorly known
$K^-$-neutron interaction. The deviation of the results from the empirical cross sections
above $p_{\rm lab} \simeq 500$ MeV in some channels may be due in part
to the fact that we do not consider the p-wave  $\Lambda(1600)$ and $\Sigma(1660) $ resonances
quantitatively in this work. As will be demonstrated below when presenting the partial-wave
amplitudes, there is, however, a strong tendency that those resonances are generated in the
$\chi$-BS(3) scheme. Giving up some of the large-$N_c$ sum rules and thereby increasing the
number of free parameters one can easily obtain a fit of much improved quality beyond
$p_{lab} = 500$ MeV. We refrained from presenting those results because its is not clear
that this procedure leads to the correct partial wave interpretation
of the total cross sections. The inelastic channel $K^-p \to \Lambda \pi \pi$, not
included in this work, is no longer negligible at a quantitative level for
$p_{\rm lab} > 300$ MeV \cite{watson}.

\begin{figure}[t]
\begin{center}
\includegraphics[width=14.0cm,clip=true]{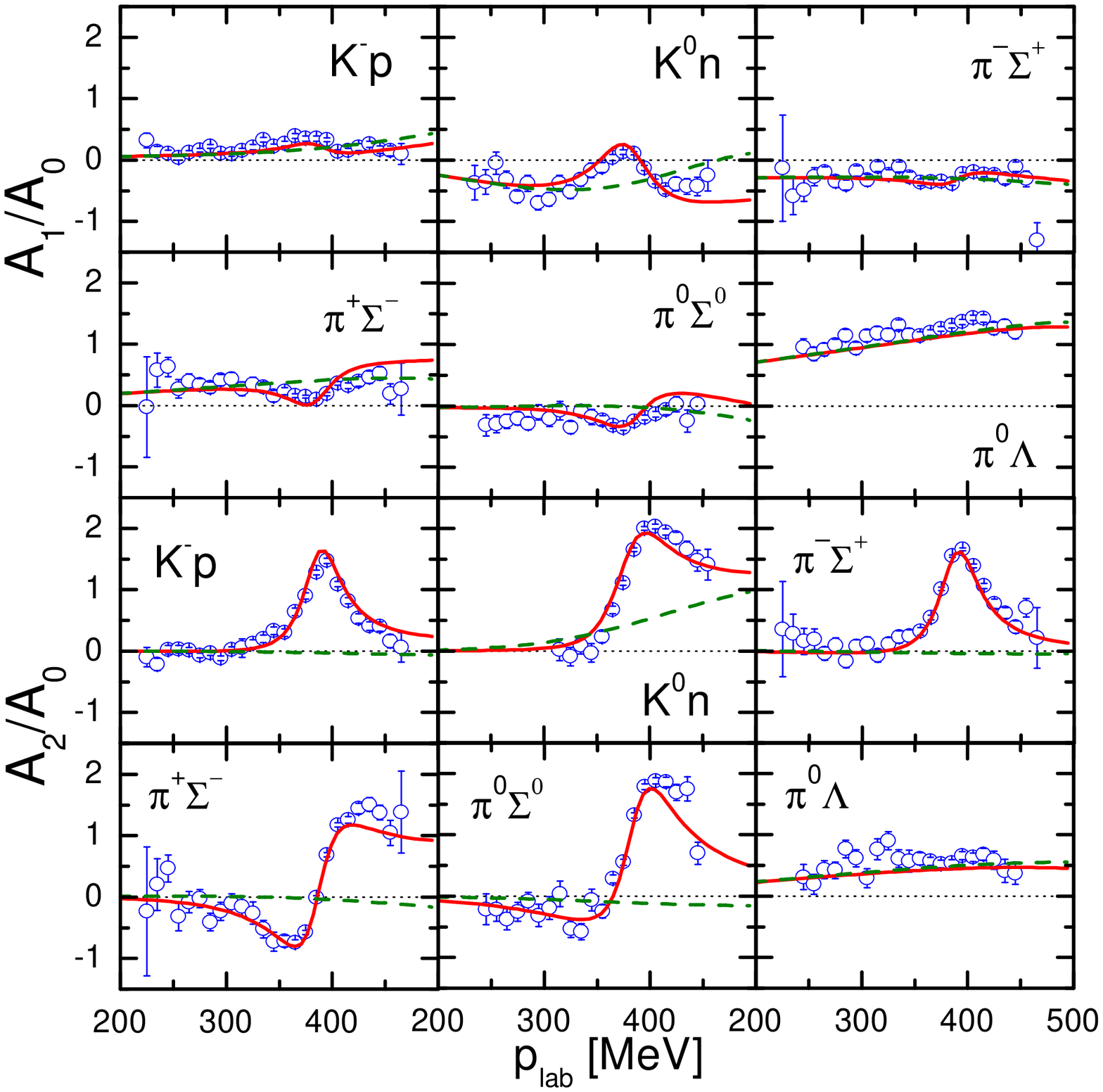}
\end{center}
\caption{Coefficients $A_1$ and $A_2$ for the $K^-p\to \pi^0 \Lambda$,
$K^-p\to \pi^\mp \Sigma^\pm$
and $K^-p\to \pi^0 \Sigma$ differential cross sections. The data are
taken from \cite{mast-pio,bangerter-piS}. The solid lines are the result of the $\chi$-BS(3) approach
with inclusion of the d-wave resonances. The dashed lines show the effect of switching off d-wave contributions.}
\label{fig:a}
\end{figure}

\begin{figure}[t]
\begin{center}
\includegraphics[width=11.0cm,clip=true]{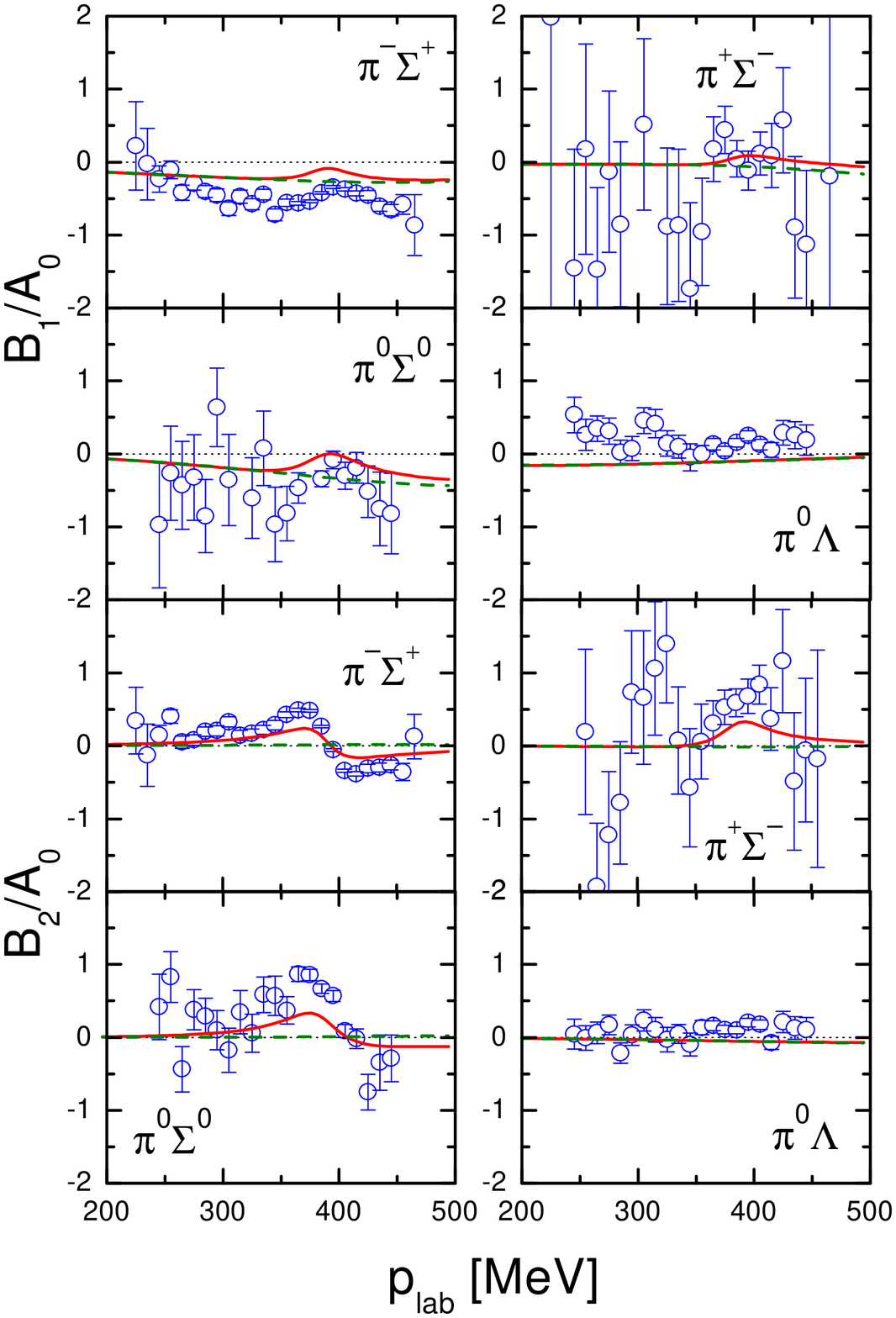}
\end{center}
\caption{Coefficients $B_1$ and $B_2$ for the $K^-p\to \pi^\mp \Sigma^\pm$,
$K^-p\to \pi^0 \Sigma$ and $K^-p\to \pi^0 \Lambda$ differential cross sections. The data are
taken from \cite{mast-pio,bangerter-piS}. The solid lines are the result of the $\chi$-BS(3) approach
with inclusion of the d-wave resonances. The dashed lines show the effect of switching off d-wave contributions.}
\label{fig:b}
\end{figure}

Further important information on the p-wave dynamics is provided by angular distributions for
the $K^-p$ reactions. The available data are represented in terms of
coefficients $A_n$ and $B_n$ characterizing the differential cross section
$d\sigma(\cos \theta , \sqrt{s}\,) $
and the polarization $P(\cos \theta , \sqrt{s}\,)$ as functions of the center of mass
scattering angle $\theta $ and the total energy $\sqrt{s}$:
\begin{eqnarray}
\frac{d\sigma (\sqrt{s}, \cos \theta )}{d\cos \theta }  &=&
\sum_{n=0}^\infty A_n(\sqrt{s}\,)\,P_n(\cos \theta ) \,,
\nonumber\\
\frac{d\sigma (\sqrt{s}, \cos \theta )}{d\cos \theta } \,P (\sqrt{s}, \cos \theta )
&=& \sum_{n=1}^\infty B_l(\sqrt{s}\,)\,P^1_n(\cos \theta ) \;.
\label{a-b-def}
\end{eqnarray}
In Fig.~\ref{fig:a} the empirical ratios $A_1/A_0$ and $A_2/A_0$ are compared with the
results of the $\chi$-BS(3) approach. Note that for $p_{\rm lab} < 300$ MeV the
empirical ratios with $n\geq 3$ are compatible with zero within their given errors.
A large $A_1/A_0$ ratio
is found only in the $K^-p\to \pi^0 \Lambda$ channel demonstrating again the importance of
p-wave effects in the isospin one channel. The dashed lines of Fig.~\ref{fig:a}, which are obtained
when switching off d-wave contributions, confirm the importance of this resonance for the angular
distributions in the isospin zero channel. The fact that the $\Lambda(1520)$ resonance appears
more important in the differential cross sections than in the total cross sections follows simply
because the tail of the resonance is enhanced if probed via an interference term. In the differential
cross section the $\Lambda(1520)$ propagator enters linearly whereas the total cross section probes
the squared propagator only. Note also the sizeable p-wave contributions at somewhat larger momenta
seen in the charge-exchange reaction of Fig. \ref{fig:a} and also in Fig. \ref{fig:totcross}.
The constraint from the ratios $B_1/A_0$ and $B_2/A_0$, presented in Fig. \ref{fig:b}, is weak due
to rather large empirical errors. New polarization data possibly with polarized hydrogen targets,
would be highly desirable.

The threshold characteristics of the $K^-p$ reaction is
constrained by experimental data for the threshold branching ratios
$\gamma, R_c$ and $R_n$ where
\begin{eqnarray}
\gamma &=& \frac{\sigma (K^-\,p\rightarrow \pi^+\,\Sigma^-)}
{\sigma (K^-\,p\rightarrow \pi^-\,\Sigma^+)}  \;, \quad
R_c = \frac{\sigma (K^-\,p\rightarrow \mbox{charged\,particles })}
{\sigma (K^-\,p\rightarrow \mbox{all})} \;,
\nonumber\\
R_n &=& \frac{\sigma (K^-\,p\rightarrow \pi^0\,\Lambda)}
{\sigma (K^-\,p\rightarrow \mbox{all\,neutral \,channels})}  \;.
\label{km:thres}
\end{eqnarray}
A further important piece of information
is provided by the recent measurement of the $K^-$ hydrogen atom state
which leads  to a value for the $K^-p$ scattering length \cite{Iwasaki}.
In Tab. \ref{tab-r-km} the empirical numbers are confronted with the present
analysis. All threshold parameters are well described within the $\chi$-BS(3)
approach. The result of \cite{Kaiser,Ramos} that the branching ratios
are rather sensitive to isospin breaking effects is confirmed. However, note that
it is sufficient to include isospin breaking effects only in the $\bar
KN$ channel to good accuracy. The empirical branching ratios are taken from
\cite{branch-rat}.  The real part of our $K^-n$ scattering length with
$\Re \,a_{K^-n \to K^- n} \simeq 0.29$ fm turns out considerably smaller than the
value of $ 0.53$ fm found in the recent analysis \cite{Ramos}.
In the Tab. \ref{tab-r-km} results for the p-wave scattering volumes are presented also.
Here isospin breaking effects are negligible. All scattering volumes but the one in the
$P_{23}$ channel are found to be small. The not too small and repulsive scattering volume
$a_{P_{23}} \simeq (-0.16+i\,0.06)$ fm$^3$ reflects the presence of the $\Sigma (1385)$ resonance
just below the $\bar K N$ threshold.  The precise values of the threshold parameters
are of crucial importance when describing  $K^-$-atom data which constitute a rather sensitive
test of the in-medium dynamics of antikaons. In particular one expects a strong sensitivity
of the level shifts on the s-wave scattering lengths.

\renewcommand{\arraystretch}{1.5}
\tabcolsep=2.4mm
\begin{table}[t]\begin{center}
\begin{tabular}{|c|c|c||c|c|c|}
\hline
& $a_{K^-p }$ [fm] & $a_{K^-n }$ [fm] & $\gamma  $ & $R_c$ & $R_n $   \\
\hline \hline
Exp. & -0.78$\pm$0.18
 & - & 2.36$\pm$ 0.04 & 0.664$\pm$0.011 & 0.189$\pm$0.015 \\
  & +$i$\,(0.49$\pm$0.37)  & & & & \\
\hline
$\chi$-BS(3) & -1.09+$i$ 0.82 & 0.29+$i$ 0.54 & 2.42 & 0.65 & 0.19  \\
\hline
 SU(2) & -0.79+$i$ 0.95 & 0.30+$i$ 0.49 & 4.58 & 0.63 & 0.32 \\
\hline
\hline
&
$a^{(\bar K N)}_{P_{01}}$ $[m_\pi^{-3}]$ & $a^{(\bar K N)}_{P_{03}}$ $[m_\pi^{-3}]$ & &
$a^{(\bar K N)}_{P_{21}}$ $[m_\pi^{-3}]$ & $a^{(\bar K N)}_{P_{23}}$ $[m_\pi^{-3}]$  \\
\hline \hline
$\chi$-BS(3)   & 0.025+$i$ 0.001 & 0.002+$i$ 0.001 & & -0.004+$i$ 0.001 & -0.055+$i$ 0.021 \\
\hline
\end{tabular}
\end{center}
\caption{$K^-$-nucleon threshold parameters. The row labelled with $SU(2)$ gives the results in the isospin limit with
$m_{K^-}=m_{\bar K^0}= 493.7$ MeV and $m_p = m_n = 938.9$ MeV. }
\label{tab-r-km}
\end{table}

\begin{figure}[t]
\begin{center}
\includegraphics[width=12cm,clip=true]{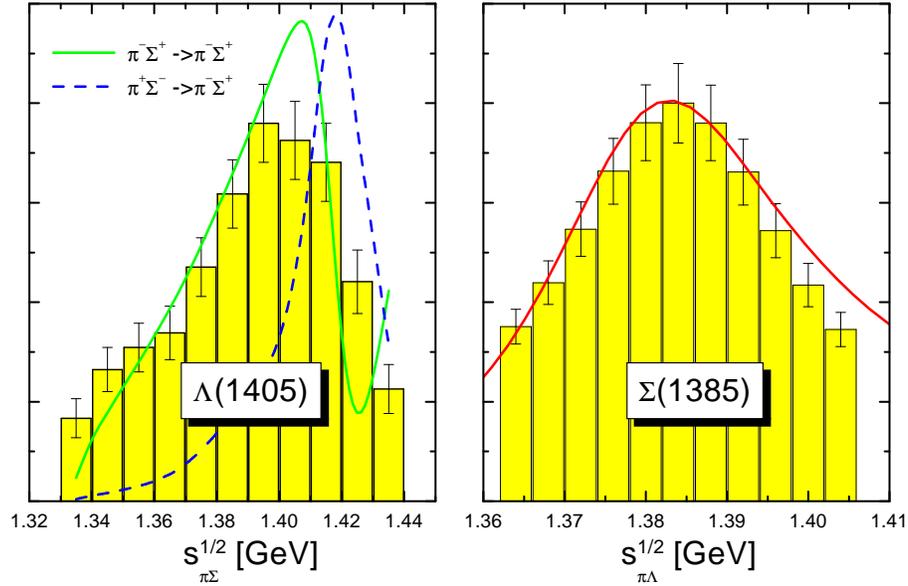}
\end{center}
\caption{$\Lambda(1405)$ and $\Sigma (1385)$ resonance mass
distributions in arbitrary units.}
\label{fig:massspec}
\end{figure}

In Fig.~\ref{fig:massspec} the $\Lambda(1405)$ and $\Sigma (1385)$ spectral
functions are shown as measured in the reactions
$K^-p\rightarrow \Sigma^+\pi^-\pi^+\pi^-$ \cite{lb-spec}
and $K^-p\rightarrow \Lambda \pi^+\pi^-$ \cite{sig-spec}  respectively.
The $\Lambda(1405)$ spectrum of \cite{lb-spec} was not included in the global fit.
Since the $\Lambda(1405)$-spectrum shows a strong energy dependence, incompatible with a
Breit-Wigner resonance shape, the spectral form depends rather strongly on the initial and
final states through which it is measured. The empirical spectrum of  \cite{lb-spec}
describes the reaction $\Sigma^+(1600) \,\pi^- \to \Lambda(1405) \to \Sigma^+ \,\pi^-$
rather than the reactions $\Sigma^\pm \,\pi^\mp \to \Lambda(1405) \to \Sigma^+ \,\pi^-$
accessible in our present scheme. In Fig. 6 the spectral form of the $\Lambda(1405)$
resulting from two different initial states $\Sigma^\pm \,\pi^\mp$ are confronted with the
empirical spectrum of \cite{lb-spec}. While the spectrum defined with respect to the initial
state $\Sigma^+ \,\pi^-$ represents the empirical spectrum reasonably well the
other choice of initial state $\Sigma^+ \,\pi^-$ leads to a significantly altered spectral
form. Hence, in a scheme that does not include the $\Sigma(1600) \pi$ state
explicitly it is not justified to use the $\Lambda(1405)$ spectrum of \cite{lb-spec} as a
quantitative constraint for the antikaon-nucleon dynamics.

Consider the mass spectrum of the decuplet $\Sigma (1385)$ state in Fig. \ref{fig:massspec}.
The spectral form, to good accuracy of Breit-Wigner form, is
reproduced reasonably well. The result for the ratio of $\Sigma (1385) \to \pi \Lambda $
over $\Sigma (13 85) \to \pi \Sigma $ of about $17\, \% $ compares well with the most
recent empirical determination. In \cite{Dionisi} that branching ratio was extracted from the
$K^-p \to \Sigma (1385) \,K\,\bar K$ reaction and found to be $20 \pm 6 \,\%$.
Note that the ratio given above was obtained from the two reaction amplitudes
$\pi \Lambda \to \pi \Lambda$ and $\pi \Lambda \to \pi \Sigma $ evaluated at
the $\Sigma (1385)$ pole. The schematic expression (\ref{decuplet-decay}) would give
a smaller value of about $ 15 \, \%$. The value for the
$\Xi (1530)$ total width of $10.8 $ MeV comfortably meets the
empirical value of $9.9^{+1.7}_{-1.9} $ MeV given in \cite{fpi:exp}.

\vskip1.5cm \section{Analyticity and crossing symmetry}

It is important to investigate to what extent the  multi-channel scattering amplitudes
are consistent with the expectations from analyticity and crossing symmetry. As discussed in
detail in chapter 4.3 the crossing symmetry constraints should not be considered in terms of
partial-wave amplitudes, but rather in terms of the forward scattering amplitudes only.
In the strangeness sector crossing symmetry is particularly important because
that sector has a large subthreshold region not directly accessible and constrained by data.
In the following the forward $\bar KN$ and $KN$ scattering amplitudes are reconstructed in
terms of their imaginary parts by means of dispersion integrals. The reconstructed
scattering amplitudes are then confronted with the original ones. It is non-trivial that
those amplitudes match even though the loop functions and effective interaction kernels are
analytic functions. One can not exclude that the coupled channel dynamics generates
unphysical singularities off the real axis which would then spoil the representation of
the scattering amplitudes in terms of dispersion integrals. Within effective field
theory unphysical singularities are acceptable, however, only far outside the applicability
domain of the approach.

The analysis performed here is analogous to that of Martin \cite{A.D.Martin}. However,
here the dispersion-integral representation is considered as a consistency check of
the theory rather
than as a predictive tool to derive the low-energy kaon-nucleon scattering amplitudes in terms
of the more accurate scattering data at $E_{\rm lab}>300$ MeV \cite{Queen}. This
way any subtle assumption on the number of required subtractions in the dispersion integral
is avoided. Obviously the dispersion integral, if evaluated for small energies, must be
dominated by the low-energy total cross sections which are not known empirically too well.
A subtracted dispersion integral is written
\begin{eqnarray}
&& \! \!T^{(0)}_{\bar K N}(s) = \frac{f^2_{KN \Lambda }}
{s-m^2_\Lambda} +  \sum_{k=1}^n\,c_{\bar K N}^{(0,k)}\,(s-s_0)^k
+\!\!\!\int_{(m_\Sigma+m_\pi)^2}^\infty \!\!\! \! \frac{d \,s'}{\pi }\,
\frac{(s-s_0)^n}{(s'-s_0)^n}\,\frac{\Im \,T^{(0)}_{\bar K N} (s')}{s'-s -i\,\epsilon}\;,
\nonumber\\
&& \! \! T^{(1)}_{\bar K N}(s) = \frac{f_{KN \Sigma }^2}
{s-m^2_\Sigma} +  \sum_{k=1}^n\,c_{\bar K N}^{(1,k)}\,(s-s_0)^k
+\!\!\! \int_{(m_\Lambda+m_\pi)^2}^\infty \!\!\! \!\frac{d \,s'}{\pi }\,
\frac{(s-s_0)^n}{(s'-s_0)^n}\,\frac{\Im \,T^{(1)}_{\bar K N} (s')}{s'-s -i\,\epsilon} \;,
\label{disp-check}
\end{eqnarray}
where $s_0 = \Lambda_{\rm opt.}^2= m_N^2+m_K^2$ is identified with the optimal matching point
of (\ref{opt-match}). Recall that the  kaon-hyperon coupling constants,
\begin{eqnarray}
&& f_{KN Y } = \sqrt{\frac{m_K^2-(m_Y-m_N)^2}{2\,m_N}}\,
\frac{m_N+m_Y}{2\,f}\,\Big( A^{(Y)}_{\bar K N}+P^{(Y)}_{\bar K N}\Big) \,,
\label{}
\end{eqnarray}
with $Y=\Lambda, \Sigma $ receive
contributions from pseudo-vector and pseudo-scalar vertices as specified
in Tab. \ref{meson-baryon:tab}. The values $f_{K N \Lambda } \simeq -12.8\,m^{1/2}_{\pi^+}$ and
$f_{K N \Sigma } \simeq 6.1\,m^{1/2}_{\pi^+}$ follow.

The forward scattering amplitude $T_{\bar KN}^{}(s)$  reconstructed in terms of the
partial-wave amplitudes $f_{\bar K N, J}^{(L)}(\sqrt{s}\,)$ of (\ref{match}) reads
\begin{eqnarray}
T^{}_{\bar KN}(s) &=&4\,\pi\,\frac{\sqrt{s}}{m_N}\,
\sum_{n=0}^\infty \,\Big(n+1 \Big)\,\Big( f_{\bar K N,n+{\textstyle{1\over 2}}}^{(n)}(\sqrt{s}\,)+
f_{\bar K N,n+{\textstyle{1\over 2}}}^{(n+1)}(\sqrt{s}\,)
\Big) \;.
\label{forward-amplitude}
\end{eqnarray}
The subtraction coefficients $c_{\bar K N}^{(I,k)}$ are
adjusted to reproduce the scattering amplitudes close to the kaon-nucleon threshold.
One must perform a sufficient number of subtractions so that
the dispersion integral in (\ref{disp-check}) is dominated by energies still within the
applicability range of our theory. With $n=4$ in (\ref{disp-check}) we indeed find that
we are insensitive to the scattering amplitudes for $\sqrt{s} > 1600$ MeV to good
accuracy. Similarly, subtracted dispersion integrals for the
amplitudes $T^{(I)}_{K N}(s)$ of the strangeness $+1$ sector are written
\begin{eqnarray}
 T^{(0)}_{K N}(s) &=& -\frac{1}{2}\,\frac{f^2_{KN \Lambda }}{2\,s_0-s-m^2_\Lambda}
+\frac{3}{2}\,\frac{f^2_{KN \Sigma }}{2\,s_0-s-m^2_\Sigma}
\nonumber\\
&+&\sum_{k=1}^n\,c_{K N}^{(0,k)}\,(s-s_0)^k+\int_{(m_N+m_K)^2}^\infty \frac{d \,s'}{\pi }\,
\frac{(s-s_0)^n}{(s'-s_0)^n}\,\frac{\Im \,T^{(0)}_{K N} (s')}{s'-s -i\,\epsilon}\;,
\nonumber\\
T^{(1)}_{K N}(s) &=& \frac{1}{2}\, \frac{f^2_{KN \Lambda }}{2\,s_0-s-m^2_\Lambda}
+\frac{1}{2}\, \frac{f^2_{KN \Sigma }}{2\,s_0-s-m^2_\Sigma}
\nonumber\\
&+&\sum_{k=1}^n\,c_{K N}^{(1,k)}\,(s-s_0)^k+\int_{(m_N+m_K)^2}^\infty \frac{d \,s'}{\pi }\,
\frac{(s-s_0)^n}{(s'-s_0)^n}\,\frac{\Im \,T^{(1)}_{K N} (s')}{s'-s -i\,\epsilon} \;.
\label{disp-check-2}
\end{eqnarray}
The subtraction coefficients $c_{K N}^{(I,k)}$ are
adjusted to reproduce the scattering amplitudes close to the kaon-nucleon threshold. The
choice $n=4$ leads to a sufficient emphasis of the low-energy $KN$-amplitudes.

\begin{figure}[t]
\begin{center}
\includegraphics[width=13cm,clip=true]{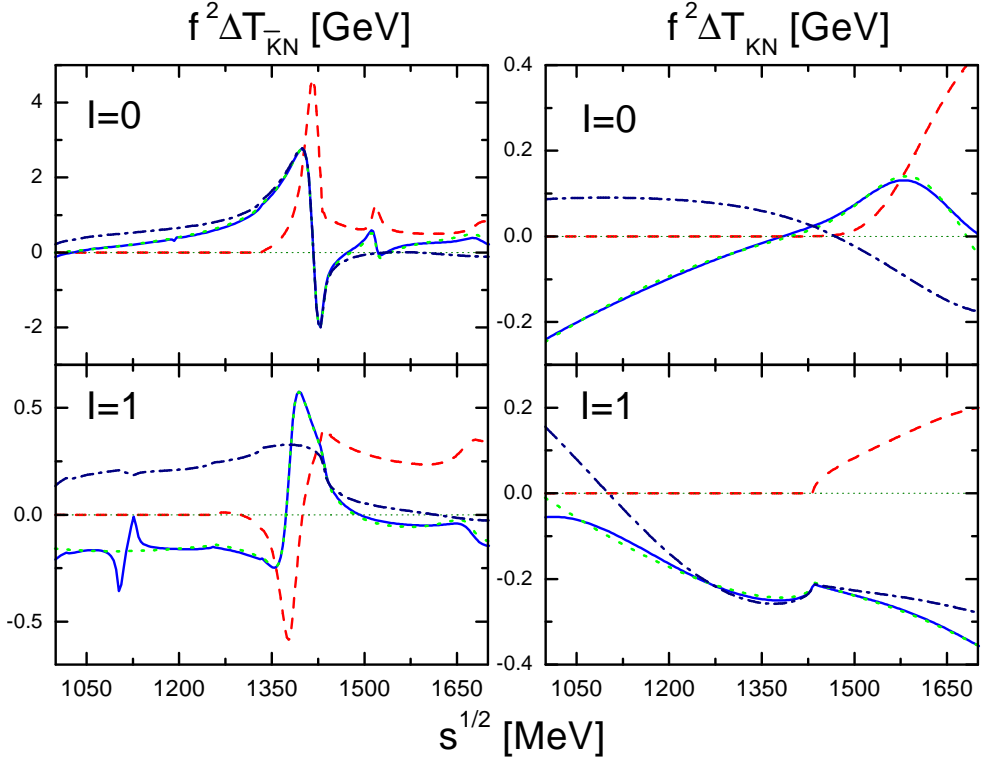}
\end{center}
\caption{Causality check of the pole-subtracted kaon-nucleon scattering amplitudes $f^2\,\Delta T^{(I)}_{K N}$
and $f^2\,\Delta T^{(I)}_{\bar K N}$. The full lines represent the real part of the forward scattering amplitudes.
The dotted lines give the amplitudes as obtained from their imaginary parts (dashed lines) in terms of the
dispersion integrals (\ref{disp-check}) and (\ref{disp-check-2}). The dashed-dotted lines give the s-wave
contribution to the real part of the forward scattering amplitudes only.}
\label{fig:causality}
\end{figure}

In Fig.~\ref{fig:causality} we compare the real part of the pole-subtracted amplitudes
$\Delta T^{(I)}_{\bar KN}(s)$ and $\Delta T^{(I)}_{KN}(s)$ with the corresponding amplitudes
reconstructed via the dispersion integral (\ref{disp-check})
\begin{eqnarray}
\Delta T_{\bar KN}^{(I)}(s) &=&  T_{\bar KN}^{(I)}(s)- \frac{f^2_{K N Y}}{s-m_H^2}
-\left( A_{\bar K N}^{(Y)}\right)^2 \frac{2\,m_N^2+m_K^2-s-m_H^2}{8\,f^2\,m_N}
\nonumber\\
&+&\left( P_{\bar K N}^{(Y)}\right)^2 \frac{(m_N+m_H)^2}{8\,f^2\,m_N}
+ P_{\bar K N}^{(Y)}\,A_{\bar K N}^{(Y)}\frac{m^2_H-m_N^2}{4\,f^2\,m_N}  \,,
\label{pole-subtr}
\end{eqnarray}
with $Y=\Lambda$ for $I=0$ and $Y= \Sigma $ for $I=1$. Whereas it
is straightforward to subtract the complete hyperon pole contribution in the $\bar K N$ amplitudes
(see (\ref{pole-subtr})), it is less immediate how to do so for the $KN$ amplitudes. Since not
all partial wave contributions are considered in the latter amplitudes the u-channel pole
must be subtracted in its approximated form as given in (\ref{u-approx-1}).
The reconstructed amplitudes agree rather well with the original amplitudes. For
$\sqrt{s} > 1200$ MeV the solid and dotted lines in Fig. \ref{fig:causality} can hardly be
discriminated. This demonstrates that the amplitudes are causal to good accuracy. Note that the
discrepancy for $\sqrt{s} < 1200$ MeV is a consequence of the approximate treatment
of the non-local u-channel exchanges  which violates analyticity at subthreshold energies to some
extent (see (\ref{prescription})). With Fig.~\ref{fig:causality} it is demonstrated that such
effects are well controlled for $\sqrt{s} > 1200$ MeV. In any case, close to
$\sqrt{s}\simeq 1200$ MeV the complete forward scattering amplitudes
are largely dominated by the s- and u-channel hyperon pole contributions absent
in $\Delta T_{\bar KN}^{(I)}(s)$.

As can be seen from Fig.~\ref{fig:causality} also, there are sizeable p-wave contributions
in the pole-subtracted amplitudes at subthreshold energies. This follows by comparing the
dashed-dotted lines, which give the s-wave contributions only, with the solid lines which
represent the complete real part of the pole-subtracted forward scattering amplitudes.
The p-wave contributions are typically much larger below threshold than above threshold.
The fact that this is not the case in the isospin zero $KN$ amplitude reflects a subtle
cancellation mechanism of hyperon exchange contributions and quasi-local two-body interaction
terms. In the $\bar K N$ amplitudes the subthreshold effects of p-waves are most dramatic in
the isospin one channel. Here the amplitude is dominated by the $\Sigma(1385)$ resonance. Note
that p-wave channels contribute with a positive imaginary part for energies larger than the
kaon-nucleon threshold but with a negative imaginary part for subthreshold energies. A negative
imaginary part of a subthreshold amplitude is consistent with the optical theorem which
relates the imaginary part of the forward scattering amplitudes to the total cross section
for energies above threshold only. The results presented here shed some doubts on the analysis
of Martin, which attempted to constrain the forward scattering amplitude via a dispersion
analysis \cite{A.D.Martin}. An implicit assumption of his analysis was, that the contribution
to the dispersion integral from the subthreshold region, which is not directly determined by
the data set, is dominated by s-wave dynamics.
As was pointed out in \cite{Hirschegg} strong p-wave contributions at subthreshold energies
should have an important effect on the propagation properties of antikaons in dense nuclear
matter. This expectation will be confirmed in the next section.

Crossing symmetry relates the subthreshold $\bar K N$ and $K N$ scattering amplitudes.
As a consequence the exact amplitude ${\rm T}^{(0)}_{\bar K N}(s)$ shows unitarity cuts
not only for $\sqrt{s}>m_\Sigma+m_\pi$ but also for $\sqrt{s}<m_N-m_K$ representing the
elastic $K N$ scattering process. For the isospin zero amplitude one expects the
following representation:

\begin{figure}[t]
\begin{center}
\includegraphics[width=12cm,clip=true]{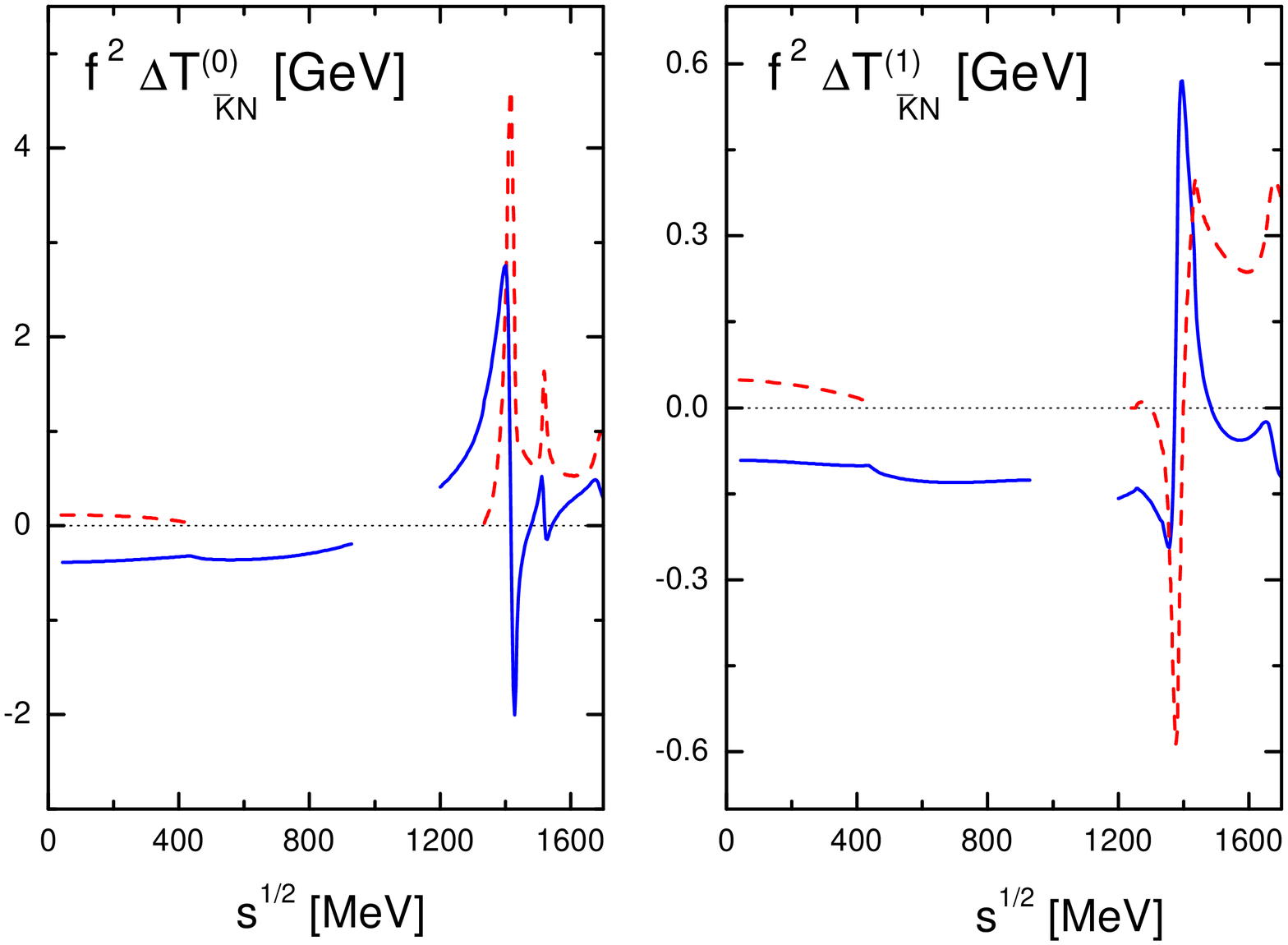}
\end{center}
\caption{Approximate crossing symmetry of the pole subtracted kaon-nucleon forward scattering amplitudes. The lines in
the left hand parts of the figures result from the $KN$ amplitudes. The lines in the right hand side of
the figures give the $\bar K N$ amplitudes.}
\label{fig:crossing}
\end{figure}

\begin{eqnarray}
\!\! {\rm T}^{(0)}_{\bar K N}(s)-{\rm T}^{(0)}_{\bar K N}(s_0) &=&
\frac{f^2_{KN \Lambda }}{s-m^2_\Lambda}-\frac{f^2_{KN \Lambda }}{s_0-m^2_\Lambda} +
\! \! \!\int_{-\infty}^{(m_N-m_K)^2} \frac{d \,s'}{\pi }\,\frac{s-s_0}{s'-s_0}\,
\frac{\Im \,{\rm T}^{(0)}_{\bar K N} (s')}{s'-s -i\,\epsilon}
\nonumber\\
&+&\int_{(m_\Sigma+m_\pi)^2}^{+\infty} \frac{d \,s'}{\pi }\,\frac{s-s_0}{s'-s_0}\,
\frac{\Im \,{\rm T}^{(0)}_{\bar K N} (s')}{s'-s -i\,\epsilon}
\;,
\label{cross-disp-check}
\end{eqnarray}
where one subtraction was performed to help the convergence of the dispersion integral.
An analogous representation holds for the isospin one amplitude.
Comparing the expressions for $T^{(0)}_{\bar K N}(s)$  in (\ref{disp-check}) and
${\rm  T}^{(0)}_{\bar K N}(s)$ in (\ref{cross-disp-check}) demonstrates that the contribution
of the unitarity cut at $\sqrt{s}< m_N-m_K$ in (\ref{cross-disp-check}) is effectively
absorbed in the subtraction coefficients $c_{\bar K N}^{(I,k)}$ of (\ref{disp-check}).
Similarly the subtraction coefficients $c_{K N}^{(I,k)}$ in (\ref{disp-check-2}) represent the
contribution to $T^{(I)}_{K N}(s)$ from the inelastic $\bar K$N scattering process. Thus,
both model amplitudes $T^{(I)}_{K N}(s)$ and $T^{(I)}_{\bar K N}(s)$
represent the exact amplitude ${\rm T}^{(I)}_{\bar K N}(s)$ within their validity
domains and therefore approximate crossing symmetry
\begin{eqnarray}
&& \Delta T^{(0)}_{\bar K N}(s) \simeq
-\frac{1}{2}\, \Delta T^{(0)}_{K N}(2\,s_0-s)+\frac{3}{2}\,  \Delta T^{(1)}_{ K N}(2\,s_0-s) \;,
\nonumber\\
&&  \Delta T^{(1)}_{\bar K N}(s) \simeq
+ \frac{1}{2}\, \Delta T^{(0)}_{K N}(2\,s_0-s)+\frac{1}{2}\, \Delta T^{(1)}_{ K N}(2\,s_0-s) \;,
\label{exp-cross}
\end{eqnarray}
is expected close to the optimal matching point $s_0=m_N^2+m_K^2$ only.
In Fig.~\ref{fig:crossing} we confront the pole-subtracted
$ \Delta T^{(I)}_{K N}$ and $  \Delta T^{(I)}_{\bar KN} $ amplitudes with the
expected approximate crossing identities (\ref{exp-cross}). Since the optimal matching point
$s_0=m_N^2+m_K^2 \simeq (1068)^2$ MeV$^2$ is slightly below the respective validity range of the
original amplitudes, one may use the reconstructed amplitudes of (\ref{disp-check}) and
(\ref{disp-check-2}) shown in Fig.~\ref{fig:causality}. This is justified, because the
reconstructed amplitudes are based on the imaginary parts of the amplitude which have support
within the validity domain of our theory only. Fig.~\ref{fig:crossing} indeed confirms that
the kaon-nucleon scattering amplitudes are approximatively crossing symmetric. Close to the
the point $s\simeq m_N^2+m_K^2$ the $K N$ and $\bar K N$ amplitudes match.
Hence, the subthreshold antikaon-nucleon scattering amplitudes are determined rather
reliably and well suited for an application to the nuclear antikaon dynamics.

\vskip1.5cm \section{Antikaons and hyperon resonances in nuclear matter}

The result section is closed with a presentation of antikaon and
hyperon resonance properties as they arise in the self consistent many-body
approach developed in chapter 5. A resonance propagator may
be identified with the appropriate antikaon-nucleon scattering amplitude of a
given partial wave. In a self consistent scheme the in-medium scattering process
is intimately related to the antikaon spectral function. According
to (\ref{kaon-final}) once the self consistent in-medium scattering process is established
the antikaon self energy follows upon a proper Fermi averaging of the in-medium scattering
amplitudes (\ref{kaon-final}). Therefore the
pertinent structures in the in-medium amplitudes already tell the characteristic features expected in the antikaon
spectral function. Most important are the s-wave $\Lambda (1405)$ and the p-wave $\Sigma (1385)$
resonances. Of central importance for the derivation of
a realistic antikaon spectral function are the subthreshold $\bar K N$ scattering amplitudes derived
in the previous sections. Since the antikaon spectral function tests the $\bar K N$ amplitudes
at subthreshold energies where they are not directly determined by empirical data it is crucial
to take $\bar K N$ amplitudes as input for the many-body calculation which are consistent with constraints set
by causality, chiral symmetry and crossing symmetry. In this work the effect of nuclear binding
and short range correlation is not yet studied. The impact of the latter effects
on the antikaon propagation properties are generally found to be of moderate importance at
nuclear saturation density \cite{Waas2,Tolos,Cieply}.

\begin{figure}[t]
\begin{center}
\includegraphics[width=16cm,clip=true]{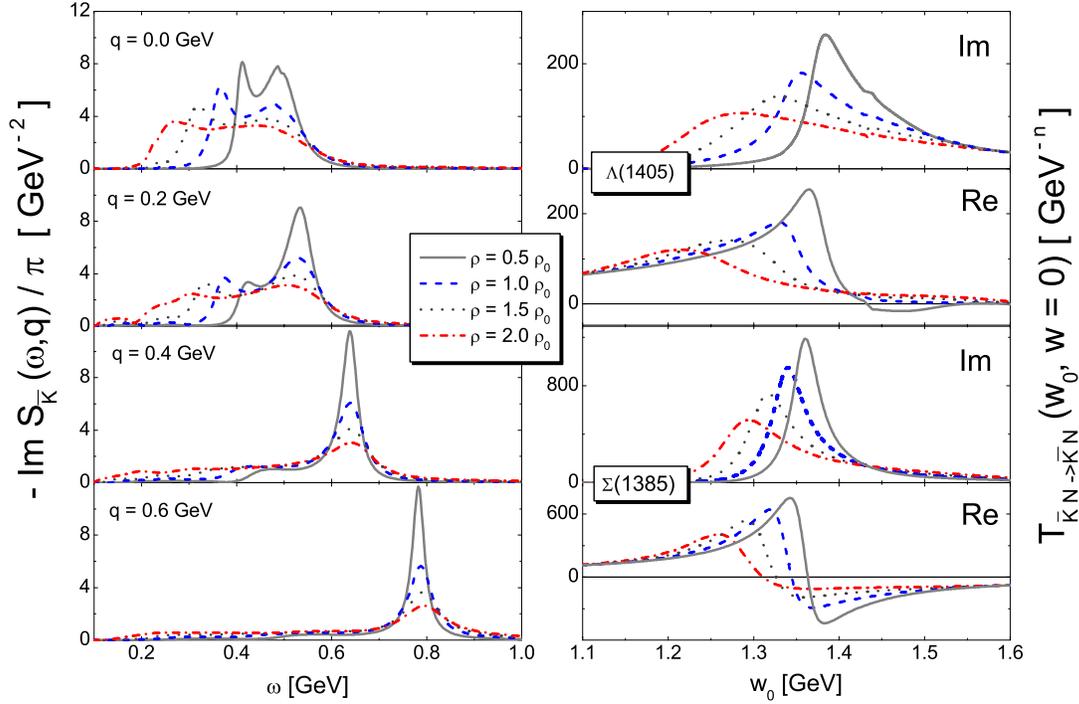}
\end{center}
\caption{The antikaon spectral function is shown in the left hand panel as
a function of the antikaon energy $\omega$, the momentum $\vec q$ and the
nuclear density with $\rho_0 = 0.17$ fm$^{-3}$. The right hand panel
illustrates the in-medium modification of the  $\Lambda(1405)$ and
$\Sigma (1385)$ hyperon resonances. It is plotted the real and imaginary
parts of the antikaon-nucleon scattering amplitudes in the appropriate channels.
The hyperon energy and momentum are $w_0$ and $\vec w =0$ respectively.}
\label{fig:spec}
\end{figure}

In Fig. \ref{fig:spec} results for the spectral functions are shown at various
isospin symmetric nuclear matter densities. They are based on the scattering amplitudes
obtained in the previous sections. The antikaon spectral function
exhibits a rich structure with a pronounced dependence on the antikaon three-momentum.
That reflects the presence of hyperon resonance states in the $\bar K N$ amplitudes.
Typically the peaks seen are quite broad and not always of quasi-particle type. For the hyperon
resonances we predict attractive mass shifts and a considerable broadening with increasing density.
At nuclear saturation density the attractive mass shifts for the $\Lambda(1405)$ and
$\Sigma (1385)$ are about 60 MeV for both states. The resonance widths increase to about
120 MeV and 70 MeV respectively. Further results for
the hyperon ground states and the $\Lambda(1520)$ resonance are reported in \cite{Lutz:Korpa}.
It is assured that the sum rule
\begin{eqnarray}
-\int \frac{d \, \omega }{\pi}\,\omega \,\Im \,S_{\bar K}(\omega,\,\vec q\,) =1 \;,
\label{k-sumrule}
\end{eqnarray}
is satisfied to good accuracy. Any violation of that sum rule (\ref{k-sumrule}) would
indicate that the microscopic theory is in conflict with constraints set by causality.
The strong in-medium modifications found for the hyperon resonance states have a simple
physical interpretation. The latter is
determined by the hyperon-resonance nucleon interaction. The self consistent approach includes in
particular the t-channel antikaon exchange contribution, which is expected to dominate that interaction
in particular when the effective antikaon mass is lowered by the nuclear environment. A dramatic
in-medium modification of the antikaon spectral function, then necessarily leads to strong effects also for the
hyperon states.

It is emphasized that here the in-medium effects for the
$\Lambda(1405)$ are reported to be significantly stronger as compared to
previous works \cite{ml-sp,ramossp,Cieply}. In the original work
\cite{ml-sp} it was was found that a self consistent evaluation,
based on s-wave interactions only, that were adjusted to reproduce
the amplitudes of \cite{Kaiser}, lead to almost zero mass shift of
the $\Lambda(1405)$ at nuclear saturation density. The large
repulsive mass shift \cite{Waas1,Waas2} predicted by the Pauli
blocking effect (see (\ref{pauli-lambda})) was compensated by the
attractive mass shift implied by a decreased effective antikaon
mass. It is reassuring that our present scheme confirms
that finding. At nuclear saturation density the in-medium change of the
antikaon-nucleon loop function is largely dominated by the last two
terms in (\ref{j-exp}). If higher partial wave contributions are switched off the
$\Lambda(1405)$ resonance does not experience any significant mass
shift at nuclear saturation density, merely the width increases
somewhat as was observed also in \cite{ml-sp,ramossp,Cieply}.

In \cite{ramossp} this result is confirmed qualitatively by a
calculation applying the amplitudes of \cite{Ramos}. The
quantitative comparison of our original calculation \cite{ml-sp} and also
\cite{Lutz:Korpa}, important for the
description of kaonic atoms \cite{Florkowski,Cieply}, is hampered
by the facts that first the subthreshold amplitudes of
\cite{Kaiser,Lutz:Kolomeitsev} and \cite{Ramos} differ significantly and second the
computation in \cite{ramossp} relies on an approximation that
amounts to neglecting the strong momentum dependence of the
antikaon self energy (see discussion in chapter 5.2). We refrain here from a detailed
comparison with the results of Tolos et al. \cite{Tolos}. Their many-body
computation includes higher partial wave contributions. However, the applied
meson exchange model \cite{Juelich:2} was confronted so far with total cross
sections data only. Thus, it remains unclear whether it describes the dynamics of higher
partial waves correctly.

\begin{figure}[t]
\begin{center}
\includegraphics[width=8cm,clip=true]{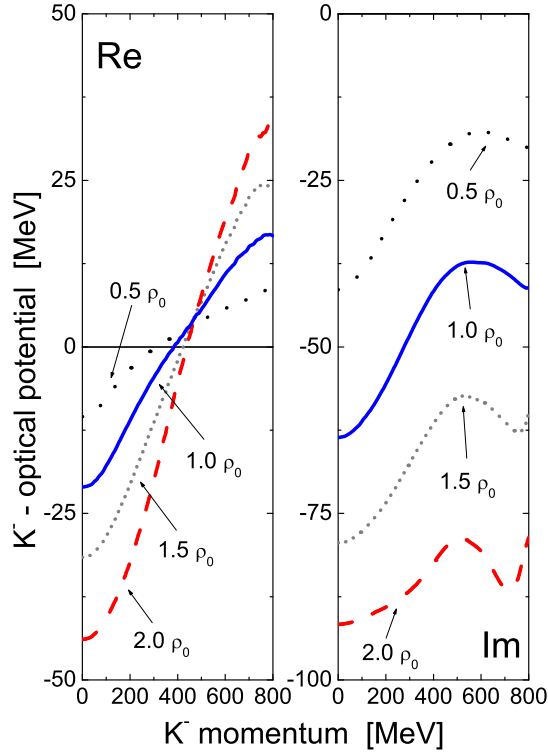}
\end{center}
\caption{Antikaon optical potential $V_{\rm opt.} ({\vec q}\,; \rho )$
for isospin symmetric nuclear matter densities. Results are shown for
$0.5\,\rho_0$, $1.0\,\rho_0$, $1.5\,\rho_0$ and $2.0\,\rho_0$ where
$\rho_0 = 0.17$ fm$^{-3}$.}
\label{fig:kaon-optical}
\end{figure}

The non-trivial dynamics implied by the presence of the hyperon resonances
reflects itself in a complicated behavior of the antikaon self energy, $\Pi_{\bar K}( \omega
, {\vec q}\,; \rho )$. This is illustrated by the antikaon nuclear optical potential
$V_{\rm opt.} (\vec q \,; \rho )$, defined by
\begin{eqnarray}
2\,E_K({\vec q}\,)\, V_{\rm opt.} ({\vec q}\, ; \rho) =\Pi_{\bar K}( \omega
=E_K({\vec q}\,), {\vec q}\,; \rho ) \;, \qquad E_K({\vec q}\, )=\sqrt{m_K^2+{\vec q}\,^2 }\,.
\end{eqnarray}
In Fig. \ref{fig:kaon-optical} the result for the optical potential is presented
as a function of the antikaon momentum $\vec q $ and the nuclear matter density.
It is pointed out that the real part of the optical potential exhibits rather moderate
attraction of about 20 MeV at nuclear saturation density and $\vec q= 0$ MeV. The attraction
is further diminished and even turns into repulsion as the antikaon momentum increases.
On the other hand we find a rather strong absorptive
part of the optical potential. This agrees qualitatively with our previous work that was
based on self consistent s-wave dynamics \cite{Florkowski} but is in striking disagreement
with mean field calculations \cite{Schaffner:Bondorf:Mishustin} that predict considerable
more attraction in the antikaon optical potential. The large attraction in the antikaon
spectral function of Fig. \ref{fig:spec} is consistent with the
moderate attraction in the optical potential of Fig. \ref{fig:kaon-optical}. It merely
reflects the strong energy dependence of the kaon self energy induced by the
$\Lambda (1405) $ and $\Sigma (1385 )$ resonances. Such important energy variations are
missed in a mean field approach. Hence, a proper treatment
of the pertinent many-body effects is required.

\begin{figure}[t]
\begin{center}
\includegraphics[width=14cm,clip=true]{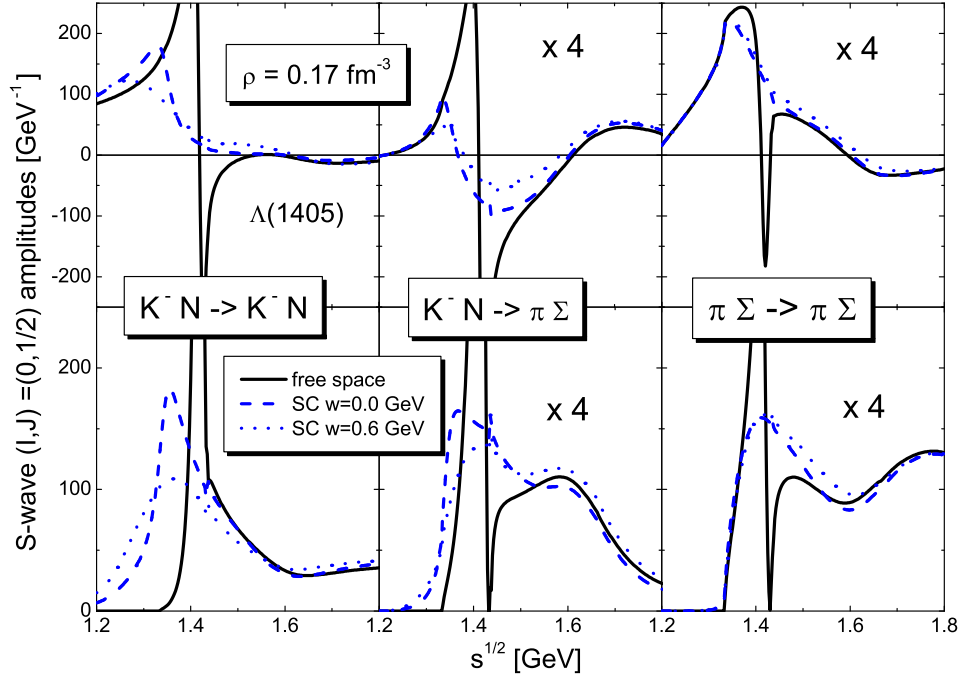}
\end{center}
\caption{S-wave reaction amplitudes in the isospin zero channel for
isospin symmetric nuclear matter $\rho = 0.17$ fm$^{-3}$. The $J=1/2$ amplitudes are the result
of a self consistent calculation. The three momentum
$\vec w $ denotes the sum of initial three momenta defined in the rest frame
of nuclear matter.}
\label{fig:cs-amplitudes-0}
\end{figure}

\begin{figure}[t]
\begin{center}
\includegraphics[width=14cm,clip=true]{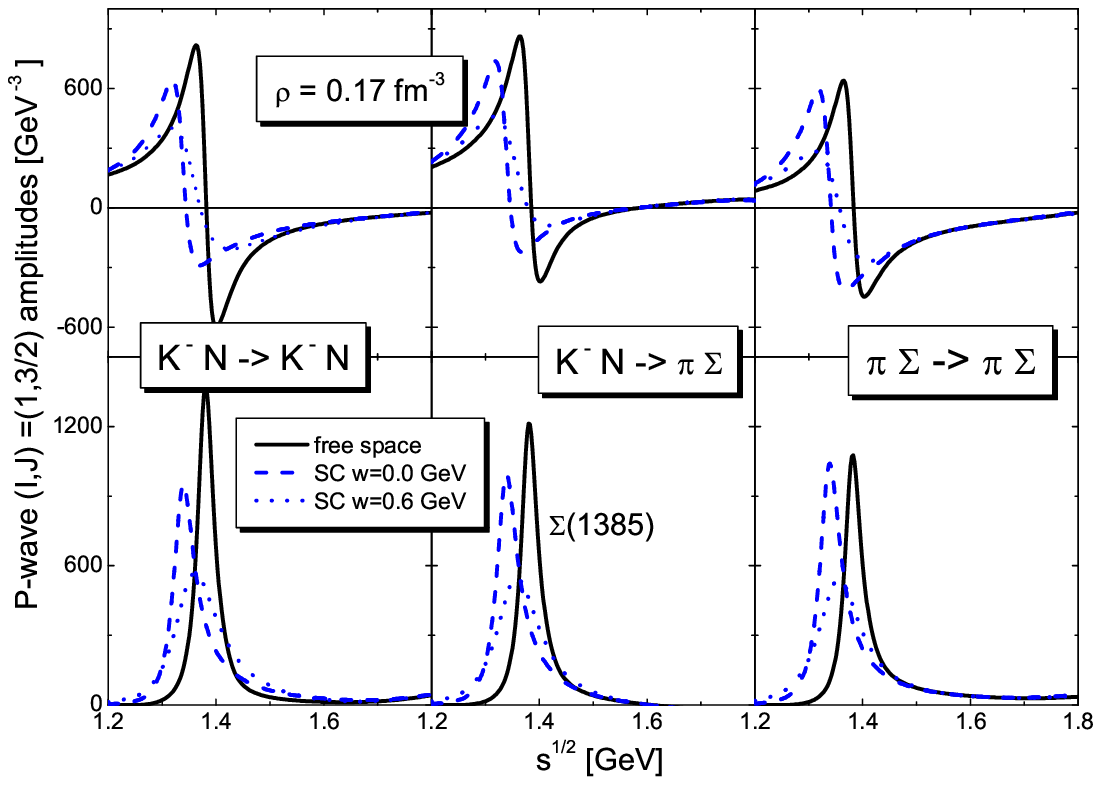}
\end{center}
\caption{P-wave reaction amplitudes in the isospin one channel for
isospin symmetric nuclear matter $\rho = 0.17$ fm$^{-3}$. The $J=3/2$ amplitudes are the result
of a self consistent calculation. The three momentum
$\vec w $ denotes the sum of initial three momenta defined in the rest frame
of nuclear matter.}
\label{fig:cs-amplitudes-1}
\end{figure}

Before analyzing the results in more detail it is useful to
comment on the claim that the in-medium modification of the $\pi
\Lambda$ and $\pi \Sigma $ channels, not considered in this work,
have an important effect on the antikaon spectral function
\cite{Ramos}. Clearly such effects probe the reaction amplitudes
$\bar K N \to \pi \Lambda, \pi \Sigma $ at subthreshold energies.
This was illustrated explicitly in chapter 5.2. Since subthreshold
amplitudes are constrained by the data set only rather indirectly
via crossing symmetry and analyticity it is not surprising that
the more recent work by Cieply et al. \cite{Cieply} does not
confirm that claim. The conflicting results reflect the
different amplitudes those computation are based on. On naive
grounds we would favor the results of Cieply at al., since we
argued consistently throughout this work that subthreshold
amplitudes should match the perturbative amplitudes as they arise
in $\chi$PT, and therefore we do not see much room for significant
effects implied by the pion dressing. An indication that this is indeed
the case follows from the form of the in-medium reaction amplitudes
$\bar K N \to \pi \Sigma $ as shown in Fig. \ref{fig:cs-amplitudes-0} and
Fig. \ref{fig:cs-amplitudes-1}. The dominant s-wave $I=0$ amplitudes
are plotted in Fig. \ref{fig:cs-amplitudes-0} at isospin symmetric
nuclear saturation density. The reaction amplitude $\bar K N \to \pi \Sigma $
is suppressed by a factor of 4 as compared to the diagonal amplitude
$\bar K N \to \bar K N $. Therefore any additional in-medium change of the
$\pi \Sigma \to \pi \Sigma $ amplitude that is induced by an in-medium pion spectral
function, manifests itself only weakly in the diagonal amplitude
$\bar K N \to \bar K N $. A larger ratio of $\bar K N \to \pi \Sigma $ to
$\bar K N \to \bar K N $ amplitudes is found for the isospin one
p-wave amplitudes that probe the $\Sigma (1385)$ resonance. This is illustrated in
Fig. \ref{fig:cs-amplitudes-1} where the various amplitudes are shown at isospin symmetric
nuclear matter. Notwithstanding a detailed investigating of the effect of the pion
dressing is eventually desirable also in the present scheme.

\begin{figure}[t]
\begin{center}
\includegraphics[width=16cm,clip=true]{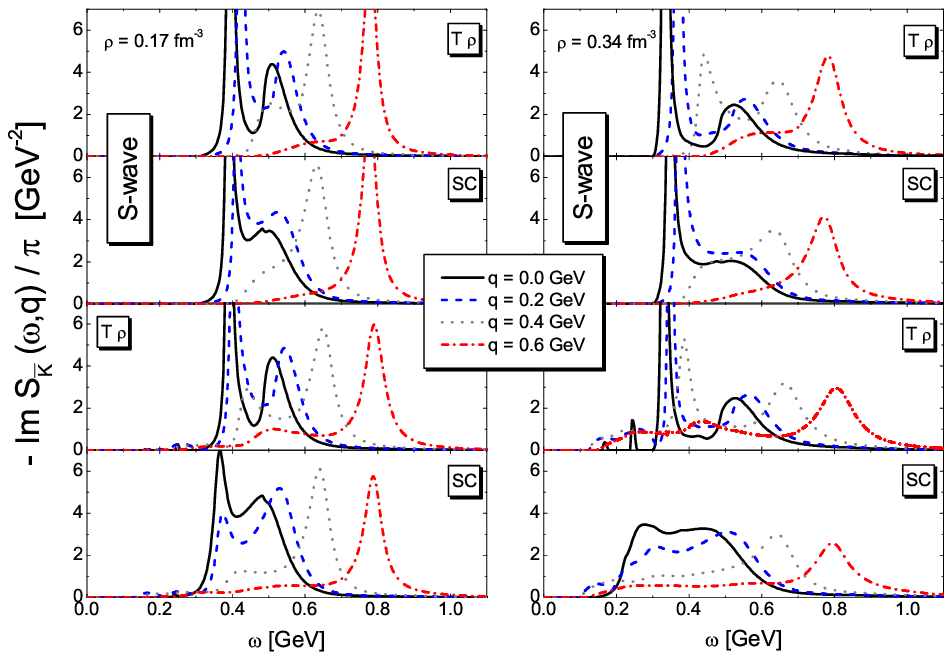}
\end{center}
\caption{Antikaon spectral function as a function of antikaon
energy $\omega$ and momentum $\vec q$. The labels 'T$\rho$' and 'SC'
refer to calculations obtained in terms of free-space and
in-medium $\bar K N$ amplitudes respectively (see (\ref{k-self})).
The first two rows give the results with only s-wave interactions
and the last two rows with all s-, p- and d-wave contributions.}
\label{fig:kaon-sp}
\end{figure}

In Fig. \ref{fig:kaon-sp} the antikaon spectral function
evaluated at symmetric nuclear densities $\rho_0$ and $2\,\rho_0$ according
to various approximation strategies is shown. In the first and
third row the antikaon self energy is computed in terms of the
free-space scattering amplitudes only. Here the first row gives
the result with only s-wave contributions and the third row
includes all s-, p- and d-wave contributions established in this
work. The second and fourth row give results obtained in the self
consistent approach developed in chapter 5 where the full result
of the last row includes all partial waves and the results in the
second row follow with s-wave contributions only. In all cases a
self consistent evaluation of the spectral function
leads to dramatic changes in the spectral function as compared to
a calculation which is based on the free-space scattering
amplitudes only. Moreover, as emphasized in the discussion of the
$\Lambda(1405)$ resonance properties, the effects of higher
partial waves are not negligible. This was anticipated in
\cite{Hirschegg,Florkowski}. As is evident upon comparing the
first and third rows of Fig. \ref{fig:kaon-sp} the p- and d-wave
contributions add quite significant attraction for small energies
and large momenta. At twice nuclear saturation density we find
most striking the considerable support of the spectral function at
small energies. That reflects in part the coupling of the antikaon
to the $\Lambda(1115)$ and $\Sigma (1185)$ nucleon-hole states.
The latter explains the peak structure of the antikaon spectral
functions at $\omega \simeq 250$ MeV seen in the 'T$\rho$'
approximation. Upon inspecting the Lindhard function of these
contributions for free-space $\Lambda(1115)$ and $\Sigma (1185)$
states \cite{Kolomeitsev} one finds that the antikaon spectral
function should be non-zero for
\begin{eqnarray}
\omega > \sqrt{m_Y^2+(|\vec q \,|-k_F)^2}-\sqrt{m_N^2+k_F^2} \,.
\label{lam-kin}
\end{eqnarray}
The strength of the spectral function is largest at $|\vec q \,|
\sim k_F$ because of the kinematical consideration
(\ref{lam-kin}). Self consistency smears out that structure
because the contribution from the dominant $\Sigma (1185)$ state
receives a finite decay width \cite{Lutz:Korpa}.

\begin{figure}[t]
\begin{center}
\includegraphics[width=16cm,clip=true]{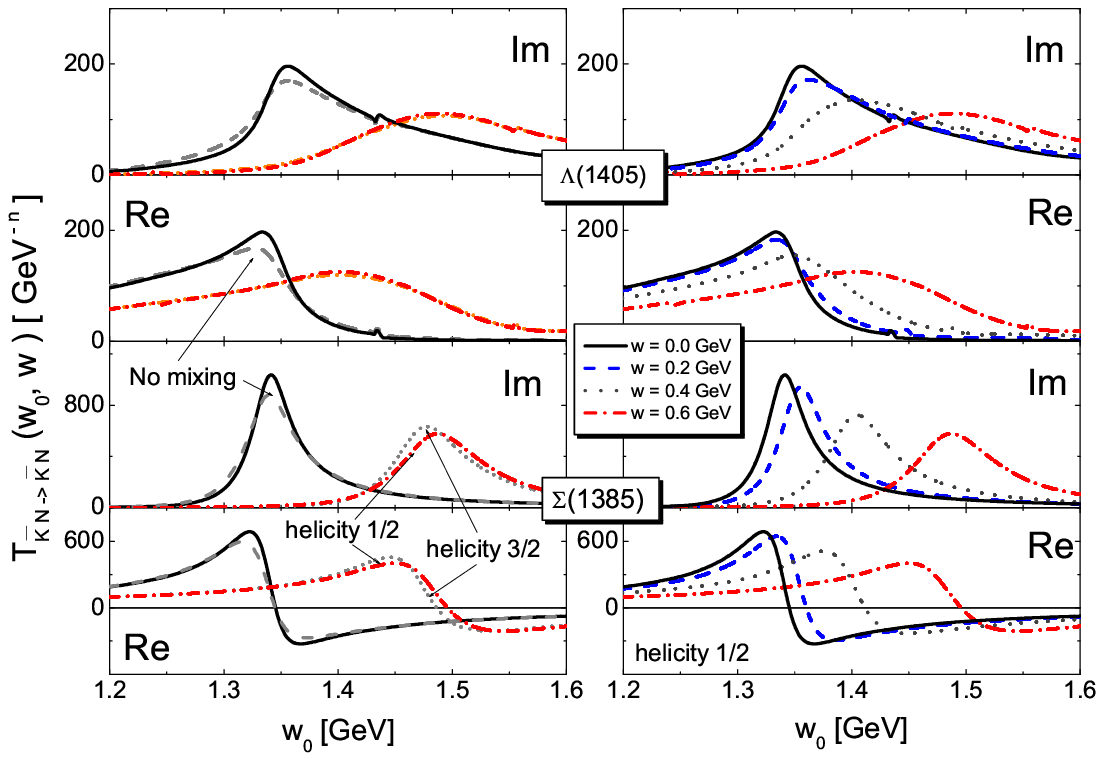}
\end{center}
\caption{Hyperon resonance spectral functions for the $\Lambda(1405)$ and
$\Sigma (1385)$ as a function of the hyperon energy $w_0$ and momentum $\vec w$
for isospin symmetric nuclear matter. It is shown the
s-wave I=0 and p-wave I=1 partial wave amplitudes at $\rho_n+\rho_p= 0.17$ fm$^{-3}$.
The left hand panels demonstrate the small effect of switching off the mixing of the partial
wave amplitudes (dashed versus full lines) and the splitting of the 4 spin 3/2 states into
helicity one half and three half modes (dotted versus dashed-dotted lines).}
\label{fig:hyperon-spec-1}
\end{figure}

\begin{figure}[t]
\begin{center}
\includegraphics[width=16cm,clip=true]{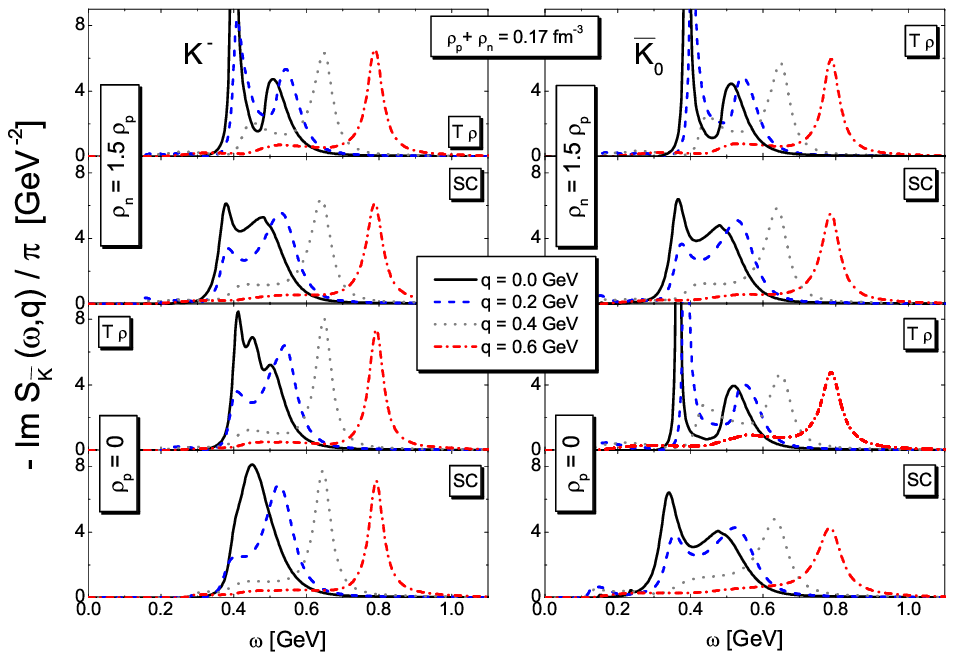}
\end{center}
\caption{Antikaon spectral functions as a function of antikaon
energy $\omega$ and momentum $q$ in asymmetric nuclear matter
with $\rho_p\!+\!\rho_n = 0.17$ fm$^{-}3$. The labels 'T$\rho$' and 'SC'
refer to calculations obtained in terms of free-space and self consistent
in-medium $\bar K N$ amplitudes respectively (see (\ref{k-self})).
The first collum gives the result for $K^{-}$, the second for the
$\bar K_0$ at $\rho_n=1.5 \,\rho_p$ and $\rho_p=0$.}
\label{fig:kaon-sp-as}
\end{figure}
In Fig. \ref{fig:hyperon-spec-1} we present the results for the propagation
properties of the $\Lambda(1405)$ and $\Sigma (1385)$ resonances as they move
with finite three momentum $\vec w$ in isospin symmetric nuclear matter. The peak
positions in the imaginary parts of the amplitudes follow in general the naive
expectation $\sqrt{(m_Y^*)^2+\vec w^2}$ with the hyperon effective mass $m_Y^*$ defined
at $\vec w=0$. However, we observe a systematic increase of the decay widths as
$\vec w$ increases. This is easily understood since the effective in-medium
$Y^* N \to Y^* N$ amplitude which is responsible for the broadening allows
for additional inelasticity as the hyperon momentum $\vec w$ increases.
An interesting phenomenon illustrated in Fig. \ref{fig:hyperon-spec-1}
is the in-medium induced mixing of partial wave amplitudes with
different quantum numbers $J^P$. At vanishing three
momentum of the meson-baryon state $\vec w = 0$, all partial wave
amplitudes decouple. In this case the system enjoys a three-dimensional rotational symmetry since
there is no three-vector available to select a particular direction.
However, once the meson-baryon pair is moving relative to the nuclear matter bulk with $\vec w \neq 0$ we find
two separate channels for a given isospin only. The system is invariant under
a subgroup of rotations only, namely those for which the rotational vector is aligned with
the hyperon momentum $\vec w$. The three-dimensional rotational symmetry is reduced and therefore
the total angular momentum $J$ is no longer a conserved quantum number. However, the total angular momentum
projection onto the $\vec w$ direction remains a conserved quantum number. The latter
defines the helicity of a hypothetical s-channel particle exchange and therefore the in-medium scattering
amplitudes decouple into an infinite tower of helicity amplitudes. Each helicity amplitude probes a well defined
infinite set of partial wave amplitudes. For instance, in the P-space,
all considered partial wave amplitudes $S_{I1}$, $P_{I1}$, $P_{I3}$ and $D_{I3}$ couple for given isospin channel $I$.
Since the angular momentum projection of the $S_{I1}$ channel onto $\vec w$ is necessarily one half, the P-space
corresponds to the first term with helicity one half. Note that reflection symmetry implies that only the absolute
value of the helicity matters here. It is evident that all partial wave amplitudes
contribute to this channel simply because any state with a given angular momentum $L$ has a
component where $\vec L \cdot \vec w $ vanishes and therefore the helicity is carried by the
nucleon spin. In the second channel, our Q-space, only the partial wave amplitudes $P_{03}$
and $D_{03}$ couple. This channel corresponds to the helicity three half term. In a computation
that considered all partial waves the helicity three half term would require all partial wave
amplitudes except the $S_{I1}$ and $P_{I1}$ waves with $J=1/2$. The fact that the
$J={\textstyle{3\over 2}}$ amplitudes $P_{I3}$ and $D_{I3}$ affect both the P- and Q-space,
or the helicity one half and helicity three half terms, is not surprising, because for those
states one would expect the nuclear medium to lift the degeneracy of the four spin modes. This
is completely analogous to the longitudinal and transverse modes of
vector mesons, which bifurcate in nuclear matter.

The splitting of the four $\Sigma (1385)$ modes is demonstrated in the left hand
panel of Fig. \ref{fig:hyperon-spec-1}. The helicity one and three half modes are
shifted by about 5 MeV for $\vec w = 600$ MeV. This is a small effect and therefore
it is justified to neglect the coupling of the partial waves for
that density to good accuracy. Indeed a run where all off diagonal loop functions
that couple waves of different angular momentum or parity are set to zero gives
results that are almost indistinguishable to those obtained in the full scheme. This
is demonstrated by the dashed lines in the left hand panel of Fig. \ref{fig:hyperon-spec-1}.
At $\vec w =0$ the results of the two computations, dashed and solid lines are quite close.
This finding has a simple interpretation. One expects sizeable effects from the in-medium
mixing of the partial-wave amplitudes only if two partial wave amplitudes that mix show both
significant strength at a given energy $w_0$. This is not
the case here. The dominant partial waves $S_{01}$ and $P_{13}$ decouple because they
carry different total isospin. Therefore it is natural to obtain small mixing effects. This may,
however, change once isospin asymmetric matter where the two isospin channels couple is
considered.

\begin{figure}[t]
\begin{center}
\includegraphics[width=16cm,clip=true]{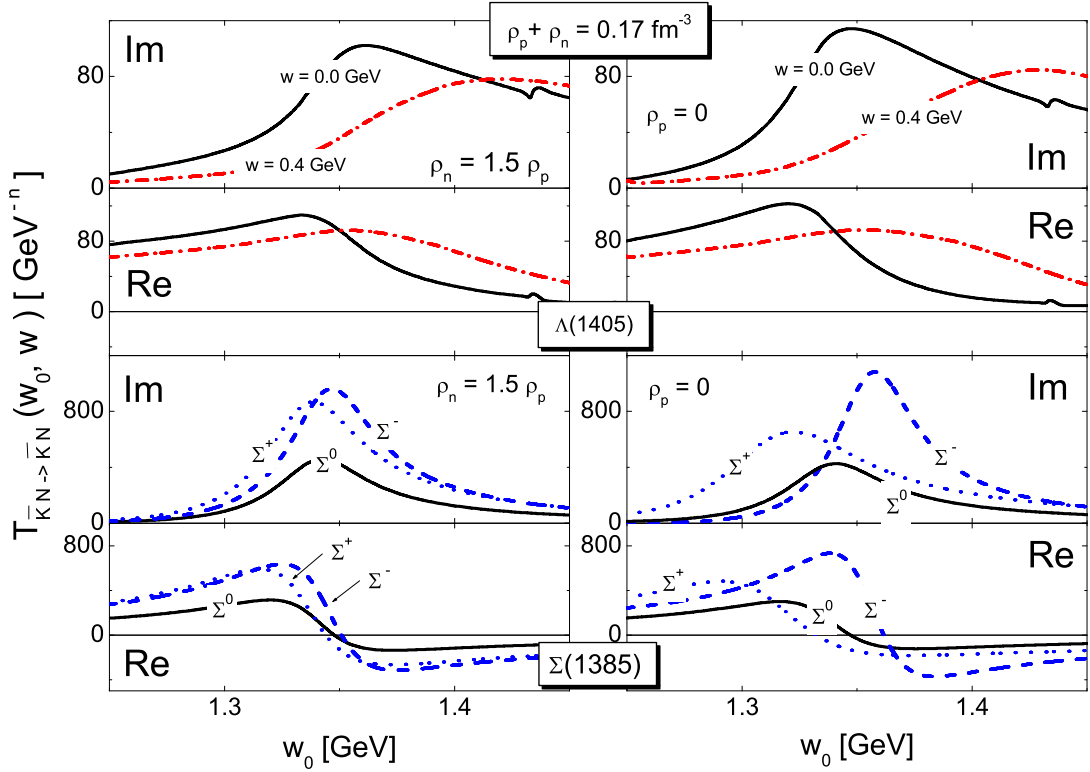}
\end{center}
\caption{$\Lambda(1405)$ and $\Sigma (1385)$ hyperon resonance propagators as a
function of hyperon energy $w_0$ and momentum $\vec w$ for asymmetric nuclear matter.
It is shown the s-wave $K^- p \to K^- p$ (for $\vec w= 0$ MeV and $\vec w= 400$ MeV) and
p-wave  $\bar K^0 n \to \bar K^0 n$, $\bar K^0 p \to \bar K^0 p$ and
$K^- n \to K^- n$ reaction amplitudes (for $\vec w= 0$ MeV) at $\rho_n+\rho_p= 0.17$ fm$^{-3}$.}
\label{fig:hyperon-sp-as}
\end{figure}

A presentation of results obtained for asymmetric nuclear matter follows.
Two cases are considered here both with $\rho_n+\rho_p= 0.17$ fm$^{-3}$.
In Fig. \ref{fig:kaon-sp-as} results for $\rho_n= 1.5\,\rho_p$, which corresponds to
the condition met in the interior of lead, and for $\rho_p=0$, which describes neutron matter,
are shown. The asymmetry breaks up the isospin doublet $(K^-, \bar K_0)$ leading to
distinct spectral functions for the charged and neutral antikaons. In all cases considered
a significant effect from the self consistency is found at small antikaon momenta. This
is illustrated by comparing the entries of Fig. \ref{fig:kaon-sp-as} labelled
with 'T$\rho$' and 'SC'. Whereas the effect of the asymmetry
is surprisingly small for the lead scenario, the spectral functions of the
charged and neutral antikaons differ strongly in neutron matter with $\rho_p=0$.
In Fig. \ref{fig:hyperon-sp-as} the corresponding properties of the hyperon resonances
are shown. Again, like one observed for the isospin
doublet $(K^-,\bar K^0)$ the isospin asymmetry of the matter breaks up the isospin
triplet state $\Sigma(1385)$ introducing a medium-induced splitting pattern.
The $\Lambda(1405)$ resonance is presented in terms of the in-medium
$K^- p \to K^- p$ s-wave amplitude, the neutral and charged $\Sigma(1385)$ states by
p-wave amplitudes $\bar K^0 n \to \bar K^0 n$, $\bar K^0 p \to \bar K^0 p$ and
$K^- n \to K^- n$. In free space the latter amplitudes are determined by
the isospin one amplitude in case of the charged hyperon states but by the mean of the
two isospin amplitudes in case of the neutral hyperon state. Thus, the isospin one component,
the $\Sigma (1385)$ couples to in free space, is smaller by a factor of two for the
neutral amplitude $\bar K^0 n \to \bar K^0 n$ as compared to the amplitudes describing
the charged hyperon states. This reflects itself in amplitudes for the $\Sigma^0(1385)$
in Fig. \ref{fig:hyperon-sp-as} that are typically smaller by a factor two as compared to
the amplitudes of the $\Sigma^\pm(1385)$. Whereas the $\Lambda(1405)$ resonance is not
affected much by the asymmetry, the splitting of the three $\Sigma (1385)$ states shows a
strong dependence on the asymmetry. For neutron matter with $\rho_n = 0.17$ fm$^{-3}$ we
find a mass difference of about 30 MeV for the charged states.

\begin{figure}[t]
\begin{center}
\includegraphics[width=9cm,clip=true]{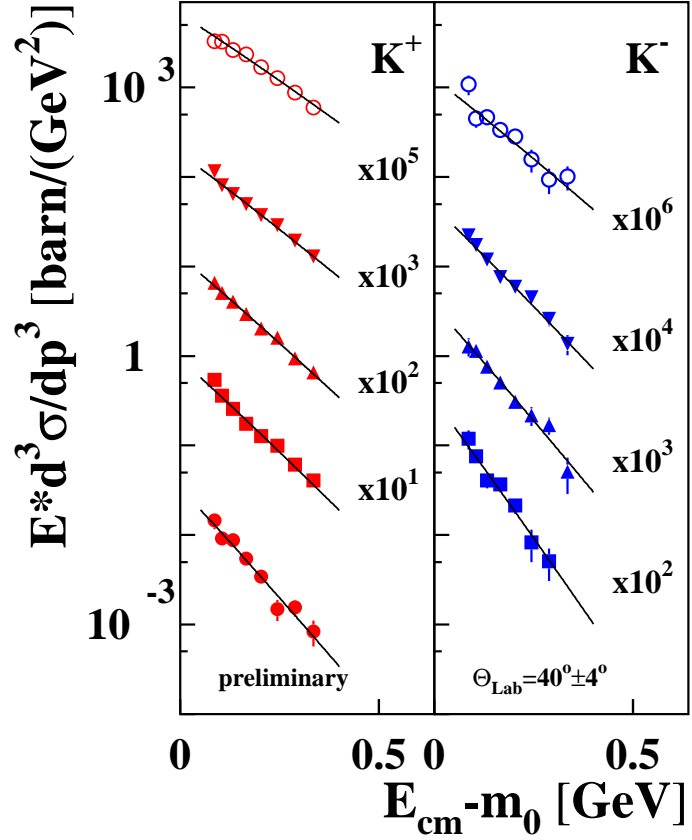}
\end{center}
\caption{Invariant cross sections for $K^+$ and $K^-$ in Au + Au collisions at
$E_{\rm beam} =$ 1.5 AGeV for five centrality classes \cite{Foerster}. The open
circles depict the most central data with decreasing centrality from the top to
the bottom of the figure. The solid lines are the result of a fit with a
Boltzmann distribution.}
\label{fig:K-spectra-Foerster}
\end{figure}

The result chapter is closed with a brief discussion of the
results presented here as they may be relevant for antikaon production in heavy
ion reactions at subthreshold energies.
Strictly speaking the results do not apply for the conditions present inside a fireball
created after the collision of two heavy nuclei. The quantitative description of such a process
requires a transport theoretical description capable to model physics of finite systems
off equilibrium. The latter requires as input
hadronic cross sections, or more precisely hadronic transition rates, that are
taken from experiment or theory. Calculations performed in the idealized world
of infinite but hot and dense nuclear matter are nevertheless a useful tool to
discuss and understand effects on a qualitative level. In certain channels
transition rates that are typically derived from free-space cross sections may be significantly
changed in a hot and dense nuclear environment. Of course one may object to
the application of our calculations to the heavy ion scenario because they are performed
at zero temperature. However, since the dominant effect found in this work is quite
insensitive to the Pauli blocking effect, it is natural to anticipate that the inclusion of
small temperatures of up to $T \simeq 80$ MeV typical for SIS energies at GSI, does
not affect the results significantly. This argument relies on the naive picture that
temperature effects may be interpreted as an effective reduction of the Pauli blocking effect.

One may argue that if the antikaon spectral function evaluated for
infinite hot nuclear matter shows considerable strength at energies smaller than the
free kaon mass, it is energetically easier to produce antikaons in a fireball than if the
effective antikaon mass distribution was not reduced. In this case one may expect to see some
type of enhancement of antikaon production rates. It is a challenge to verify the above
simple but physical interpretation of the observed kaon and antikaon yields in terms of
transport model simulations. In Fig. \ref{fig:K-spectra-Foerster} preliminary
results of the KaoS collaboration \cite{Foerster} for the kaon and antikaon transverse
spectra for Au+Au at $E_{\rm beam}$=1.5 AGeV are shown. The spectra shown are typical
for most observed particle multiplicities. They are fitted well by a simple
Boltzmann distribution
\begin{eqnarray}
\sim \exp \big( - (E_{cm}-m_K)/T \big)\,,
\label{}
\end{eqnarray}
in terms of an effective temperature parameter $T$.
The effective temperatures in Fig. \ref{fig:K-spectra-Foerster} of the $K^+$ are about
20 MeV larger than the ones for the $K^-$. It is important to realize that the
Boltzmann type behavior of the spectra does not imply that the kaons and antikaons are
necessarily equilibrated. Estimates for the kaon mean free path, which are based on the
empirical kaon-nucleon cross section, suggest that the kaons do not have a chance to reach
equilibrium conditions in a typical heavy ion collision at SIS energies. That may be
different for antikaons as will be discussed in some detail below.

\begin{figure}[t]
\begin{center}
\includegraphics[width=5cm,clip=true]{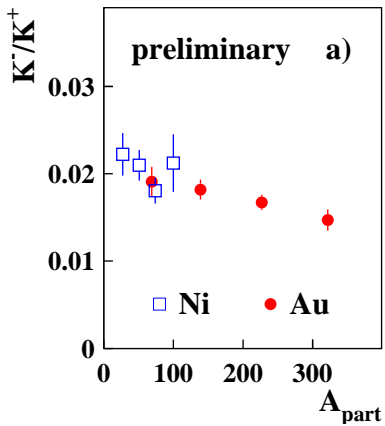}
\end{center}
\caption{It is shown the ratio of the $K^-/K^+$ yields in the Ni + Ni and Au + Au collisions at
$E_{\rm beam} =$ 1.5 AGeV as a function of the number of nucleons $A_{\rm part}$,
that participate in the reaction \cite{Foerster}. The data are taken
at $\theta_{\rm lab}= 40^\circ$. }
\label{fig:K-ratio-Foerster}
\end{figure}

Since at subthreshold energies the mean
energy available by one nucleon is not sufficient to produce an antikaon in an elementary
nucleon-nucleon collision by definition, one visualizes the production process to occur
in a series of successive reactions
\cite{Cassing:Mosel,Li:Brown,Aichelin:Hartnack,Aichelin:Oeschler:Hartnack},
\begin{eqnarray}
&& N\,N \to N \,\Delta \to N\,N \,\pi \,, \quad
N\,N \to N \,Y\,K \,,\quad
\pi \,Y \to \bar K \,N \,,
\label{}
\end{eqnarray}
where $Y = \Lambda(1115), \Sigma (1185)$. Here we do not
intend to provide a complete listing of reactions included in transport model
simulations rather we would like to discuss the main effects qualitatively. A heavy
ion reaction produces, besides photons and leptons, dominantly pions, the lightest
hadronic degrees of freedom available. For SIS energies at GSI
this production is thought to be driven by the isobar production process
$N\,N \to N \,\Delta \to N \,N \pi $. The kaon production is determined by the primary
$N\,N \to N \,Y\,K $ reaction far above threshold but by the secondary reaction
$\pi \,N \to K \,Y$ at subthreshold energies. Because of strangeness conservation the antikaon
production threshold in nucleon-nucleon collisions is much larger than the
threshold of kaon production. The former is determined by the $N\,N \to N\,N\,K\,\bar K$
reaction leading to
\begin{eqnarray}
\sqrt{s}^{(\bar K)}_{\rm thres}-\sqrt{s}^{(K)}_{\rm thres} = m_K +m_N-m_\Lambda \simeq
320 {\rm MeV} \,.
\end{eqnarray}
It is therefore plausible that antikaons are dominantly
produced in the secondary $\pi \,Y \to \bar K \,N$ reaction in particular at subthreshold
energies. Consequently these reactions received considerable attention in the study of
subthreshold antikaon production \cite{Cassing:Mosel,Li:Brown,Schaffner}. A clear hint that
the antikaon production in nucleus-nucleus collisions is indeed driven by the secondary
$\pi\, Y \to \bar K\, N$ reaction follows from the centrality dependence of the measured
ratio $K^-/K^+$  \cite{Foerster} as shown in Fig. \ref{fig:K-ratio-Foerster}. The number of
nucleons, $A_{\rm part}$ that participate in a Ni on Ni or Au on Au collision
is a measure for the inverse size of the impact parameter. Since both, the kaon and antikaon
yields depend strongly on the impact parameter \cite{Menzel:KaoS} the empirical observation
that the ratio of the yields is almost insensitive to it signals that the production
mechanisms of the kaons and antikaons must be strongly related.

\begin{figure}[t]
\begin{center}
\includegraphics[width=14cm,clip=true]{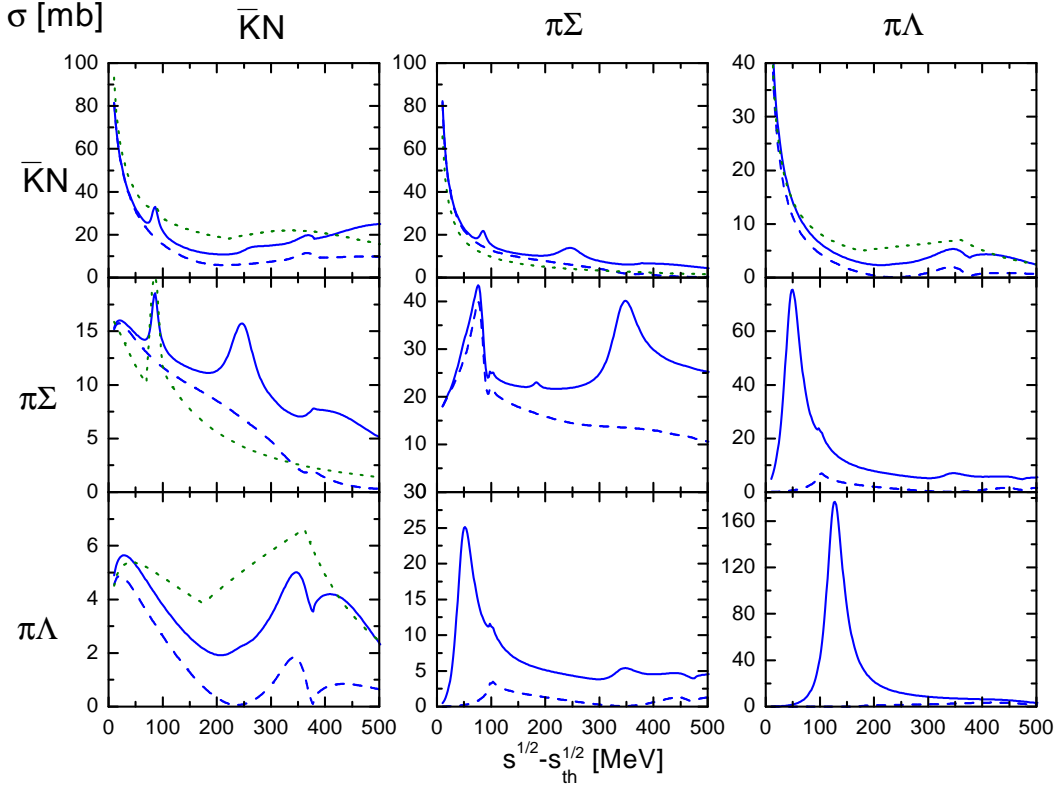}
\end{center}
\caption{Total cross sections
$\bar K N \to \bar K N$, $\bar K N \to \pi \Sigma $, $\bar K N \to \pi \Lambda $ etc relevant for subthreshold
production of kaons in heavy-ion reactions. The solid and dashed lines give the results of the $\chi$-SB(3) approach
with and without p- and d-wave contributions respectively. The
dotted lines correspond to the parameterizations given in \cite{Brown-Lee}.}
\label{fig:cross-pred}
\end{figure}

The importance of the $\pi \,Y \to \bar K \,N$ reactions suggests that it may be useful
to investigate the latter within the $\chi$-BS(3) scheme. In Fig. \ref{fig:cross-pred}
the isospin averaged cross sections of the channels
$\bar K N, \pi \Sigma $ and $\pi \Lambda$ as obtained in the chiral coupled channel
analysis \cite{Lutz:Kolomeitsev} are shown. The results
in the first row repeat in part the content of Fig. \ref{fig:totcross} only that
here the cross sections are confronted with typical parameterizations used in transport model
calculations. The cross sections in the first column are determined by detailed balance from
those of the first row. Uncertainties are present nevertheless, reflecting the large empirical
uncertainties of the antikaon-nucleon cross sections close to threshold. The remaining four
cross sections in Fig. \ref{fig:cross-pred} are true predictions of the $\chi$-BS(3) approach.
It should be emphasized that the results are quantitatively reliable only for
$\sqrt{s}< 1600$ MeV. It is remarkable that nevertheless the cross sections agree with the
parameterizations in \cite{Brown-Lee} qualitatively up to much higher energies except in the
$\bar K N  \leftrightarrow \pi \Sigma $ reactions where they overshoot those parameterizations
somewhat. Besides some significant deviations of our results from \cite{Cugnon,Brown-Lee} at
$\sqrt{s}-\sqrt{s_{\rm th}} <$ 200 MeV, an energy range where the results are quantitatively
reliable, we find interesting the sizeable cross section of about 30 mb for the elastic
$\pi \Sigma $ scattering. As demonstrated by the dotted line in Fig. \ref{fig:cross-pred},
which represent the $\chi$-BS(3) approach with s-wave contributions only, the p- and d-wave
amplitudes are of considerable importance for elastic $\pi \Sigma $ scattering.
A further striking effect is seen in the elastic $\pi \Lambda $ cross section that
has a maximum of about 160 mb at $\sqrt{s} \simeq 1385$ MeV. That reflects a strong
coupling of the $\pi \Lambda $ state to the p-wave $\Sigma (1385)$ resonance. This resonance
cannot be seen in the $\bar K N $ reactions because it is below the antikaon nucleon threshold.
It is evident that the large cross sections found for the $\pi Y \to \pi Y$ reactions will help
significantly to drive hyperons produced in nucleus-nucleus collisions towards a
statistical phase space distribution. As it will be argued in detail below the presence of
the $\Sigma (1385)$ resonance will play a crucial role for antikaon production in nucleus-nucleus
collisions at subthreshold energies.

So far most transport model simulations of antikaon production rely on rough approximations.
For instance the complicated antikaon spectral function is substituted by a quasi-particle
ansatz parameterized in terms of an effective antikaon mass that drops basically linearly
with the nuclear density \cite{Cassing:Mosel,Li:Brown}. Technically this is implemented by
incorporating an appropriate mean field modifying the trajectories of the
antikaon between collisions in the fireball. The width of the quasi particle is
then modelled by including the inelastic cross sections like $\bar K N \to \pi Y$. For cross
sections involving antikaons in initial or final states simple substitution rules like e.g.
$\sqrt{s} \to \sqrt{s} - \Delta m_{\bar K}$ are applied. Such computations confirm
the expectation that the antikaon yield is quite sensitive to an attractive mean field of
the antikaon \cite{Cassing:Mosel,Li:Brown}. Runs where an attractive mean field and the absorption
cross sections are both switched off or on are reported to be fairly close to the measured
multiplicities \cite{Cassing:Mosel,Li:Brown}. Neglecting, however the attractive mean field
but including the antikaon absorption cross sections underestimates the antikaon yield up to a
factor 5 \cite{Cassing:Mosel,Li:Brown}.

A more sophisticated simulation was performed by Schaffner, Effenberger and
Koch \cite{Schaffner} in which the consequences of two effects were studied.
First, the effect of in-medium modifications of the $\pi Y \to \bar K N$
reactions was studied. And second, a momentum dependence of the antikaon
potential was considered. The momentum dependence was chosen such that at
moderate antikaon momenta of about 300-400 MeV the attractive mean field was basically
switched off. This was motivated by microscopic many-body evaluations of the antikaon
self energy which predicted that type of behavior \cite{ml-sp}. The in-medium modification
of the production cross sections $\pi Y \to \bar K N$ were assumed to be dominated by
the Pauli blocking effect \cite{Koch} that shifts the $\Lambda(1405)$ resonance from
$\sqrt{s}= 1405$ MeV to larger energies, about $\sqrt{s}\simeq 1490$ MeV at nuclear
saturation density. As a consequence the authors report an enhancement factor
for the $\pi \Sigma \to \bar K N$ reaction of about 20 close to $\sqrt{s} =1490$ MeV.
Both effects, the attractive mean field and the increased
$\pi Y \to \bar K N$ cross section gave rise to a similar effects in the antikaon
yield. Together the two mechanism predict an enhancement factor of about 4 as compared
to a computation that applies the empirical cross sections and discards any mean
field effects. However, as was pointed out by Schaffner, Effenberger and
Koch \cite{Schaffner}, once the momentum dependence of the attractive mean field was
incorporated together with the fact that in their model the enhancement of the
cross section disappears already at moderate temperatures $T \simeq 80$ MeV, the
results are quite close to the reference calculation
with no medium effects. All together the empirical antikaon yields are not described, they
are underestimated by about a factor 3-4.

In a more recent work by Aichelin, Hartnack and Oeschler
\cite{Aichelin:Hartnack,Aichelin:Oeschler:Hartnack} a
further interesting aspect was emphasized. Since the antikaon production is
driven by the $\pi Y \to \bar K N$ reaction the total antikaon yield should
be sensitive to the kaon production mechanism. A repulsive mass shift for the
kaon can therefore compensate for the effect implied by an attractive mean field
for the antikaon. It is argued, however, that the computation
by Aichelin and Hartnack \cite{Aichelin:Hartnack} most likely overemphasizes this effect due
to a too large mass shift for the kaon (about 70 MeV at
2 $\rho_0$ \cite{Schaffner:Bondorf:Mishustin}). This objection
is based on a recent calculation based on the kaon-nucleon s- and p-wave phase shifts
\cite{Lutz:Kolomeitsev:Korpa} that demonstrates that a kaon produced with
momenta of about 400 MeV at 2 $\rho_0$ the repulsive mass shift is only
about 45 MeV, smaller than the value of 70 MeV used in \cite{Aichelin:Hartnack}.
The conclusions of Aichelin, Hartnack and Oeschler
\cite{Aichelin:Hartnack,Aichelin:Oeschler:Hartnack} depend also
sensitively on the magnitude of the attractive mean fields for the nucleons and hyperons
as well as on the poorly known $N\,\Delta \to N \,Y \,K$ reaction rates. In order to
reduce such uncertainties it would be useful to describe not only total yields of kaons,
antikaons and hyperons but at the same time achieve quantitative agreement with the
azimuthal emission pattern \cite{Li:Ko:Li,Fuchs:flow,Bratkovskaya}. The
mean fields used in
\cite{Aichelin:Hartnack,Aichelin:Oeschler:Hartnack} for the hyperons show a strong non-linear
behavior in the nuclear
density \cite{Schaffner:Bondorf:Mishustin}. In particular at twice saturation density the
hyperons' mean fields are zero approximatively. We would conclude from the self consistent
antikaon dynamics \cite{Lutz:Korpa}, that the $\Lambda(1115)$ should experience an
attractive mass shift of about 20-40 MeV at twice saturation density instead \cite{Lutz:Korpa}.
We do not find any sign of a strong non-linear behavior in density.
Such a strong non-linear density behavior in the hyperon self energy may be questioned.
It is a consequence of a mean field picture that relies on a scalar iso-scalar $\sigma$
degree of freedom. If, on the other hand, the physics of the nucleon self energy is driven by the
many-body dynamics of pions instead as suggested in \cite{Lutz:Friman:Appel}, a strong
non-linear behavior for the $\Lambda$ ground state is not expected. The non-linearities
should be significantly reduced because the analogous (reducible) diagrams that
drive the nuclear saturation mechanism in \cite{Lutz:Friman:Appel} do not exist in case
of the $\Lambda $. Even for the $\Sigma $ ground state, which permits the analogous
reducible diagrams of \cite{Lutz:Friman:Appel}, one would expect significantly smaller
effects simply because only one of the two baryon lines is affected in strangeness zero
nuclear matter. Thus, one may expect that the hyperon self energies are
dominated by the antikaon t-channel exchange. The latter contribution was studied
in \cite{Lutz:Korpa} but no indication for a strong non-linear density dependence was found
despite the non-trivial form of the antikaon spectral function in nuclear matter. Of course
it is impossible at this stage to make rigorous estimates of the hyperon self energy at twice
nuclear saturation density simply because the hyperon-nucleon interaction is at present not
too well constrained by the data set \cite{Nijmegen:NY,Juelich:NY}.
It is interesting to observe, however, that
a Brueckner-Hartree-Fock evaluation \cite{Schultze:Baldo} based on the Nijmegen soft-core
hyperon-nucleon interaction \cite{Nijmegen:NY} suggests about 30 MeV attraction for the
$\Lambda(1115)$ mean field at twice nuclear saturation density. As a consequence in
nucleus-nucleus collisions the net effect of increased kaon mass and decreased hyperon mass
would then tend to cancel in the total kaon yield. An investigation of further observable
quantities like the azimuthal emission pattern ad mid rapidity, which discriminate among
the various effects, reveals that a small repulsive mean field for the kaon is favored by
the data set \cite{Shin:KaoS,Li:Ko:Li,Fuchs:flow,Bratkovskaya}.
Aichelin, Hartnack and Oeschler
\cite{Aichelin:Hartnack,Aichelin:Oeschler:Hartnack} affirm that the $K^-/K^+$ ratio is
insensitive to the details of the kaon production and therefore should be considered as
a superior observable to test the antikaon dynamics.
In their work \cite{Aichelin:Oeschler:Hartnack} the possible importance of
in-medium modified cross sections is studied. Within their
scenario, which does not consider a momentum dependence of the antikaon mean field,
they observe an enhancement of the $K^-/K^+$ ratio by a factor 2 if the
$\pi Y \to \bar K N$ cross sections are enlarged by a factor three. However, the
impact parameter dependence of that ratio is affected strongly
by this ad-hoc procedure leading to a dependence that appears incompatible
with the behavior shown in Fig. \ref{fig:K-ratio-Foerster}. It is argued that the
antikaons are produced at a late stage of the collision even though the difference
of absorption and production rates shows a clear maximum in the high density initial phase.
Increasing the $\pi Y \to \bar K N$ cross sections only affects the antikaon absorption
process at the late stage of the fireball expansion since the typical pions and hyperons
do not have sufficient kinetic energy to produce antikaons anymore. This would lead to a drop
of the $K^-/K^+$ ratio with increasing impact parameter. Moreover, their claim, that
including an attractive mean field for the antikaon into the simulation does not affect the
total antikaon yield, remains puzzling. This contradicts the typical enhancement factors 2-4
as were obtained in previous simulations \cite{Cassing:Mosel,Li:Brown,Cassing:review,Schaffner}.

\begin{figure}[t]
\begin{center}
\includegraphics[width=14cm,clip=true]{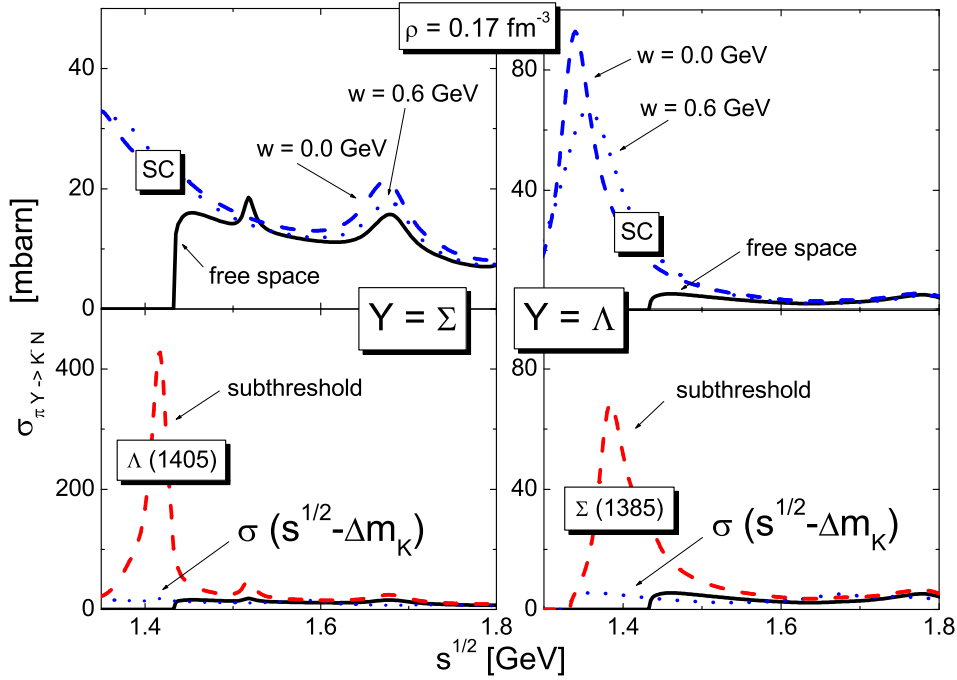}
\end{center}
\caption{The upper panels show the pion induced cross sections of
antikaons obtained in a self consistent many-body evaluation (dashed for $\vec w=0$ MeV
and dotted lines for $\vec w=600$ MeV) at nuclear saturation density as
compared to the free-space cross sections (full lines). The lower panels give the results
of schematic evaluations. The dotted lines are the free-space cross section shifted in
$\sqrt{s}$ by $\Delta m_{\bar K}=-100$ MeV. The dashed lines follow with computations that
are based on the subthreshold free-space amplitudes as predicted by the $\chi$-BS(3) approach
together with a final state phase space evaluated with a reduced kaon mass.}
\label{fig:in-medium-cross}
\end{figure}

The above discussion implies that it is important to obtain an improved
understanding of the $\pi Y \to \bar K N$ cross sections as they may change in a
nuclear environment \cite{Schaffner}. In Fig. \ref{fig:in-medium-cross} the
result of a computation based on the self consistent scheme introduced in this work
are presented. In the upper panel we show the isospin averaged $\pi \Sigma \to \bar K N$
and $\pi \Lambda \to \bar K N$ cross sections where the in-medium scattering amplitudes
(see e.g. Figs. \ref{fig:cs-amplitudes-0},\ref{fig:cs-amplitudes-1}) are used
together with the antikaon spectral function defining the
final-state phase space (see Fig. \ref{fig:spec}). Comparing the solid with the dashed and dotted
lines one realizes a dramatic enhancement for the $\pi \Lambda$ reaction but a much
more moderate enhancement for the $\pi \Sigma $ reaction. The dashed and dotted lines
correspond to the situation where the sum of initial three-momenta is 0 MeV and 600 MeV
respectively. To explain the source of this effect the figure shows in its lower
panel the same cross sections evaluated in two different schematic ways. The dotted
lines give the result obtained according to the prescription
$\sqrt{s} \to \sqrt{s} - \Delta m_{\bar K}$ as commonly applied in transport
model simulations. It is used $\Delta m_{\bar K} =- 100$ MeV typically for the amount of
attraction found for the antikaon at saturation density. The cross sections remain as
small as they are in free-space. A striking enhancement is shown by the dashed lines in
the lower panel. Here we use the free-space scattering amplitude and evaluate the cross section
with the final state phase space determined by a reduced kaon mass of 394 MeV. As a result
this cross section probes the scattering amplitudes at subthreshold energies a kinematical
region where they are not directly constrained by the scattering data. Since the $\chi-$BS(3)
scheme predicts considerable strength in the subthreshold amplitudes from the s-wave
$\Lambda(1405)$ and p-wave $\Sigma (1385)$ resonances the former cross sections are
dramatically enhanced as compared to the free-space cross sections
\footnote{The results for the $\pi Y \to \bar K N$ transition rates may change somewhat
once the in-medium spectral functions for the pion, nucleon and hyperons are used. One would
expect that the effect of the attractive shifts in the mass distributions of the nucleons and
hyperons tend to cancel in the transition rates. The inclusion of the pion spectral function
will most likely broaden the $\Sigma(1385)$ resonance peak in the $\pi \Lambda \to \bar K N$
transition rate. This issue deserves further studies.}. Comparing the lower
with the upper panel demonstrates that the latter cross section, though providing a
simple physical interpretation of the enhancement, do not adequately reproduce the
full computation as it arises in a self consistent framework. In particular the large
cross section of the $\pi \Sigma \to \bar K N$ reaction predicted by the free-space amplitudes
is significantly reduced in the self consistent scheme. Here one should note that
the s-wave and p-wave final-state phase space factors probe the antikaon spectral function
in different ways. Thus, the net result is a combined effect depending on the in-medium
amplitude and a projection of the antikaon spectral function that depends on the angular
momentum.

The moderate enhancement for the $\pi \Sigma \to \bar K N$ reaction confirms the results of our
previous work \cite{ml-sp}, in which it was pointed out that a self consistent
antikaon dynamics does not generate the large enhancement factor predicted by a scheme that
considers Pauli blocking only \cite{Koch,Schaffner}. Since an enhancement factor that is driven
by the Pauli blocking effect will eventually disappear once a finite temperature is allowed
for, it can not explain enlarged cross sections to be used in heavy ion reactions in any case.
The striking effect induced by the $\Sigma (1385)$ resonance in the $\pi \Lambda \to \bar K N$
reaction is novel. It is the result of our detailed analysis of the kaon and antikaon
scattering data which predicts that there is a strong coupling of the $\Sigma (1385)$
resonance to the $\pi \Lambda$ and $\bar K N$ channels at subthreshold energies. The
self consistent many-body computation presented here suggests that this strong coupling
persists in the nuclear medium giving rise to the large enhancement factor found for the
in-medium $\pi \Lambda \to \bar K N$ reaction. This effect should have important consequences
for the antikaon yield in nucleus-nucleus collisions at SIS energies.

As demonstrated in
\cite{Aichelin:Oeschler:Hartnack} the ad-hoc increase of the $\pi Y \to \bar K N$
cross section away from its free-space limit does increase the total antikaon yield.
This demonstrates that transport model simulations of the antikaon yield that are based
on free-space cross sections do not reach equilibrium conditions. Of course beyond
some critical enhancement factor one would expect the yield to saturate implying that
phase space is populated statistically. To put it in another way an enhancement factor
40 will not increase the yield by a factor of 40. However, we argue that a strong
enhancement factor is rather welcome since that implies that the enhancement should
be strong enough exceeding the critical value required for driving the antikaons towards
equilibrium conditions even at smaller density or equivalently larger impact parameter.
Therefore we would expect that the impact parameter dependence of the $K^-/K^+$ yield
should be weak as seen in Fig. \ref{fig:K-ratio-Foerster} and predicted by
the statistical model of Cleymans, Oeschler and Redlich \cite{Cleymans:Oeschler:Redlich}.
Here we would deviate from the line of arguments put forward by Aichelin, Hartnack
and Oeschler who argued that with their ad-hoc enhancement the impact parameter
dependence of the $K^-/K^+$ yield decreases with increasing impact parameter. Since
the unscaled cross sections predict a flat behavior in their simulation, more consistent
with Fig. \ref{fig:K-ratio-Foerster}, one may favor the free-space cross sections.
This conclusion would be based, however, on a simple energy independent enhancement factor
of the $\pi \,Y \to \bar K \,N$ transition rate. In contrast Fig. \ref{fig:in-medium-cross}
suggests that even at a rather late stage of the fireball expansion the pions and hyperons need
very little kinetic energy to produce a virtual antikaon. That should lead to a flattening
of the number of antikaons with increasing reaction time contradicting the results shown
by Aichelin, Hartnack and Oeschler \cite{Aichelin:Oeschler:Hartnack}.

It would be very useful to incorporate the dynamics of the $\Sigma (1385)$ into
transport model simulations. If it can be confirmed that that the description of
the antikaon yield requires an in-medium enhancement of the
$\pi\, Y \to \bar K \, N$ rates we would interpret this as a clear signal for an attractive
shift in the antikaon mass distribution \footnote{One may
argue that an attractive mean field of the nucleon gives rise to a similar enhancement. However,
our results predict that this effect would be annihilated by the attractive mass shift
of the $\Sigma (1385)$ resonance.}. Only if the in-medium $\bar K N$ phase space has
significant overlap with the $\Sigma(1385)$ resonance the $\pi\, \Lambda \to \bar K \, N$
rate can be enhanced significantly.

\cleardoublepage

\chapter{Summary and  outlook}
\label{k7}
\markboth{\small CHAPTER \ref{k7}.~~~Summary and  outlook}{}

In this work the relativistic chiral SU(3) Lagrangian was applied to the
meson-baryon scattering processes. Within the $\chi$-BS(3) approach
a unified description of pion-nucleon, kaon-nucleon and antikaon-nucleon scattering
was established. A large amount of scattering data and also the empirical axial vector coupling
constants for the baryon octet ground states are described. We derived the Bethe-Salpeter
interaction kernel at chiral order $Q^3$ and then computed the scattering amplitudes by
solving the Bethe-Salpeter equation within a systematic on-shell reduction scheme that
preserves covariance, analyticity and two-body unitarity. Moreover, the number of
colors ($N_c$) in QCD was considered as a large parameter performing a
systematic $1/N_c$ expansion of the interaction kernel. This lead to a significant
reduction of the number of parameters.
The analysis provides first reliable estimates of poorly known s- and p-wave parameters.
An important result of the work is that the strength of all quasi-local 2-body interaction
terms are consistent with the large-$N_c$ sum rules of QCD. Further intriguing results
were obtained for the meson-baryon coupling constants. The chiral $SU(3)$ flavor symmetry
was found to be an extremely useful and accurate tool. Explicit symmetry breaking effects
are quantitatively important but sufficiently small to permit an evaluation within chiral
perturbation theory. Two essential ingredients for a successful application of the chiral
Lagrangian to the meson-baryon dynamics were derived. First it was found that the explicit
s- and u-channel decuplet contributions are indispensable for a good fit the data set.
Second, it is crucial to employ the relativistic chiral Lagrangian.
It gives rise to well defined kinematical structures in the quasi-local 4-point interaction
terms which imply a mixing of s-wave and p-wave parameters. Only in the heavy-baryon mass
limit, not applied in this work, the parameters decouple into the s-wave and p-wave sector.
A projector formalism was constructed which decouples the Bethe-Salpeter equation into
covariant partial wave equations. This formalism permits the application of dimensional
regularization crucial for a proper renormalization of the scattering equation. A
novel minimal chiral subtraction prescription within the dimensional regularization
scheme was developed that complies  manifestly with the chiral counting rules.
A consistency check of the forward scattering amplitudes was performed by confronting
them with their dispersion-integral representations. The analysis showed that the scattering
amplitudes are compatible with their expected analytic structure.
Moreover it was demonstrated that the kaon-nucleon and antikaon-nucleon scattering amplitudes
are approximatively crossing symmetric in the sense that the $K N$ and $\bar K N$ amplitudes
match at subthreshold energies. An important test of the analysis could be provided by
new data on kaon-nucleon scattering e.g. from the DA$\Phi$NE facility \cite{DAPHNE}. In
particular further polarization data possibly obtained with a polarized hydrogen or
deuteron target would be extremely useful.

The results for the $\bar K N$ amplitudes have interesting consequences for antikaon
propagation in dense nuclear matter as for example probed in heavy ion collisions \cite{Senger}.
According to the low-density theorem \cite{dover,njl-lutz} an in-medium antikaon spectral
function with support at energies smaller than the free-space kaon mass, probes the
antikaon-nucleon scattering amplitudes at subthreshold energies. The required amplitudes
are well established in this work. In particular sizeable contributions from p-waves
not considered systematically so far \cite{Kolomeitsev,ml-sp,ramossp} were found.
The microscopic $\chi$-BS(3) dynamics was applied to antikaon and hyperon resonance propagation
in cold nuclear matter in a manifestly covariant and self consistent manner. For the antikaon
spectral function a pronounced dependence on the three-momentum of the antikaon is
predicted. The spectral function shows typically a rather wide structure invalidating a
simple quasi-particle description. For instance at $\rho = 0.17$ fm$^{-3}$ the strength starts
at the quite small energy, $\omega \simeq $ 200 MeV. Furthermore,
at nuclear saturation density attractive mass shifts for the $\Lambda(1405)$, $\Sigma (1385)$
and $\Lambda(1520)$ of about 60 MeV, 60 MeV and 100 MeV are predicted. The hyperon states are
found to show at the same time an increased decay width of about 120 MeV
for the s-wave $\Lambda(1405)$, 70 MeV for the p-wave $\Sigma (1385)$ and 90 MeV for the
d-wave $\Lambda(1520)$ resonances. It would be interesting to explore whether it is feasible to
confirm the strong in-medium modifications
of hyperon resonance by suitable experiments at ELSA or MAMI.

The importance of the p-wave $\Sigma (1385)$ resonance for the antikaon production in heavy
ion collisions as studied at GSI was pointed out. The presence of the latter resonance
implies a substantially enhanced $\pi \Lambda \to \bar K N $ reaction rate in nuclear matter.
Contrary to naive expectations only a small enhancement of the $\pi \Sigma \to \bar K N $
reaction was derived even though it couples to the s-wave $\Lambda (1405)$ resonance. It was
argued that the incorporation of the $\Sigma (1385)$ resonance into transport model simulations
of nucleus-nucleus collisions at SIS energies should help to obtain a quantitative and
microscopic understanding of the empirical antikaon yields.

The analysis is expected to pave the way for a microscopic description of kaonic atom data.
The latter are known to be a rather sensitive test of the antikaon-nucleon dynamics \cite{Gal}.
An extended domain of applicability for the microscopic chiral theory is foreseen once additional
inelastic channels are considered systematically. Also an application of the covariant and self
consistent many-body dynamics, developed in this work, to the propagation properties of pions,
vector-mesons and nucleon resonances is being pursued. Here  interesting in-medium effects due
to mixing of the s-, p- and d-wave nucleon resonances are expected. Before a quantitative
application of the results to heavy-ion reactions it is necessary to extend the approach to
finite temperatures.

\cleardoublepage

\cleardoublepage
\pagestyle{headings}

\newpage

\pagestyle{empty}

\vskip5.0cm

\centerline{\bf \large Acknowledgements}

\vskip1.2cm

I would like to thank E. Kolomeitsev and C. Korpa for enjoyable and very fruitful collaboration.
The writing of my habilitation thesis depended crucially on being part of a very stimulating and
supportive environment present at GSI. The project profitted significantly by the constant support
and encouragement of W. N\"orenberg. Many thanks also to H. Feldmeier, B. Friman, H. van Hees and
J. Knoll for their never ending readiness to discuss and help.

Particular encouragement I received by the many discussions with P. Senger and Ch. Sturm from the KaoS
collaboration. Also I would like to express my gratitude to W. Florkowski, M. Soyeur and G. Wolf who I
collaborate with on projects not part of this thesis.

Thanks also to the frequent visitors of the theory group, K. Redlich and D. Voskresensky with who
I had many useful discussions.

\end{document}